\documentclass[a4paper,fleqn,usenatbib]{mnras}

\pdfoutput=1

\usepackage[T1]{fontenc}
\usepackage{ae,aecompl}

\usepackage{graphicx}	% Including figure files
\usepackage{amsmath}	% Advanced maths commands
\usepackage{amssymb}	% Extra maths symbols
\usepackage{pdflscape}

\title[Subgroups of stars associated with open clusters]{Structural properties of subgroups of stars associated with open clusters}

\author[Gregorio-Hetem \& Hetem]{
Jane Gregorio-Hetem$^{1}$\thanks{E-mail: gregorio-hetem@usp.br (JGH)},
Annibal Hetem$^{2}$
\\
$^{1}$Universidade de S\~{a}o Paulo, IAG, Rua do Mat\~{a}o 1226, 05508-090 
S\~{a}o Paulo, SP, Brazil\\
$^{2}$UFABC Federal University of ABC, Av. dos Estados, 5001, 09210-580 Santo Andr\'{e}, SP, Brazil\\
}

\date{Accepted 2024 July 27. Received 2024 July 27; in original form 2024 January 30.}

\pubyear{2024}

\begin{document}
\label{firstpage}
\pagerange{\pageref{firstpage}--\pageref{lastpage}}
\maketitle

\begin{abstract}
Recent studies have identified star clusters with multiple components based on accurate spatial distributions and/or proper motions from {\it Gaia}~DR3, utilising diverse diagnostics to improve our understanding of subgroup evolution. These findings motivated us to search for subgroups among the objects examined in our previous work, which employed fractal statistics. The present study considers seven open clusters that exhibit significant dispersion in age and/or proper motion distributions, suggesting that they would not be single clusters.
For characterizing the stellar groups, we calculate the membership probability using Bayesian multi-dimensional analysis by fitting the observed proper motion distribution of the candidates.  A probability distribution  is also used to determine the distance of the cluster, which is obtained from the mean value of the distance 
modes. 
The photometry from {\it Gaia}~DR3 is compared with evolutionary models to estimate the cluster age and total mass.
 In our sample, double components are found only for Markarian~38 and NGC~2659. The other five clusters are confirmed as being single. The structural parameters, such as  
 $\mathcal{Q}$,  $\Lambda_{\rm MSR}$ and  $\Sigma_{\rm LDR}$
 are compared with results from N-body simulations 
to investigate how the morphology of the stellar clustering evolves.
The new results, for a more complete sample of cluster members, provide a better definition of the distribution type (central concentration or substructured region) inferred from the  $\overline{m}  -  \overline{s}$  plot.

 \end{abstract}


%%%%%%%%%%%%%%%%%%%%%%%%%%%%%%%%%%%%%%%%%%%%%%%%%%

%%%%%%%%%%%%%%%%% BODY OF PAPER %%%%%%%%%%%%%%%%%%
 %----------------------------------------Sect. 1
\section{Introduction}
\label{sec:intro}

In the studies of stellar groupings, mass segregation and high surface density of sources are the properties typically expected in young star clusters. However, many stellar groups showing more or less concentrated clustering have been observed  probably due to different conditions for star formation and dynamical evolution,
 such as NGC~2264 \citep{gonz17}, NGC~2548 \citep{belen16},  NGC~6231 \citep{kuhn17}, NGC~3105 \citep{davi17}, and  NGC~346 \citep{schmeja09}, among other examples.
The formation of bound clusters or dispersed stellar groups depends on the initial physical condition of turbulence effects in a molecular cloud \citep{elm08}. For instance, recently formed clusters tend to follow the fractal structure of the clouds where they are found embedded \citep[e.g.][]{elm18}. 

Exploring the relationship between stellar groups and their natal star-forming regions hints at the hierarchical structure of star distributions and cluster evolution. 
For instance, to systematically search for reliable small-scale structures in star-forming regions that could be related to the star formation process, \cite{gonz21} developed a robust procedure to statistically analyse different regions having a varied sample of initial conditions. Their method was tested for synthetic and observed clusters, presenting successful results for regions with high degree of structures, where significant small-scale substructures were detected. For concentrated regions, they find a main structure surrounded by smaller ones. \cite{gonz21}  argue that multi-scale analysis is needed to disentangle the complexity of the region. 
In this work, we are particularly interested in clusters showing a low degree of structure, which could be evidence that structural characteristics may not change within the early stages of cluster evolution. 

\citet[][hereafter HGH19]{HGH19} explored the fractal structure of a large sample of open clusters  noticing some objects that seem to belong to larger groups due to the possible presence of separate components having slightly different values of parallax and proper motion. 
NGC~2659 is one of those considered in the literature as a single cluster  \citep{Dias21,CantatGaudin18b,Poggio21}. However, in a study of the 
interaction of adjacent open clusters found in the Galaxy, which have spatial projected separation lower than 50 pc,   \citet{Song22} suggest 
NGC~2659 is a binary cluster.

Among the open clusters studied by HGH19, there are several objects for which it was suggested the possible presence of more than a single component.
This hypothesis is due to the large deviations in the mean values of astrometric and dynamical parameters,
as well as differences when comparing with results from the literature, meaning that subgroups appeared mixed and considered a single cluster. 
To explore the changes in structural parameters, we  have selected a sample of clusters to be analysed in light of the possible presence of subgroups. 

The paper is organized as follows. In Sections 2 and 3, we respectively present some of the literature results related to our sample and describe the data used to revisit previous studies.
The analysis presented in Sect. \ref{sec:groups} is based on characterizing stellar groups that depend on identifying the cluster membership, mode of distance, 
mass, and age. Section \ref{sec:distrib} is dedicated to the statistical analysis of the surface density distribution. In  Sects. \ref{sec:compare} and \ref{sec:conclus}, we respectively present a comparison with previous results and summarize our conclusions. Finally, two appendices are included to present additional results (supplementary material available only in digital form).

 %----------------------------------------Sect. 2
\section{Sample selection}
\label{sec:selection}

For the revisiting analysis proposed here, we selected seven clusters from HGH19 with a larger number of members, 
which are  listed in Table \ref{tab:data}. We have searched the recent literature for the parameters to compare with our previous results (HGH19). 
Most of the parameters are presented by Cantat-Gaudin and collaborators, whose works are 
dedicated to studying the structure and history of the Milky Way by characterising open clusters and studying the Galactic 
disc using the distance, age, and interstellar reddening for stellar clusters identified with {\it Gaia} astrometry 
\citep[e.g.][]{CantatGaudin18a,CantatGaudin18b,CantatGaudin20}.

%%%%%%%%%%%%%%%% literature table 1
\begin{table*}
%\centering
\caption{Coordinates, distance, astrometric parameters, age, and size found in the literature for the selected sample of open clusters.}
\label{tab:data} 
\begin{tabular}{c|c|c|c|c|c|c|c|c|c|c|c}
\hline 
\hline
cluster	& $\alpha$ 	& $\delta$	 	&	d &N	&$\varpi$	&$\mu_{\alpha^\star}$ &$\mu_{\delta}$ &log(Age)	&$R$	&	 $R_{50}$	 &	Reference	\\
                 &  deg              &   deg              &  pc        & &mas      & mas yr$^{-1}$           & mas yr$^{-1}$    &  yr & deg        & deg               &            \\
                 \hline
Col205 	&	135.123	&	-48.983	&	2044	&	143& 0.49 	&	 -4.81 	&	 3.94 	&	 5.7$\pm$2.1 	&	0.10	&	$\dots$	&	\tiny{HGH19}	\\
             	&	135.091	&	-48.985	&	1400	&	99& 0.46 	&	 -4.67 	&	 3.93 	&	 6.95             	&	$\dots$	&	0.048	&	\tiny{ \citet{Dias21}}	\\
             	&	135.119	&	-48.984	&	1953	&	102& 0.48 	&	 -4.80 	&	 3.92 	&	         $\dots$            	&	$\dots$	&	0.047	&	\tiny{\citet{CantatGaudin18b}}	\\
             	&	135.119	&	-48.984	&	2394&	80& 0.54 	&	 -4.80 	&	 3.97 	&	 6.66           	&	$\dots$	&	0.047	&	\tiny{\citet{Poggio21}}	\\	
\hline																																									
IC~2602	&	 	160.596	 	&	 	 -64.419	 	&	154	&	140& 6.49	&	-17.56	&	10.73	&	6.7$\pm$0.9	&	1.97	& $\dots$&	\tiny{HGH19}	\\
	&	 	160.613	 	&	 	 -64.426	 	&	152	&	311& 6.60	&	-17.58	&	10.70	&	7.5	&$\dots$&	 	1.449	 	&	\tiny{\citet{CantatGaudin18b}; B19}	\\
	&	160.515	&	-64.444	&	151	& 318 &	$\dots$& 	-17.69	&	10.69	&	7.65	&$\dots$&1.404& \tiny{\citet{Pang22}}\\
		&	 	160.605	 	&	 	 -64.399	 	&	$\dots$&	30-99& 6.70	&	$\dots$&	$\dots$&	8.0	&	$\dots$&	 		 $\dots$&	\tiny{\citet{Yen18}}	\\
\hline																					
Mrk38	&	 	273.820	 	&	 	 -19.004   	&	1892	&	167 & 0.53	&	0.22	&	-1.75	&	6.55$\pm$0.9	&	0.05	&	$\dots$&	\tiny{HGH19}	\\
	&	 	273.812	 	&	 	 -19.005	 	&	1770	&	26& 0.56	&	0.85	&	-2.26	&	7.0	&	$\dots$&	 	0.041	 	&	\tiny{\citet{Dias21}}	\\
	&	 273.819	 	&	  -18.997	 &	1678	 &	27& 0.57	&	0.84	&	-2.28	&	7.0	&	$\dots$&	 	0.044	 	&	\tiny{\citet{CantatAnders20}, P14}	\\
	             	&	273.819    	&	-18.997 	&	1802&	17&  0.56 	&	 0.81 	&	-2.31 	&	 7.2           	&	$\dots$	&	0.044	&	\tiny{\citet{Poggio21}}	\\
\hline																					
NGC~2168	&	 	92.267	 	&	 	24.310  	&	887	&	127 & 1.13	&	2.22	&	-2.93	&	7.3$\pm$1.2	&	0.86	&$\dots$&\tiny{	HGH19}	\\
	&	 	92.263	 	&	 	24.334	 	&	821	&	1215& 1.13	&	2.30	&	-2.90	&	8.2	&	$\dots$&	 	0.316	 	&	\tiny{\citet{Dias21}}	\\
	&	 	92.272	 	&	 	24.336	 	&	862	&	1325& 1.13	&	2.31	&	-2.90	&	8.6	&	$\dots$&	 	0.319	 	&	\tiny{\citet{CantatGaudin18b}, B19}	\\
	&	92.302	&	24.360	&$\dots$&	 1239& 1.12	&	0.62	&	-4.06	&	8.3	&$\dots$&	$\dots$&	\tiny{\citet{CantatGaudin18a}}	\\
\hline																					
NGC~2659	&	 	130.662	 	&	 	 -44.975 	 	&	2024	&	233& 0.49	&	-3.99	&	3.59	&	6.5$\pm$1.41	&	0.12	&$\dots$&\tiny{	HGH19}	\\
	&	 	130.633	 	&	 	 -44.999	 	&	1815	&	91& 0.43	&	-5.36	&	5.03	&	7.6	&	$\dots$&	 	0.038	 	&	\tiny{\citet{Dias21}}	\\
	&	 	130.634	 	&	 	 -44.999	 	&	2080	&	97& 0.45	&	-5.34	&	5.03	&	7.4	&	$\dots$&	 	0.042	 	&	\tiny{\citet{CantatGaudin18b}, B19}	\\
		             	&	130.634     	&	-44.999   	&	2095&	82&  0.47 	&-5.31 	& 5.07 	&	 7.6           	&	$\dots$	&	0.042	&	\tiny{\citet{Poggio21}}	\\

\hline																					
NGC~3532	&	 	166.489	 	&	 	 -58.723 	 	&	488	&	262& 2.05	&	-10.41	&	4.99	&	6.6$\pm$0.9	&	0.54	& $\dots$&	\tiny{HGH19}	\\
	&	 	166.412	 	&	 	 -58.722	 	&	477	&	1762& 2.06	&	-10.37	&	5.19	&	8.6	&	$\dots$&	 	0.441	 	&	\tiny{\citet{Dias21}}	\\
	&	 	166.417	 	&	 	 -58.707	 	&	477	&	1889& 2.07	&	-10.38	&	5.17	&	$\dots$&	$\dots$&	 	0.536	 	&	\tiny{\citet{CantatGaudin18b}}	\\
	&	166.389	&	-58.702	&	478	& 2559&	2.10	&	-10.40      	&	5.23	&	8.6	&	$\dots$&	0.887&	\tiny{\citet{Pang22}}	\\
\hline																					
NGC~6494	&	 	269.251	 	&	 	 -18.969	 	&	740	&	170& 1.35	&	0.55	&	-1.78	&	6.7$\pm$0.75	&	0.52	&$\dots$& \tiny{HGH19}	\\
	&	 	269.241	 	&	 	 -18.969	 	&	691	&	694 & 1.36	&	0.28	&	-1.80	&	8.5	&	$\dots$&	 	0.293	 	&	\tiny{\citet{Dias21}}	\\
	&	 	269.237	 	&	 	 -18.987	 	&	674	&	789& 1.35	&	0.28	&	-1.81	&	8.7	&	$\dots$&	 	0.292	 	&	\tiny{\citet{CantatGaudin18b}, B19}	\\
	&	269.227	&	-19.000	&$\dots$&	 53& 1.23	&	1.17	&	0.09	&	8.5	&$\dots$&$\dots$&	\tiny{\citet{CantatGaudin18a}}	\\
\hline	
\hline
\end{tabular} 
\tiny{Notes: Results from \citet{CantatGaudin18b} are presented along with  log(Age) adopted from  \citet[][B19]{Bossini19}, 
when available. In the case of Mrk38, log(Age) is given by \citet[][P14]{Piatti14} complementing data from \cite{CantatAnders20}. 
For IC2391, the results from \citet{Dias21} are complemented by coordinates from \citet{Gaia18b}. The size of the cluster is indicated by 
$R_{50}$ represents the radius of the area that contains 50 percent of members, excepting the results from \citet{Pang22} that correspond to the half-mass radius.}
\end{table*}

Our sample is also included in the list of clusters revisited by  \citet{Dias21}, for which updated parameters are presented 
based on an isochrone fitting code \citep{Monteiro20} to the {\it Gaia} DR2 photometry, taking into account the nominal errors
to derive distance, age, and extinction of each cluster. 
\citet{Bossini19}  also used the method of isochrone fitting to estimate visual extinction and age for part of our sample.

 \cite{Poggio21}  used {\it Gaia} EDR3 data to map the segments of the nearest spiral arms in the Milky Way, 
based on the overdensity distribution of young upper main sequence stars, open clusters, and classic Cepheids. 
They used a list of open clusters from \cite{CantatGaudin20}, but their analysis was restricted to the 
intrinsically bright objects, selecting the open clusters with more than five members 
with absolute magnitude $M_G \leq$ 0. 
We notice that four of the objects in our sample (Collinder~205, 
NGC~2659, NGC~2168, NGC~6494) coincide with the bright clusters present in the list of  \cite{Poggio21}.

A machine learning code (SytarGO) was used by \citet{Pang22} to study the hierarchical star formation process by analysing
the three-dimensional morphology of young clusters, which revealed spatial and kinematic substructures. Two of our clusters appear in the sample studied by \citet{Pang22}: IC~2602 and NCG~3532; however, the morphology is indicated only for the second one, the ``halo" type.

%----------------------------------------Sect. 2.1 
\subsection{Previous results}
\label{sec:individual}
In this section, we summarize individual comments for our sample to complement the information provided in Table 1. Similar information could not be included for all clusters due to varying availability in the literature searched.

\noindent{$\bullet$ \bf Collinder 205}: The distance that was previously estimated by us for Collinder~205 (Col205 hereafter) is about 2~kpc (HGH19) in agreement with the value found by \citet{CantatGaudin18b}, but is considerable  larger than d~$\sim$ 1.4~kpc found by different works \citep{Kharchenko05, Dib18, Dias21} and smaller than d~$\sim$ 2.4~kpc presented by \cite{Poggio21}. There are also some discrepancies in our previous result of cluster age that tends to be younger than the ages found by other works \citep[e.g.][]{CantatGaudin20}.
Col205 was included in the Galactic mapping survey of chemical abundances  of open clusters  \citep{Ray22}. As expected for clusters near the Sun, the metallicity ([Fe/H] = -0.07 $\pm$ 0.19) and all other elements are within the normal of solar neighborhood cluster mean abundances. This result is consistent with [Fe/H] = -0.169 found by \cite{Dias21}.

\noindent{$\bullet$  \bf IC~2602}:  Density maps were obtained by \citet{Richer21} with the method of fitting  Gaussian functions to kernel density estimates, aiming to test in the solar vicinity the structural and dynamical early evolution theories,
IC~2602 is one of the studied objects, which was considered a benchmark in this kind 
of test because it is one of the nearest clusters with ages $\sim$ 50 Myr. 
\citet{Bravi18} used the {\it Gaia}-ESO Survey \citep[see][and references there in]{Bragaglia22}, products to estimate the gravity index, the lithium 
 equivalent width, and the metallicity to identify candidate members for IC~2602. They also used the radial velocities to derive the cluster membership
  probabilities and the intrinsic velocity dispersion that changes from $\sim$0.48 km s$^{-1}$ to $\sim$0.20  km s$^{-1}$  depending on the number of members considered. 
    
\noindent{$\bullet$  \bf Markarian~38}: The estimates of proper motion and distance of the cluster Markarian~38 (hereafter Mrk38) presented by \cite{Poggio21} are in excellent agreement with those from other works \citep{CantatAnders20, Dias21}. An estimation of the morphological aspect is given by \cite{Hu21b}, indicating that the ellipticity $\sim$ 0.13) is measured for both the core and the overall shape of the cluster. According to the study of morphological evolution of a sample of open clusters presented by \cite{Huetal21}, the overall shape of clusters becomes more elliptical as they grow older, while their core remains circular. This seems not to be the case for Mrk~38, whose core tends to be more elliptical.

\noindent{$\bullet$  \bf NGC~2659}: \cite{Casado21} has explored the spatial distribution of {\it Gaia} sources aiming to identify possible double and multiple groups of star clusters, such as their “Group 18” that contains 8 open clusters. Four of these clusters were previously identified as a group by \cite{Liu19}, including  NGC~2659, which was suggested to constitute a pair with UBC~482. According to \cite{Song22}, this pair has similar values of proper motion and parallax but are $1\degr$ apart in the space distribution, corresponding to $\sim$ 50 pc of projected separation. However, their distances exceed 250 pc \citep{Poggio21}. For 82 members in NGC~2659 and 79 in UBC~482, \cite{Song22} found ages of ~44 Myr and ~27 Myr, respectively.

\noindent{$\bullet$  \bf NGC~3532}: The adopted age in the literature for NGC~3532 is $\sim$ 300 Myr \citep{Dobbie09},
but it was two orders of magnitude larger than the value estimated by HGH19. This cluster was included in the three-dimensional kinematic 
study by  \citet{Jack22} to estimate the membership probabilities of open clusters. The morphology of NGC~3532  was studied by \cite{Huetal21}  
giving the shape parameters corresponding to the ellipticity of the core $e_{core} = 0.074 \pm 0.031$ and of the cluster overall $e_{all} = 0.120 \pm 0.045$. 
\cite{Jadhav21} identified the high mass ratio of binaries found in a sample of 23 open clusters to estimate their fraction and trace their radial segregation. Their results show that NGC~3532 is one of the clusters with the lowest fraction of high mass-ratio binaries compared to the rest of 
the population. These results agree with \cite{Li20}, arguing that NGC~3532 is not a binary-rich cluster, and its binary mass ratio follows a nearly uniform distribution.

\noindent{$\bullet$  \bf NGC~6494}: \cite{Cordoni}  infer a core radius $r_c$ = 1.8 pc and $A_V$ = 1.28 mag considering 1273 members in NGC~6494.  
\citet{tarricq22} analysed a sample of 870 members and obtained
$r_c$ = 2.55 pc, where the core has low eccentricity, while the halo has an elongated distribution, such 
as a tail-like structure with semi-major axis = 16.5 pc and a semi-minor axis =  6.29 pc. A comparable 
result was found by \citet{Hu21b}, estimating for NGC~6494 a value of 0.15 for the ellipticity of the core.
\citet{Rain} and \citet{Bossini19} have adopted extinction correction, respectively E(B-V) = 0.27 and $A_V$ = 0.85 mag that coincide with the values used by us. The age estimated by HGH19 is lower than the range of  380 Myr to 479 Myr  \citep[e.g.][]{Spina, Poggio21, Rain}.

The examples above are the primary sources used for parameter comparison (see Table \ref{tab:data}). 
Our criteria for sample selection were based on the large differences found among our previous results and the literature, suggesting that discrepancies in the parameters and/or large uncertainties could indicate more than one structure in these selected clusters, possibly not noticed by HGH19.  
The data used in our analysis are described as follows.

 %----------------------------------------New Sect. 3
\section{Data sets}
\label{sec:datasets}

%----------------------------------------Sect. 3.1
\subsection{Optical data}
\label{sec:data}

The astrometric and photometric data was obtained  from {\it Gaia} DR3 \citep[][]{Gaia23} by adopting 
query ranges defined on the basis of parameters estimated by HGH19: J2000 equatorial coordinates ($\alpha$, $\delta$),
parallax ($\varpi$), and proper motion ($\mu_{\alpha^{\star}}$, $\mu_{\delta}$), where $\mu_{\alpha^{\star}}  \equiv \mu_{\alpha}\cos\delta$.  
Table  \ref{tab:data} gives the adopted position and size of the clusters, compared with other results from the literature. 

Using the positions given in Table \ref{tab:data}, we searched for all the sources in the area defined by a radius 10 percent larger than the cluster radius (HGH19). 
To avoid {\it Gaia} sources showing low quality of  the astrometric solution,  the selection was  restricted to objects having
RUWE{\footnote{Re-normalised unit weight error (see details in the technical \\
note GAIA-C3-TN-LU-LL-124-01).}} $< 1.4$. 

To estimate individual mass and age of the cluster members, we constructed Colour-Magnitude Diagrams (see Sect. \ref{sec:age}) with  photometric data at bands 
$G$ ($\sim$ 639 nm), $G_{BP}$ ($\sim$ 518 nm), and $G_{RP}$ ($\sim$ 782 nm) from {\it Gaia} DR3. 
The observed apparent magnitudes were corrected for extinction based on reddening $E(G_{BP}-G_{RP})$  that is derived from $A_0$,  the 
line-of-sight extinction available for some of the {\it Gaia} sources. $A_0$ is one of the stellar parameters inferred by the algorithm Aeneas that
 is part of the package GSP-Phot (General Stellar Parametrizer from Photometry)\footnote{Details are found in the {\it Gaia} DR3 documentation 
 (Sect. 11.3.3)  at https://gea.esac.esa.int/archive/documentation/GDR3/ .}.  The method performed by Aeneas consists of a simultaneous fitting of 
 the observed parallax, BP/RP spectra, and apparent magnitude ({\it Gaia} G band) using the technique of Bayesian posterior maximisation. 

In the case of  sources with no reddening information given by the {\it Gaia} catalogue, we estimate the colour excess $E(B-V)$ 
based on the colour-magnitude diagram by fitting the observed unreddened magnitudes to the distribution of magnitudes that were corrected by the algorithm 
Aeneas. The value of  $E(B-V)$  that gives the best fitting was used to infer a mean value for visual extinction ($A_V$) by adopting $R=\frac{A_V}{E(B-V)}=3.1$ \citep{Savage}. 
We checked the validity of the mean values of  $A_V$  by inspecting the extinction maps provided by \cite{Dobashi},  discussed in Sect. \ref{sec:av}.
The conversion between  $A_V$ and the expected extinction at the {\it Gaia} photometric bands was adopted
from \cite{Casagrande}.

%%%%%%%--------------------------------- Fig.1 
\begin{figure*}
\includegraphics[width=4.8cm]{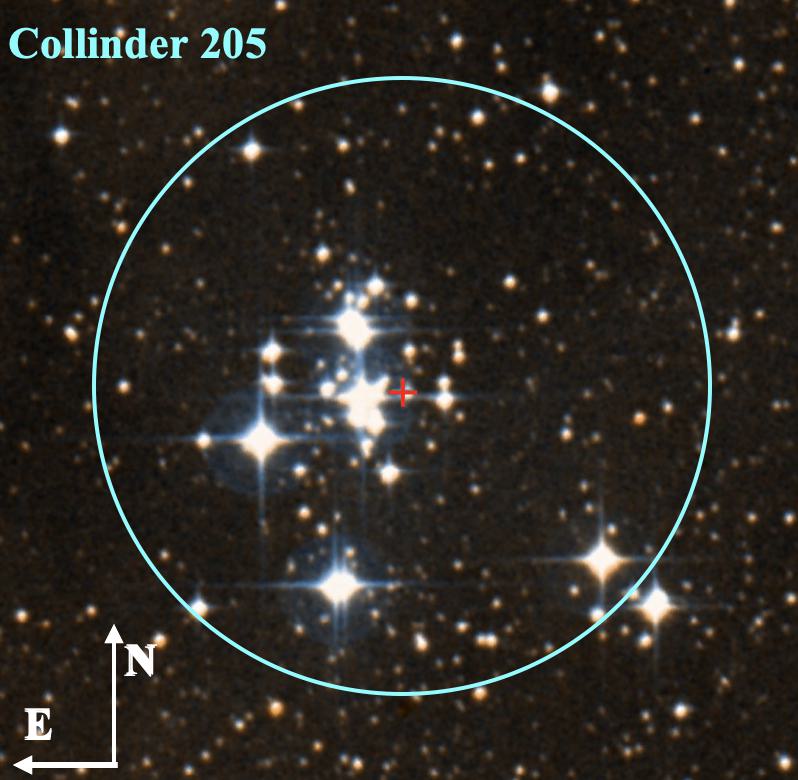}
\includegraphics[width=6.3cm]{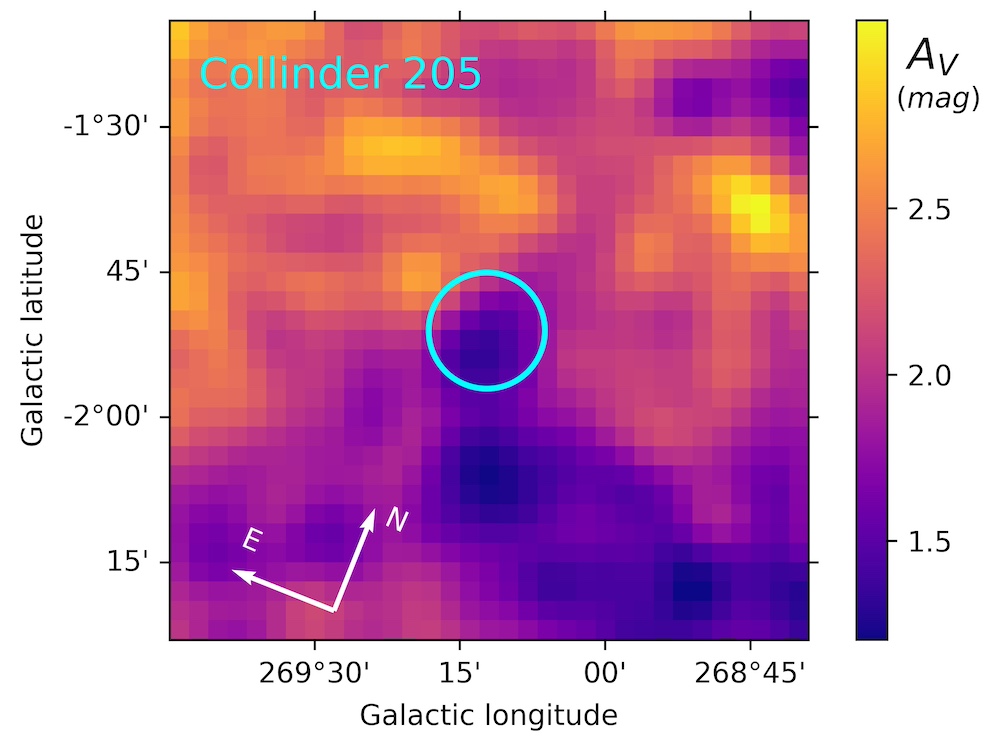}
\includegraphics[width=6.3cm]{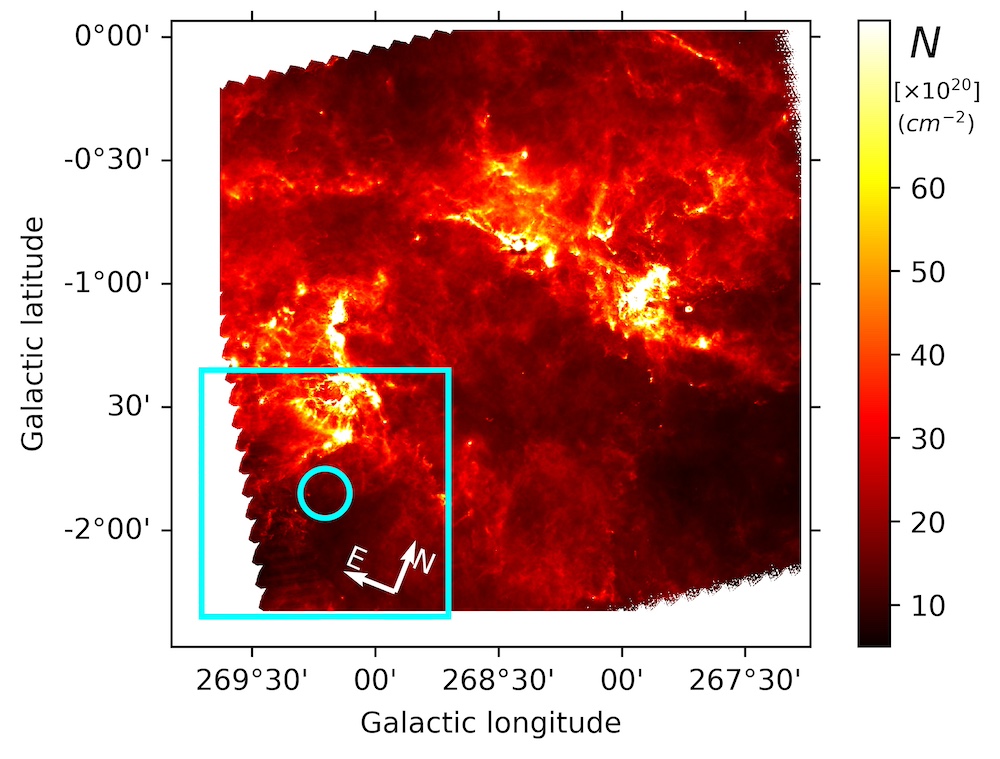}
\caption{The field-of-view of Col205 (cluster radius = $6 \arcmin$, cyan circle). {\it Left}: Optical image (DSS coloured, SIMBAD database) centered  at equatorial coordinates  given in Table 1. 
{\it Middle}: Visual extinction map presented in galactic coordinates, obtained by \citet{Dobashi} based on DSS images. The arrows show the equatorial coordinates direction
 , and the circle is the same as the left panel. {\it Right:} Integrated column density map of hydrogen molecules estimated from Hi-GAL based maps of dust continuum emission 
 ({\it Herschel}-PPMAP). The cyan square approximatively corresponds to the same area seen in the middle panel.}
\label{fig:dss}
\end{figure*}

%%%% -----------------------------------------end Fig.1 

%----------------------------------------Sect. 3.2
\subsection{Infrared photometry and mapping}
\label{sec:ir}

We checked  in our sample the occurrence of circumstellar dust emission to identify the 
cluster members 
that are disc-bearing stars. Large infrared excess is expected for Young Stellar Objects (YSOs) that still 
are embedded in their natal cloud (Class 0 objects), as well as for pre-Main Sequence  stars with large amounts of 
circumstellar dust (Class I and Class II). 
Considering that protostellar discs only survive up to a few tens of Myr, members of older 
clusters are expected to be Class III objects (without discs). The classification based on the emission at near- and mid-infrared 
confirms the evolutive status of the clusters and could help identify the presence of mixed 
populations (different ages) in the sample. In Appendix \ref{sec:wise} (supplementary material), we describe the search for disc-bearing sources performed by fetching from public catalogues the infrared counterparts of the Gaia sources.
A cross-matching between the position of the sources was performed on the 2MASS \citep[The Two Micron All Sky Survey,][]{Cutri03,Skrutskie06}, 
and All-WISE \citep[Wide-field Infrared Survey Explorer,][]{Wright10,Cutri13} catalogues. The observed magnitudes at  2MASS (JHK) and WISE (W1: 3.35~$\mu$m; W2: 4.6~$\mu$m, W3: 11.6 ~$\mu$m) bands were fetched only for the sources showing flags indicating good photometric quality  
and data that are unaffected by known artifacts. 

Aiming to inspect the dust distribution in the direction of the clusters, we searched for the continuum images at  70-500~$\mu$m obtained by the {\it Herschel} Infrared Galactic Plane Survey \citep[Hi-GAL,][]{Molinari10}. We found only two   $\sim 2 \degr$ strip tiles covering objects of our sample: Fields 264 and 269,  respectively, containing the clusters NGC~2659 and Collinder~205. 
A third cluster, Mrk38, is found near the southeast corner of Hi-GAL Field 11 but outside the image (see Fig. \ref{fig:Adss},  supplementary material). In this case, it is not possible to directly infer the local
 infrared emission. However, we could verify the lack of dust emission at this corner of Field 11 and around its neighborhood (Hi-GAL Field 13). Inspecting the far-infrared maps gives us some clues that the dust distribution is in good agreement with the visual extinction maps, as discussed in the following.

%----------------------------------------Sect. 3.3
\subsection{Visual extinction maps}
\label{sec:av}

The dust distribution in the Galaxy is related to the $A_V$ maps and traces the gas distribution in molecular clouds. 
 
Among the existing dust-based maps, there are two  of our 
 particular interest: the visual extinction maps from \cite{Dobashi}, and the column density mapping
 derived from the Hi-GAL survey \citep{Marsh17} using the point process mapping 
 (PPMAP)\footnote{https://www.astro.cf.ac.uk/research/ViaLactea/ .} 
 described by \citet{Marsh15}. In this case, the  {\it Herschel} images of dust continuum emission are used to produce image cubes of differential column density as a function of dust temperature and position.
 
In this work we use the $A_V$ maps\footnote{https://darkclouds.u-gakugei.ac.jp.} produced from the optical DSS\footnote{Digitized Sky Survey - STScI/NASA.} 
 images \citep{Dobashi}, in spite of their lower resolution ($\sim  6 \arcmin$~pixel$^{-1}$) than the extinction maps based on 2MASS data \citep[$\sim$ 1$\arcmin$~pixel$^{-1}$,][]{Dobashi11}. This choice was made to avoid unreal ($A_V < 0$) values that could occur in some regions mapped by using near-infrared photometry, which is not the case in the DSS maps. 

Since we are interested in a rough estimation 
 of $A_V$, to be used to check the reddening correction (see Sect. \ref{sec:age}), we 
 can consider the comparison  between dust extinction maps produced from different data sets as a 
 confirmation that the position of the clusters coincides (or not) with regions showing low levels of dust 
 concentrations. 

Aiming to illustrate  the use of the $A_V$ maps as a diagnosis of the presence of dust concentration, 
in Fig. \ref{fig:dss} we compare the maps from \cite{Dobashi} with the {\it Herschel} PPMAP, presented in the form of 2D map of integrated line-of-sight column density of 
hydrogen molecules (in units of 10$^{20}$ ~cm$^{-2}$). 
In Table \ref{tab:result1}, we present the range of extinction values inferred by visual inspection of the 
$A_V$ maps. The values of E(B-V) used to correct the reddening (as discussed in Sect. \ref{sec:data}) 
are also presented in Table \ref{tab:result1}, which are in good agreement with the estimates from 
the $A_V$ maps, excepting NGC~2659, which shows $E(B-V)$ that is 0.19 mag larger than the value expected from the extinction map.

%----------------------------------------Sect. 4
\section{Characterization of stellar groups}
\label{sec:groups}

The optical data extracted from the {\it Gaia} DR3 catalog (see Sect. \ref{sec:data}) were used for three 
purposes: to analyze the vector-point diagram (VPD)  
constructed using proper motion ($\mu_{\alpha^{\star}}$, $\mu_{\delta}$) of the stars to identify the members of different groups; using the parallax ($\varpi$) measured for the
members to estimate the  distance mode; and 
determining cluster age and individual stellar masses from isochrones fitting in the optical colour-magnitude diagram.  
The following sections are dedicated to describing the adopted methodology for identifying structures, the characterization of the clusters, 
and their respective stellar content.

%----------------------------------------Sect. 4.1

\subsection{Membership probability}
\label{sec:membership}

In the literature, the search for substructures has been conducted using methods and techniques based on spatial and kinematic analyses to identify discrete structures \citep[e.g.,][]{kuhn14, gonz21, Buckner22b, Buckner24}, spatial stellar associations \citep[e.g.,][]{Buckner19},  or dynamical analyses  \citep[e.g.,][]{kuhn19, Buckner20, Buckner24}, among other approaches. 
Young star clusters that are no longer embedded in their natal molecular cloud \citep[10-50 Myr, e.g.][]{Bica} are not expected to exhibit overdensities in their spatial distribution, which requires different criteria to identify which stars are physically related members confidently. Common velocities and distance are the main characteristics that stars inherit from the original cloud, giving the cluster a coeval movement until the tidal disruption that destroys the cluster \citep{LadaLada03}.

In this work, we adopt the method from  \cite{Sanders71} for identifying cluster members by their proper motion modeled according to a 
Probability Distribution Function  (PDF), following the formalism presented by \citet{Dias14}. It was adopted  as a segregation 
procedure based on a mixed bivariate density function model 
for proper motions, considering that the cluster and the field are independent, as suggested by \citet{Uribe94}. 

We adopted a PDF model that considers a sum of normal distributions, whose number of components can be two (cluster$+$field) or three (2 subclusters$+$field). The choice depends on visual inspection of proper motion VPD. Considering that candidates are projected against the same area, they do not present large differences in spatial distributions that could help distinguish subclusters. By this way, the only other parameter considered is parallax. The resulting PDFs are defined by the mean, standard deviation, and correlations for proper motion and parallax.

In summary, the 
membership probability is defined by using the maximum likelihood method that depends on the contrast between cluster members and 
field-stars. 
Accurate membership probabilities can be achieved by using  Bayesian 
 multi-dimensional analysis and performing a global optimization procedure  based on the  Cross-Entropy technique 
 \citep{Rubinstein97}, to fit the observed distribution of proper motions and to obtain the probability of a given star 
belonging or not to the cluster. We also used a genetic algorithm to improve the first guess of the parameters set.  HGH19 gives details on the membership probability calculation. 

%%%%%%%--------------------------------- Fig.2 
\begin{figure*}
\includegraphics[width=\columnwidth]{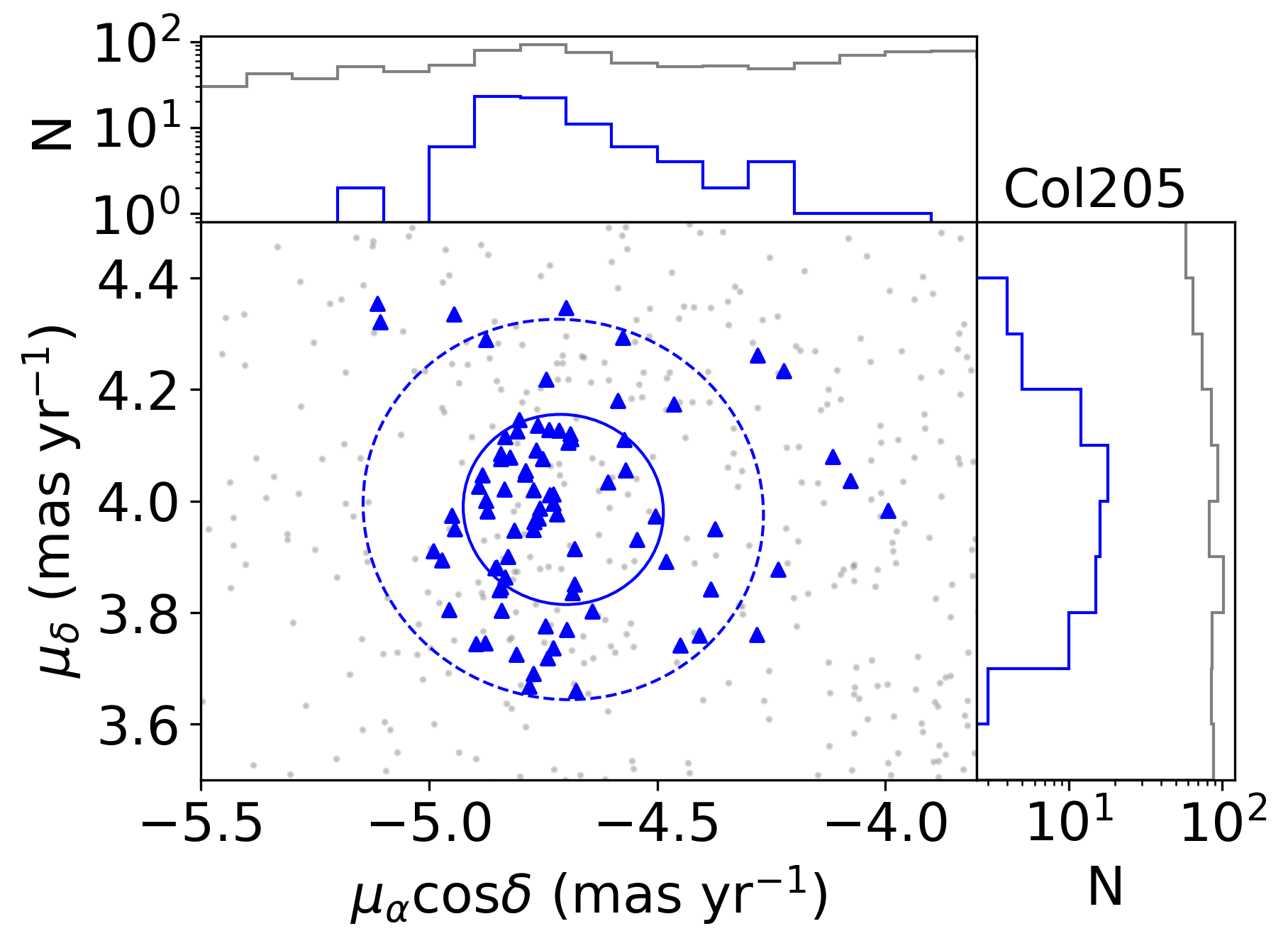}
\includegraphics[width=\columnwidth]{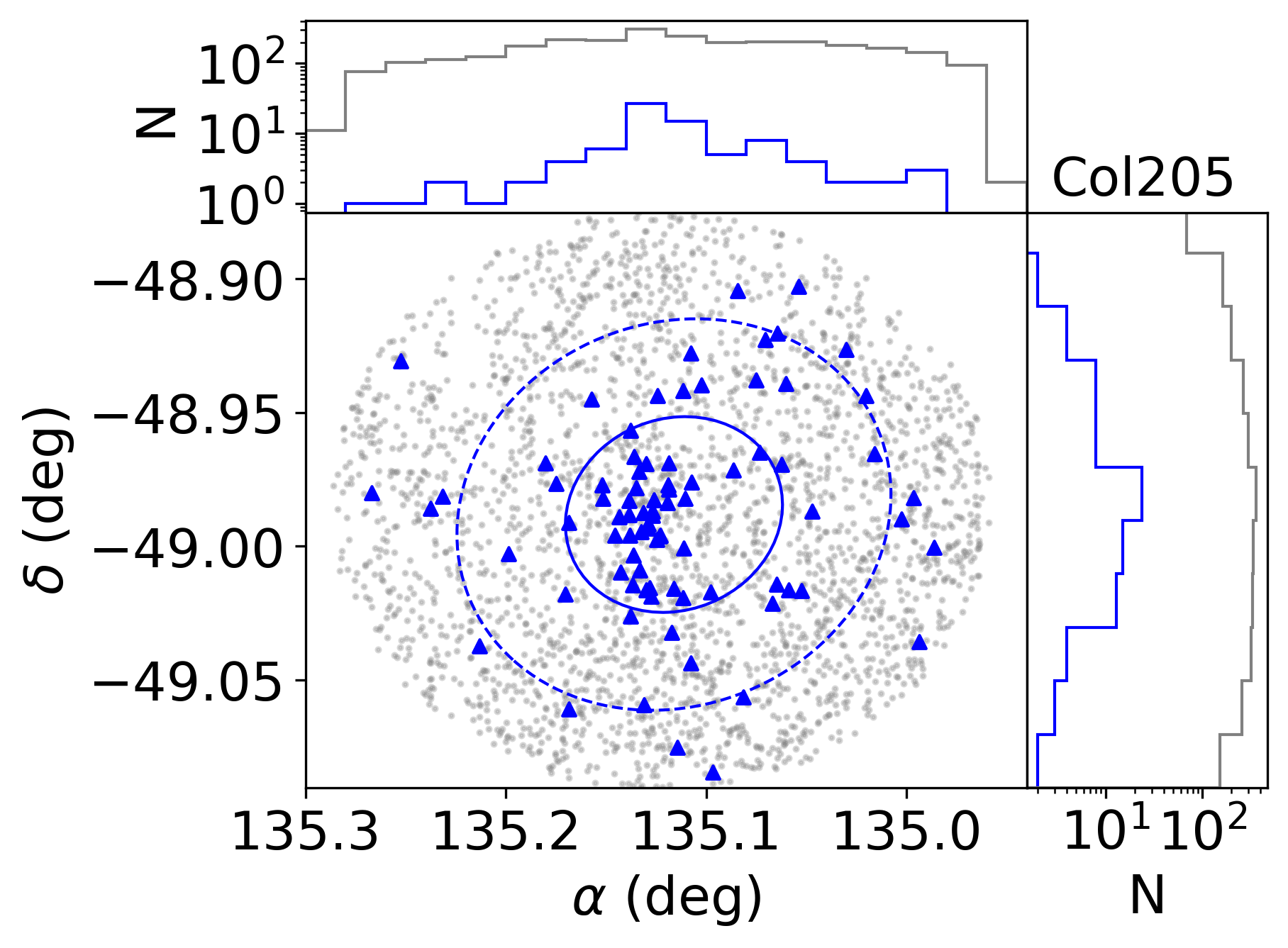}
\includegraphics[width=\columnwidth]{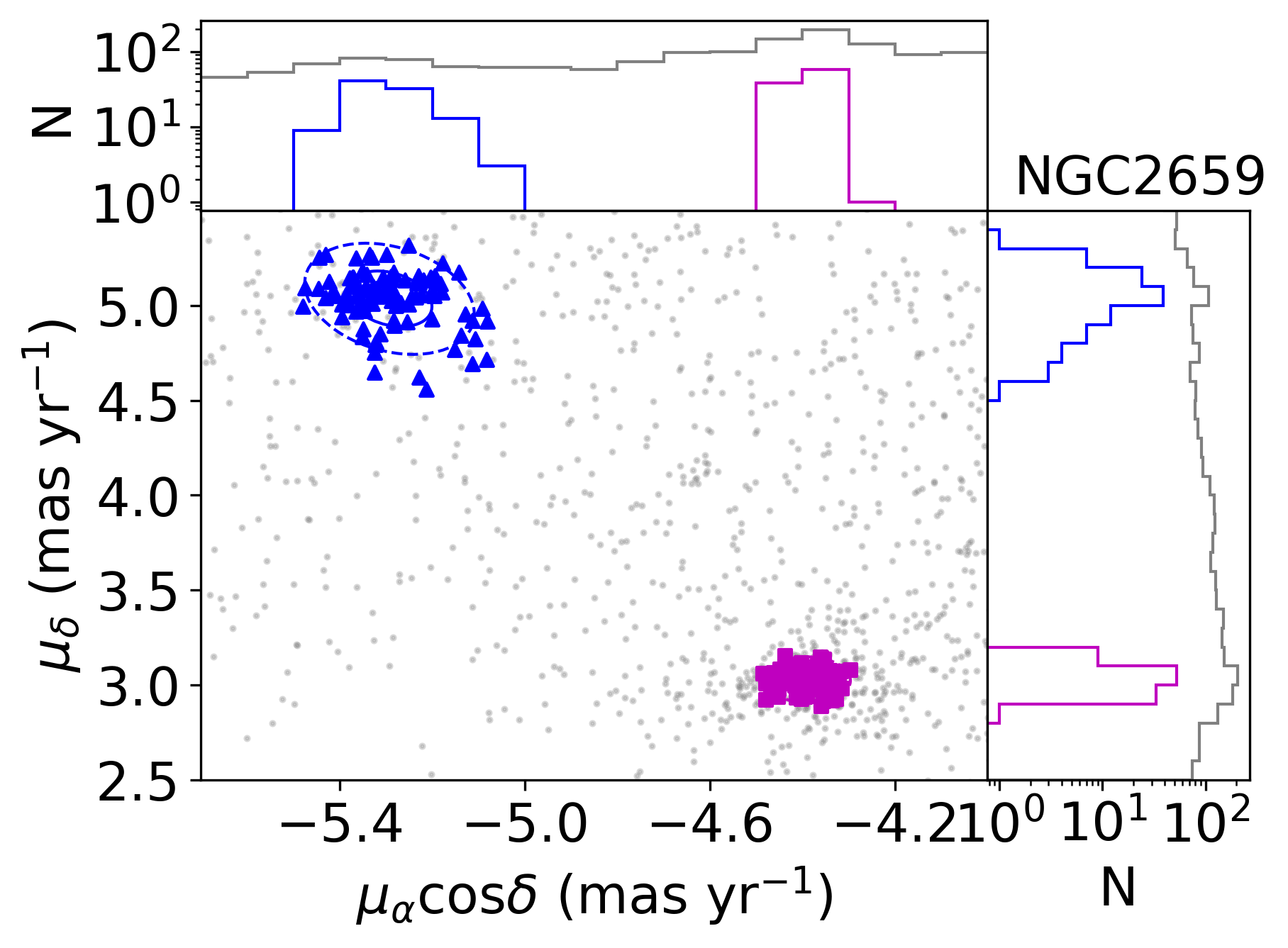}
\includegraphics[width=\columnwidth]{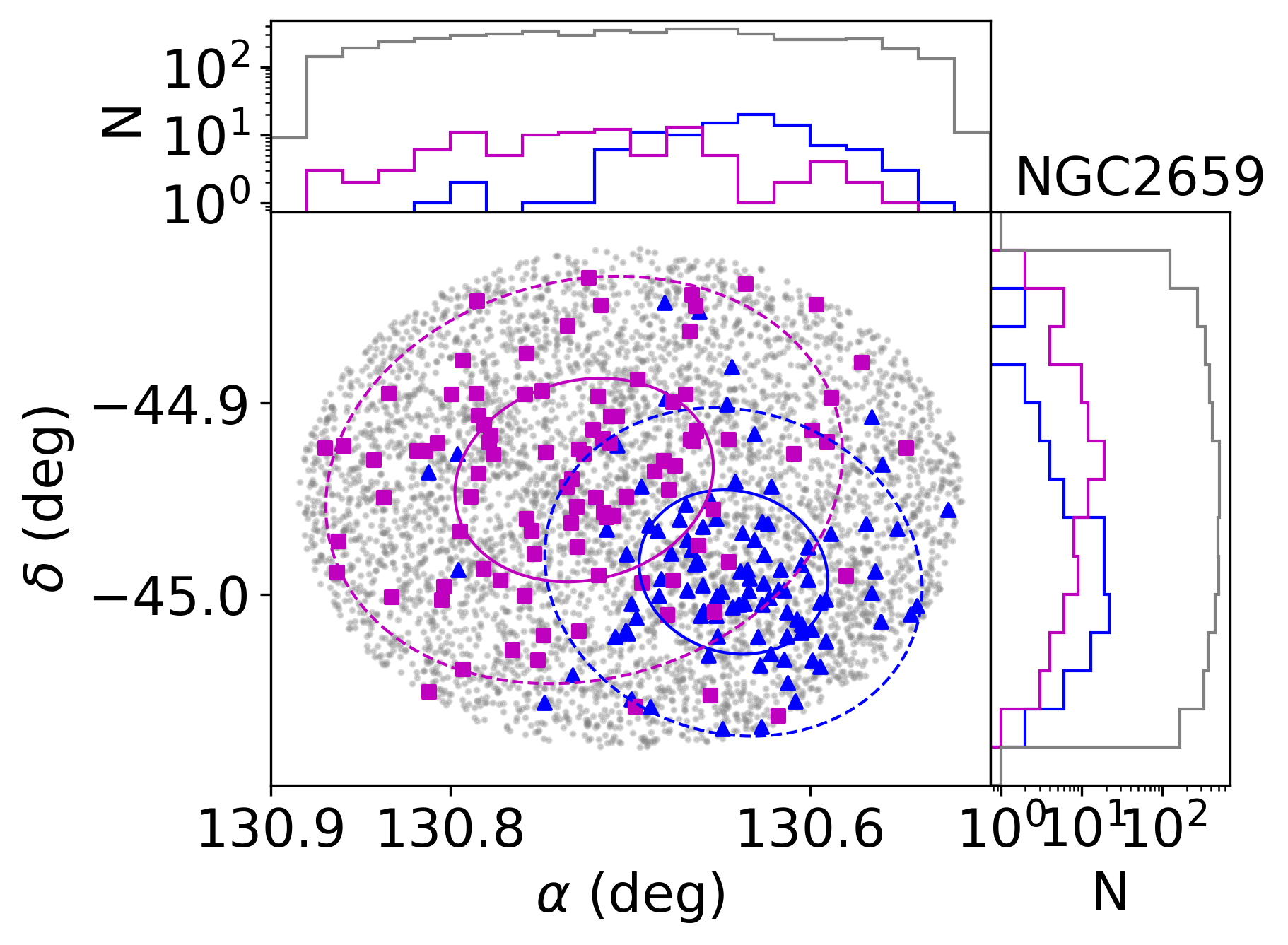}
\includegraphics[width=\columnwidth]{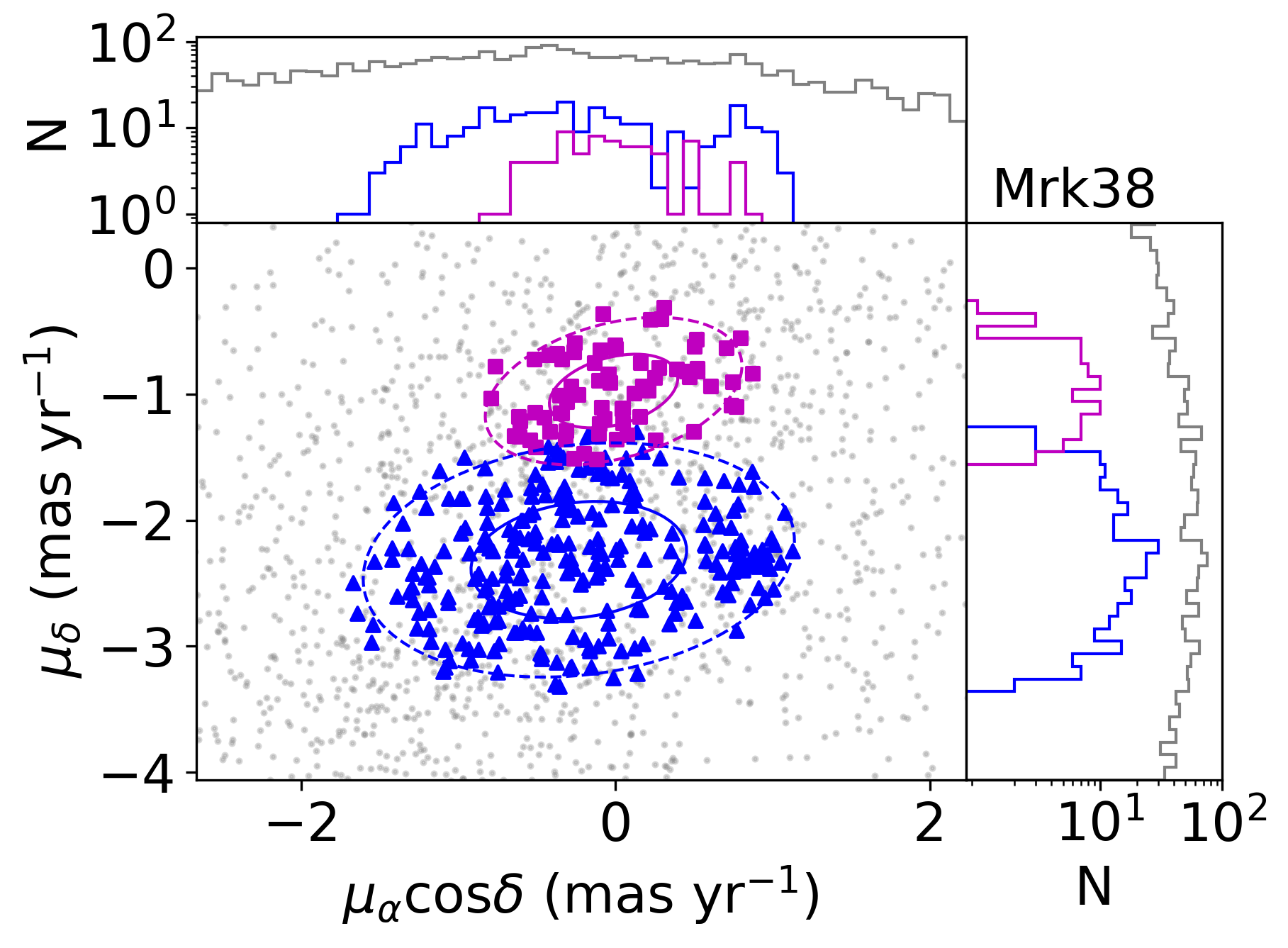}
\includegraphics[width=\columnwidth]{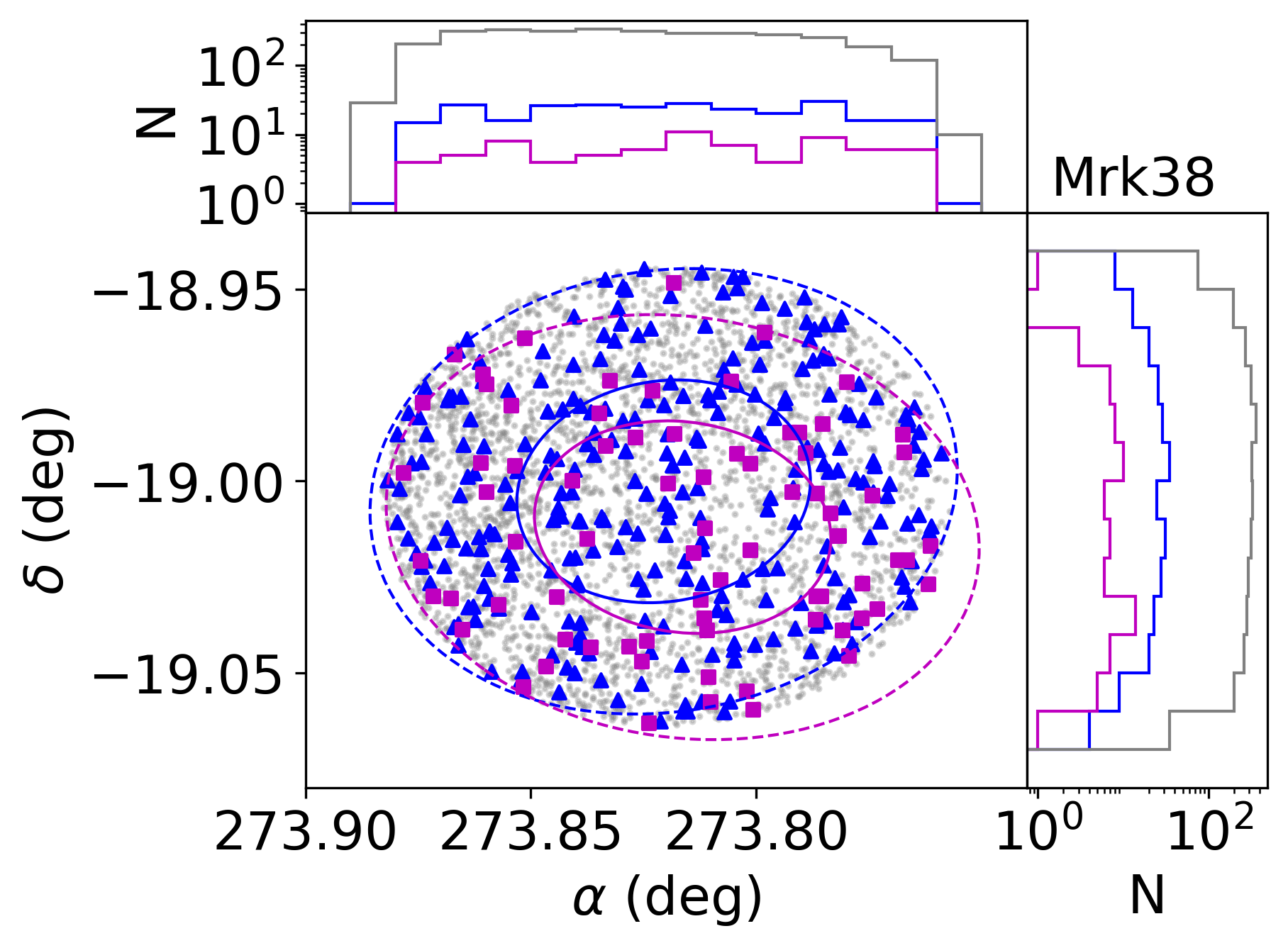}
\caption{Vector point diagram of proper motion ({\it left panel}) and the spatial distribution ({\it right panel}) for all the selected {\it Gaia} sources (grey dots) and the $P_{50}$ 
members (blue triangles). Magenta symbols are used for NGC~2659b and Mrk38b. The ellipses indicate the standard deviation of the mean values  (1$\sigma$: full lines, 2$\sigma$: dashed lines). 
The  histograms show the distribution of number of members and field stars.}
\label{fig:member}
\end{figure*}
%%%% -----------------------------------------end Fig.2

Our analysis is restricted to the samples of stars with 50 percent  or more of membership probability, which will be hereafter referred to as $P_{50}$ members or simply members. The mean values for the parameters estimated for the clusters are presented in Table \ref{tab:result1}.
Among the objects we studied, only for Mrk38 and NGC~2659 we found a second group in the same studied region. In these cases of double groups, the main cluster is denoted by index {\it a} and the secondary group by {\it b}.  

The distribution of members in the proper motion VPD and their position in the plot of equatorial coordinates are shown in Fig. \ref{fig:member} for 
Collinder 205 (hereafter Col205), a representative of single clusters{\footnote{The figures corresponding to the other clusters of the sample are given in Appendix \ref{sec:figs}  (supplementary material)}.},  and the two targets that were found bimodal. The main criteria for choosing the single or two-component
clusters are based on the separation of groups showing different proper motions. This separation is also evidenced by the mean values of proper 
motion and distances shown for each subgroup in  Table \ref{tab:result1}. 
Despite the clear separation on proper motion, the spatial distribution of the members of the two groups associated with NGC 2659 roughly covers the same projected area, with a slight trend of subgroup  {\it a} being more concentrated in the SW region, while subgroup  {\it b} tends to 
the NE region.  On the other hand, both components of Mrk38 occupy the same projected area without any separation in the plot of equatorial 
coordinates.

\begin{figure}
\includegraphics[width=\columnwidth]{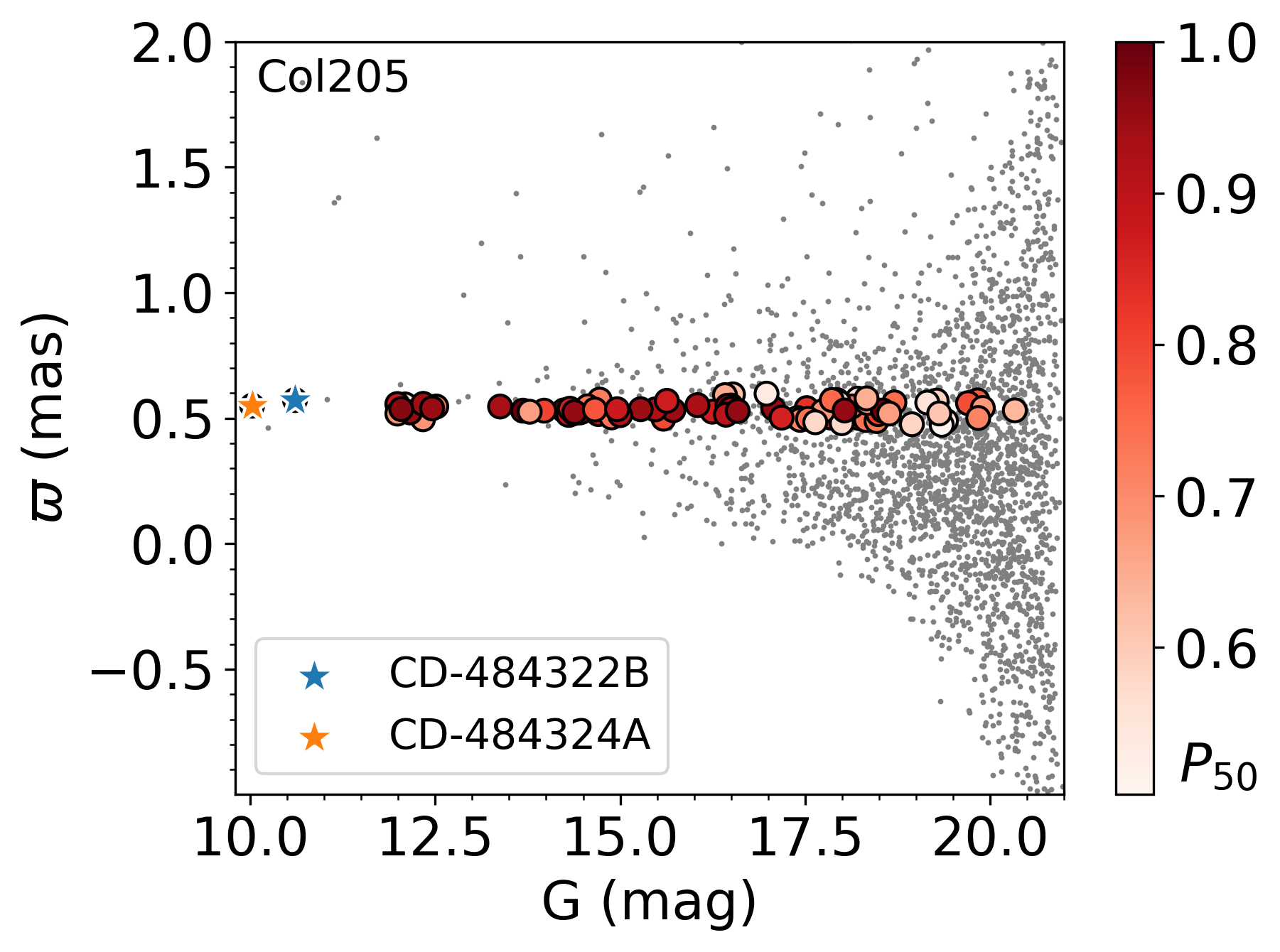}
\includegraphics[width=\columnwidth]{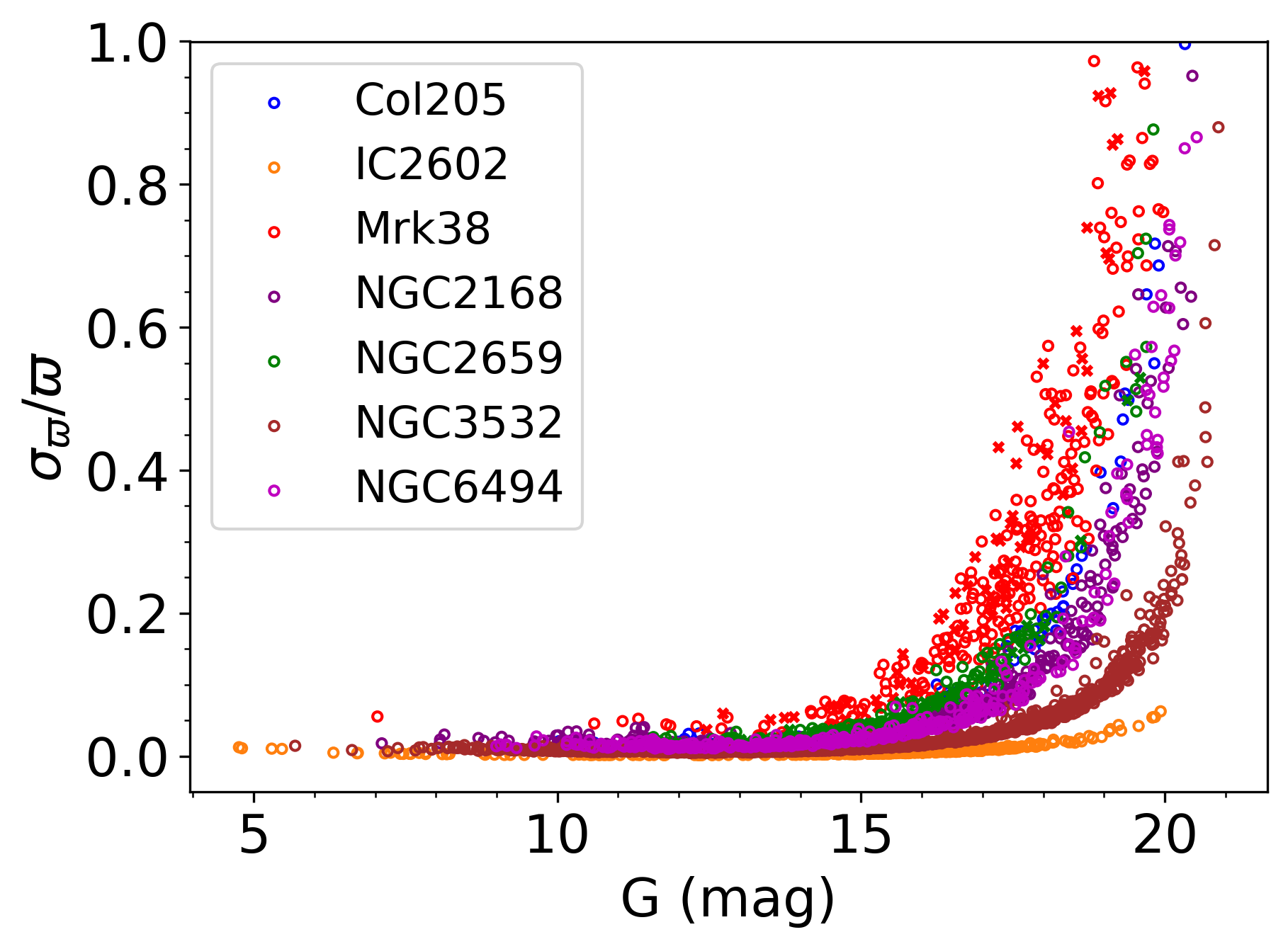}
\caption{{\it Top:} Plot of parallax as a function of G magnitude and the membership probability (colour-bar scale), highlighting two of the brightest stars of Col205. 
{\it Bottom:} The  fractional parallax uncertainty ($f  =  \sigma_{\varpi} / \varpi $)
as a function of G magnitude. Members of the secondary group of double clusters are indicated by $\times$. Faint objects  ($G \sim$ 18 mag, depending on distance)  have $f > $ 0.3, which occurs for a minor part of the members, except Mrk38.}
\label{fig:G_pxl}
\end{figure}

Another criterion confirming the cluster membership is based on the mean value of the parallax ($\varpi$).  In  Fig. \ref{fig:G_pxl} (top panel), we present the distribution of 
$\varpi$ as a function of apparent magnitude in band G, using different colour grades to indicate membership of Col205. 
It can be noted that the most 
 probable members show a narrow distribution around $\varpi \sim $ 0.5 mas, mainly for the brighter sources ($G < 17$ mag). 
 Fainter objects, for which the membership probability is lower, show a dispersion around the mean value of $\varpi$. 
 To infer the distances of the clusters, we adopted a statistical methodology based on a parallax distribution, which is described below.

%%%%%%%--------------------------------- Fig.3 + old 4 Parallax as a function of G mag + Parallax S/N
%%%% -----------------------------------------end Fig.3

%%%%%% Sect. 4.2 Distance 
\subsection{The most probable distance}
\label{sec:moda}

We discuss here the procedure to improve the distance determination based on {\it Gaia} parallax of objects at kilo-parsec scale distances. 
In this case, the simple method of inversion of $\varpi$ does not give accurate results, mainly for objects having high fractional parallax uncertainty, 
 defined as $f = \sigma_{\varpi} / \varpi $. 
 According to \citet{Luri18},  the estimation by the inversion of the parallax tends to overestimate the distance modulus, which is a poor approximation due to systematics and correlations in the {\it Gaia} astrometric solution.
 
For instance,   the use of model fitting to obtain the PDF by an automated Bayesian approach (BASE-9) was adopted by \citet{Bossini19} to infer 
the distance modulus and estimate the age for 268 open clusters. They found a median offset of -0.11 mag for the difference between the distance moduli derived from the analysis 
of BASE-9 results and the inversion of the median parallax. 

\citet{Navarete19} also used a Bayesian inference method to estimate the distance to the W3 complex (d = 2.14 kpc). They derived the PDF by adopting the exponentially 
decreasing space density suggested by  \citet{Bailer15}. The distances found for W3  and its substructures, considering the median value of the PDF, are about 5 percent different 
from those obtained by the simple inversion of parallaxes. For a more extreme case, the massive stellar cluster Westerlund~1 (Wd~1), \citet{Navarete22} 
revisited the distance estimation by using {\it Gaia} EDR3 and considering new cluster members.  They inferred an improved distance of 4.06 kpc for Wd~1, adopting the parallax 
method suggested by  \citet{CantatGaudin18b}  when dealing with cluster members with large uncertainties on the individual parallaxes. In these cases, the distance is assumed to be the same for all the cluster members and is calculated by following a maximum-likelihood procedure \citep[see Eq. 1 from][]{CantatGaudin18b}.

In Fig. \ref{fig:G_pxl} (bottom panel), we show the plot of  $f$ as a function of $G$ magnitude, aiming to evaluate, among the objects of our sample, the occurrence of 
observed parallaxes with large fractional parallax uncertainty. We verified that only the faint ($G > 18$ mag) members of our clusters may present high 
parallax uncertainty  $(f > 0.3)$, which occurs for less than 20 percent of the members, excepting for Mrk38a and Mrk38b that have 
more than 50 percent of their members showing $f > 0.3$.

\begin{figure}
\includegraphics[width=0.8\columnwidth]{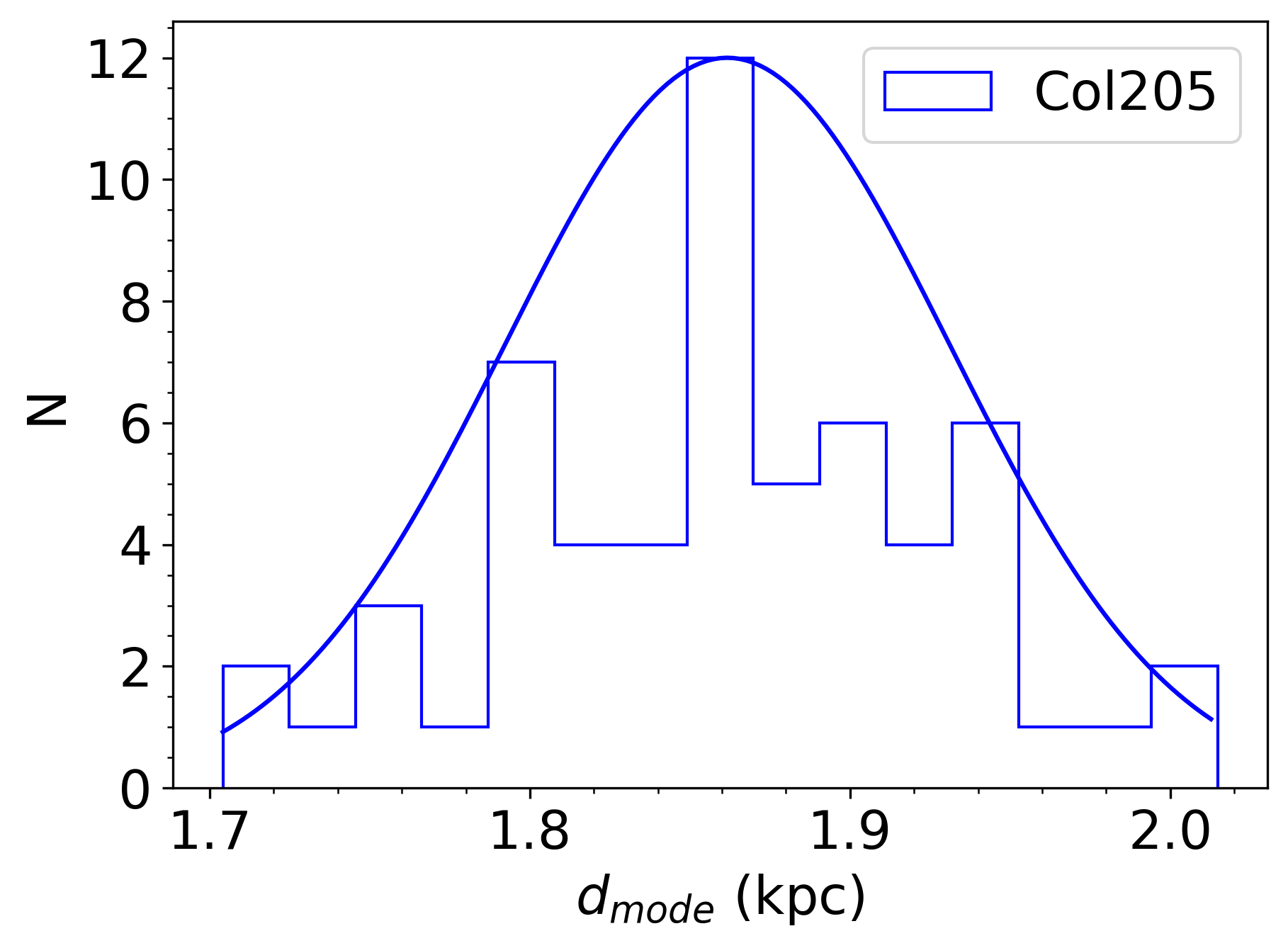}
\includegraphics[width=0.8\columnwidth]{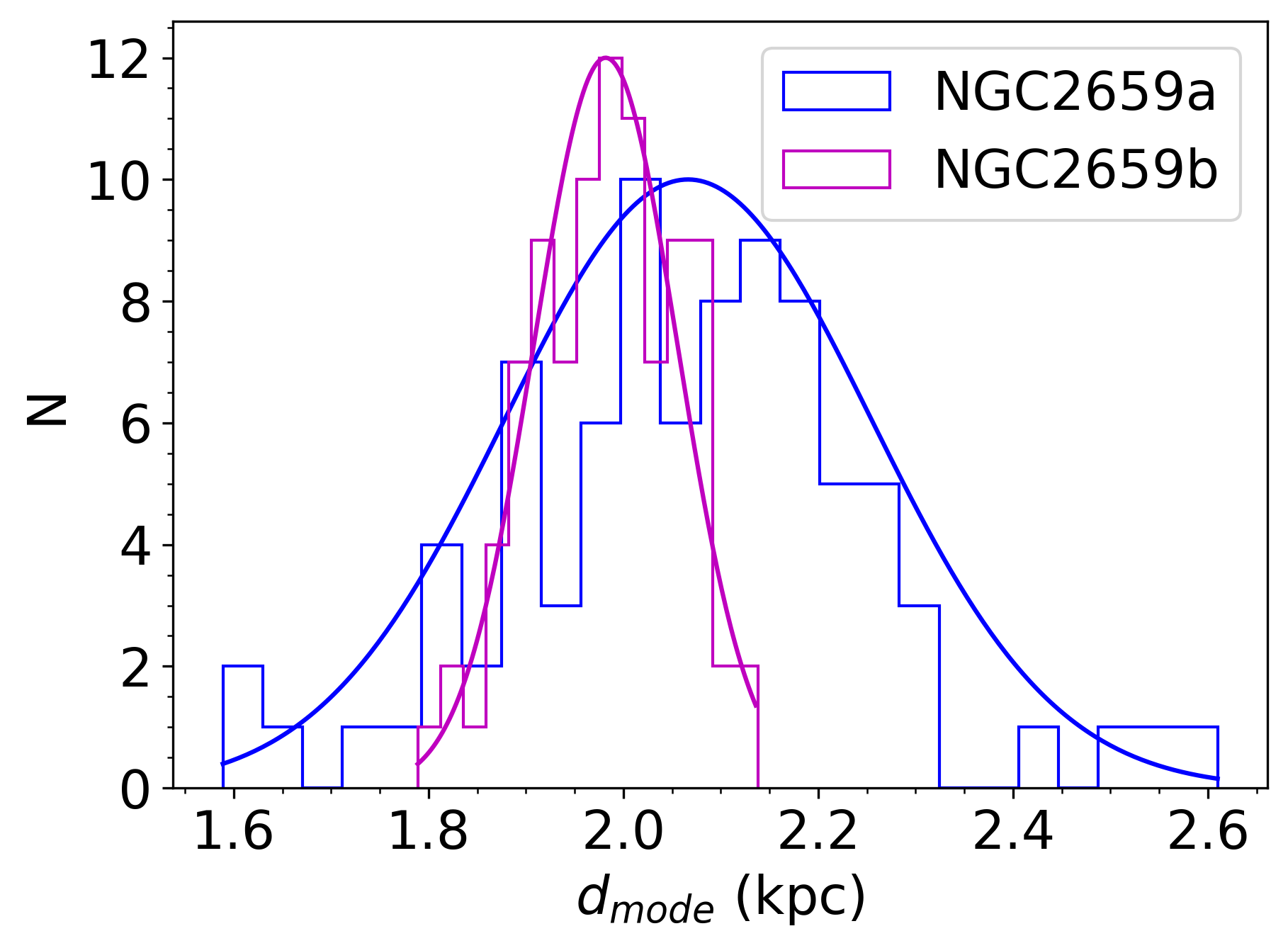}
\includegraphics[width=0.8\columnwidth]{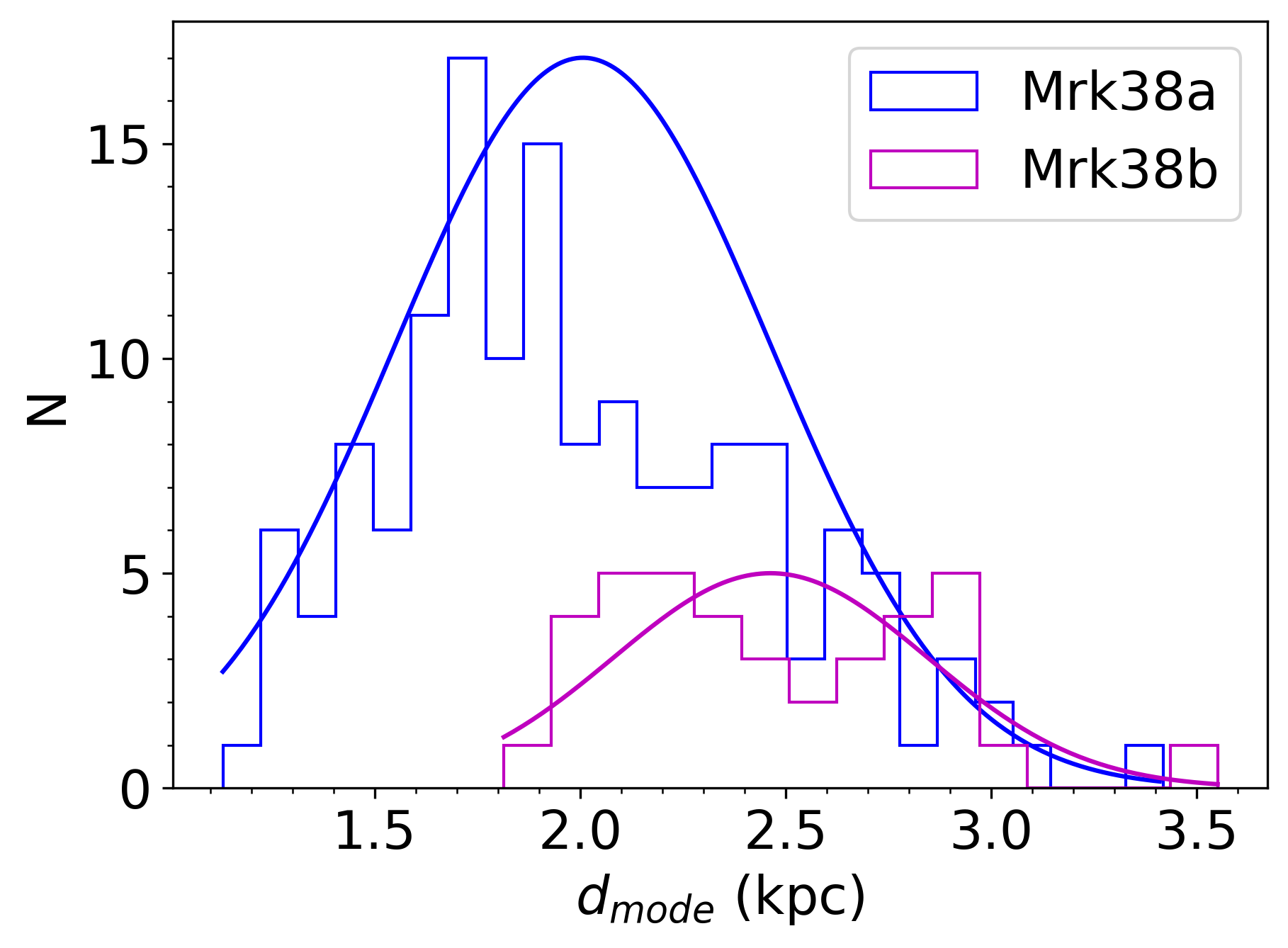}
\caption{Distance mode histogram obtained for Col205, NGC~2659 and Mrk38, excluding $P_{50}$ members that have  fractional parallax uncertainty $f  < 0.3$. The curves show the Gaussian fitting used to determine the mean value adopted as the cluster distance.}
\label{fig:mode}
\end{figure}

Considering that most objects in our sample do not show extreme cases of high fractional parallax uncertainty, we calculate the
distance for each cluster member  by using a simple PDF to estimate the true distance $d = 1/ \varpi_{\rm true}$ based on the observed 
parallax ($\varpi$). For the individual distance estimation, we adopt  the probability distribution suggested by  \citet{Luri18}, assuming that the observed parallax is normally distributed around the true parallax:
\begin{eqnarray}
p(\rho | \varpi_{\rm true})  =  \frac{1}{(2 \pi  ~\rho^4 ~\sigma_{\varpi}^2)^{0.5}} ~{\rm exp}( -\frac{(1/\rho - \varpi_{\rm true})^2}{2  \sigma_{\varpi}^2}),
\end{eqnarray}
\noindent{where $\rho = 1/\varpi$ and $\sigma_{\varpi}$ is the uncertainty on the individual observed parallax}.  The mode of the 
distribution was obtained for each member of the cluster, and finally, we calculated the mean value of the modes that
is adopted as the distance of the whole cluster. 
Figure \ref{fig:mode} displays histograms of the distribution of distance modes, whose mean values are presented in Table \ref{tab:result1}. Only the members with low fractional parallax uncertainty $(f < 0.3)$ were considered in the calculation
of the mean value of distance modes.

%%%%%% Sect. 4.3 	Colour-magnitude diagram (extinction, age, mass) 
\subsection{Colour-Magnitude diagrams}
\label{sec:age}

\begin{figure*}
\includegraphics[width=0.578\columnwidth]{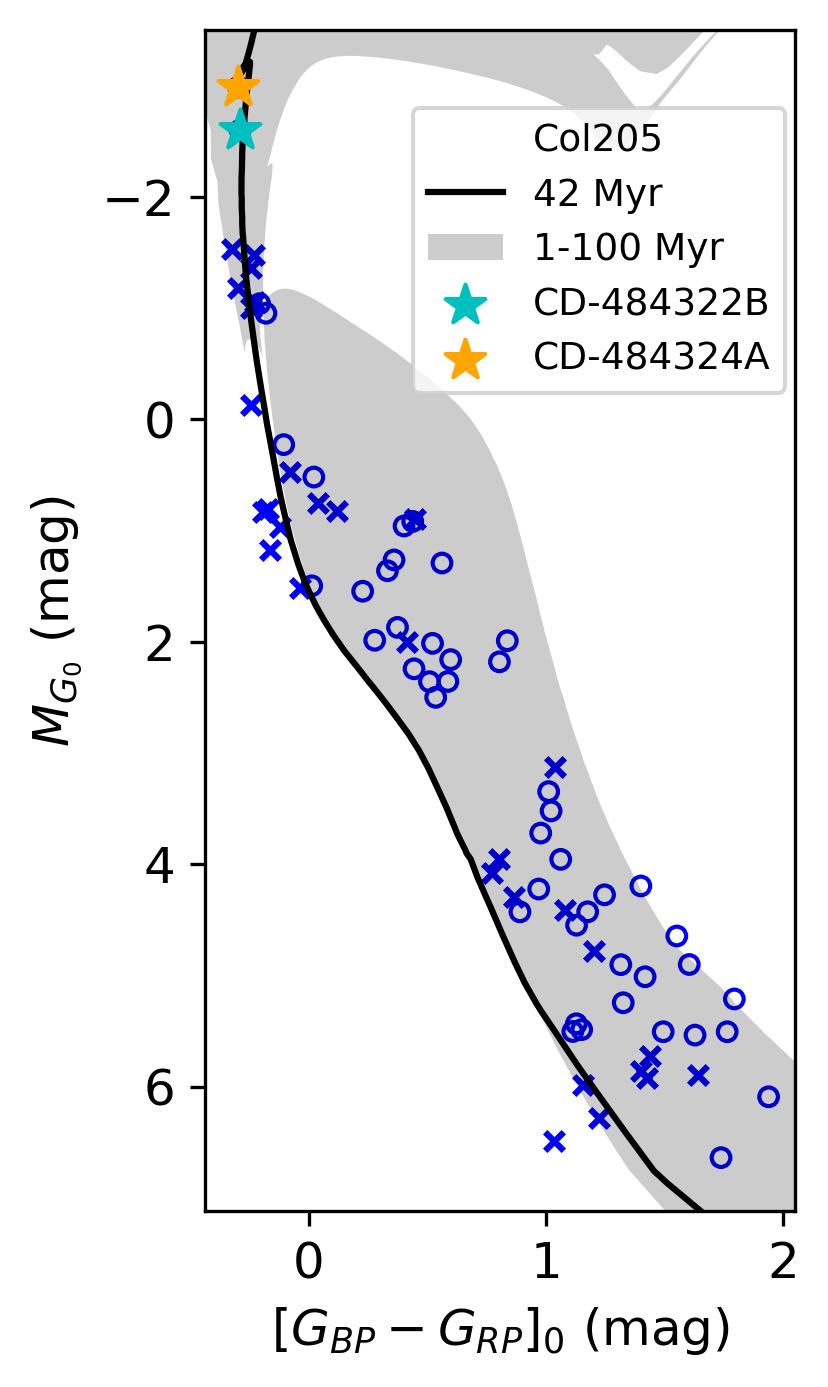}
 \includegraphics[width=0.6\columnwidth]{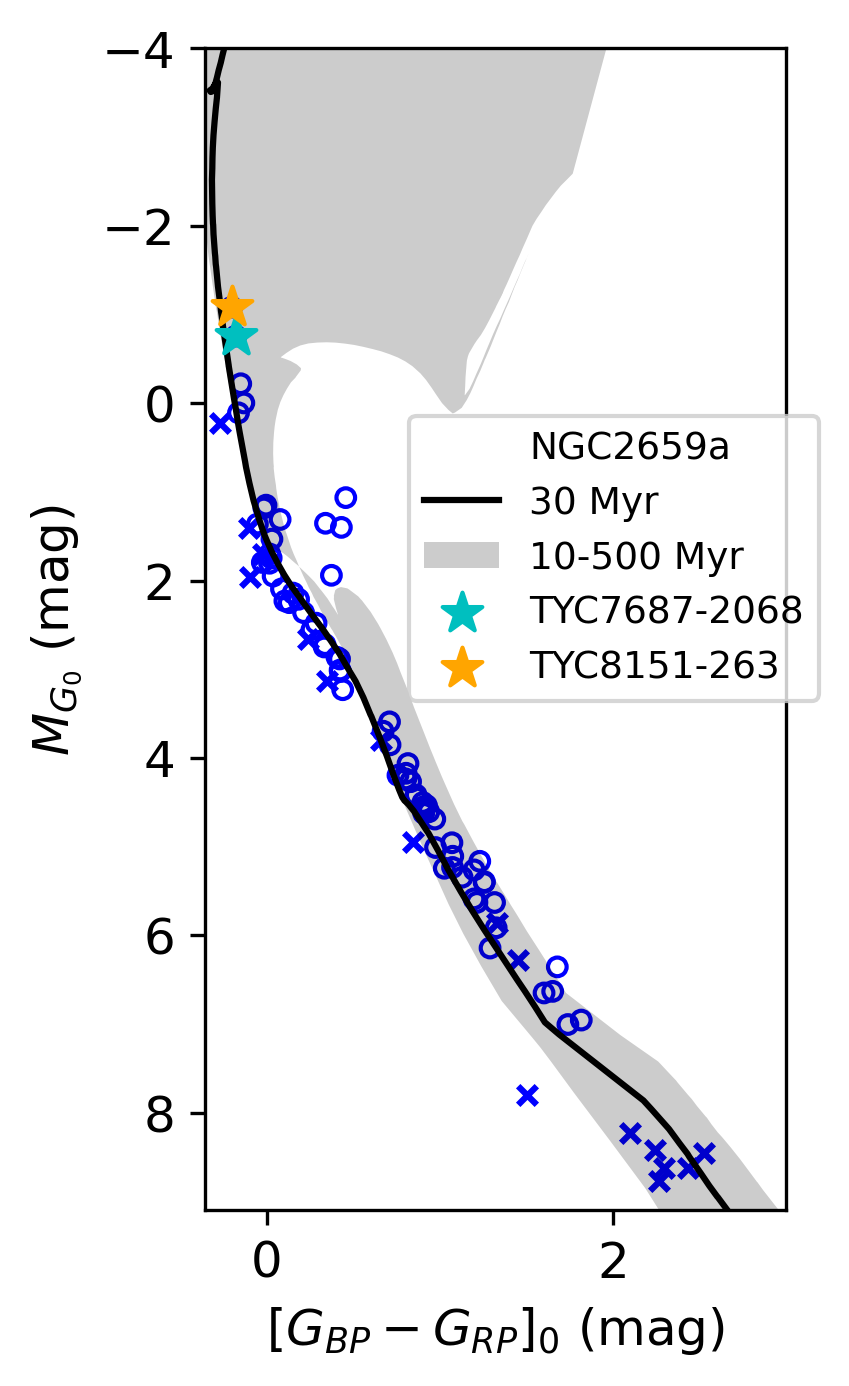}  
  \includegraphics[width=0.6\columnwidth]{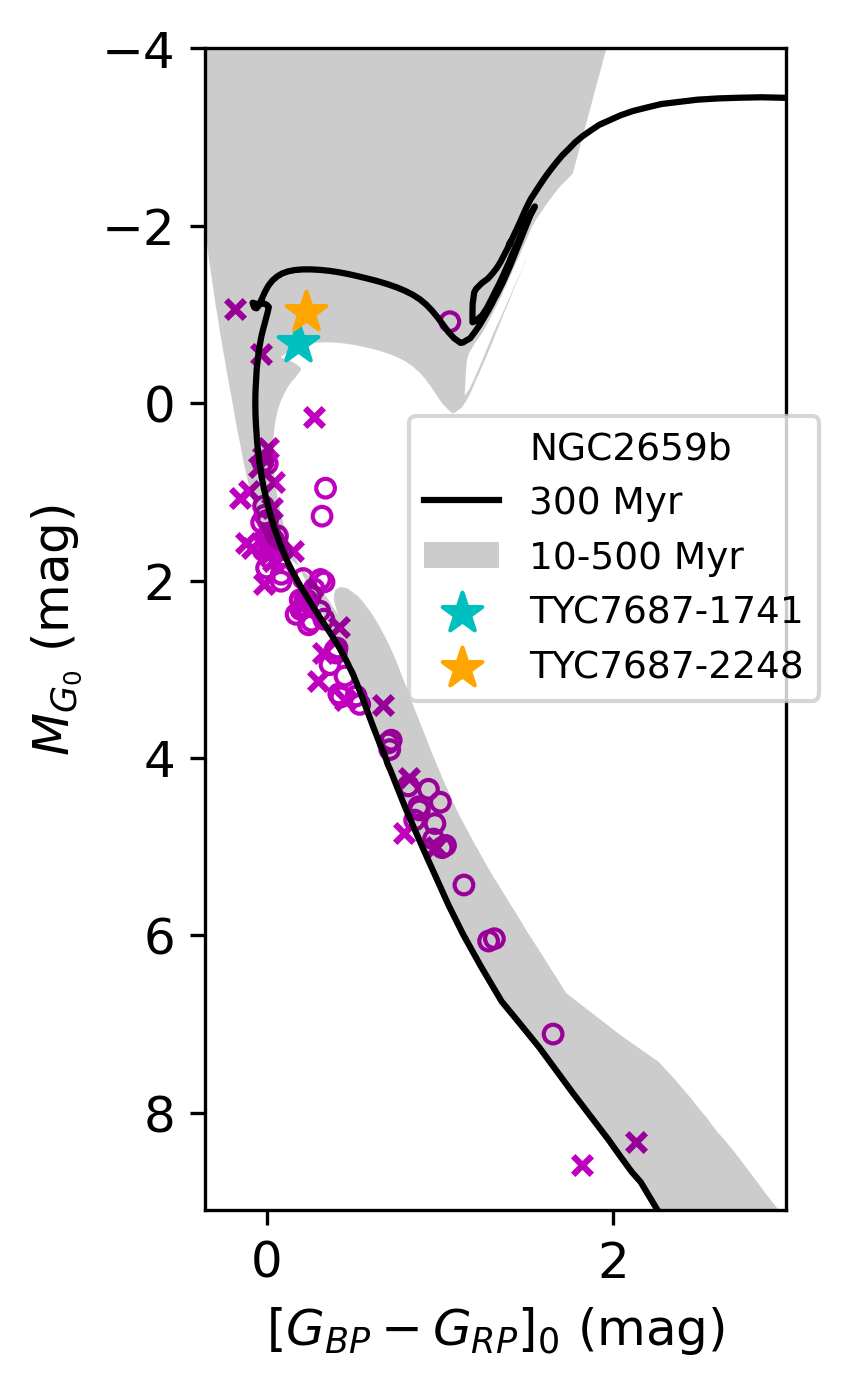}  
 \caption{Colour-Magnitude Diagram for $P_{50}$ members compared with isochrones from PARSEC models that best fit the cluster age. 
 The {\it Gaia} photometric data are unreddened by using the extinction correction estimated by the algorithm Aeneas (open circles). For the sources without Aeneas correction 
 ($\times$ symbols),  we adopted a mean value for E(B-V) (see Sect. \ref{sec:data}). }
    \label{fig:cormag}
\end{figure*}

We used the photometric data from {\it Gaia} DR3 to compare the observed colours and magnitudes of the cluster members 
with theoretical isochrones that give us an estimation of stellar mass and age.
The isochrones were adopted from 
{\it PARSEC}{\footnote{Version v1.2S+COLIBRI PR16 of {\it PARSEC} models available 
on http://stev.oapd.inaf.it/cgi-bin/cmd.}} evolutive models \citep{Bressan12, Marigo17}, 
which were plotted 
 in the $(M_G)_0 \times [G_{BP} - G_{RP}]_0$ diagram, where the absolute magnitude was estimated by adopting for all the stars the same value of distance inferred for the cluster (see Sect. \ref{sec:moda}). The observed apparent magnitudes were corrected for extinction following the procedure described in 
 Sect. \ref{sec:data}. Figure  \ref{fig:cormag} shows the distribution of {\it Gaia} sources in the colour-magnitude diagram, compared with the {\it PARSEC} isochrone 
 that provides a good fitting of the observed data, which was adopted as the cluster age.  
 The uncertainty on the age is indicated by the grey area shown in Fig. \ref{fig:cormag}, which is defined by lower and upper isochrones encompassing the distribution of data points. 
 
 %%%%%%%--------------------------------- Fig.4 (old 5) Distance mode histogram
%%%% -----------------------------------------end Fig.4

Stellar mass for each cluster member was estimated by adopting a simple interpolation method based on algebraic fitting inside the regions delimitated by the 
intersection between the theoretical lines (isochrones and evolutionary tracks). By localizing the observed position of a given star on the colour-magnitude diagram, the method 
infers the value of the mass by interpolating the theoretical values from the two nearest evolutionary tracks. 
The sum of the individual masses obtains the total observed mass of the cluster.

\begin{figure}
\includegraphics[width=\columnwidth]{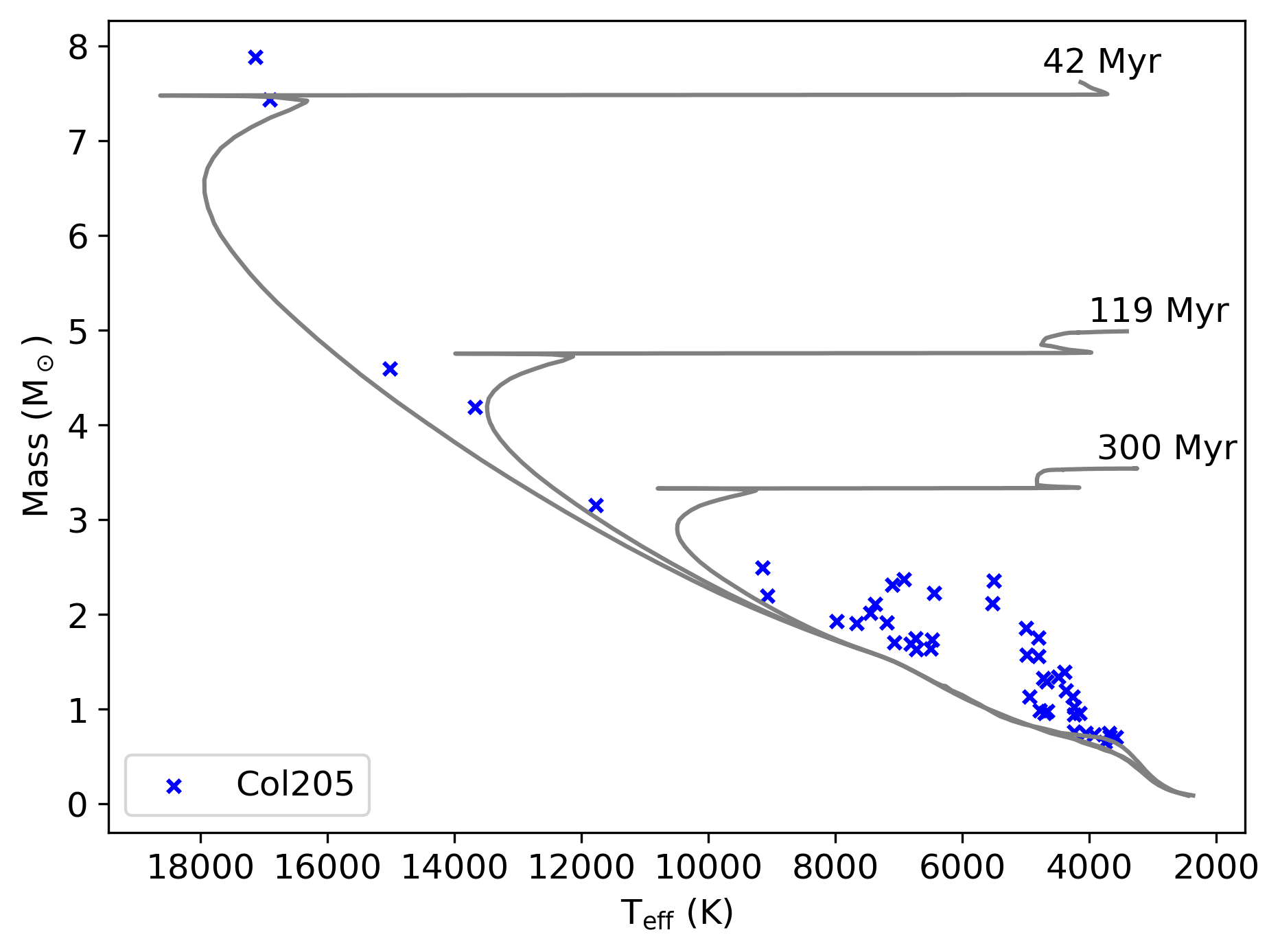}
 \caption{Individual stellar mass estimated in this work as a function of effective temperature obtained from {\it Gaia} DR3, compared with three isochrones from PARSEC models 
 that illustrate the agreement of parameters obtained from different databases.}
    \label{fig:mteff}
\end{figure}

To validate the results derived from the colour-magnitude diagrams, in Fig. \ref{fig:mteff}, we plot the individual masses (estimated by us) 
 as a function of effective
 temperature (provided by {\it Gaia}) for Col205. Fig. \ref{fig:Amteff}   (supplementary material) shows the plots for the other single clusters of our sample.  
 As an illustration,   the theoretical lines corresponding to three examples of isochrones from stellar models are also plotted. 
It can be noted a good mass-temperature correlation of low-mass stars ($<$ 2 M$_{\odot}$) following a main sequence, while some massive stars having low temperature coincide with the Red Giants region, as shown by the isochrones of 119 Myr (two members of NGC~2168), and 300 Myr (NGC~6494).

The good agreement among the parameters, obtained from different databases ({\it Gaia}, PARSEC) and the individual masses estimated by us, gives confidence on the observed mass of the cluster. 

\begin{figure}
\includegraphics[width=\columnwidth]{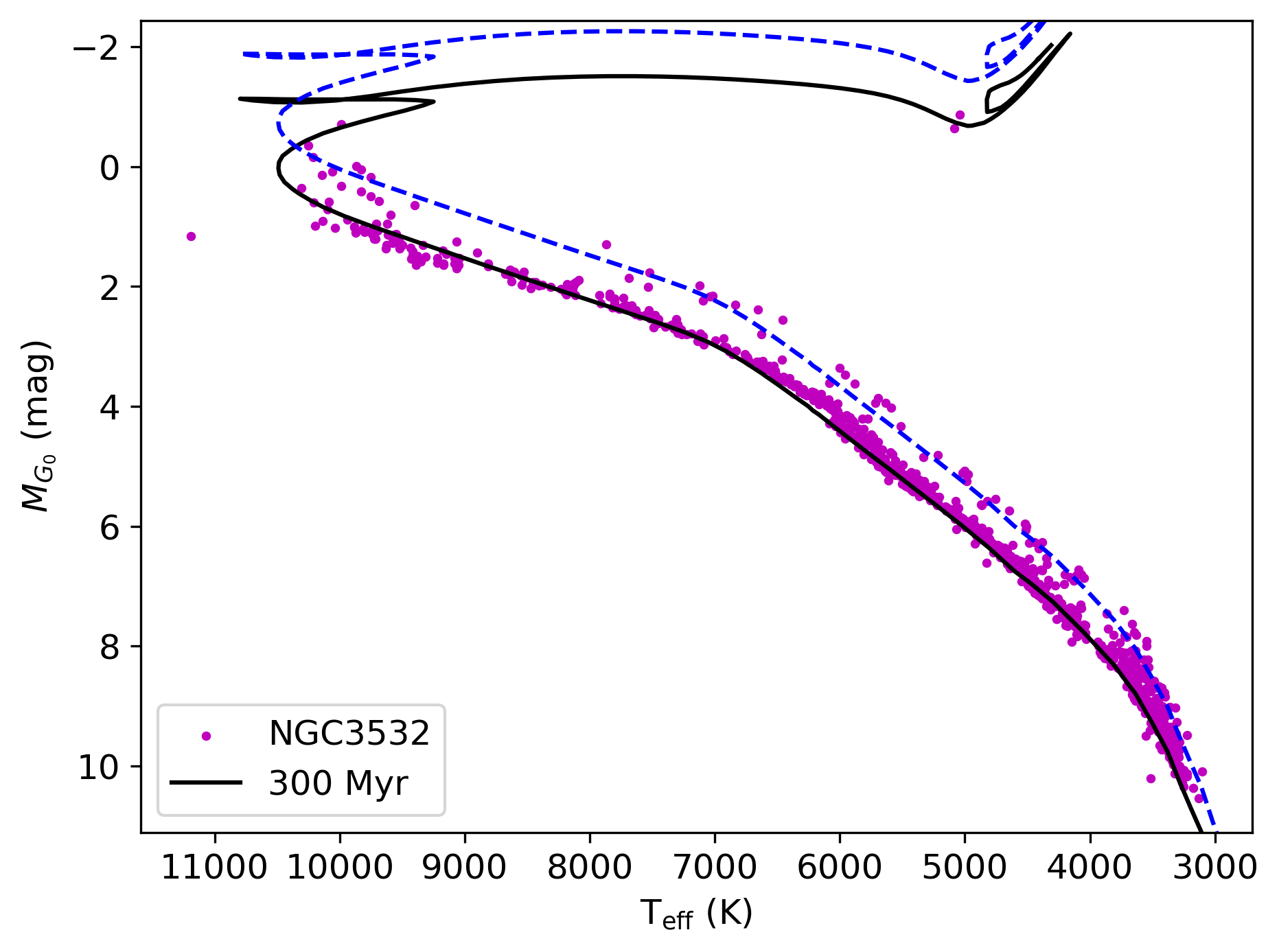}
 \caption{Distribution of absolute magnitude G as a function of effective temperature obtained from {\it Gaia} DR3 for the cluster NGC~3532 showing a dispersion due to the presence of binary stars. The PARSEC isochrone corresponding to the cluster age (full line) is also plotted with an offset of 0.75 mag (dashed line) to illustrate the good fitting of the binary population.}
    \label{fig:gteff}
\end{figure}

A similar confirmation is obtained when comparing the isochrone corresponding to the cluster age with the distribution of observed G magnitude as a function of the effective 
temperature ($T_{\rm eff}$) shown in Fig. \ref{fig:gteff}. It is interesting to note the expected dispersion due to the presence of binary stars. In the case of NGC~3532, the separation of 
binaries is very clear and can be fitted by the same isochrone, but using an offset of $\Delta G = 0.75$ mag. According to the simulations performed by  \cite{donada}  using 
evolutive models that include unresolved binaries in the Main Sequence population, this offset best fits the secondary distribution in the $G \times [BP-RP]$ 
diagram.

%%%%%--------------------------------- Fig. 5 (old 6) color-magnitude diagram 
%%%%%%%%------------------end Fig. 5

%%%--------------------------Table 2
\begin{table*}
 \caption{Parameters estimated for the $P_{50}$ members of the clusters, and  $A_V$ ranges obtained from the visual extinction maps.}
\label{tab:result1}
\scriptsize{
\begin{tabular}{lccccccccccc}
\hline
Cluster & $\alpha$ (2000) & $\delta$ (2000) & $\mu_\alpha$cos$\delta$  & 
    $\mu_\delta$  & $d$ & $N_{\rm P50}$ & $R_{\rm P50}$ & $M_{\rm P50}$  & age & $E(B-V)$ &  $A_V$ map \\ 
 & deg & deg & mas yr$^{-1}$ & mas yr$^{-1}$ & pc &  &pc & $M_{\odot}$ &Myr & mag & mag\\ 
\hline
Col205 & 135.116$\pm$0.054 & -48.988$\pm$0.036 & -4.71$\pm$0.22 & 3.98$\pm$0.17 & 
   1862$\pm$70 & 83 & 3.26$\pm$0.15 &170   & 42 & 0.89 &  1.5 - 2.5 \\ 
IC2602 & 160.899$\pm$2.101 & -64.523$\pm$0.778 & -17.75$\pm$0.94 & 10.64$\pm$0.97 & 
    152$\pm$3 & 295 & 8.17$\pm$0.14 & 197 & 119 & 0.10& 0 - 1.0  \\ 
Mrk38a & 273.821$\pm$0.032 & -19.003$\pm$0.029 & -0.23$\pm$0.69 & -2.32$\pm$0.46 & 
    1984$\pm$455 & 271 & 1.77$\pm$0.01 & 265  & 20 & 0.34 &  0 - 1.0 \\ 
Mrk38b & 273.816$\pm$0.033 & -19.012$\pm$0.028 & -0.01$\pm$0.41 & -0.98$\pm$0.29 & 
    2461$\pm$382 & 75 &  2.01$\pm$0.05 & 76  & 250 &0.34 &  0 - 1.0\\ 
NGC2168 & 92.273$\pm$0.3 & 24.342$\pm$0.289 & 2.22$\pm$0.17 & -2.9$\pm$0.15 & 
    861$\pm$29 & 845 & 13.7$\pm$0.2   &1056 & 119 &0.28 &  0 - 0.5 \\ 
NGC2659a & 130.643$\pm$0.052 & -44.988$\pm$0.043 & -5.29$\pm$0.09 & 5.04$\pm$0.15 & 
    2066$\pm$189 & 98 &  4.06$\pm$0.18 & 140  & 30 &0.51&  0.75 - 1.0 \\ 
NGC2659b & 130.726$\pm$0.071 & -44.94$\pm$0.053 & -4.39$\pm$0.05 & 3.03$\pm$0.06 & 
    1979$\pm$78 & 96 & 4.64$\pm$0.14  & 174  & 300 &0.51 &  0.75 - 1.0 \\ 
NGC3532 & 166.46$\pm$0.484 & -58.705$\pm$0.243 & -10.42$\pm$0.4 & 5.22$\pm$0.38 & 
    477$\pm$11 & 1262 &  6.68$\pm$0.03  & 1218 & 300 &0.07&  0.5 - 1.0\\ 
NGC6494 & 269.252$\pm$0.244 & -18.989$\pm$0.233 & 0.35$\pm$0.15 & -1.84$\pm$0.16 & 
    741$\pm$21 & 424 &   7.02$\pm$0.06 & 530 & 250 & 0.45& 1.5 - 2.0 \\ 
\hline
\end{tabular}}
\end{table*}
%%%-------- End Table 2

%%-----------------Fig.6 (old 7) Mass x Teff
%%-----------------end Fig.6

%%----------------- new Fig.7 G x Teff

%%-----------------end new Fig.7

%%---------
%%%%%% Sect. 4.4 	Mass Function and Sample incompleteness
\subsection{Mass function}
\label{sec:mf}

The results for the sample containing only $P_{50}$ members give accurate mean values of astrometric parameters, distance, and age of the cluster. However, the incompleteness of the sample has an impact on the correct estimation of parameters depending on total mass, such as crossing time and dynamical age, as well as  for  the cluster size and core radius derived from the fitting of the surface density distribution (see Sect. \ref{sec:core}).

Part of the problem can be solved by enlarging the sample at least to counting with a similar number of members found in the published catalogues (see Table \ref{tab:data}). For some clusters (Mrk38, NGC~2659, and NGC~6494), the counting of $P_{50}$ members are close to the cataloged values. For the other clusters, we increased the list of considered members by choosing lower limits of membership probability. The adopted limit was $P= 31$ percent for Col205 and IC~2602, $P=20$ percent for NGC~2168, and $P=4.5$ percent for NGC~3532,
which defined the sample containing the total number of stars that we consider as observed members ($N_{\rm obs}$). These observed members were included in determining the individual mass based on the colour-magnitude diagram.

Observational bias due to {\it Gaia} detection limit must also be taken into account when discussing the sample 
completeness mainly for clusters at large distances \citep[see][and references therein]{Buckner24}. 
We derived the mass function of the clusters to estimate the potential contribution of faint low-mass stars that may not have been detected by
 {\it Gaia}. The histogram of observed mass distribution was used in 
the fitting of the mass function  $\xi (m) \propto m^{- (1+ \chi)}$ adopted from \cite{Kroupa}. Following the method described by \cite{SS12}, we use the slope of the observed mass distribution 
($\chi$ given in Table \ref{tab:tab3})
 to obtain  the number of lacking faint stars ($N_{MF}$), below the limit of detection.  $N_{MF}$ is estimated by integrating the Initial Mass Function suggested by Kroupa in the range of low-mass stars, assuming 
 $\chi= 0.3 \pm 0.5$ for $m < 0.5$ $M_{\odot}$. Table \ref{tab:tab3} gives the total number of stars
  $N_{\rm tot} = N_{\rm obs} + N_{MF}$ and the total mass of the cluster 
  $M_{\rm tot} = M_{\rm obs} + M_{MF}$, where  
  $M_{\rm obs}$ is the sum of individual mass derived from the colour-magnitude diagram  of the observed stars,
  and $M_{MF}$ was estimated by integrating the mass function in the range of low masses.

%%%%%% Sect. 4.5 	Comparing Table 2 with Table 1
\subsection{Comparison with previous results}
\label{sec:comp}
When increasing the lists of sources with more objects that have membership probability lower than 50 percent, the samples are comparable to the maximum number of members reported in the literature (see Table  \ref{tab:data}), showing completeness that varies from 72 percent to 100 percent for Col205, IC~2602, NGC~2168, NGC~3532. 
For the lists that remained with $P_{50}$ members only, the number of objects is between the minimum and the maximum values reported in the literature, varying in the range of 42 -  60 percent of the maximum values, respectively, for NGC2659a and NGC6494. On the other hand, for Mrk38a, the list of $P_{50}$ objects is 0.62 larger than the maximum value found in the literature.

Concerning the age estimation obtained from the isochrone fitting, which strongly depends on the massive and intermediate-mass stars, the inclusion of more objects (mainly low-mass stars) does not change the age obtained for the P50 members. Comparing the results given in Table \ref{tab:result1} with those of Table \ref{tab:data}, we find a good agreement (more than 0.75 of the maximum value from the literature) for the clusters IC~2602, Mrk38a, NGC~2659a, and NGC~3532. On the other hand, for NGC~2168 and NGC~6494, we found ages more compatible with results from \cite{Dias21} and \cite{CantatGaudin18b} that reported lower values when compared with \cite{Bossini19}. In the case of Col205, our estimation (42 Myr) is almost 5 times larger than the previous results.

As discussed in Sect. \ref{sec:astrom}, an enlarged list of possible members ($P > 10$ percent) was analysed for NGC~2659a, but there was no change in the parameters that were determined for the constrained sample ($P_{50}$ members). The main impact of the incompleteness is on the estimation of total mass (see Sect. \ref{sec:mf}). However, since the cluster mass is not provided in the public catalogues we analysed (Table  \ref{tab:data}), it cannot be compared with our derivations.

%----------------------------------------Sect. 5
%\section{Surface density distribution}
\section{Spatial Distribution}
\label{sec:distrib}

In the literature, several methods exploring the surface density distribution have been adopted 
to better understand the formation and evolution of 
substructures in the distribution of young stars, investigating if they are linked (or not) to the mass segregation 
in their original clouds. Recently, statistical tests have been performed to quantify the kinematic 
substructures of star-forming regions based on N-body simulations of artificial data points 
distributions  \citep[e.g.][]{blaylock,becky}.

Different examples of works are found covering structures ranging from large-scale distributions to the smallest ones.  
For instance, \cite{Pouteau}  used the ALMA 1.3 mm and 3 mm continuum images to investigate the relation of core distributions and mass segregation with 
the density and kinematics of the gas of star-forming clouds. 
In nearby clusters, such as NGC~1333, the method to estimate mass segregation was applied to investigate the distribution of very low-mass stars to understand the origin of planetary-mass stars \citep{Parker23}.
Considering more distant regions, the estimation of parameters related to the fractal statistics have been explored for extragalactic objects, for instance, a large sample of star clusters 
of the Magellanic Clouds  that were studied by numerical simulations \citep{Daffern}  and based on observational  VISCACHA data\footnote{VISCACHA: VIsible Soar photometry of 
star Clusters in tApii and Coxi HuguA’ Survey \citep{Maia, Bruno}.} \citep{JoaoF, Jimena}. 

Here, we infer the geometric structure of our sample by analysing the stars' spatial distribution that hints at how the stellar clustering's morphology evolves. 
This section is dedicated to the morphology diagnosis that uses the density profile fitting to estimate the size of the cluster and its core radius;  the 
parameters related to the fractal statistics; local surface density, and mass segregation.  

The methodology for fitting the surface density profile and calculating fractal parameters is adopted from HGH19, whose description is summarized below.

%----------------------------------------Sect. 5.1 old radial Density profile fitting 
\subsection{Cluster size and core radius}
\label{sec:core}
\begin{figure*}
\includegraphics[width=0.68\columnwidth]{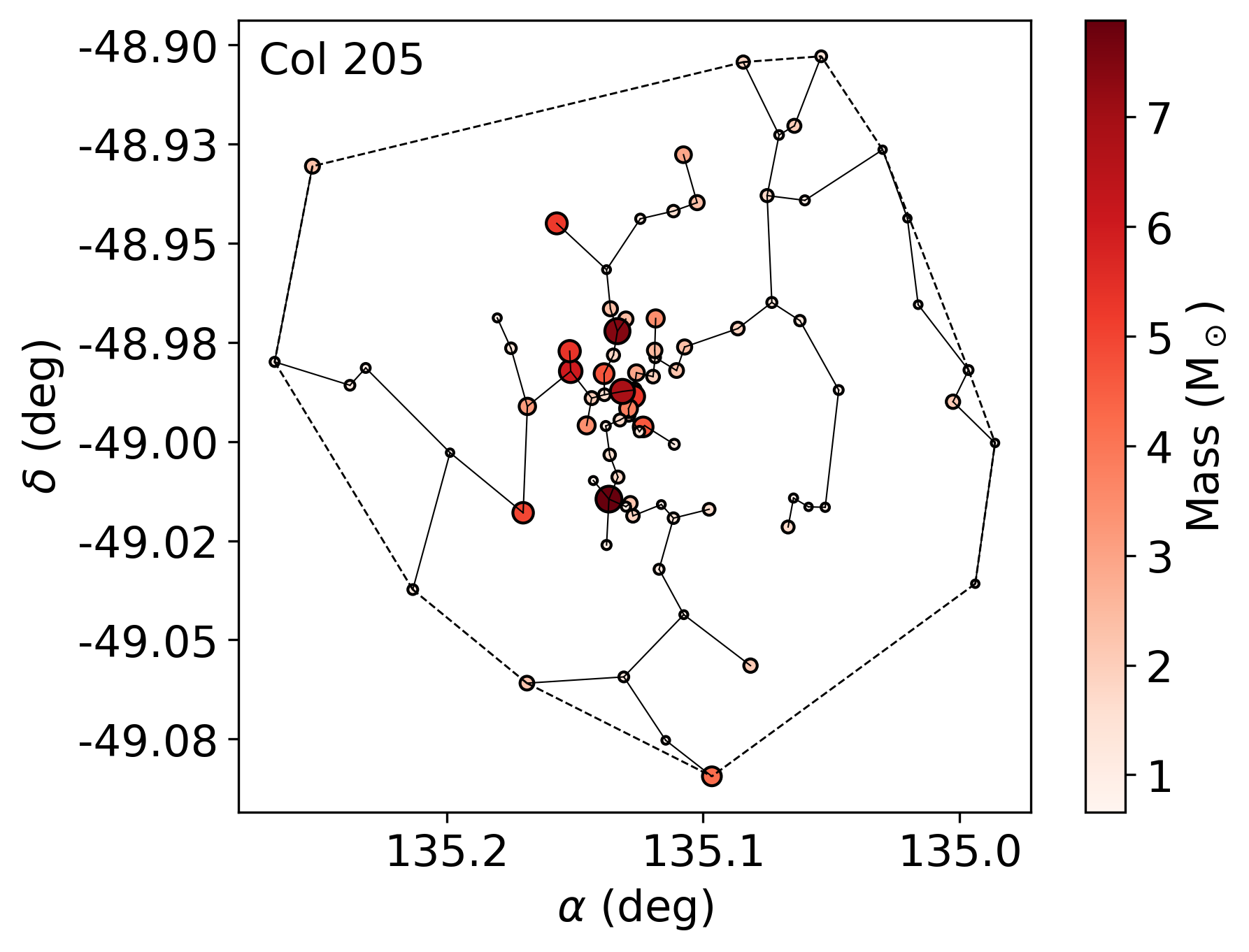}
\includegraphics[width=0.68\columnwidth]{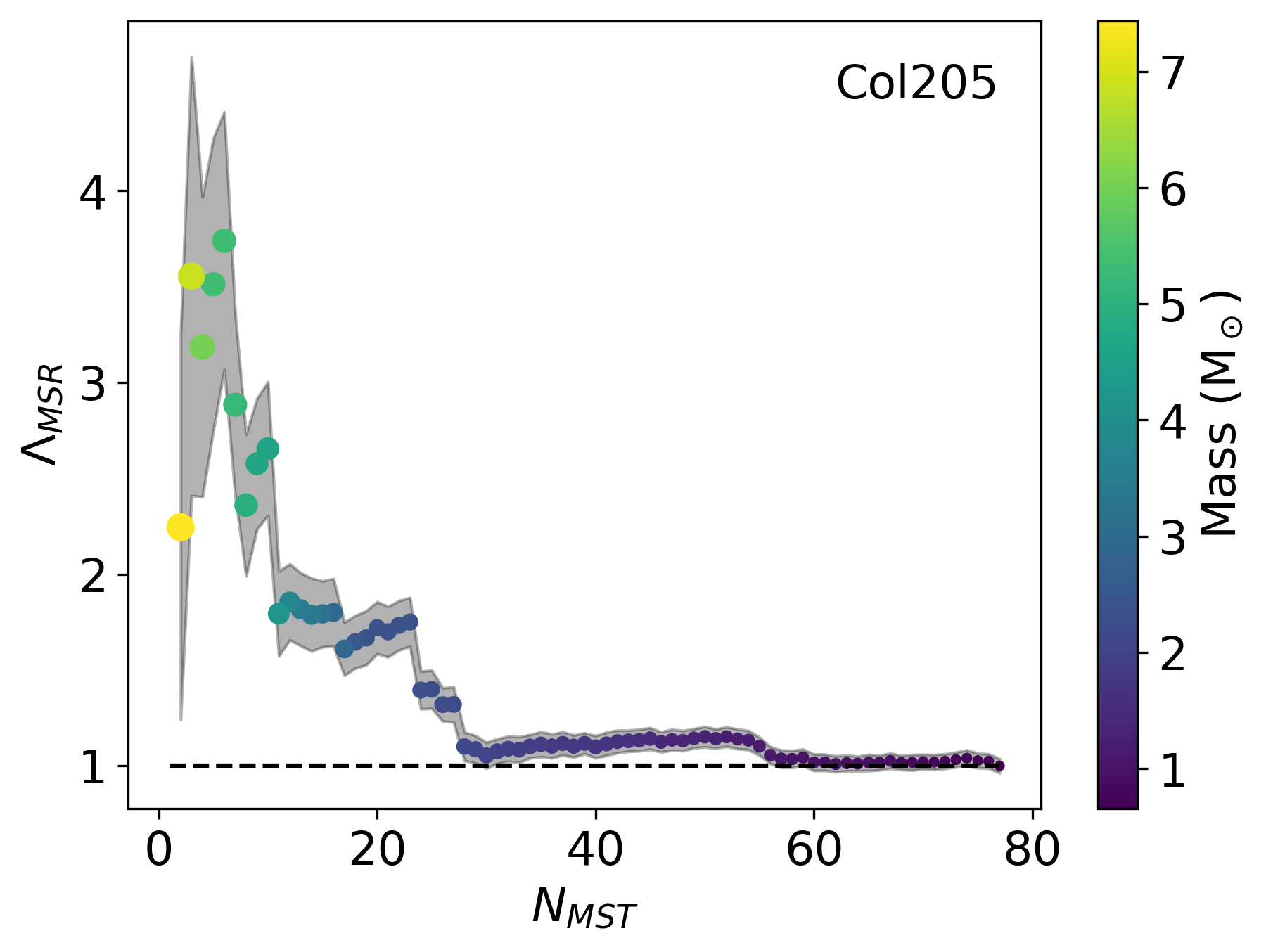}
\includegraphics[width=0.68\columnwidth]{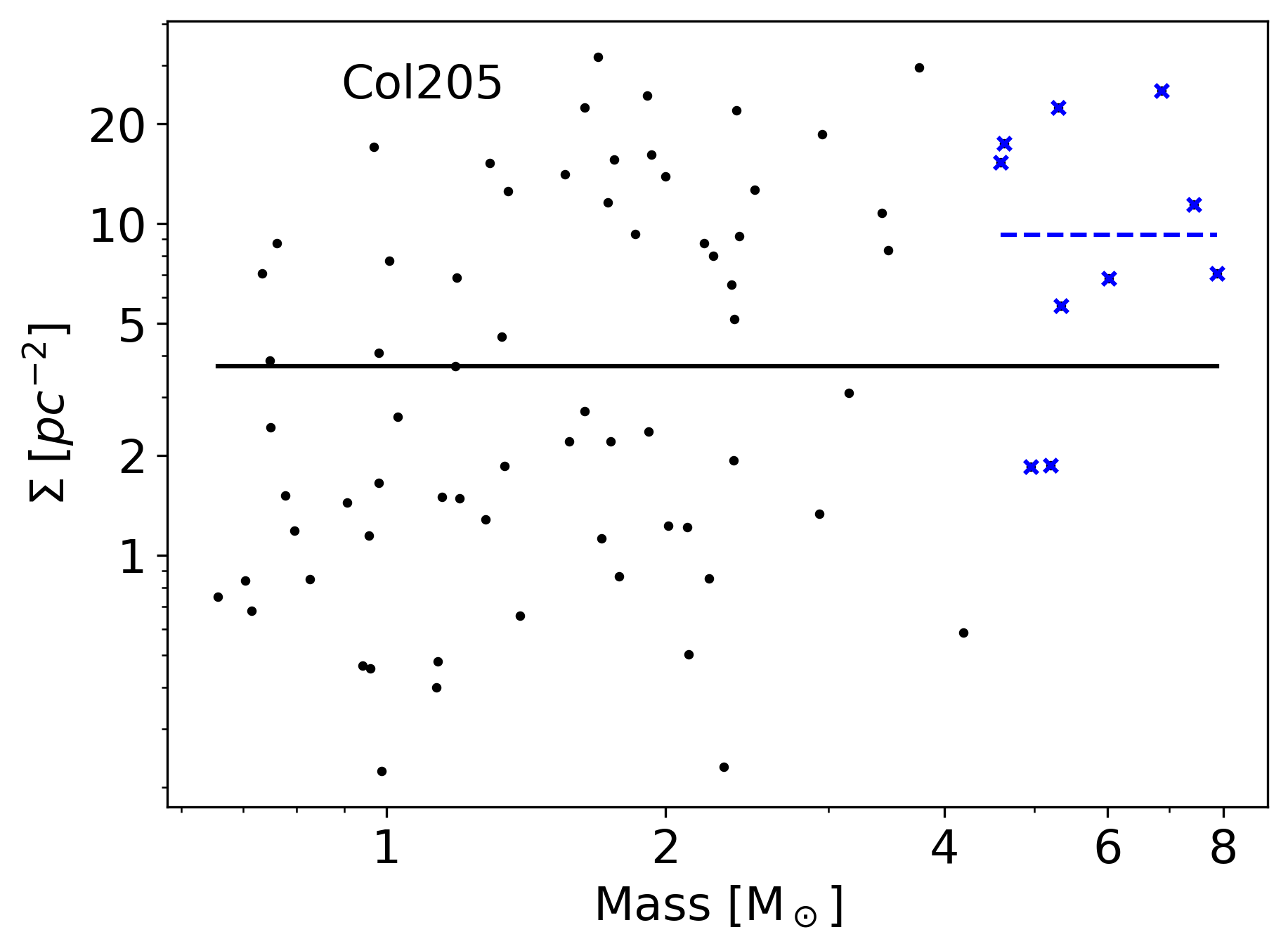}
\caption{Results for Col205 from the analysis of surface density distribution.  {\it Left}: Minimal spanning tree (MST) and convex hull area showing objects with different colours and symbol sizes according to their stellar mass.
{\it Middle}: The mass segregation ratio $\Lambda_{\rm MSR}$ as a function of the number of stars $N_{\rm MST}$. 
The occurrence of mass segregation is found above the limit $\Lambda_{\rm MSR}$ = 1 (dashed line).
{\it Right}: The $\Sigma - m$ plot comparing the local surface density with the stellar mass. The full line shows  $\tilde\Sigma_{all}$, the mean value obtained for all the cluster members, and a dashed line is used to indicate $\tilde\Sigma_{10}$ that is estimated for the 10 most massive stars, which are denoted by blue $\times$.}
\label{fig:Lambda}
\end{figure*}

The mean values of equatorial coordinates  and respective standard deviations 
($\alpha \pm \sigma_{\alpha}, \delta \pm \sigma_{\delta}$) given in Table \ref{tab:result1} roughly correspond to the center of the 2D projected spatial distribution of the cluster members.  
As indicated by the different values of  $\sigma_{\alpha}$ and $\sigma_{\delta}$, some clusters may display an elongated 
distribution, which area is better represented by a convex hull. Figure \ref{fig:Lambda} shows an example of this geometric distribution that defines the minimal spanning tree (MST), 
the smallest network of lines connecting the set of data points without forming closed loops. The sum of the edge lengths in the MST is minimized, and the area of the convex hull contains all 
points projected on the cluster plane.

The cluster size was estimated using two different methods based on different lists of objects that provide minimum and maximum values for the cluster radius.

First, we studied only the distribution of the most probable members ($P_{50}$).
Following the methodology proposed by \citet{gower}, we constructed the MST adopting 
the algorithm from \citet{Kruskal}.  The lower value for the radius ($R_{P50}$)  given in Table \ref{tab:result1} is defined by the radius of the circle having the same area as the convex hull. The standard deviation of $R_{\rm P50}$ was estimated using the Bootstrap method \citep{efron}. 

A second method was adopted to estimate the total radius of the cluster ($R_{\rm tot}$) that is achieved
 by fitting the density profile of the region containing the projected distribution of both cluster members and field stars.
The observed density profile is determined by counting the number of stars as a function of their distance to the center of the cluster. We adopted the  King-like radial density profile (RDP)  model suggested by \cite{Bonatto09} that is adapted from  the surface brightness profiles proposed by \cite{King}: 
\begin{equation}
\sigma (r) = \sigma_{\rm bg} + \frac{\sigma_0}{1+\big(\frac{r}{r_c}\big)^2} ,
\label{eq:king}
\end{equation}

\noindent{where $\sigma_0$ is the density at the center of the cluster and  $\sigma_{\rm bg}$ is the background  density. The radius of the core ($r_c$) is derived from  $\sigma (r_c) = \sigma_0 / 2$.} 

The  Levenberg–Marquardt method \citep{press} was adopted for the RDP fitting based on maximum-likelihood statistics, with 
goodness-of-fitting function given by  $\chi^2$.

The maximum value for the cluster radius ($R_{\rm tot}$) is defined by the point where the cluster stellar density reaches the background density (see Table \ref{tab:tab3}). 

Based on the parameters derived from the RDP fitting, we 
also calculate the density-contrast ($\delta_c$) parameter that quantifies how compact the cluster is. 
According \cite{Bonatto09}, this parameter is given by 
\begin{equation}
    \delta_c  = {1+\left(\frac{\sigma_0}{\sigma_{bg}}\right)} ,
	\label{eq:delta}
\end{equation}

\noindent{where compact clusters are expected to have $7 < \delta_c < 23$. Table \ref{tab:tab4} gives the values of density-contrast calculated for our clusters. According to this criterion, only  NGC~2659a is considered compact.}

%%%-------- NEW Table 3
\begin{table}
 \caption{Results from the fitting of the mass function and the radial density profile.}
\label{tab:tab3}
\scriptsize{
\begin{tabular}{lcccccc}
\hline
Cluster & $N_{\rm obs}$ & $M_{\rm obs}$ & $\chi$ & $N_{\rm tot}$ & $M_{\rm tot}$ & 
    $R_{\rm tot}$ \\ 
 &  & M$_\odot$ &  &  & M$_\odot$ & pc \\ 
\hline
Col205 & 108 & 218 & 1.12 & 943 & 1052$\pm$17 & 8.1$\pm$0.1 \\ 
IC2602 & 317 & 199 & 0.54 & 326 & 232$\pm$20 & 9.1$\pm$0.4 \\ 
Mrk38a & 271 & 265 & 1.59 & 421 & 469$\pm$14 & 4.6$\pm$1.9 \\ 
Mrk38b & 75 & 76 & -0.46 & 83 & 102$\pm$3 & 4.6$\pm$1.9 \\ 
NGC2168 & 1281 & 1490 & 1.00 & 4173 & 4406$\pm$160 & 22.9$\pm$0.1 \\ 
NGC2659a & 98 & 140 & -0.13 & 219 & 275$\pm$14 & 8.5$\pm$0.2 \\ 
NGC2659b & 96 & 174 & -0.78 & 163 & 239$\pm$15 & 10.3$\pm$0.3 \\ 
NGC3532 & 1847 & 1463 & 0.49 & 2151 & 2083$\pm$145 & 17.8$\pm$0.1 \\ 
NGC6494 & 424 & 530 & -0.39 & 554 & 713$\pm$47 & 7.5$\pm$0.1 \\ 
\hline
\end{tabular}}
\end{table}

%%%-------- End NEW Table 3

%%%%--------------------------------- Fig. 8 (old 8+10+ 11) Lambda x N  + Sigma x Mass
%%-----------------------end 

%---------------------------------------- NEW Sect. 5.2 
\subsection{Crossing time and dynamical age}
\label{sec:dyn_age}

The radius  obtained from the fitting of the King's profile  corresponds to an area larger than the distribution of $P_{50}$ members and possibly encompasses low-mass field stars. 
The same can be said about the estimation of $M_{\rm tot}$  that is considered an upper limit for the total mass of 
the cluster. Due to the differences in the lists of objects, we calculated two values for crossing time and dynamical 
age, giving a range of expected values. These ranges are presented in Table \ref{tab:tab4}, where the first value 
corresponds to the calculation using the minimum estimation for mass and radius ($M_{P50}$, $R_{P50}$) and 
the second value is a result from the use of the maximum parameters ($M_{\rm tot}$, $R_{\rm tot}$).

Based on the mass $M$ and radius $R$ adopted for the cluster, we used  the expression
$T_{\rm cr} = 10(R^3/GM)^{1/2}$ to estimate the crossing time. The ratio of cluster age to crossing time is used to quantify the dynamical age expressed by the parameter $\Pi$. According to  \cite{Gieles}, the expanding objects have $\Pi<1$ and can be distinguished from bound star clusters with $\Pi >1$.

Five objects of our sample 
show $\Pi < 1$ that corresponds to unbound clusters (Col205, IC~2602, Mrk38a, NGC~2168, and NGC~2659a), in agreement with previous results (HGH19). On 
the other side, the new results for NGC~3532 and NGC~6494 indicate $\Pi >1$, leading to a new 
classification that suggests these are bound star clusters. As discussed in Sect. \ref{sec:conclus}, the 
main difference from previous results is the larger number of members that are considered in the
 present work, which increased the total mass of the cluster, as well as the new isochrone fitting that gives an older age for these objects. 
  
The range of $\Pi$ values presented by Mark38b and NGC~2659b indicates they are  unbound clusters. Since HGH19 did not 
individually analyse these subgroups, we compared them with their main companions. Both cases show considerable differences in 
dynamical age, suggesting that Mark38b and NGC~2659b are
stellar groups respectively distinguished from Mrk38a and NGG~2659a.

%%%-------- NEW Table 4
\begin{table*}
 \caption{Results based on surface density distribution, crossing time, dynamical age, and fractal statistics.}
\label{tab:tab4}
\scriptsize{
\begin{tabular}{lccccccccccc}
\hline
Cluster & $n$ & $r_{\rm c}$ & $\delta_{c}$ & $T_{\rm cr}$ & $\Pi$ & $\bar m$ & 
    $\bar s$ & $\mathcal Q$ & $\Lambda_{\rm MSR}$ & $\Sigma_{\rm LDR}$ & p-value \\ 
 & pc$^{-2}$ & pc &  & Myr &  &  &  &  &  &  &  \\ 
\hline
Col205 & 9.5$\pm$0.6 & 0.90$\pm$0.08 & 4.63 & 67 -- 105 & 0.6 -- 0.4 & 
    0.67$\pm$0.06 & 0.74$\pm$0.05 & 0.90$\pm$0.07 & 2.65$\pm$0.35 & 2.48$\pm$0.34 & 
    0.158 \\ 
IC2602 & 20.8$\pm$3.0 & 1.23$\pm$0.37 & 3.75 & 248 -- 270 & 0.5 -- 0.4 & 
    0.62$\pm$0.03 & 0.88$\pm$0.03 & 0.70$\pm$0.03 & 1.66$\pm$0.23 & 2.16$\pm$0.29 & 
    0.006 \\ 
Mrk38a & 7.6$\pm$3.0 & 2.16$\pm$1.87 & 1.26 & 22 -- 69 & 0.9 -- 0.3 & 
    0.69$\pm$0.02 & 0.95$\pm$0.02 & 0.73$\pm$0.03 & 1.21$\pm$0.15 & 1.30$\pm$0.20 & 
    0.333 \\ 
Mrk38b & 7.6$\pm$3.0 & 2.16$\pm$1.87 & 1.26 & 49 -- 147 & 5.0 -- 1.7 & 
    0.78$\pm$0.05 & 0.99$\pm$0.03 & 0.79$\pm$0.06 & 0.92$\pm$0.10 & 0.85$\pm$0.15 & 
    0.748 \\ 
NGC2168 & 69.1$\pm$1.7& 1.74$\pm$0.08 & 5.42 & 234 -- 246 & 0.5 -- 0.5 & 
    0.57$\pm$0.02 & 0.55$\pm$0.01 & 1.04$\pm$0.03 & 2.05$\pm$0.31 & 2.04$\pm$0.37 & 
    0.090 \\ 
NGC2659a & 18.7$\pm$3.5 & 0.95$\pm$0.24 & 9.04 & 103 -- 222 & 0.3 -- 0.1 & 
    0.65$\pm$0.05 & 0.71$\pm$0.04 & 0.92$\pm$0.06 & 1.95$\pm$0.26 & 1.50$\pm$0.24 & 
    0.229 \\ 
NGC2659b & 8.5$\pm$9.7 & 0.37$\pm$0.29 & 4.64 & 113 -- 317 & 2.7 -- 0.9 & 
    0.72$\pm$0.05 & 0.83$\pm$0.04 & 0.87$\pm$0.05 & 1.21$\pm$0.14 & 1.11$\pm$0.20 & 
    0.646 \\ 
NGC3532 & 162$\pm$8 & 1.12$\pm$0.12 & 3.09 & 74 -- 249 & 4.1 -- 1.2 & 
    0.64$\pm$0.01 & 0.83$\pm$0.01 & 0.77$\pm$0.02 & 1.36$\pm$0.18 & 1.60$\pm$0.31 & 
    0.095 \\ 
NGC6494 & 136$\pm$20 & 0.24$\pm$0.05 & 1.91 & 121 -- 115 & 2.1 -- 2.2 & 
    0.65$\pm$0.02 & 0.77$\pm$0.02 & 0.85$\pm$0.03 & 2.11$\pm$0.27 & 2.12$\pm$0.33 & 
    0.030 \\ 
\hline
\end{tabular}}
\end{table*}

%%%-------- End NEW Table 4

%----------------------------------------Sect. 5.3 	Fractal statistics (m_bar, s_bar, Q)
\subsection{Fractal statistics}
\label{sec:Q}

The fact that most young clusters are found in a concentrated hierarchy of clusters within clusters makes their geometric distributions well-represented by fractals. In fractal star clusters, the level of substructures can be inferred using statistical analysis and measuring the $\mathcal{Q}$-parameter for observed clusters or simulations of artificial data point distributions  \citep[see][for instance]{delgado13,jaffa17}.

The method for determining the fractal parameters $\overline{m}$, $\overline{s}$, and $\mathcal{Q}$ was first proposed by \citet{Cart04}, 
aiming to describe the geometrical structure of points distribution and statistically quantifies the substructures. Measurements on the MST allow us to obtain
$\mathcal{Q} = \frac{\overline{m}}{\overline{s}}$, where  $\overline{m}$ is the mean edge length that is related to the 
surface density of the points distribution,  and $\overline{s}$ is the mean separation of the points. 

Generally, an initially fractal star-forming region is expected to evolve to become smoother and more centrally concentrated. However, different initial conditions or differences in establishing the borders of the considered region will cause significant changes in the position of the pair of parameters  $\overline{m} - \overline{s}$ in a plot that is often used to separate smooth distributions from substructured regions \citep{Daffern}.

Inferences of $\overline{m}$  and $\overline{s}$  depend on the total number of considered points $N$, which correspond in this work to the number of $P_{50}$ members, and are normalised by $A_ {N} $, the area of the convex hull  \citep{schkl}.

These fractal parameters are useful to distinguishing fragmented ($\mathcal{Q} <  0.8$)  from smooth distributions  ($\mathcal{Q} > 0.8$),
small-scale fractal subclustering can be quantitatively distinguished from distributions with large-scale radial clustering. 

 Table \ref{tab:tab4} gives the fractal parameters obtained for our sample, which are discussed in Sect. \ref{sec:structure} based on the comparative analysis with previous results. In this case, the study considers the expected distribution in the $\overline{m}-\overline{s}$ plot indicating different types of structures, according to numerical simulations \citep{Parker18}. 

Our new results better indicate the cluster type distribution, probably due to the larger number of members considered here compared to HGH19.  The results are not well defined only in the case of  NGC~2659b, 
but the value  $\mathcal{Q} = 0.87$ indicates a radial concentration, while
Mrk38a ($\mathcal{Q} = 0.73$) is near the homogeneous boundary between fractals and radial profiles.

 An additional analysis of the cluster type distribution was performed by calculating the fractal dimension for the clusters that have 
 $\mathcal{Q} < 0.8$.  Following \cite{Canavesi}, for instance, we adopted a simple definition of the Box-Counting dimension from \cite{Feder}. 
 In summary, this method considers several cubes
($N_{\delta (F)}$)  in a $F \in {\delta}^n$ set with a $\delta$ size. If $N_{\delta (F)}$ intersects $F$, the box-counting dimension $D_b$ 
is defined by 
the slope of  $\log(N_{\delta (F)})$ {\it versus}  $- \log (\delta)$.
We found fractal box dimension $D_b  > 2.6$ for IC~2602, Mrk38a, and NGC~3532, indicating they do not show high levels of substructures, tending to smooth distributions as expected for regions having fractal dimension 
$D \sim 3$. For Mrk38b, which has $\mathcal{Q} \sim 0.8$, the type of distribution remains undefined.

For clusters with $\mathcal{Q} > 0.8$, we adopted the radial distribution  $n \propto r^{-\alpha}$  in order to estimate $\alpha$
through the fitting of star counts in intervals of radius $r+dr$ \citep{Cart04}. In this case, we found 
$1.2 < \alpha < 2.4$ corresponding to intermediary 
distributions, in between uniform density profile ($\alpha = 0$) and centrally concentrated distribution  ($\alpha = 3$).

%----------------------------------------Sect. 5.4	Mass segregation & local surface density
\subsection{Distribution of massive stars}
\label{sec:mass}

Analysing the properties of the massive star as a function of age, surface density, and structure is useful in 
order to understand better the origin and evolution of mass segregation of stellar clusters 
\citep[e.g.][]{Dib18,maurya,kim,nony21}

In this section, we discuss two forms to quantify if massive stars are concentrated (or not) in a projected distribution by calculating 
 the mass segregation ratio ($\Lambda_{\rm MSR}$) and plotting  the local surface density ($\Sigma$) as a function of mass.

The parameter  $\Lambda_{\rm MSR}$ \citep{Allison09} is used to determine if 
massive stars are closer to each other. It is defined by: 
\begin{equation}
\Lambda_{\rm MSR} = \frac{\langle l_{\rm average} \rangle}{l_{\rm subset}},
    	\label{eq:lambda}
\end{equation}

\noindent{where $\langle l_{\rm average} \rangle$ is the average (median) length of the MST measured for sets of $N_{\rm MST}$ random stars,  
and $l_{\rm subset}$ is the same measure made for  the subset of  $N_{\rm MST}$ most massive stars.  
In this work, we use a subset of $N_{\rm MST} = 10$.
We adopted the standard deviation from the dispersion associated with the roughly Gaussian distribution around $\langle l_{\rm average}\rangle$ to represent the uncertainty on the estimate of the mass segregation ratio. The cases of  
 mass segregation are indicated by $\Lambda_{\rm MSR} >$ 1. } 

Another parameter to be considered is $\Sigma_{\rm LDR}$, commonly used to indicate if the massive stars are in regions of higher 
surface density than those where low-mass stars are found. 
The calculation of the  local stellar surface density  \citep{Masch} for an individual star $i$ is expressed by:
\begin{equation}
    \Sigma_i = \frac{N-1}{\pi r^2_{i,N}},
	\label{eq:LSD}
\end{equation}

\noindent{where $r_{i,N}$ is the distance between a given star and  its $N^{th}$ nearest neighbour}.  
In this work, we set $N=10$.
Following \citet{Parker14}, we adopted the local density ratio:

\begin{equation}
   \Sigma_{\rm LDR} = \frac{\tilde{\Sigma}_{\rm subset}}{\tilde{\Sigma}_{\rm all}},
\end{equation}

\noindent{where $\tilde{\Sigma}_{\rm subset}$ is the average measured for a given subset of stars, and  $\tilde{\Sigma}_{\rm all}$ corresponds to
the average calculated for all the stars of the cluster. If the subset contains massive stars concentrated in dense regions, its local
surface density is expected to be higher than the whole sample. In this case, the mass segregation is indicated by 
$\Sigma_{\rm LDR} > 1$. With this method, it is possible to quantify if massive stars are found in dense regions, which is  different from the mass segregation  measured by 
$\Lambda_{\rm MSR}$ \citep{Parker14, Parker15}.} 

Following the same method  to infer  the standard deviation of $R_{\rm P50}$ (see Sect. \ref{sec:core}), the 
$\Sigma_{\rm LDR}$ uncertainties were estimated using the bootstrap technique \citep{efron}. The method creates a simulated data set considering a variation of each parameter (positions) within a confidence level corresponding to a given distribution. Then, the standard deviation of this set of values is calculated. 

An example of the results of  $\Lambda_{\rm MSR}$  and $\Sigma$ compared with the mass 
distribution is shown in  Fig. \ref{fig:Lambda}. The parameters related to the fractal statistics and the geometric 
structure of the clusters are given in Table \ref{tab:tab4}.
The resulting $\Sigma_{\rm LDR}$, obtained from the ratio of the values corresponding to the horizontal lines in 
Fig. \ref{fig:Lambda} (right panel) shows that our clusters tend not to have massive stars concentrated  in regions
of high surface density, which is indicated by $\Sigma_{\rm LDR} \sim 1$.

Aiming to quantify the significance of the deviation from the median for all stars, we calculate the p-values using a Kolmogorov-Smirnov test 
for the cumulative distribution of the radial distances from the center of all stars and the 10 most massive stars. Following \cite{Parker15}, we adopted a p-value < 0.1 as the significance threshold. 
If $\Sigma_{\rm LDR} \sim 1$ and p-value > 0.1, the distributions show no significant difference, indicating that the most massive stars are not 
mass segregated. This is confirmed for Mrk38a,b and NGC~2659a,b. However, Col205 and NGC~3532 are too close to the adopted threshold, making it difficult to gauge whether the differences are significant. On the other hand, for IC~2602, NGC~2168, and NGC~6494, the parameters 
$\Sigma_{\rm LDR} > 2$ and p-value < 0.1 are clear signature of mass segregation.

In particular for NGC~6494, \citet{tarricq22} found mass segregation ratio  $\Lambda_{\rm MST}$ = 1.37 that is lower than our result but 
still compatible (see Table \ref{tab:tab4} and Fig. \ref{fig:ALambda2}).

%----------------------------------------Sect. 6
\section{Comparative analysis}
\label{sec:compare}

%----------------------------------------Sect. 6.1
\subsection{Astrometric results}
\label{sec:astrom}

The mean values of position found for the cluster members (see Table \ref{tab:result1}) were compared with  previous results 
from the literature (see Table \ref{tab:data}) by calculating the offset on equatorial coordinates ($\alpha$, $\delta$) with respect the values obtained in this work. 
The conversion of angular measurements to linear dimensions, which gives  $\Delta {\rm pos}$ (position offset in parsec), was made using the mean value of the distance mode.

The comparison of proper motion was made by calculating the offset of tangential  velocity 
$\Delta \mu = (\Delta \mu_{\alpha^\star}^2 + \Delta \mu_{\delta}^2)^{0.5}$,  given in mas yr$^{-1}$ that was converted 
into km s$^{-1}$ in the same way we used for the position offset.

In the case of double clusters, we adopted the results for Mrk38a and NGC~2659a as a reference, meaning that these main groups and the single clusters present offset equal zero in this comparative analysis. 

%%-----------------Fig. 9 (old 12) Off set
\begin{figure*}
\includegraphics[width=\columnwidth]{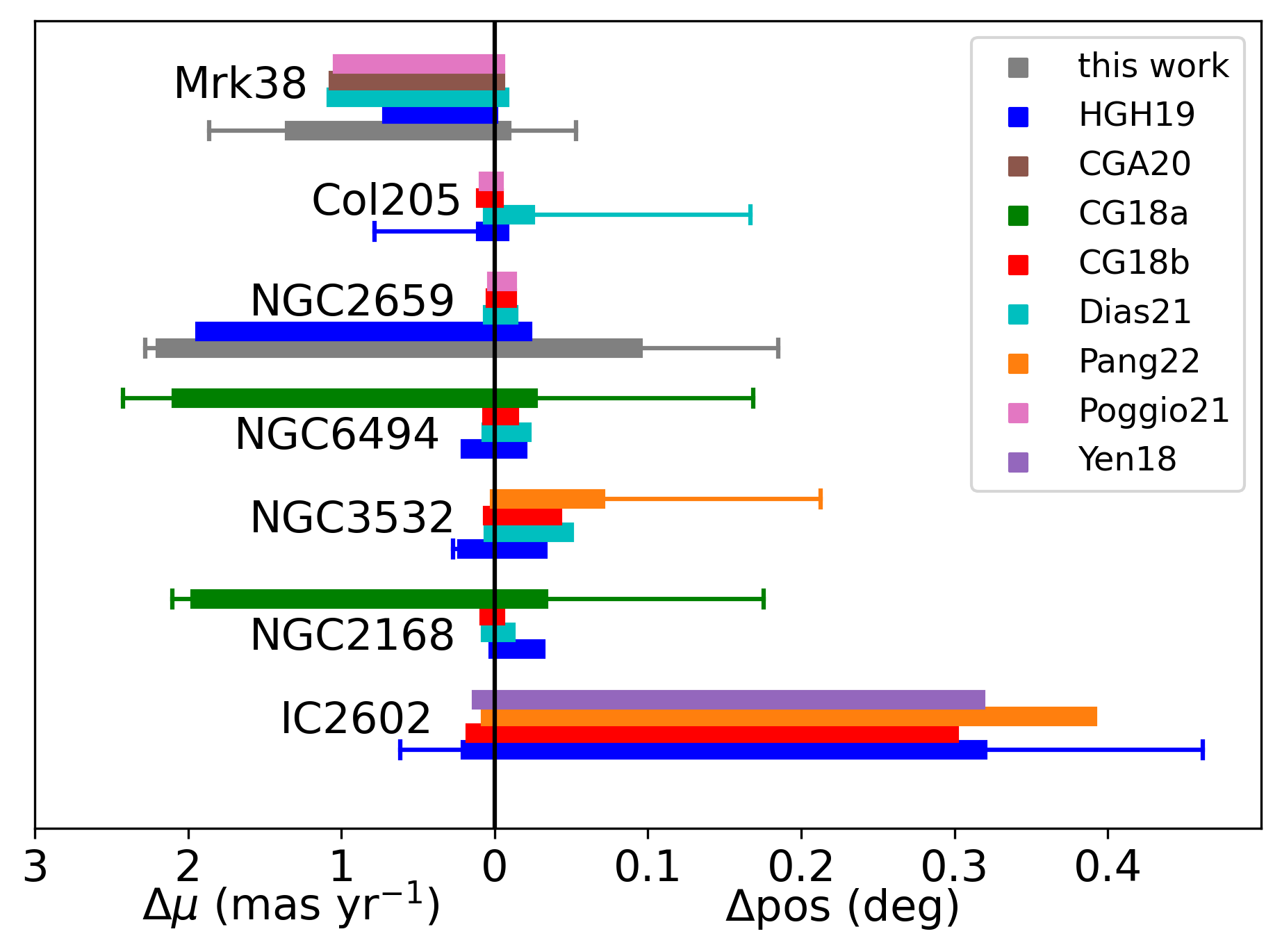}
\includegraphics[width=\columnwidth]{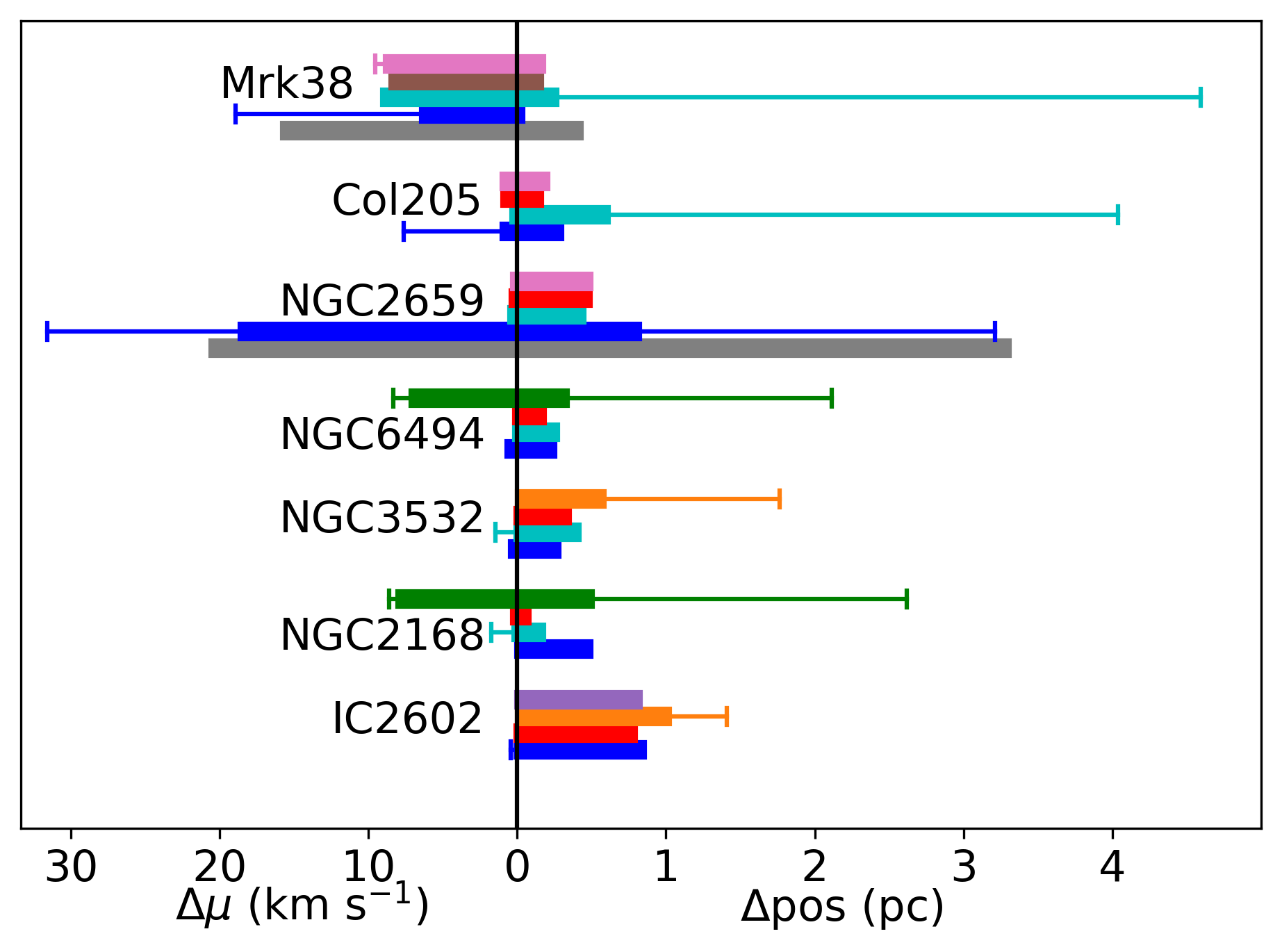}
\caption{The spatial distribution and velocities estimated in this work compared with the literature. The uncertainties are represented by error bars corresponding to the largest error or offset found for each cluster.
{\it Left:} The coloured bars
 indicate different works (see Table \ref{tab:data}) and 
the offsets of their respective values on the position ($\Delta {\rm pos}$) and  tangential velocity ($\Delta \mu$), given in angular units.  In the case of double clusters, the grey bars represent the offset of results for the secondary group (Mrk38b and NGC~2659b), respectively, to the main clusters (Mrk38a and NGC~2659a), which were adopted as referential. {\it Right:}  The same as left panel, with offsets given in linear units.}
    \label{fig:offset}
\end{figure*}
%%-----------------end Fig. 9

%%%--------Table 5
\begin{table}
 \caption{Comparison of our results with the literature.}
\label{tab:tab5}
\begin{tabular}{lcccc}
\hline
\hline
Cluster	&	$\Delta {\rm pos}$ &$\Delta {\rm pos}$ &	$\Delta \mu$ &	  $\Delta \mu$ 	\\		        
                 &      (deg)                 &      (pc)                   &  mas yr$^{-1}$  & km s$^{-1}$ \\
\hline
Col205	&	0.009$^c$	          &	0.17$^e$             &	0.06$^d$	&	0.43$^d$	\\
                 &      0.025$^d$          &    0.62$^d$	           &	0.11$^e$   &      1.1$^{cef}$  \\
 \hline
IC2602	&	0.302$^e$ 	&	0.80$^e$	            & 0.08$^g$	&	0.06$^g$ \\	
                 &      0.392$^g$         &    	1.03$^g$              &  0.21$^c$    &      0.15$^c$	            \\
 \hline
Mrk38a	&	0.001$^c$	         &  0.05$^c$	   &	0.72$^c$	      &	 6.5$^c$\\	
                &      0.010$^b$          &   0.51$^b$       &    1.36$^b$        &    18.3$^b$         \\
 \hline	
NGC2168	&	0.006$^e$		&	0.09$^e$	&	0.03$^c$ &	0.12$^c$	\\
                &      0.034$^i$         &     	0.51$^i$    &     1.98$^i$      &     8.0$^i$       \\
 \hline	
NGC2659a&	0.014$^{ef}$ 	&	0.45$^d$.  &	0.04$^f$	&	0.39$^f$	\\
                &     0.095$^b$          &       3.31$^b$	 &      2.20$^b$  &     20.7$^b$    \\
 \hline
NGC3532	&	0.034$^c$       &	0.29$^c$	&	0.02$^g$	&	0.05$^g$	\\
                &       0.071$^g$       &       0.59$^g$    &   0.24$^c$    &    0.54$^c$           \\
 \hline	
NGC6494	&	0.015$^e$	 &	0.19$^e$	&	0.07$^e$	&	0.24$^{de}$ 	\\
                &       0.027$^i$	  &   0.34$^i$     &     2.09$^i$     &  7.2$^i$      \\
\hline	
\hline
\end{tabular} \\
\tiny{Notes: For each cluster, the minimum and maximum offsets are in the top and bottom lines, respectively. (b) Secondary cluster in this work; (c) HGH19; (d) \citet{Dias21}; (e)  \citet{CantatGaudin18b}; (f) \citet{Poggio21}; (g) \citet{Pang22}; 
(h) \citet{Yen18}; (i) \citet{CantatGaudin18a}.}
\end{table}
%%%------------------Tab.5 end	

The estimated ranges quantifying the differences and similarities between our results and those in the literature are shown in Table \ref{tab:tab5}. The table provides the minimum and maximum values of position and velocity offsets, along with their respective references.
Figure \ref{fig:offset} displays the offsets $\Delta {\rm pos}$ and  $\Delta \mu$  for the entire sample compared with the literature, identifying individual works by coloured bars.

First, we discuss the single clusters. Compared with our new results, the offsets are very low, excepting the comparison with \cite{CantatGaudin18a} data  (green bars) for NGC~2168 and NGC~6494, which are based on UCAC4\footnote{The fourth U.S. Naval Observatory CCD Astrograph Catalogue.}. A possible explanation for 
these higher velocity offsets  is the difference in the proper motion used from a different database,  which in some 
cases may present larger uncertainties when compared with the {\it Gaia} data. Another possible cause of discrepancies between our results and the literature is related to different estimates of the 
cluster distance. This problem is avoided in the  left panel of Fig. \ref{fig:offset}  that displays $\Delta {\rm pos}$ as a function 
of  $\Delta \mu$, both given in angular measurements.
Again, the points showing large offset in velocity are NGC~2168 and NGC~6494, due to the values adopted from UCAC4.

Proceeding with the same evaluation for double clusters, we found larger discrepancies than expected. For Mrk38, the separation is larger on 
proper motion. 
On the other hand, the position we found for Mrk38a is in good 
agreement with the literature, while Mrk38b shows  negligible $\Delta {\rm pos}$ in comparison with its pair, corresponding to minimal differences on the projected distribution.

Noticeable discrepancies are also found for NGC~2659 compared with HGH19, which results are roughly in between the values obtained for both groups in the present work, mainly for NGC2659b that has the 
largest distance from its pair.
The discrepancies of both components of NGC~2659 in comparison with
HGH19 indicate that our previous results were obtained from a mixing of members of both groups. It can be also noted that our present results for NGC~2659a are in better 
agreement with literature \citep{Dias21, CantatGaudin18b, Bossini19}.

\begin{figure}
\includegraphics[width=\columnwidth]{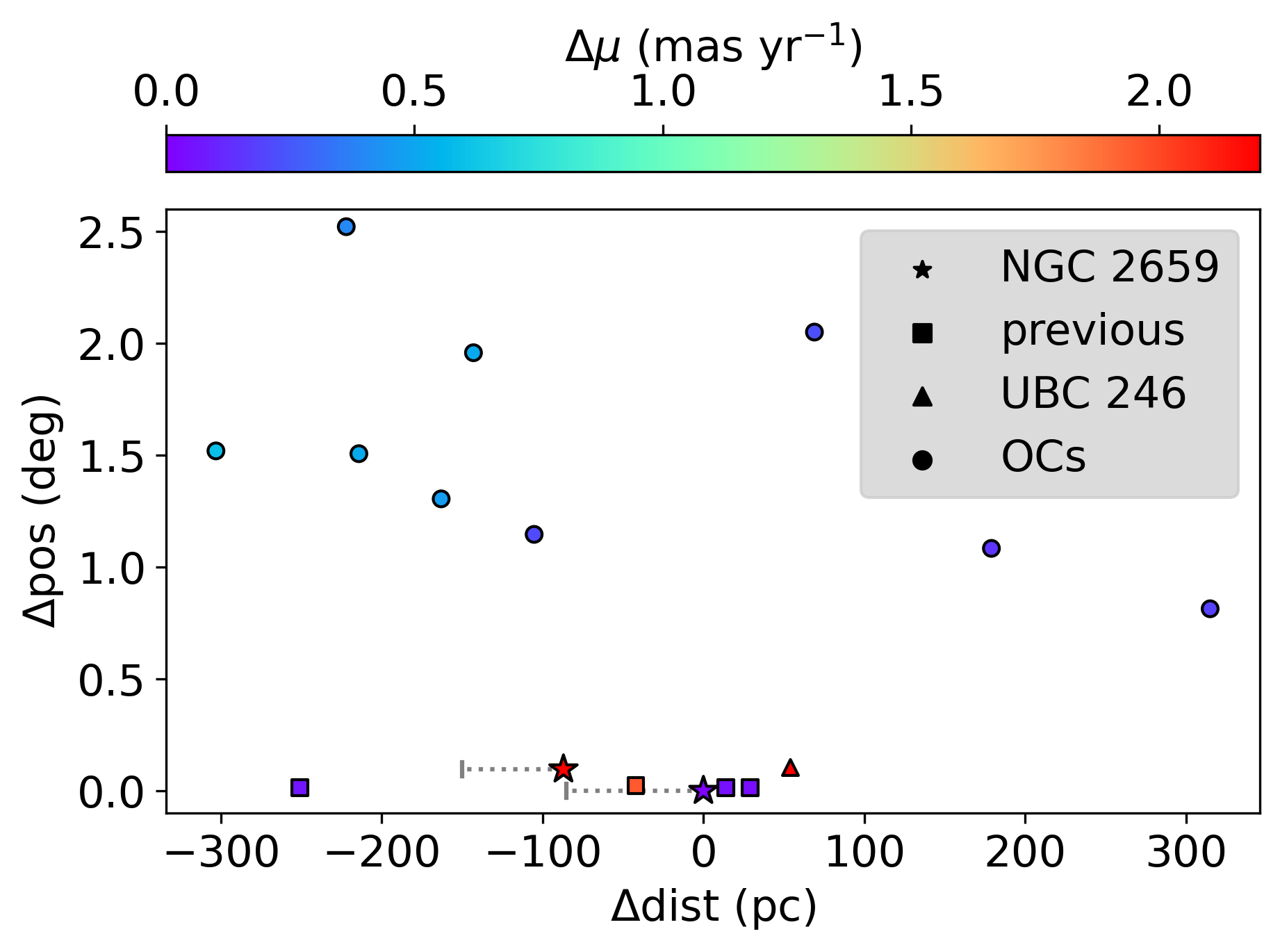}
\caption{The offsets in position and velocity for NGC~2659a (blue $\star$) and NGC~2659b (red $\star$ ), compared with previous results from literature, and other open clusters (OCs) suggested to be companions groups, such as UBC~246, for instance.  The grey dotted lines show the distance offset obtained for a larger sample that includes additional members (P $<$ 50 percent)}.
    \label{fig:offset2}
\end{figure}

We have checked the possible coincidence of NGC~2659b with other open clusters (OCs) previously suggested as part of the same group. Figure \ref{fig:offset2} compares results only for NGC~2659 and other OCs that could have similar properties. 
These OCs were reported in the literature as candidates to possible companions of NGC~2659, which could constitute a double or multiple cluster.
This plot shows the offsets in position as a function of differences in cluster distance ($\Delta {\rm dist}$) when compared with our present results.
In this case, offsets with positive values indicate more distant objects (background), while negative values are used for lower distances (foreground). 
The symbols have different colours representing the offset of tangential velocity. The blue symbols indicate similarity in velocity compared with NGC2659a, 
as it is noted for the results from the literature \citep[shown by blue squares][]{Dias21, CantatGaudin18b, Bossini19}, while red symbols correspond to velocities similar to the results for NGC2659b ($\Delta \mu > 2$ ~mas ~yr$^{-1}$, where $\star$ indicates this work, and the red square is from HGH19).

The OCs \citep[e.g.][]{Liu19,Casado21} are represented by blue circles indicating similarity in velocity compared with NGC2659a. However, three of them have larger distances ($\Delta {\rm dist}>$ 100 pc):   Gull5, UBC~482, and Cas61. Other five OCs have lower distances ($\Delta {\rm dist}<$ -70 pc): LP58; Pismis 8; SAI92; Rupr71;  and NGC2645. The last candidate, Cas28, 
has projected position larger than 1 degree in the sky, which is also observed for all the other clusters. Due to these large discrepancies, it is more conservative not considering they belong to the same group.  

%%-----------------new Fig. 10 (old 9 right) Off set
%%-----------------end new Fig. 10

Finally, UBC~246, also named Pismis~9, was suggested to form a pair with NGC~2659a  \citep{giorgi}, which has a projected position and tangential velocity similar to NGC2659b. However, its distance is larger at more than 150 pc \citep[e.g.][]{CantatGaudin20, Poggio21}. 
Even when considering the uncertainty in estimation, the extent of their separation remains too significant to align with the same subgroup.
It must be kept in mind that our selection of 96 members is more restrictive than other works that consider samples with at least three times more objects scattered around the mean spatial position, which naturally span in larger ranges of distances. To be more conclusive about the similarities and differences between NGC~2659b and UBC~246, it is necessary to include additional objects, candidates showing membership lower than 50 percent, to have the same basis of comparison with other works.  

 \cite{Poggio21} reported a  list of 270 members for UBC~246. To achieve a similar sample, for NGC~2659b, we selected the objects that have 
membership probability $P>0.1$ percent and  fractional parallax uncertainty $f <0.5$. 
The mean values (and respective standard deviation)\footnote{($\alpha,\delta$)=(130.712(0.006), -44.935(0.054)) deg, $\varpi$=0.509(0.052) mas, $\mu_{\alpha^{\star}}$= -4.379(0.101) mas yr$^{-1}$, 
$\mu_{\delta}$ = 3.015(0.134)~mas~yr$^{-1}$, d = 1916(218) pc.}  for this enlarged 
sample are in good agreement with the literature for UBC~246 \citep[mainly][]{Poggio21}, except for the slightly different distance: d = 2120 pc \citep{CantatGaudin20} or d = 2025 pc \citep{tarricq22}, but still within the uncertainties. This confirms that NGC~2659b is indeed the same cluster previously cataloged as UBC~246.

When comparing NGC~2659b with NGC~2659a (using a sample of 154 objects that have $P>10$ percent, $f<0.5$), the differences between 
the mean values for each parameter (presented above) remain comparable with those obtained for the samples restricted to $P_{50}$ members. The
new results show less than 0.5 percent for differences in position and proper motion, while the standard deviations, as expected, are almost 50 
percent larger than the previous results ($P_{50}$ members). The main difference, but not significant, occurs for the new values of distance mode, 
which change the offset 
from $\Delta {\rm dist} = 2066 -1979 = 87$ pc (measured for samples of $P_{50}$ members) to $\Delta {\rm dist} = 1980 - 1916 = 64$ pc.
An illustration of these differences is shown by dotted lines linked to the star points representing NGC~2659a,b in Fig. \ref{fig:offset2}.
These  results indicate that enlarging the sample (by a factor of $\sim$3 in the case of NGC2659b) does not give different parameters that were 
found for  the list of the most probable members. Despite the similarity in the projected distribution of both groups, NGC~2659b tends to be in a 
region  more than 0.05 deg to the NE away from NGC~2659a. There is a difference of more than 50 pc in distances, where NGC~2659b is in the 
foreground, and significant differences in proper motion and age confirm that this cluster is not a subgroup of NGC~2659a.

%%%% Section 6.2
\subsection{Structure}
\label{sec:structure}

The early evolution of star clusters and their original structure can be explored by comparing the fractal 
parameters obtained from the observed surface density distribution with theoretical simulations.  In a 
recent study of the NGC 2264 star-forming region, \cite{Parker22} used the $\mathcal{Q}$-parameter to 
quantify the spatial distribution of stars for two subclusters centered around the stars S~Mon and IRS~1/2.
Both groups have $\mathcal{Q} \sim 0.8$, meaning they have neither a substructured nor a centrally 
concentrated distribution. According to different age estimates in the literature, the star formation 
activity seems to have started first in the S Mon region ($\sim$ 5 Myr) and more recently ($\sim$2 Myr) 
for the group IRS 1/2 \citep[e.g.][]{schoettler22}. This is one example of studies using the$\mathcal{Q}$-parameter, comparing it with the mass segregation ratio and the local density, similar to the analysis performed in the present work.

\begin{figure}
\includegraphics[width=\columnwidth]{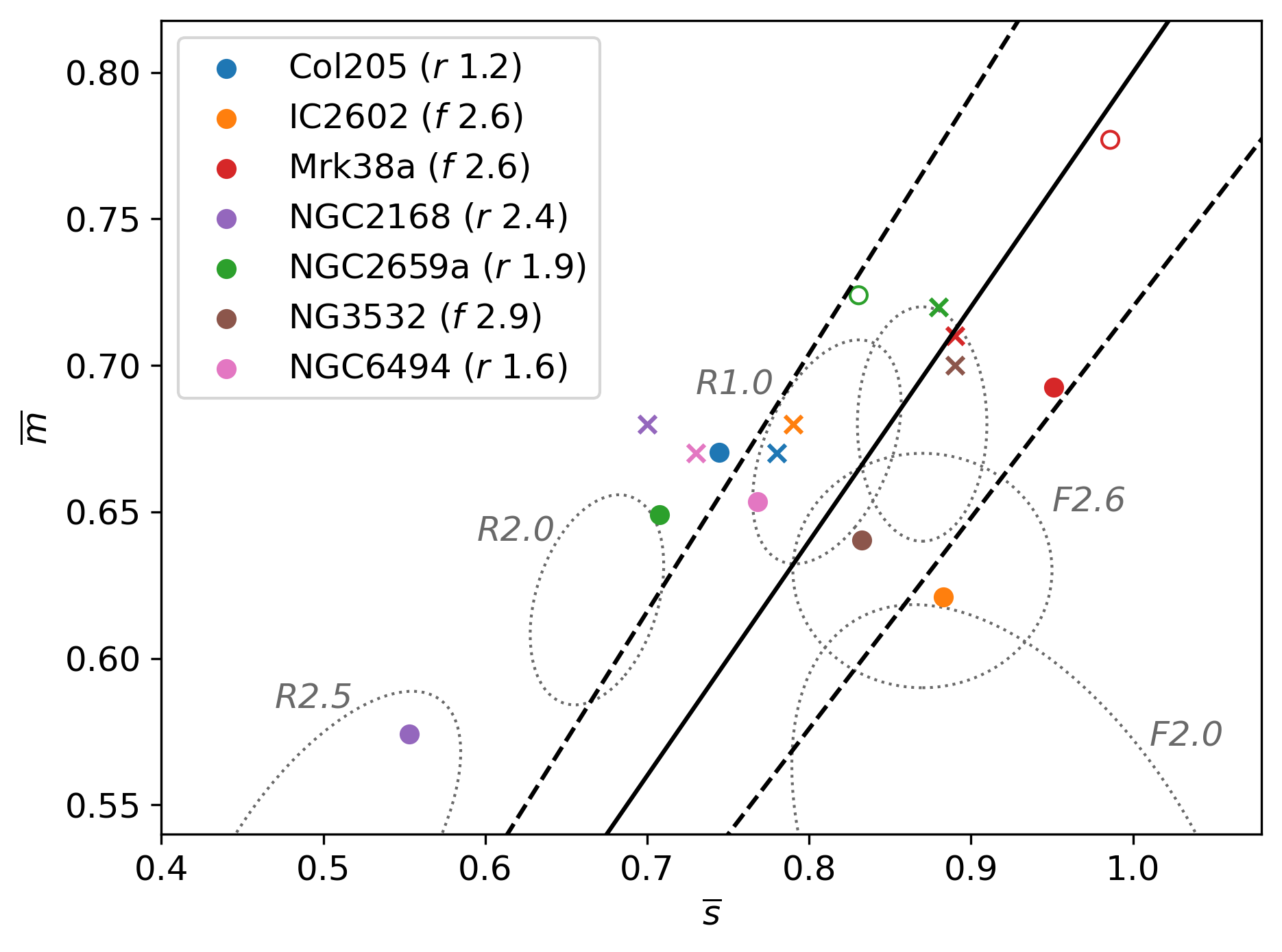}
\caption{ Plot of $\overline{m}$ and $\overline{s}$  comparing results from HGH19 (shown by $\times$) with those obtained here (circles, with open symbols in the case of Mrk~38b and NGC~2659b).
The full line indicates $\mathcal{Q}=0.8$, with 10 percent deviation  (dashed lines). 
The dotted ellipses show the simulations by \citet[]{Parker18} defining the locus of regions with radial distributions (from R0 to R2.5) or 
fractals (F2.0, F2.6, and F3.0). In both types of structure, the smooth distributions (R0 and F3.0) coincide in the same area (unlabeled ellipse).
Results from our estimation of fractal dimension ({\it f}) or slope of radial profile ({\it r}) are indicated in the figure's legend.}
\label{fig:ms}
\end{figure}

Here, we aim to investigate the variation in the fractal parameters if single or multiple structures are considered 
in the distribution of  our cluster sample. Figure \ref{fig:ms} shows the locus of different regions defined by numerical 
simulations from \citet{Parker18} using convex hull area normalization for synthetic star-forming regions that 
contain 300 points. We adapted ellipses covering the span of points from the results for each adopted 
geometry (fractal dimension or radial density profile) to display the loci of six parameters used in the simulations
\citep[see original distributions of points in Fig. A3(b) from][]{Parker18}. Overimposed on these ellipses,  we 
plot the values of $\overline{m}$  and $\overline{s}$  obtained in this work compared with previous results from HGH19.
Above the line of  $\mathcal{Q} = 0.8$ is found the loci of simulations with RDP described by $n \propto r^{-\alpha}$, 
where $\alpha$ = 0 indicates uniform density profile, 
while $\alpha$ = 2.9 corresponds to centrally concentrated distributions. On the other hand, the regions with 
 $\mathcal{Q} <$ 0.8 correspond to simulations  that adopt geometries varying from very substructured fractal 
 (fractal dimension $D = 1.6$) to smoother  distributions ($D = 3$).

 It can be noted in Fig. \ref{fig:ms} the improvement we obtained in the accuracy of the fractal parameters,  
 probably due to the larger number of members
 considered in this work for most of our clusters, compared with previous results (HGH19). 
 This means that the position of the clusters in the  $\overline{m} - \overline{s}$  plot coincides with well-defined 
 regions representing RDP distributions in the case of Col205, 
 NGC~6494 ($\alpha$ = 1), NGC~2659a ($\alpha$ = 2), and  NGC~2168 ($\alpha$ = 2.5). On the other hand,
 IC~2602 and NGC~3532  have $\mathcal{Q} < 0.8$, coinciding with the simulated fractal region F2.6. 
 As discussed in Sect. \ref{sec:Q}, the legend of Fig. \ref{fig:ms} also shows the results from calculating 
 the fractal box-dimension ($D_b$) or the slope ($\alpha$) of the RDP, respectively labeled by ``f" or ``r" aiming to compare with 
 the simulated regions directly.

\begin{figure*}
\includegraphics[width=\columnwidth]{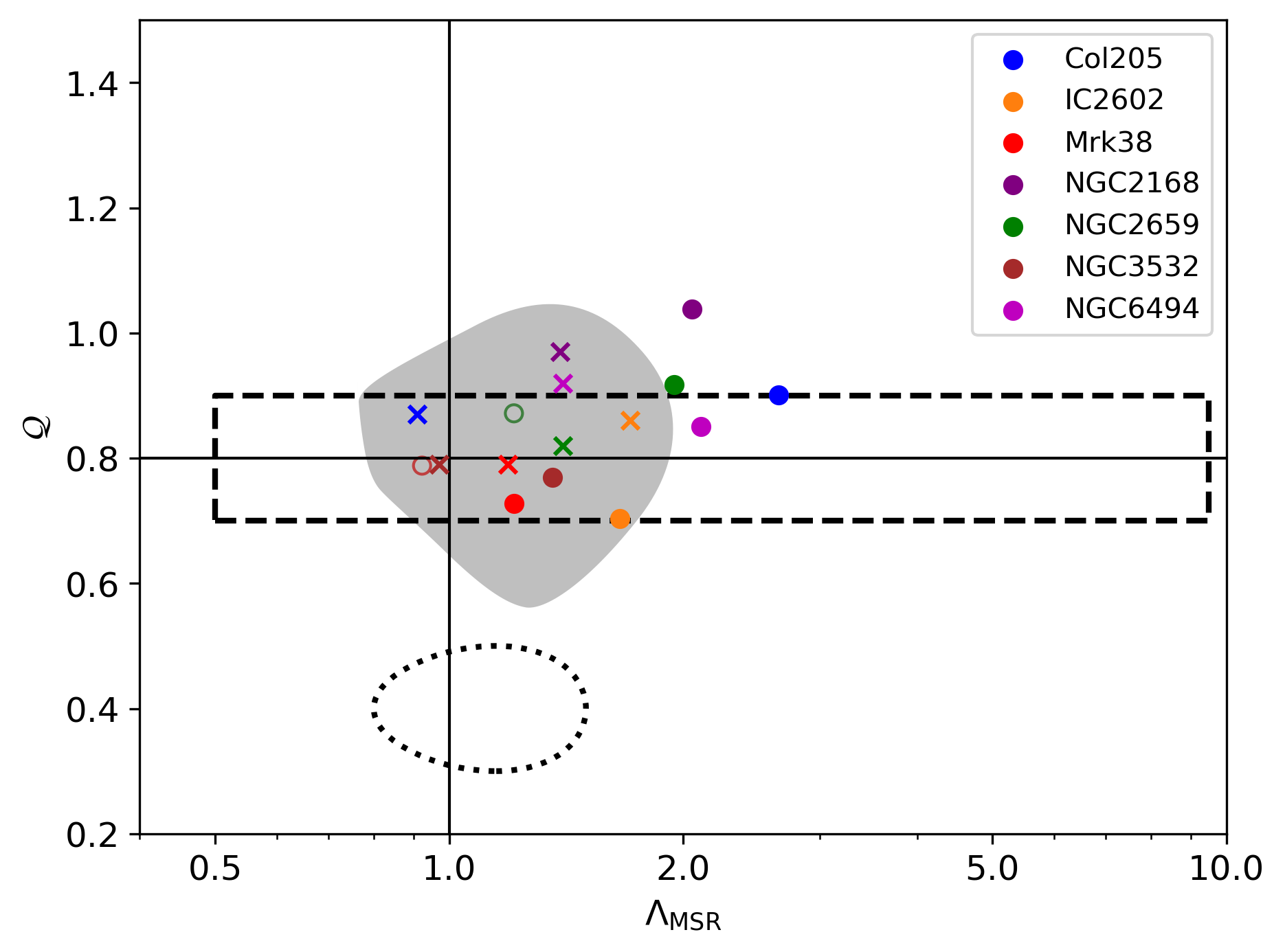}
\includegraphics[width=\columnwidth]{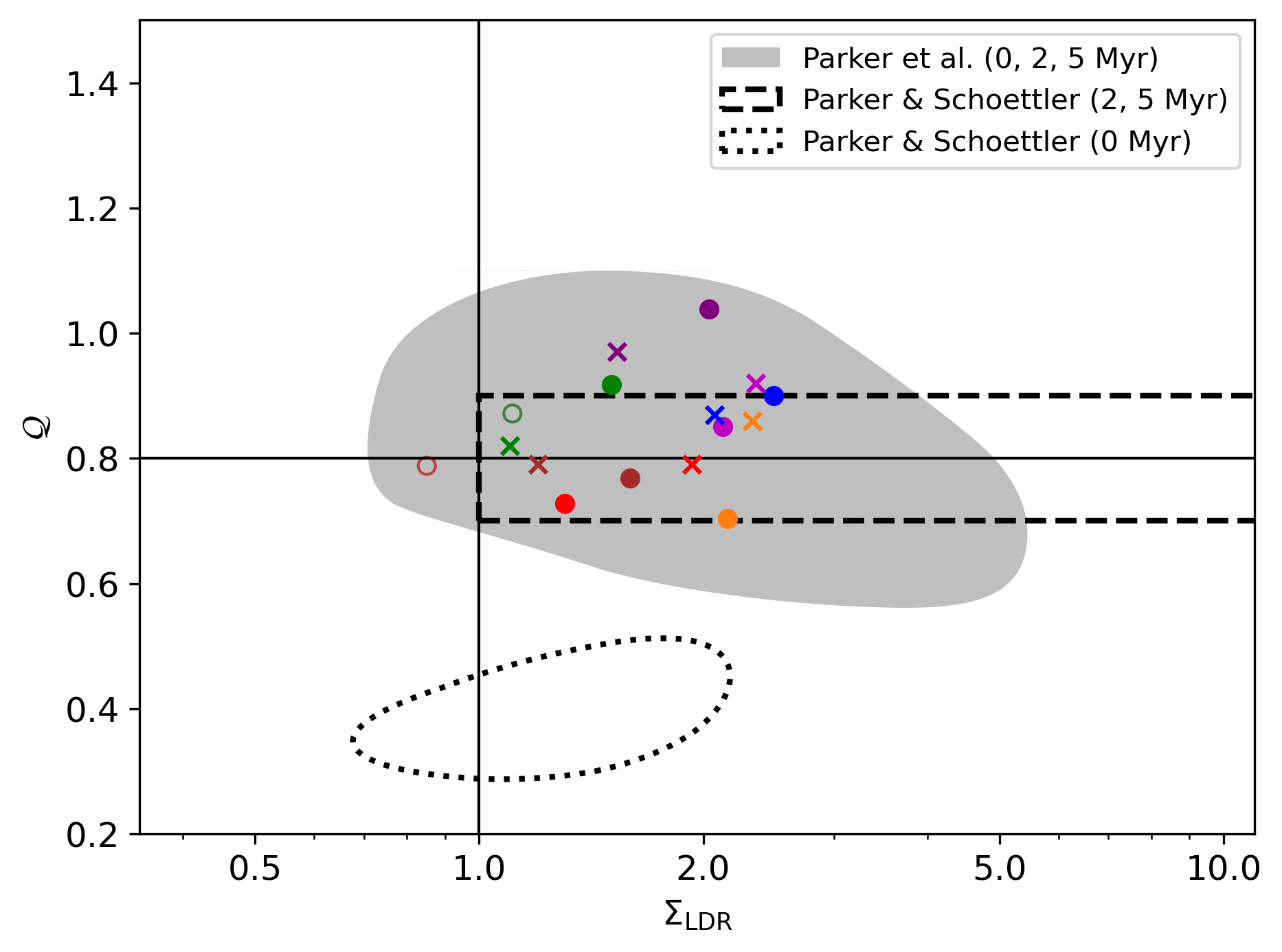}
\caption{Comparison of $\mathcal{Q}$ with mass segregation ratio  (left panel)  and local surface density ratio 
(right panel). The legends are the same for both panels. Results from HGH19 are shown 
by $\times$, and open 
circles indicate the secondary group of double clusters. 
The highlighted  areas represent smoothed distributions of 
points from N-body simulations of star-forming regions under different initial conditions. The hatched grey area 
shows simulations from {\citet{Parker14}}, and the areas delimited by dotted or dashed  lines roughly correspond to
the points distribution obtained by  {\citet{Parker22}}, which reaches up to $\Sigma_{\rm LDR} \sim$ 40 (not shown in the plot).}
\label{fig:QLambSig}
\end{figure*}

A last diagnosis of our analysis is obtained from the plots of the  $\mathcal{Q}-$parameter against the 
mass segregation ratio and the local density shown in Fig. \ref{fig:QLambSig}. In these plots, the 
areas corresponding to the N-body simulations from \citet{Parker22} are displayed, representing results 
for a highly substructured star-forming region under subvirial{\footnote{ $\alpha_{\rm vir}$ = 0.3, where $\alpha_{\rm vir}$ is the ratio of  potential energy to kinetic energy.}} initial conditions. 
Previous results (HGH19) are also plotted for comparison, with principal differences noted for $\Lambda_{\rm MSR}$ tending to be larger in the present work.
$\Sigma_{\rm LDR}$ seems to show lower values, but for all the objects of our sample, both parameters 
remain in the 1.0 to 2.5 range. These values coincide with the interval expected by the simulations
corresponding to the early evolution of the clusters (age = 2 and 5 Myr) but
 do not reach the maximum values found by  \citet{Parker22} for these adopted initial conditions. No significant changes were noted in the 
$\mathcal{Q}$ range, since our sample coincides with the range expected by 
the simulations, except for Mrk38b, which shows inverse mass segregation.

%%%%%%%--------------------------------- Fig.10   m_barra x s_barra 
%%%% -----------------------------------------end Fig. 10

 Comparing $\mathcal{Q}$, $\Lambda_{\rm MSR}$ and $\Sigma_{\rm LDR}$ obtained from simulations for the 
 initial conditions adopted by \citet{Parker22}, it can be concluded that the dynamical evolution erases the initial 
 structure (see in Fig.  \ref{fig:QLambSig} the area corresponding to the distribution of points 
at 0 Myr). After 2 Myr, their levels of substructure decrease significantly, and their levels of mass 
 segregation increase. Although the ages of our sample are much older, their  values of  $\mathcal{Q}$ 
 and mass segregation suggest that most of our clusters would be consistent with those initial conditions. 
 However, as discussed in HGH19, they would also be consistent with other conditions. For instance, we also 
 compare our results with simulations of synthetic regions from \citet{Parker14} 
 using $\alpha_{\rm vir}$ = 1.5 and $D = 3$. 
 Fig.  \ref{fig:QLambSig}   shows a shaded area corresponding to the smoothed span of 
 the points resulting from these simulations at three ages (0, 2 and 5~Myr). In this case, the choice of initial 
 conditions is related to the simulated distribution of points coinciding with the same range of 
 parameters found for our clusters.

%%%%--------------------------------- Fig. 11 QxLambda, QxSigma
%%%%------------end Fig. 11

%-----------------------------------Sect. 7
\section{Discussion}
\label{sec:conclus}
 
We selected seven OCs from a sample previously studied by us to investigate the possible presence of double- or multiple structures. The choice of the objects was based on the large dispersion of their ages, which were estimated by using near-infrared photometry (HGH19). In the present work, the characterization of the cluster members was revisited in the light of {\it Gaia} DR3 astrometric and photometric data that allowed a significant increase in the list of studied stars, corresponding to a more complete population when compared with our previous results. Near- to mid-infrared data and visual extinction maps were used to inspect the circumstellar and interstellar environments that could be related to dust emission from a protoplanetary disc or the presence of a surrounding molecular cloud. However, our sample does not show large amounts of dust emission, considering that only a few members of the younger clusters seem to be disc-bearing stars, and the $A_V$ levels in the direction of the clusters are lower than 1 mag for most of them.

We considered members of the clusters the stars with a membership larger than 50 percent ($P_{50}$), estimated by fitting a Probability Distribution Function to the observed proper motion. Only members presenting low fractional uncertainty of parallax ($f < 0.3$) were used to estimate the mean values of the astrometric parameters, as well as in the calculation of the most probable distance of the cluster that is determined from the distribution of the distance mode of the members.
In total, we analysed 9 stellar groups, five of which are single clusters: Col205, IC~2602, NGC~2168, NGC~3532, and NGC~6494, which do not show any subgroup in their spatial distribution neither in proper motion. The other four groups are related to two candidates that possibly are pairs of clusters Mrk~38a,b and NGC~2659a,b, whose components are mainly distinguished when comparing the proper motion and the distance of the subgroups, while their position overlaps in the same projected region.

The colour-magnitude diagram constructed with the {\it Gaia} photometry was corrected from reddening using the extinction automatically 
estimated by the code Aeneas of the GSP-Phot package. For the {\it Gaia} sources lacking this correction, we have adopted a mean value 
of E(B-V) 
that agrees with the $A_V$ maps and provides a good fit for the isochrone corresponding to the cluster age. Individual stellar masses, determined from interpolation between evolutive tracks, 
were compared with the effective temperature, confirming that the mass estimates are reliable mass. 
A comparison of the astrometric results with the literature was performed by measuring offsets in position ($\Delta {\rm pos}$) and proper motion calculated in the form of tangential velocity
 ($\Delta {\mu}$), where our present results were adopted as references (that means zero points). 
We found low values for these offsets in the case of the single clusters that show $\Delta {\rm pos} < $ 1 pc and $\Delta {\mu} < 1 $ km s$^{-1}$ suggesting an excellent agreement with the literature.

The surface stellar distribution of subgroup Mrk38b entirely coincides with the main cluster Mrk~38a ($\Delta {\rm pos} \sim$ 0.1 deg), but the offset on proper motion corresponds to more than 10 km s$ ^{-1}$. This subgroup is about 500 pc more distant than the distance we found for the main cluster. These differences suggest Mrk38b is a distinguished stellar group, which is older (250 Myr) and more scarce ($n = 2.4$ stars pc$^{-2}$) than  Mrk38a (age= 30 Myr, $n = 14$ pc$^{-2}$) that is the main group previously known as a single cluster.
Due to unusual nature of Mrk38b, as a second cluster in the direct line of sight, we examined the possibility that it is instead an asterism caused by a combination of the Gaia missions limited ability to detect (fainter) stars more distant than Mrk38b and/or large parallax uncertainties of identified members. We performed a test of eliminating sources with a parallax uncertainty f$>$0.2 and re-calculated the membership probabilities. In this case, the number of sources is insufficient for the algorithm to discern two groups. Only one group is found, showing an elongated distribution in the considered parameter space. However, through visual inspection, it is still possible to identify a few members coinciding with the ranges of proper motion and parallax found for Mrk38b.

The largest offsets are found for NGC~2659b, separated from the main cluster by 3.3 pc and  tangential velocity offset $>$ 20 km s$^ {-1}$.  In Fig. \ref{fig:offset} (right panel), we compared the main cluster NGC~2659a with other OCs from the literature that were suggested to be part of the same group due to their similar proper motion. However, all of these OCs have shown large differences in distance ($>$ 100 pc) or projected position ($>$ 1 degree), which are too distant from each other to be considered components of the same stellar group. NGC2659b shows differences that indicate it is not a substructure of NGC~2659a. NGC~2659b   has projected position and proper motion coinciding with the cluster UBC~246, but the distance we estimated is lower than the results from the literature. 

The calculation of the structural parameters was based on the surface density distribution by constructing the MST that is used to estimate the size of the cluster and fractal parameters, which were compared with N-body simulations from the literature. For instance, the 
$\overline{m} - \overline{s}$ plot (see Fig. \ref{fig:ms}) indicates that clusters with 
$\mathcal{Q} >$ 0.8 (corresponding to radial distribution): Col~205, NGC~2168, NGC~2659a, and NGC~6494 occupy the areas coinciding with simulations that define the locus of centrally concentrated regions (R = 1 to 2.5). NGC~2659b also has $\mathcal{Q} >$ 0.8, but its 
$\overline{m}$ and  $\overline{s}$ parameters do not coincide with the loci predicted by the simulations, just appearing near the region of smooth uniform distribution (R=0).
Only  IC~2602 and NGC~3532 do not show radial distribution, but low levels of substructures are found since they coincide with simulated fractal regions (F=2.6). Mrk38a and Mrk38b appear out of the simulated regions, but Mrk38a ($\mathcal{Q} <$ 0.8) seems to be tending to the region of smooth fractal distribution (F=3), while Mrk38b ($\mathcal{Q} \sim$ 0.8) has undefined distribution.
These results are consistent with those calculated in Sect. \ref{sec:Q}.

The total mass of the clusters was used to estimate the crossing time and the dynamical age ($\Pi$) that indicates if the cluster is gravitationally bound. Half of our clusters are considered unbound since they have $\Pi < 1$. 
A possible exception is Mrk38a, which has $\Pi \sim 1$ and could be a bound young cluster whose age is near its crossing time.
Larger values of dynamical ages were found for the oldest clusters of our sample: Mrk38b, NGC2659b, NGC~3532, and NGC~6494. 
There is a $\Lambda_{\rm MSR} >$ 1 trend for most of the clusters, except for the subgroup Mrk38b. However, the mass segregation is only evident for half of the sample, which has 
$\Lambda_{\rm MSR} \gtrsim$ 2. The other clusters are consistent with $\Lambda_{\rm MSR} \sim$ 1 (within the errors).
On the other hand, based on the values of  $\Sigma_{\rm LDR} \sim 1$ we conclude that massive stars tend not to be concentrated in regions of high 
surface density for most of our clusters, excepting IC~2602 and NGC~6530 that show mass segregation signature 
($\Sigma_{\rm LDR} > $ 2 and  p-value $<$ 0.1).

The diagram comparing the mass segregation parameters as a function of $\mathcal{Q}$ also displays the results from N-body simulations considering different initial conditions. These simulations indicate the data points' distribution during the artificial clusters' early evolution. The plots in Fig. \ref{fig:QLambSig} show that most of our clusters present the same distribution found for simulations at 2 and 5 Myr. Only Col205, NGC~2168, and NGC2659a appear out of the areas predicted by the simulations. 
Their high values of $\mathcal{Q}$ indicate central concentration that is confirmed by their position in the 
$\overline{m}  - \overline{s}$  plot tending to RDP distributions (R2 to R2.5 regions in Fig. \ref{fig:ms}). The comparison with the simulations (highlighted areas in Fig. \ref{fig:QLambSig}) indicates that most of our clusters tend to smooth distribution ($\mathcal{Q} \sim 0.8$) or a radial concentration   ($\mathcal{Q} > 0.8$) as they are evolving. One possibility is that these clusters could have lost their original substructures if they had been formed in highly substructured regions. Or, on the contrary, they retained their original geometry that possibly had low levels of substructures.

\section*{Acknowledgements}
We warmly thank Professor Vera Jatenco-Pereira for the valuable suggestions on the text of this manuscript.
We thank the anonymous referee for the constructive comments and suggestions.
We acknowledge support from FAPESP (2020/15245-2; 2023/08726-2).
This work has made use of data from the European Space Agency (ESA) mission
{\it Gaia} (\url{https://www.cosmos.esa.int/gaia}), processed by the {\it Gaia}
Data Processing and Analysis Consortium (DPAC,
\url{https://www.cosmos.esa.int/web/gaia/dpac/consortium}). Funding for the DPAC
has been provided by national institutions, in particular the institutions
participating in the {\it Gaia} Multilateral Agreement.
This research has made use of the SIMBAD astronomical database \citep[][]{Simbad}, and
the VizieR catalogue access tool \citep[][DOI : 10.26093/cds/vizier]{Vizier}
operated at CDS, Strasbourg, France.
This publication makes use data products from the Two Micron All Sky Survey, which is a joint project of the University of Massachusetts and the Infrared Processing and Analysis Center/California Institute of Technology, funded by the National Aeronautics and Space Administration and the National Science Foundation.
This publication makes use of data products from the Wide-field Infrared Survey Explorer, which is a joint project of the University of California, Los Angeles, and the
Jet Propulsion Laboratory/California Institute of Technology, funded by the National Aeronautics and Space Administration.

\section*{Data Availability}

The data underlying this article are available in machine-readable form at the CDS (VizieR On-line Data Catalogue),
which correspond to the tables containing individual members' parameters and results. 

%Figures presented in the Appendices are available in the online supplementary material. 



%\bibliographystyle{mnras}

\begin{thebibliography}{99}

\bibitem[\protect\citeauthoryear{Allison et al.}{2009}]{Allison09}
Allison R. J., Goodwin S. P., Parker R. J. et al.,
%Portegies Zwart, S. F., de Grijs, R., Kouwenhoven, M. B. N., 
2009, \mnras, 395, 1449

\bibitem[\protect\citeauthoryear{Arnold et al.}{2022}]{becky}
Arnold B., Wright N. J., Parker R. J., 2022, \mnras, 515, 2266

\bibitem[\protect\citeauthoryear{Bailer-Jones}{2015}]{Bailer15}
Bailer-Jones C. A. L., 2015, PASP, 127, 994

\bibitem[\protect\citeauthoryear{Blaylock-Squibbs et al.}{2022}]{blaylock}
Blaylock-Squibbs G. A., Parker R. J., Buckner A. S. M., Güdel M., 2022, \mnras, 510, 2864

\bibitem[\protect\citeauthoryear{Bica et al.}{2019}]{Bica}
Bica E., Pavani D. B., Bonatto C., Lima E. F.,  2019, \aj, 157, 12

\bibitem[\protect\citeauthoryear{Bonatto \& Bica}{2009}]{Bonatto09}
Bonatto C., \& Bica E., 2009, \mnras, 397, 1915

\bibitem[\protect\citeauthoryear{Bossini et al.}{2019}]{Bossini19}
Bossini D., Vallenari A., Bragaglia A. et al., 2019, \aap, 623, A108

\bibitem[\protect\citeauthoryear{Bragaglia et al.}{2022}]{Bragaglia22}
Bragaglia A.,  Alfaro E. J.,  Flaccomio E. et al., 2022, \aap, 659, A200

\bibitem[\protect\citeauthoryear{Bravi et al.}{2018}]{Bravi18}
Bravi L., Zari E., Sacco G. G. et al., 2018, \aap, 615, A37.

\bibitem[\protect\citeauthoryear{Bressan et al.}{2012}]{Bressan12}
Bressan A., Marigo P., Girardi L., Salasnich B., Dal Cero C., Rubele S., Nanni A., 2012, \mnras, 427, 127

\bibitem[\protect\citeauthoryear{Buckner et al.}{2019}]{Buckner19}
Buckner A. S. M.  et al.,  2019, \aap, 622, A184

\bibitem[\protect\citeauthoryear{Buckner et al.}{2020}]{Buckner20}
Buckner A. S. M.  et al.,  2020, \aap, 636, A80

\bibitem[\protect\citeauthoryear{Buckner et al.}{2022b}]{Buckner22b}
Buckner A. S. M., Khorrami, Z., González M., Lumsden S. L., Clark P., Moraux E., 2022b, \aap, 659, A72

\bibitem[\protect\citeauthoryear{Buckner et al.}{2024}]{Buckner24}
Buckner A. S. M., Naylor T., Dobbs C. L., Rieder, S., Bending T. J. R., 2024, \mnras, 527, 5448

\bibitem[\protect\citeauthoryear{Canavesi \& Hurtado}{2020}]{Canavesi}
Canavesi, T., \& Hurtado, S., 2020, BAAA Vol. 61B, 
%eds V\'asques,A.M., Benaglia P., Iglesias F.A., \& Sgr\'o M.A.  

\bibitem[\protect\citeauthoryear{Cantat-Gaudin \& Anders}{2020}]{CantatAnders20}
Cantat-Gaudin T., Anders F., 2020, \aap, 633, A99

\bibitem[\protect\citeauthoryear{Cantat-Gaudin et al.}{2018a}]{CantatGaudin18a}
Cantat-Gaudin T., Vallenari A., Sordo R. et al., 2018a, \aap, 615, A49

\bibitem[\protect\citeauthoryear{Cantat-Gaudin et al.}{2018b}]{CantatGaudin18b}
Cantat-Gaudin T., Jordi C., Vallenari A. et al., 2018b, \aap, 618, A93

\bibitem[\protect\citeauthoryear{Cantat-Gaudin et al.}{2020}]{CantatGaudin20}
Cantat-Gaudin T., Anders F.,  Castro-Ginardi A. et al., 2020, \aap, 640, A1

\bibitem[\protect\citeauthoryear{Cardelli et al.}{1989}]{Cardelli89}
Cardelli J. A., Clayton G. C., Mathis J. S., 1989, \apj, 345, 245

\bibitem[\protect\citeauthoryear{Cartwright \& Whitworth}{2004}]{Cart04}
Cartwright A., Whitworth A.~P., 2004, \mnras, 348, 589

\bibitem[\protect\citeauthoryear{Casado}{2021}]{Casado21}
Casado J., 2021, Astronomy Reports, 65, 755

\bibitem[\protect\citeauthoryear{Casagrande \& VandenBerg}{2018}]{Casagrande}
Casagrande L., VandenBerg D. A., 2018, \mnras, 479, L102

\bibitem[\protect\citeauthoryear{Cordoni et al.}{2023}]{Cordoni}
Cordoni G., Milone A. P., Marino A. F. et al., 2023, \aap, 672, A29

\bibitem[\protect\citeauthoryear{Cutri et al.}{2003}]{Cutri03}
Cutri R.M. , Wright E.L., van Dyk, S. et al., 2003,  
VizieR On-line Data Catalog II/246:  2MASS All-Sky Catalog of Point Sources

\bibitem[\protect\citeauthoryear{Cutri et al.}{2013}]{Cutri13}
Cutri, R. M., Wright, E L., Conrow, T. et al., 2013, 
VizieR Online Data Catalog II/328: AllWISE Data Release, 
originally published in: IPAC/Caltech.

\bibitem[\protect\citeauthoryear{Daffern-Powell \& Parker}{2020}]{Daffern}
Daffern-Powell E.C.,  Parker R. J., 2020, \mnras, 493, 4925

\bibitem[\protect\citeauthoryear{Davidge}{2017}]{davi17}
Davidge T. J., 2017, \apj, 837, 178

\bibitem[\protect\citeauthoryear{Delgado et al.}{2013}]{delgado13}
Delgado A. J.,  Djupvik A. A.,  Costado M. T., Alfaro E. J., 2013, \mnras,  435, 429

\bibitem[\protect\citeauthoryear{Dias et al.}{2014}]{Dias14}
Dias W.S., Monteiro H., Caetano T.~C., L\'{e}pine J.~R.~D., Assafin M., Oliveira, A.~F., 2014, \aap, 564, 79

\bibitem[\protect\citeauthoryear{Dias et al.}{2021}]{Dias21}
Dias W. S., Monteiro H., Moitinh, A., L\'epine J.R.D., Carraro G., Pauzen E., Alessi B., Villela L., 2021, \mnras, 504, 356

\bibitem[\protect\citeauthoryear{Dias et al.}{2020}]{Bruno}
Dias B. et al., 2020, in Bragaglia A., Davies M., Sills A., Vesperini E.,
eds, Proceedings of the International Astronomical Union , Volume
14 , Symposium S351: Star Clusters: From the Milky Way to the Early
Universe , Vol. 14. p. 89–92, doi:10.1017/S174392131900694X

\bibitem[\protect\citeauthoryear{Dib et al.}{2018}]{Dib18}
Dib S., Schmeja S., Parker R., 2018, \mnras, 473, 849

\bibitem[\protect\citeauthoryear{Dobashi et al.}{2005}]{Dobashi}
Dobashi K., Uehara H., Kandori R. et al., 2005, \pasj, 57, S1

\bibitem[\protect\citeauthoryear{Dobashi}{2011}]{Dobashi11}
Dobashi K., 2011, \pasj, 63, S1-S362

\bibitem[\protect\citeauthoryear{Dobbie et al.}{2009}]{Dobbie09}
Dobbie P. D., Napiwotzki R., Burleigh M. R. et al., 2009, \mnras, 395, 2248

\bibitem[\protect\citeauthoryear{Donada et al.}{2023}]{donada}
Donada J., Anders F., Jordi C. et al., 2023, \aap, 675, A89

\bibitem[\protect\citeauthoryear{Efron}{1979}]{efron}
Efron B., 1979, Bootstrap Methods: Another Look At the Jackknife, The Annals of Statistics 7, 1-26

\bibitem[\protect\citeauthoryear{Elmgreen}{2008}]{elm08}
Elmegreen B. G., 2008, \aap, 672, 1006

\bibitem[\protect\citeauthoryear{Elmgreen}{2018}]{elm18}
Elmegreen B. G., 2018, \aap, 853, 88

\bibitem[\protect\citeauthoryear{Fedele et al.}{2010}]{Fedele}
Fedele D., van den Ancker M. E., Henning Th. et al., 2010, \aap, 510, A72

\bibitem[\protect\citeauthoryear{Feder}{2013}]{Feder}
Feder J., 2013, Fractals, Springer Science \& Business Media

\bibitem[\protect\citeauthoryear{Fernandes et al.}{2019}]{Bia}
Fernandes B., Montmerle T.,  Santos-Silva T.,  Gregorio-Hetem J., 2019, \aap, 628, A44

\bibitem[\protect\citeauthoryear{Fűrész et al.}{2006}]{furesz}
Fűrész G., Hartmann L. W., Szentgyorgyi A. H. et al., 2006, \apj, 648, 1090

\bibitem[\protect\citeauthoryear{Gaia Collaboration et al.}{2018b}]{Gaia18b}
Gaia Collaboration, Babusiaux C., van Leeuwen F., Barstow,M.~A., et al., 2018b, \aap 616, A10

\bibitem[\protect\citeauthoryear{Gaia Collaboration et al.}{2023}]{Gaia23}
Gaia Collaboration, Vallenari A., Brown, A.~G.~A., Prusti T. et al., 2023, \aap, 674, A1

\bibitem[\protect\citeauthoryear{Gieles \& Portegies Zwart}{2011}]{Gieles}
Gieles M., Portegies Zwart S. F., 2011, \mnras, 410, L6

\bibitem[\protect\citeauthoryear{Giorgi et al.}{2023}]{giorgi}
Giorgi E. E., Pera M. S., Perren G. I. et al., 2023, Boletín de la Asociación Argentina de Astronomía, 64, 90
%Giorgi, E. E., Pera, M. S., Perren, G. I., Vazquez, R. A., Cruzado, A., 2023, 

\bibitem[\protect\citeauthoryear{González \& Alfaro}{2017}]{gonz17}
González M., Alfaro E. J., 2017, \mnras, 465, 1889 

\bibitem[\protect\citeauthoryear{González et al.}{2021}]{gonz21}
González M., Joncour I.; Buckner A. S. M. et al., 2021, \aap, 647, A14

\bibitem[\protect\citeauthoryear{Gouliermis et al.}{2014}]{gouliermis14}
Gouliermis D. A., Hony S., Klessen R. S., 2014, \mnras, 439, 3775

\bibitem[\protect\citeauthoryear{Gower \& Ross}{1969}]{gower}
Gower J.C., Ross G.J.S., 1969, Appl.~Stat., 18, 54

\bibitem[\protect\citeauthoryear{Gregorio-Hetem}{2008}]{GH08}
Gregorio-Hetem J., 2008, The Canis Major Star Forming Region,
Handbook of Star Forming Regions, Volume II: The Southern Sky ASP Monograph Publications, 
Vol. 5. Edited by Bo Reipurth, p.1, ISBN: 978-58381-671-4

\bibitem[\protect\citeauthoryear{Gregorio-Hetem et al.}{2021}]{GH21}
Gregorio-Hetem J., Lefloch B., Hetem A. et al., 2021, \aap, 654, A150

\bibitem[\protect\citeauthoryear{Hernández et al.}{2008}]{Hernandez08}
Hernández J., Hartmann L., Calvet N. et al., 2008, \apj, 686, 1195

\bibitem[\protect\citeauthoryear{Hetem \& Gregorio-Hetem}{2019}]{HGH19}
Hetem A., Gregorio-Hetem J., 2019, \mnras, 490, 2521

\bibitem[\protect\citeauthoryear{Hu et al.}{2021a}]{Huetal21}
Hu Q., Zhang Y., Esamdin A., Liu J.,  Zheng X., 2021a, \apj, 912, 5

\bibitem[\protect\citeauthoryear{Hu et al.}{2021b}]{Hu21b}
Hu Q., Zhang Y., Esamdin A., 2021b,  \aap, 656, A49

\bibitem[\protect\citeauthoryear{Jackson et al.}{2022}]{Jack22}
Jackson R. J., Jeffries R. D.,  Wright N. J. et al., 2022, \mnras, 509, 1664

\bibitem[\protect\citeauthoryear{Jadhav et al.}{2021}]{Jadhav21}
Jadhav,V. V., Roy K., Joshi N., Subramaniam A., 2021, \aj, 162, 264

\bibitem[\protect\citeauthoryear{Jaffa et al.}{2017}]{jaffa17}
Jaffa S.E., Whitworth A.P., Lomax O., 2017, \mnras, 466, 1082

\bibitem[\protect\citeauthoryear{Kharchenko et al.}{2005}]{Kharchenko05} 
Kharchenko N.V., Piskunov A. E., Roeser S. et al., 2005,  \aap, 438, 1163

\bibitem[\protect\citeauthoryear{Kim et al.}{2021}]{kim}
Kim S., Lim B., Bessell M. S. et al., 2021,  \aj, 162, 140

\bibitem[\protect\citeauthoryear{King}{1962}]{King}
King I., 1962, \aj, 67, 471

\bibitem[\protect\citeauthoryear{Koenig \& Leisawitz}{2014}]{K14} 
Koenig X.P., Leisawitz D.T., 2014, \apj, 791, 131

\bibitem[\protect\citeauthoryear{Kroupa}{2001}]{Kroupa} 
Kroupa P., 2001, \mnras, 322, 231

\bibitem[\protect\citeauthoryear{Kruskal}{1956}]{Kruskal}
Kruskal J.~B.~J., 1956, Proc. Amer. Math. Soc., 7, 48

\bibitem[\protect\citeauthoryear{Kuhn et al.}{2014}] {kuhn14} 
Kuhn M. A., Feigelson E. D., Getman K. V.,  et al., 2014, \apj, 787, 107

\bibitem[\protect\citeauthoryear{Kuhn \& Feigelson}{2017}]{kuhnfeig17} 
Kuhn M.A.,  Feigelson E. D., 2017, in the Handbook of Mixture Analysis, edited 
by S. Fr\"uwirth-Schnatter, G. Celeux, and C. P. Robert (Chapman \& Hall/CRC), ISBN 9780367732066 

\bibitem[\protect\citeauthoryear{Kuhn et al.}{2017}] {kuhn17} 
Kuhn M. A., Getman K. V., Feigelson E. D. et al., 2017, \aj, 154, 214

\bibitem[\protect\citeauthoryear{Kuhn et al.}{2019}] {kuhn19} 
Kuhn M. A., Hillenbrand L. A., Sills A., Feigelson E. D., Getman K. V., 2019, \apj, 870, 32

\bibitem[\protect\citeauthoryear{Lada \& Lada}{2003}]{LadaLada03}
Lada C.~J., Lada E.~A., 2003, \araa, 41, 57

\bibitem[\protect\citeauthoryear{Li et al.}{2020}]{Li20}
Li L., Shao Z., Li Z.-Z. et al., 2020, \apj, 901, 49

\bibitem[\protect\citeauthoryear{Liu \& Pang}{2019}]{Liu19}
Liu L., Pang X., 2019, \apjs, 245, 32 

\bibitem[\protect\citeauthoryear{Liu  et al.}{2021}]{Liu21}
Liu J., Fang M., Tian H. et al., 2021, \apjs, 254, 20

\bibitem[\protect\citeauthoryear{Luri et al.}{2018}]{Luri18}
Luri X., Brown A. G. A., Sarro L. M. et al., 2018, \aap, 616, A9

\bibitem[\protect\citeauthoryear{Maia et al.}{2019}]{Maia}
Maia F. F., Dias B., Santos Jr J. F. et al., 2019, \mnras, 484, 5702

\bibitem[\protect\citeauthoryear{Marigo et al.}{2017}]{Marigo17}
Marigo P., Girardi L., Bressan et al., 2017, \apj, 835, 77

\bibitem[\protect\citeauthoryear{Marsh et al.}{2015}]{Marsh15}
Marsh K. A., Whitworth A. P., Lomax O., 2015, \mnras, 454, 4292

\bibitem[\protect\citeauthoryear{Marsh et al.}{2017}]{Marsh17}
Marsh K. A., Whitworth A. P., Lomax O. et al., 2017, \mnras, 471, 2730

\bibitem[\protect\citeauthoryear{Maschberger \& Clarke}{2011}]{Masch}
Maschberger T., Clarke C. J., 2011, \mnras, 416, 541

\bibitem[\protect\citeauthoryear{Maurya et al.}{2020}]{maurya}
Maurya J., Joshi Y. C., Gour A. S., 2020, \mnras, 495, 2496

\bibitem[\protect\citeauthoryear{Minniti et al.}{2010}]{Minniti10}
Minniti D., Lucas P. W., Emerson J. P. et al., 2010, New Astron., 15, 433

\bibitem[\protect\citeauthoryear{Molinari et al.}{2010}]{Molinari10}
Molinari S., Swinyard B.,  Bally J. et al., 2010, PASP, 122, 314

\bibitem[\protect\citeauthoryear{Monteiro et al.}{2020}]{Monteiro20}
Monteiro H., Dias W. S., Moitinho A., Cantat-Gaudin T., L\'epine J. R. D.,
Carraro G., Paunzen E., 2020, \mnras, 499, 1874

\bibitem[\protect\citeauthoryear{Navarete et al.}{2019}]{Navarete19}
Navarete F., Galli Ph.A.B., Damineli A., 2019, \mnras, 487, 2771

\bibitem[\protect\citeauthoryear{Navarete et al.}{2022}]{Navarete22}
Navarete F.,  Damineli A., Ramirez A. E., Rocha D. F., Almeida L. A., 2022, \mnras, 516, 1289

\bibitem[\protect\citeauthoryear{Nony et al.}{2021}]{nony21}
Nony T., Robitaille J. F-., Motte F. et al., 2021, \aap, 645, A94

\bibitem[\protect\citeauthoryear{O'Donnell}{1994}]{Odonnell}
O'Donnell J. E., 1994, \apj, 422, 158

\bibitem[\protect\citeauthoryear{Ochsenbein et al.}{2000}]{Vizier}
Ochsenbei, F.; Bauer P.; Marcout J., 2000, A\&AS, 143, 23

\bibitem[\protect\citeauthoryear{Pang et al.}{2022}]{Pang22}
Pang, X., Tang S-Y., Li Y. et al., 2022, \apj 931, 156

\bibitem[\protect\citeauthoryear{Parker}{2018}]{Parker18}
Parker R.~J., 2018, \mnras, 476, 617

\bibitem[\protect\citeauthoryear{Parker et al.}{2014}]{Parker14}
Parker R.~J., Wright N.~J., Goodwin S.~P., Meyer M.~R., 2014, \mnras, 438, 620

\bibitem[\protect\citeauthoryear{Parker \& Goodwin}{2015}]{Parker15}
Parker R.~J., Goodwin S.~P.,  2015, \mnras, 449, 3381

\bibitem[\protect\citeauthoryear{Parker \& Alves de Oliveira}{2023}]{Parker23}
Parker R.~J., Alves de Oliveira  C., 2023, \mnras, 525, 1677

\bibitem[\protect\citeauthoryear{Parker \& Schoettler}{2022}]{Parker22}
Parker R. J.,  Schoettler C., 2022, \mnras, 510, 1136

\bibitem[\protect\citeauthoryear{Piatti}{2014}]{Piatti14}
Piatti A.~E., 2014, \mnras, 445, 2302

\bibitem[\protect\citeauthoryear{Poggio et al.}{2021}]{Poggio21}
Poggio E., Drimmel R., Cantat-Gaudin T. et al., 2021, \aap,  651, A104

\bibitem[\protect\citeauthoryear{Pouteau et al.}{2023}]{Pouteau}
Pouteau Y., Motte F., Nony T. et al., 2023, \aap,  674, A76

\bibitem[\protect\citeauthoryear{Press et al.}{1992}]{press}
Press W. H., Teukolsky S. A., Vetterling W. T., Flannery B. P., 2007, 
Numerical recipes: the art of scientific computing, 3 edn. Cambridge University Press, Cambridge, UK

\bibitem[\protect\citeauthoryear{Rain et al.}{2021}]{Rain}
Rain M. J., Ahumada J. A., Carraro G., 2021, \aap, 650, A67

\bibitem[\protect\citeauthoryear{Ray et al.}{2022}]{Ray22}
Ray A., Frinchaboy P., Donor J. et al., 2022, \aj, 163, 195 

\bibitem[\protect\citeauthoryear{Richer et al.}{2021}]{Richer21}
Richer H. B., Caiazzo I., Du H. et al., 2021, \apj, 912, 165

\bibitem[\protect\citeauthoryear{Rodríguez et al.}{2023}]{Jimena}
Rodrígues M. J.,  Feinstein C., Baume G. et al., 2023, \mnras, 519, 3357

\bibitem[\protect\citeauthoryear{Rubinstein}{1997}]{Rubinstein97}
Rubinstein R.~Y., 1997, Eur. J. Operat. Res., 99, 89

\bibitem[\protect\citeauthoryear{Saito et al. }{2012}]{Saito12} 
Saito R. K., Hempel M., Minniti,D. et al., 2012, \aap, 537, A107

\bibitem[\protect\citeauthoryear{Sanders}{1971}]{Sanders71}
Sanders W. L., 1971, \aap, 14, 226

\bibitem[\protect\citeauthoryear{Santos et al.}{2020}]{JoaoF}
Santos J. F. C. J. et al., 2020, \mnras, 498, 205

\bibitem[\protect\citeauthoryear{Santos-Silva \& Gregorio-Hetem}{2012}]{SS12}
Santos-Silva T., Gregorio-Hetem J., 2012, \aap, 547, A107

\bibitem[\protect\citeauthoryear{Santos-Silva et al.}{2022}]{Thais}
Santos-Silva T.,  Perottoni H. D., Almeida-Fernandes F. et al., 2021, \mnras, 508, 1033

\bibitem[\protect\citeauthoryear{Savage \& Mathis}{1979}]{Savage}
Savage B. D., Mathis J. S., 1979, \araa, 17, 73

\bibitem[\protect\citeauthoryear{Schmeja \& Klessen}{2006}]{schkl}
Schmeja S.; Klessen R.~S., 2006, \aap, 449, 151

\bibitem[\protect\citeauthoryear{Schmeja et al.}{2009}]{schmeja09}
Schmeja S., Gouliermis D.A.,  Klessen  R.~S., 2009, \apj, 694, 367

\bibitem[\protect\citeauthoryear{Schoettler et al.}{2022}]{schoettler22}
Schoettler C., Parker R. J., de Bruijne J., 2022, \mnras, 510, 3178

\bibitem[\protect\citeauthoryear{Skrutskie et al.}{2006}]{Skrutskie06}
Skrutskie M.F.,  Cutri R.M., Stiening R. et al., 2006, \aj, 131, 1163

\bibitem[\protect\citeauthoryear{Song et al.}{2022}]{Song22}
Song F., Esamdin A., Hu Q., Zhang M. 2022, \aap, 666, A75

\bibitem[\protect\citeauthoryear{Spina et al.}{2021}]{Spina}
Spina L., Ting Y. -S., De Silva G. M. et al., 2021, \mnras, 503, 3279

\bibitem[\protect\citeauthoryear{Tarricq et al.}{2022}]{tarricq22}
Tarricq Y., Soubiran C., Casamiquela L. et al., 2022, \aap, 659, A59

\bibitem[\protect\citeauthoryear{Uribe \& Brieva}{1994}]{Uribe94}
Uribe A., Brieva E., 1994, \apss, 214, 171

\bibitem[\protect\citeauthoryear{Vicente et al.}{2016}]{belen16}
Vicente B., Sánchez N., Alfaro E. J., 2016, \mnras, 461, 2519

\bibitem[\protect\citeauthoryear{Wenger et al.}{2000}]{Simbad}
Wenger M., Ochsenbein F., Egret D. et al., 2000, A\&AS, 143, 9

\bibitem[\protect\citeauthoryear{Wright et al.}{2010}]{Wright10}
Wright E.L., Eisenhardt P.R.M., Mainzer A.K. et al., 2010, \aj, 140, 1868

\bibitem[\protect\citeauthoryear{Yen et al.}{2018}]{Yen18}
Yen S.X., Reffert S., Schilbach E., Röser S., Karchenko N.V., Piskunov A.E.,  2018, \aap, 615, A12


\end{thebibliography}


\clearpage

%%%%%%%%%%%%%%%%% APPENDICES %%%%%%%%%%%%%%%%%%%%%
\appendix

%%%%%%%%%%%%%%%%% APPENDIX A %%%%%%%%%%%%%%%%%%%%%

%%----------------------------------------------------------------
\section{Circumstellar emission}
\label{sec:wise}

The identification of cluster members showing  infrared (IR) emission is based on the {\it WISE}  colours diagram using photometry at 3.4 $\mu$m, 4.6 $\mu$m, and 12 $\mu$m that is commonly adopted to 
distinguish between disc-bearing stars and pre-main sequence (PMS) stars without discs. 
 \citet{K14} proposed limits on the [3.4 $\mu$m $-$ 4.6 $\mu$m] $\times$ [4.6 $\mu$m$-$12 $\mu$m] diagram defining the expected locus for Class I and Class II objects, which are PMS with the presence of protoplanetary disc, as indicated by their significant IR-excess. On the other hand,  field stars and Class III PMS objects lacking dust discs do not show IR emission and are well separated in the {\it WISE} colours diagram.

Following the methodology described by \citet{GH21},  we extracted from the {\it AllWISE} catalogue only the sources with good photometry quality at band $W3$ (12 $\mu$m) and avoiding contamination from fake detections, according with the filters proposed by \citet{K14}: $0.45 < W3_{r\chi^2} <1.15$ and $W3_{snr} > 5$, where $r\chi^2$ and $snr$ correspond to the photometric error and signal-to-noise ratio, respectively.

\begin{figure}
\includegraphics[width=0.9\columnwidth]{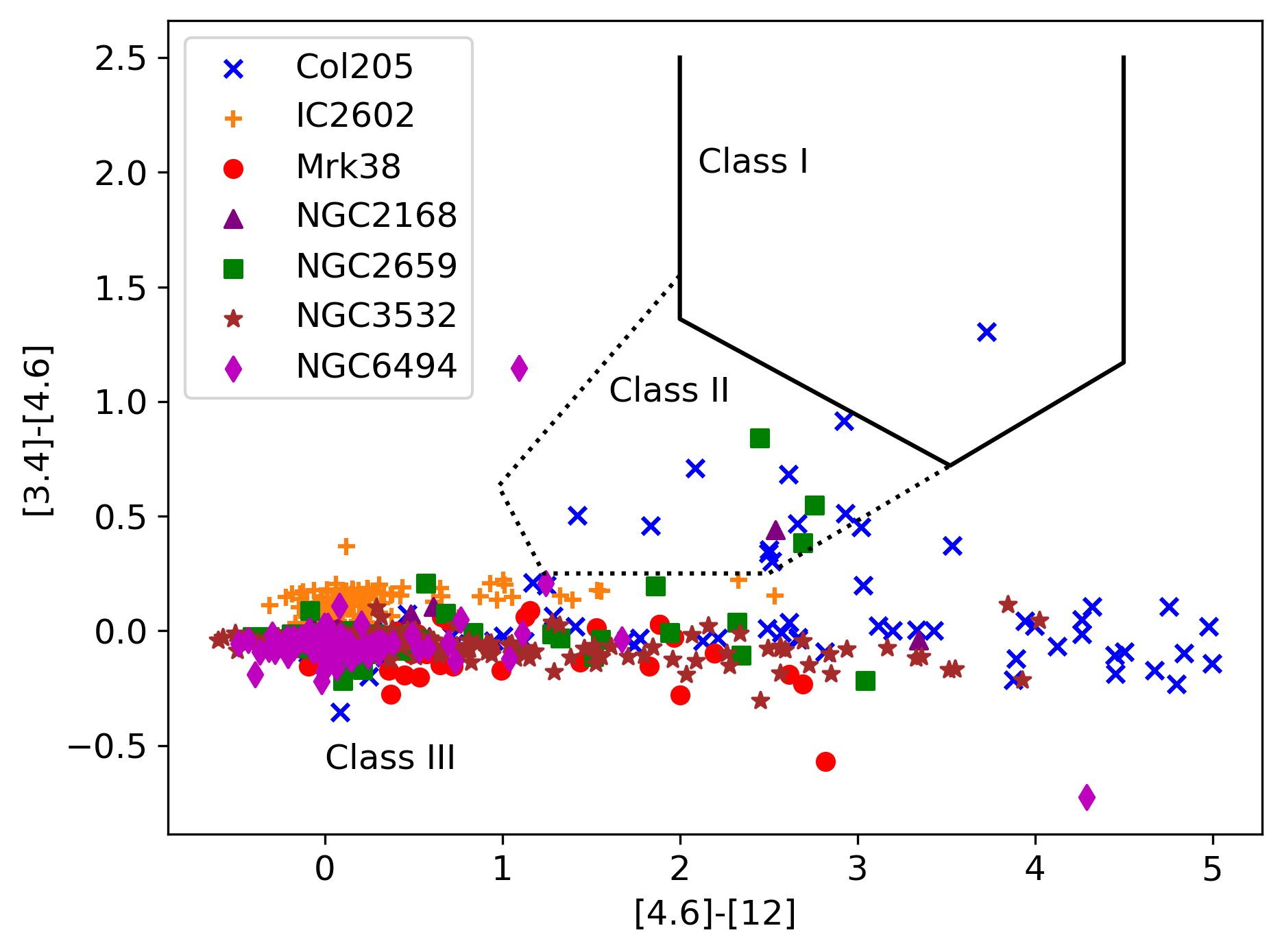}
\includegraphics[width=0.9\columnwidth]{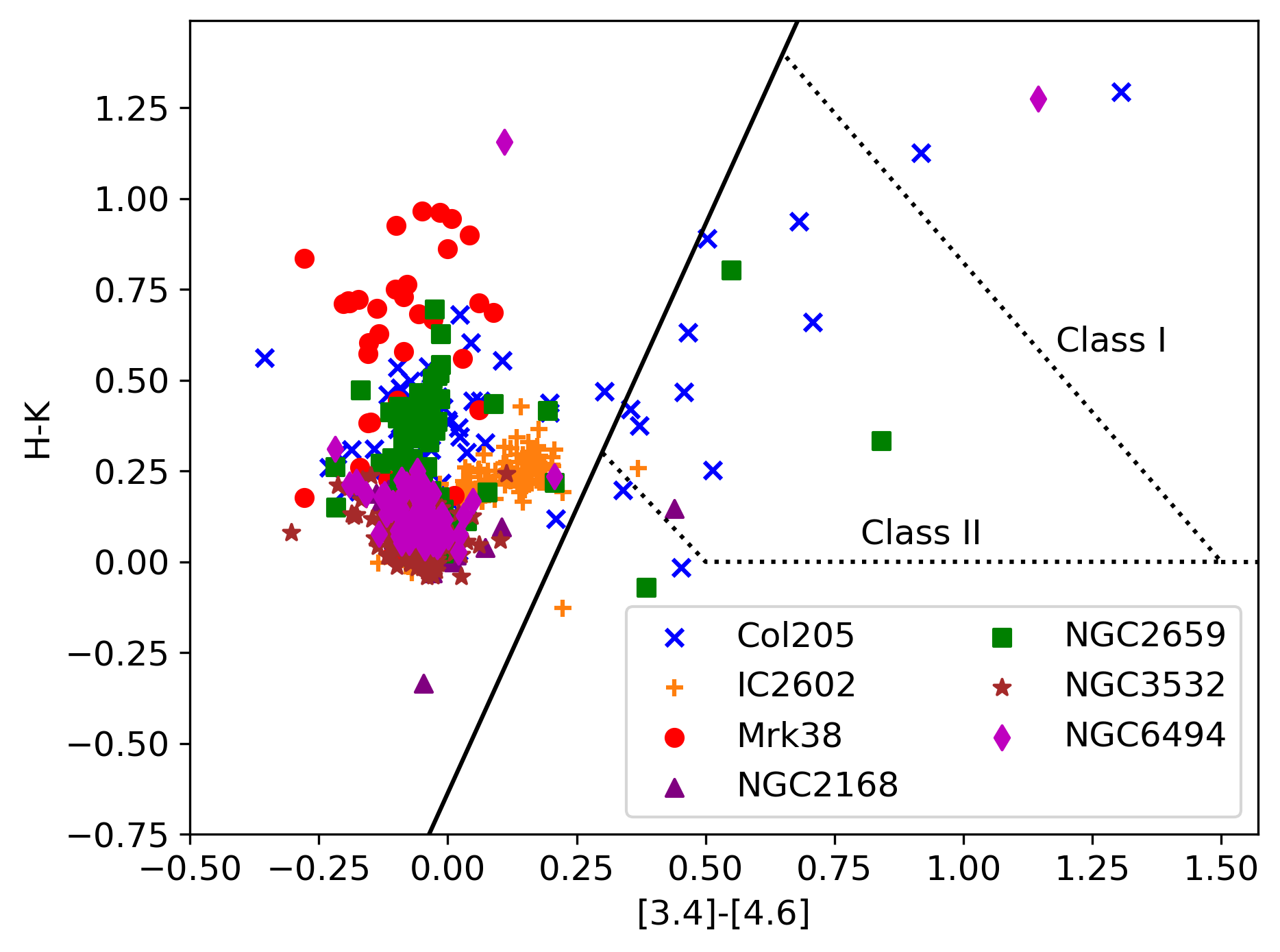}
\caption{{\it Top}: {\it WISE} colour-colour diagram displaying the expected locus for disc-bearing stars (Class I and Class II sources). Most of the members of our clusters are found in the region of Class III objects, with no dust emission from protoplanetary discs.
{\it Bottom}: The separation of objects from Classes I, II and III shown by {\it 2MASS} and {\it WISE} colours.}
\label{fig:wisebox}
\end{figure}

In Fig. \ref{fig:wisebox} (top panel), we show the {\it WISE} colours diagram for all the cluster members
 that were selected from the {\it AllWISE} catalogue.  As expected for the older clusters, most objects do not exhibit an IR-excess and are located within the region of Class III or field stars, indicating the lack of protoplanetary disc.
 
Only 13 objects belonging to the youngest clusters of our sample appear in the region of disc-bearing stars (Class I and Class II). Nine of them are members of Col205, which corresponds to a significant fraction of disc-bearing stars (11 percent), considering the number of members we studied. For young clusters at this age (42 Myr), it is observed a lower fraction of disc-bearing stars \citep[e.g.][]{Hernandez08,Fedele}. In the case of NGC2659a (30 Myr old), we found only 3 Class II objects, meaning a fraction of 3 percent of stars with disc, more compatible with other clusters of the same age.  
The other Class II object is member of NGC~2168, whose low fraction (0.1 percent) of disc-bearing stars is in agreement with the age 119 Myr found for this cluster

The bottom panel of Fig. \ref{fig:wisebox}  displays a similar diagram showing the [H-K] colour from {\it 2MASS} as a function of [3.4 $\mu$m $-$ 4.6 $\mu$m], which confirms the low-level of IR emission for most members of the studied clusters.
Once the diagrams in Fig.  \ref{fig:wisebox} are constructed for cluster members, we suggest 
 that the lacking-disc stars of our sample are Class III objects. That means, once they have a high membership probability, they can be distinguished from MS field stars.

%%%%%%%--------------------------------- Fig.A
%%%% -----------------------------------------end Fig.A

%%%%%%%%%%%%%%%%% APPENDIX B %%%%%%%%%%%%%%%%%%%%%
\section{Additional figures}
\label{sec:figs}

%%%%%%%--------------------------------- Fig.B1 
\begin{figure}
\centering
\includegraphics[width=0.7\columnwidth]{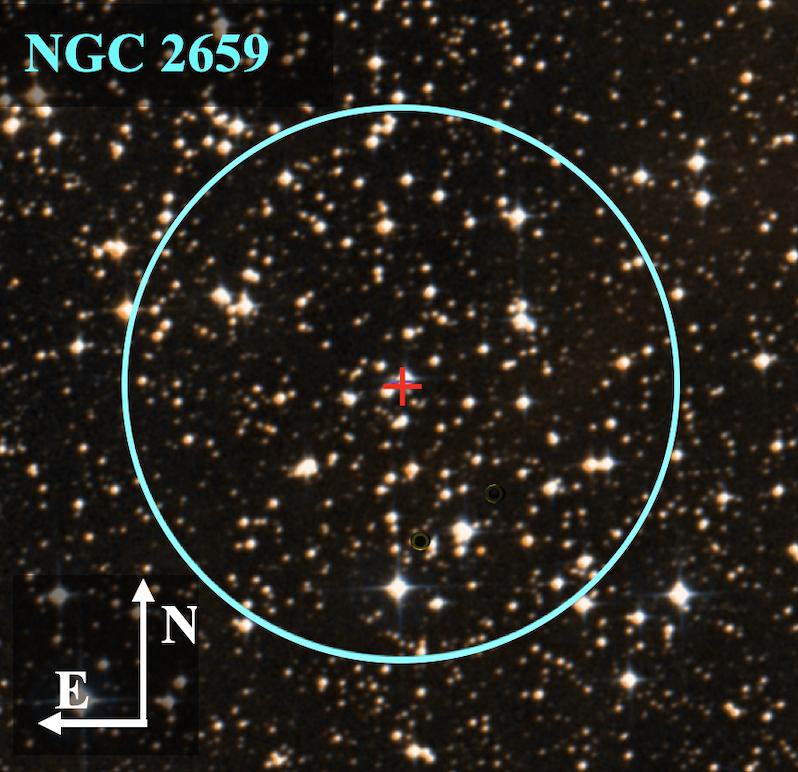}
\includegraphics[width=\columnwidth]{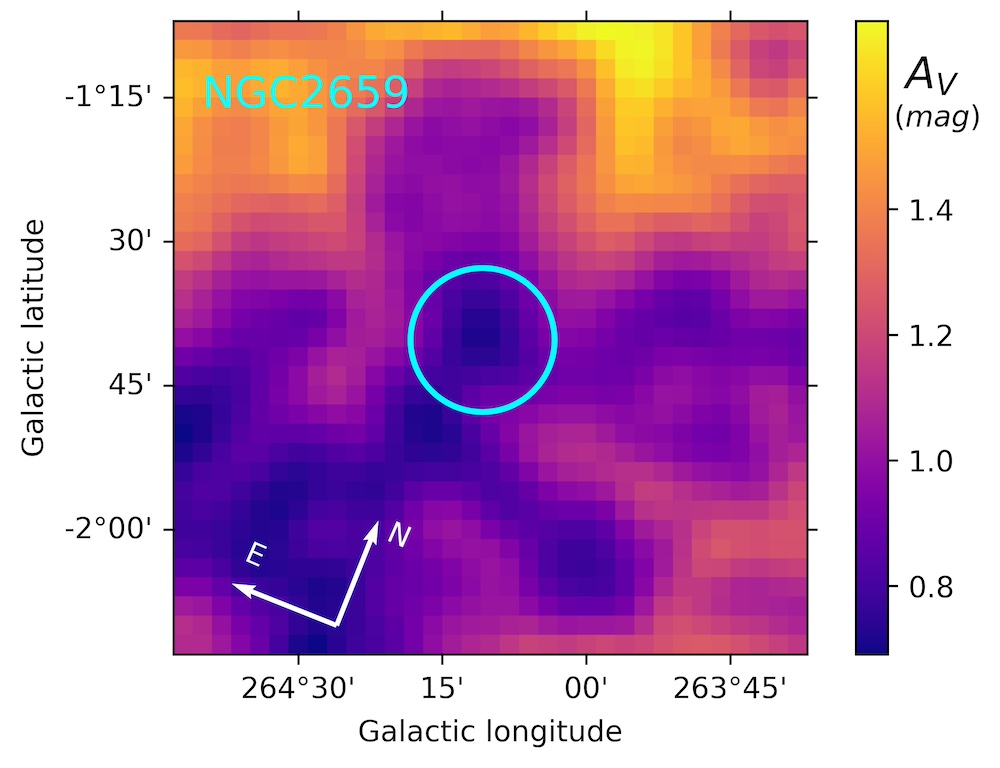}
\includegraphics[width=\columnwidth]{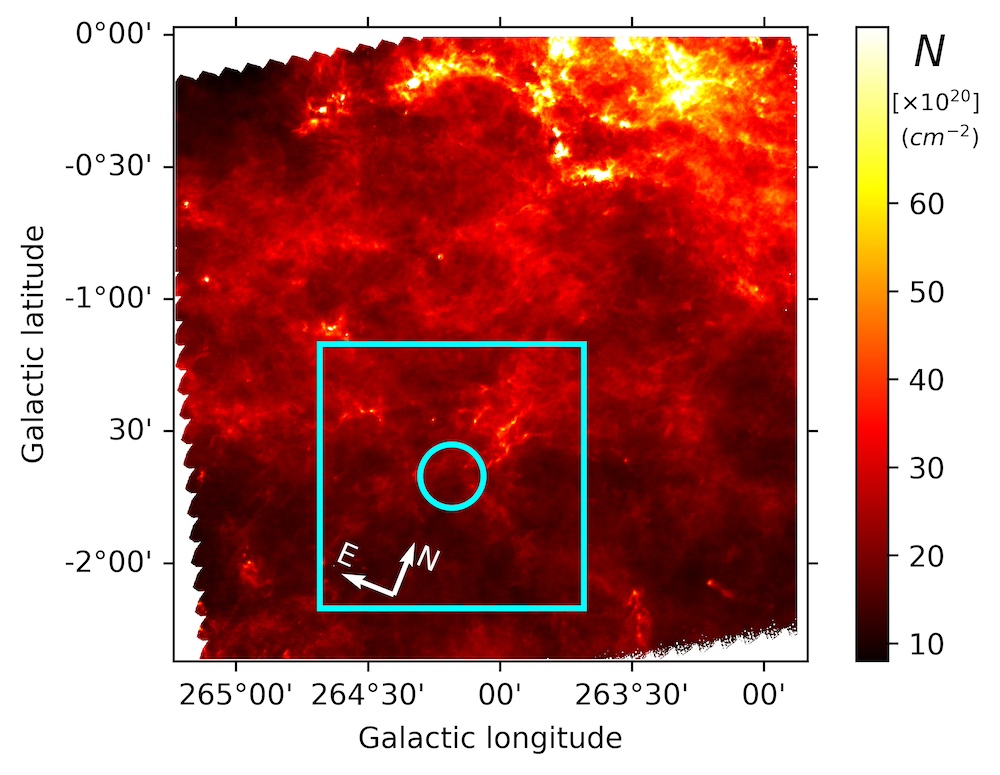}
\caption{The same as Fig. \ref{fig:dss} for NGC~2659. {\it Top}: DSS optical image. {\it Middle}: Visual extinction map. ({\it Bottom}: 
Integrated column density map of hydrogen molecules estimated from 
 ({\it Herschel}-PPMAP). The cyan square approximatively corresponds to the same area seen in the middle panel.}
\label{fig:Adss}
\end{figure}
%%%% -----------------------------------------end Fig.B1 

%%%%%%%--------------------------------- Fig.B2 
\begin{figure}
\centering
\includegraphics[width=0.7\columnwidth]{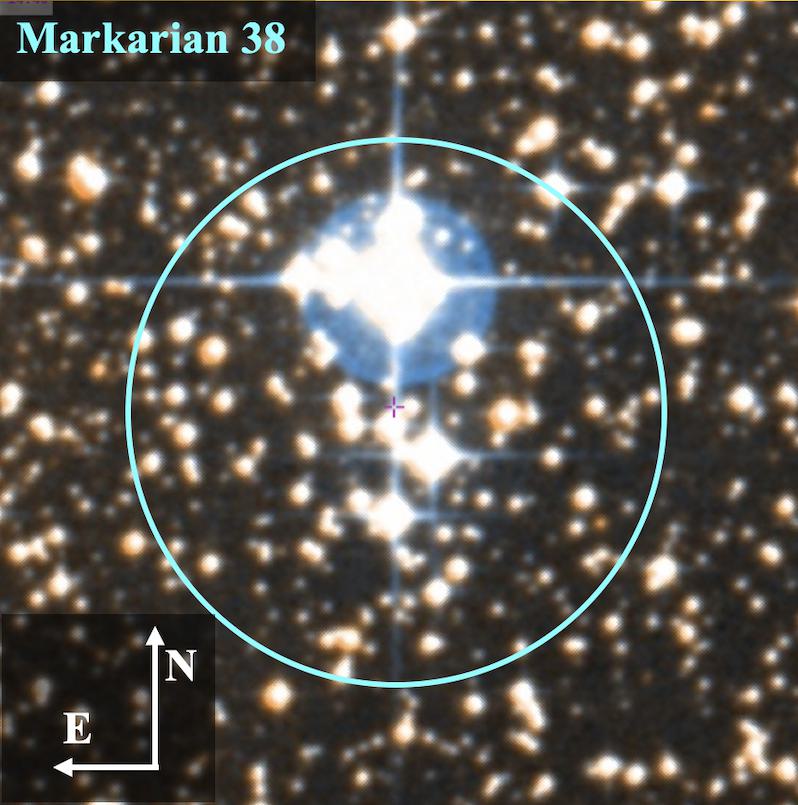}
\includegraphics[width=\columnwidth]{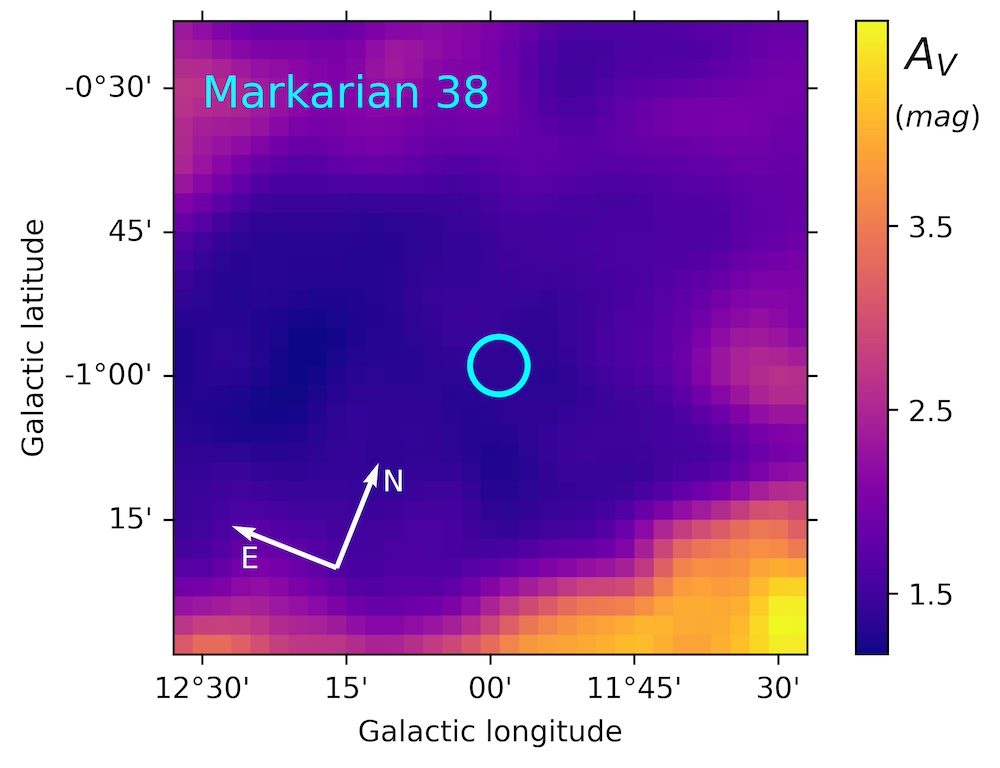}
\includegraphics[width=\columnwidth]{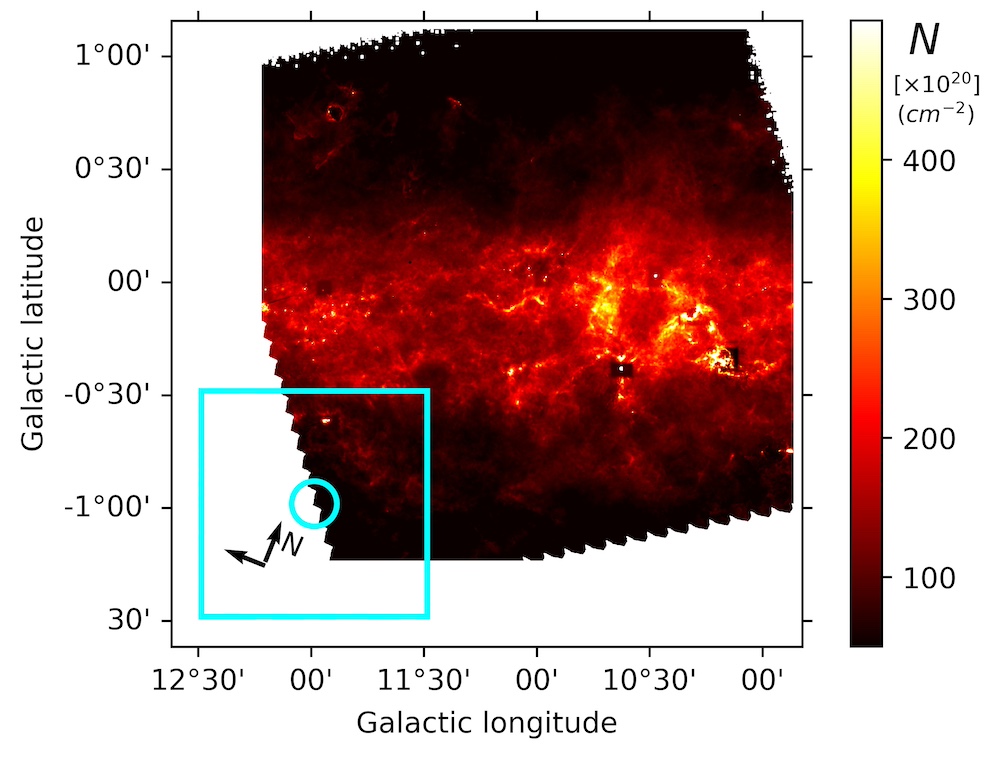}
\caption{The same as Fig. \ref{fig:Adss} for Mrk38.}
\label{fig:Adss2}
\end{figure}
%%%% -----------------------------------------end Fig.B2 

%%%%%%%--------------------------------- Fig.B3 
\begin{figure*}
\includegraphics[width=0.9\columnwidth]{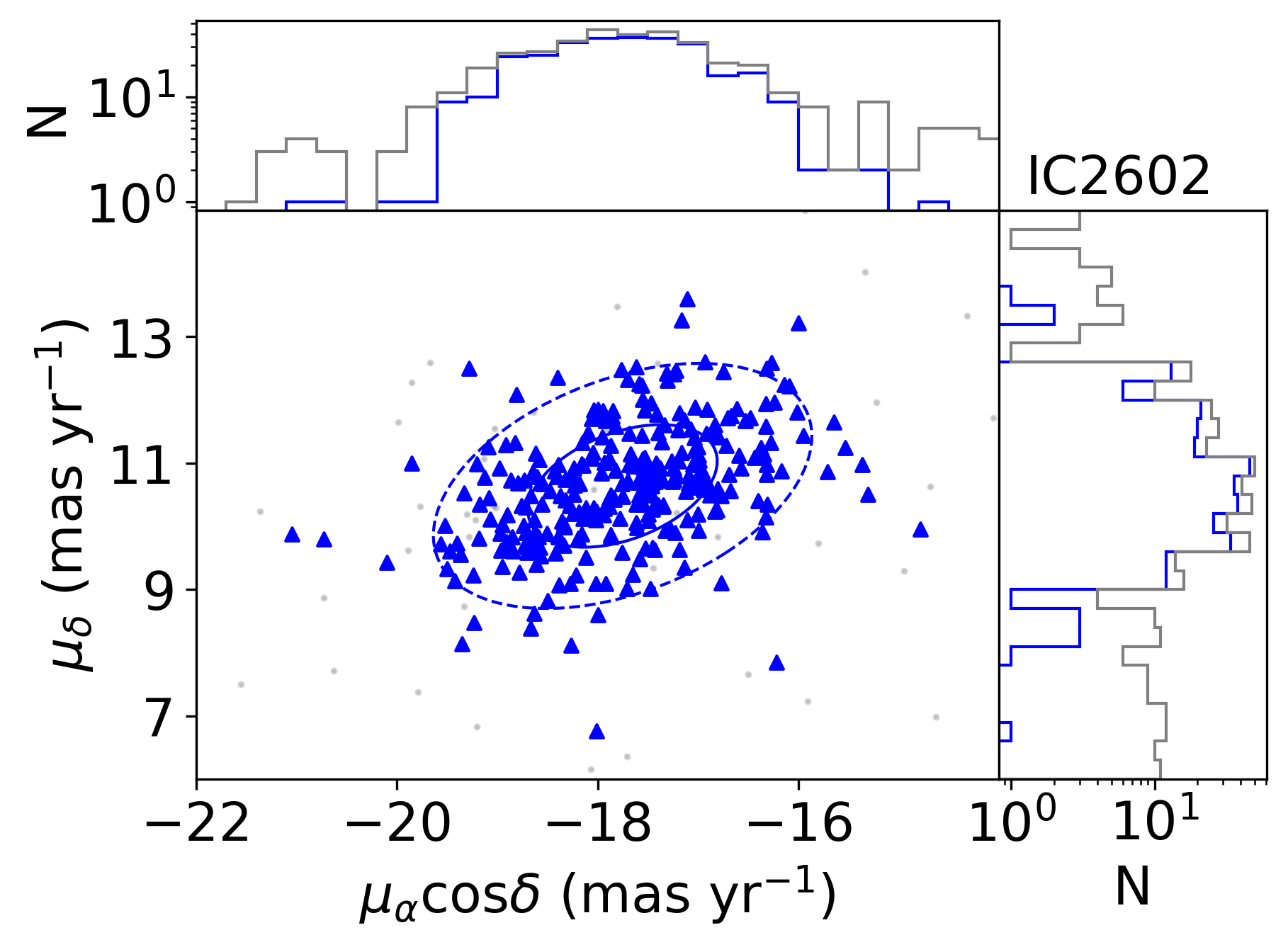}
\includegraphics[width=0.9\columnwidth]{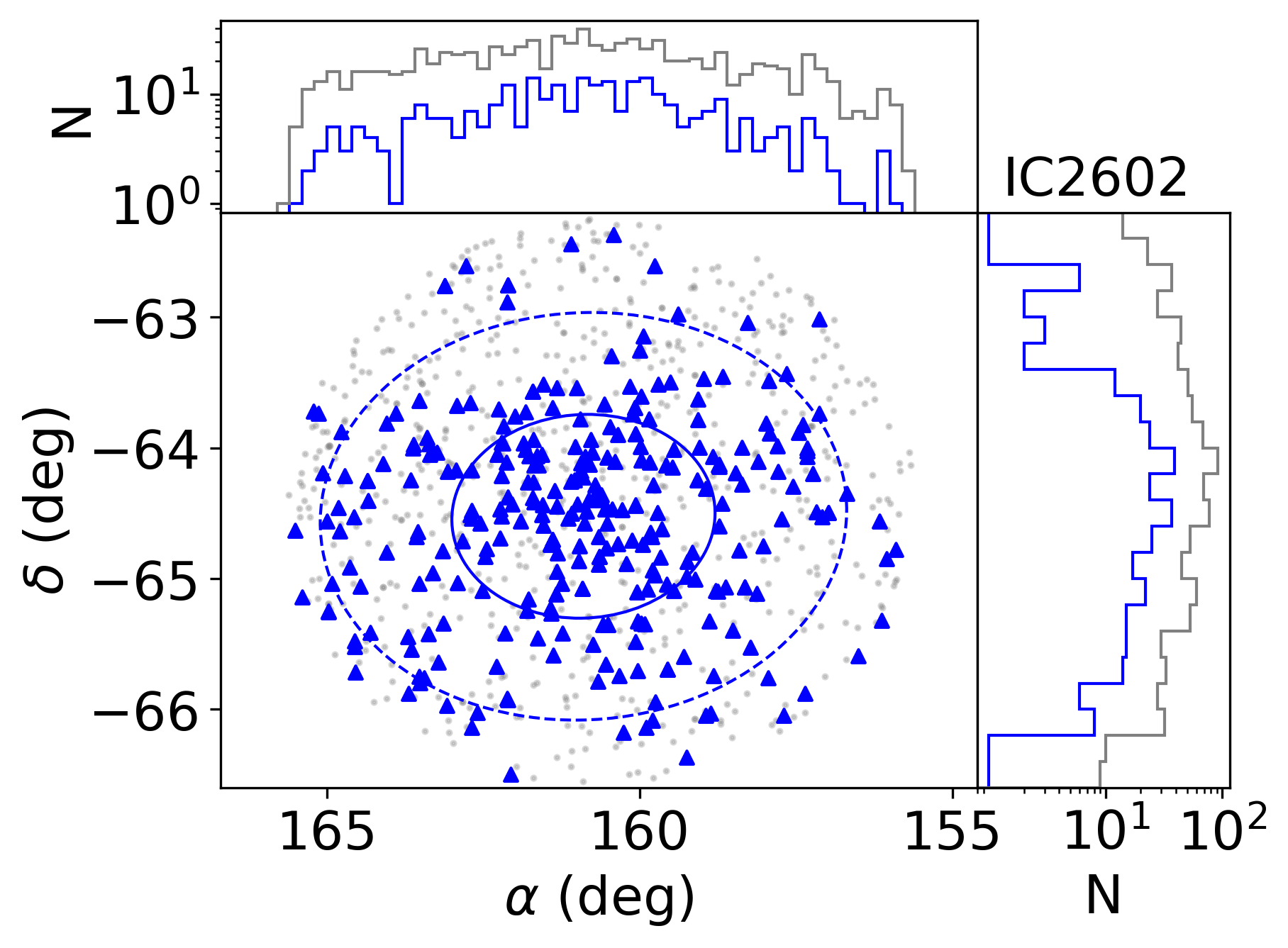}
\includegraphics[width=0.9\columnwidth]{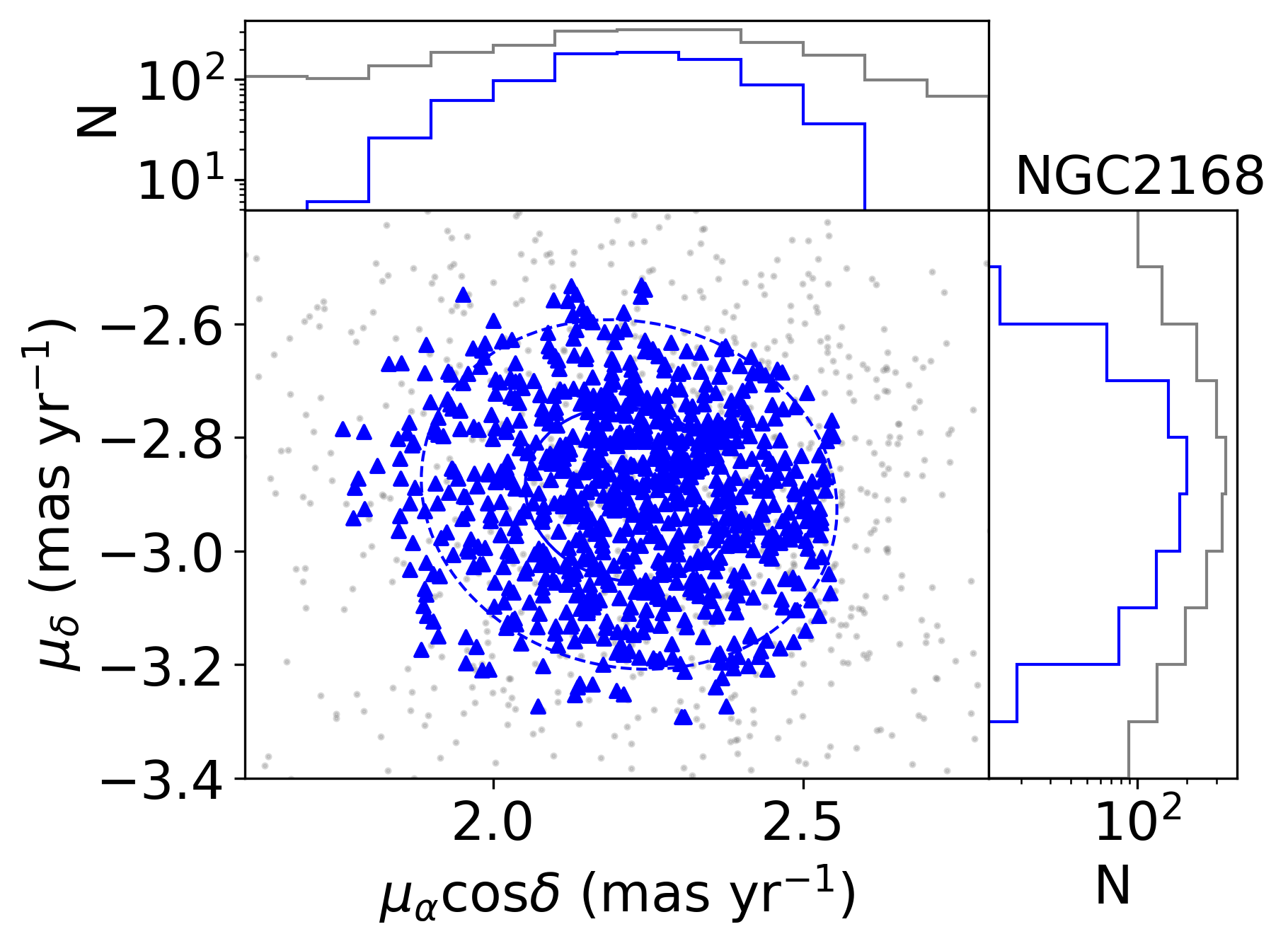}
\includegraphics[width=0.9\columnwidth]{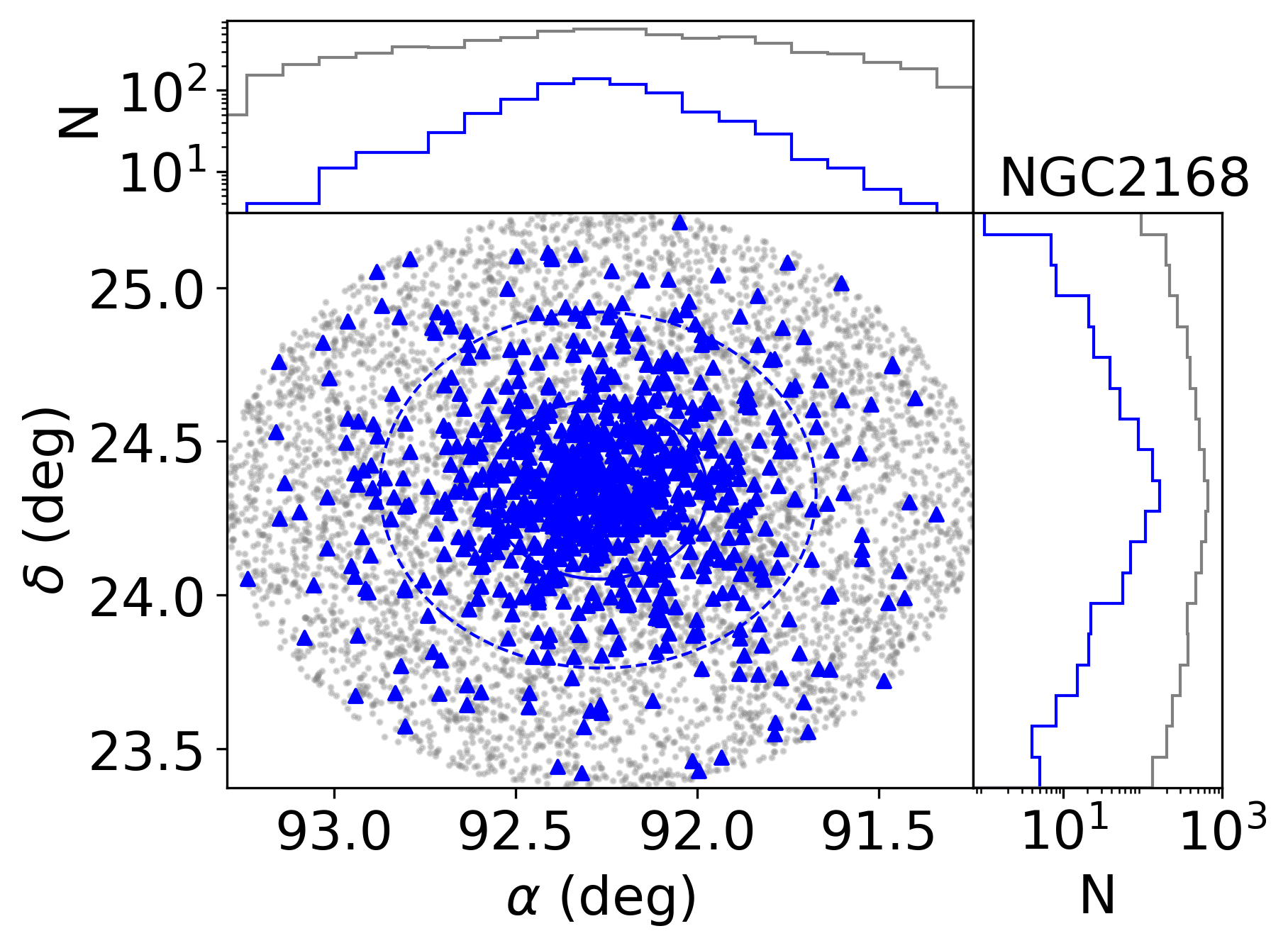}
\includegraphics[width=0.9\columnwidth]{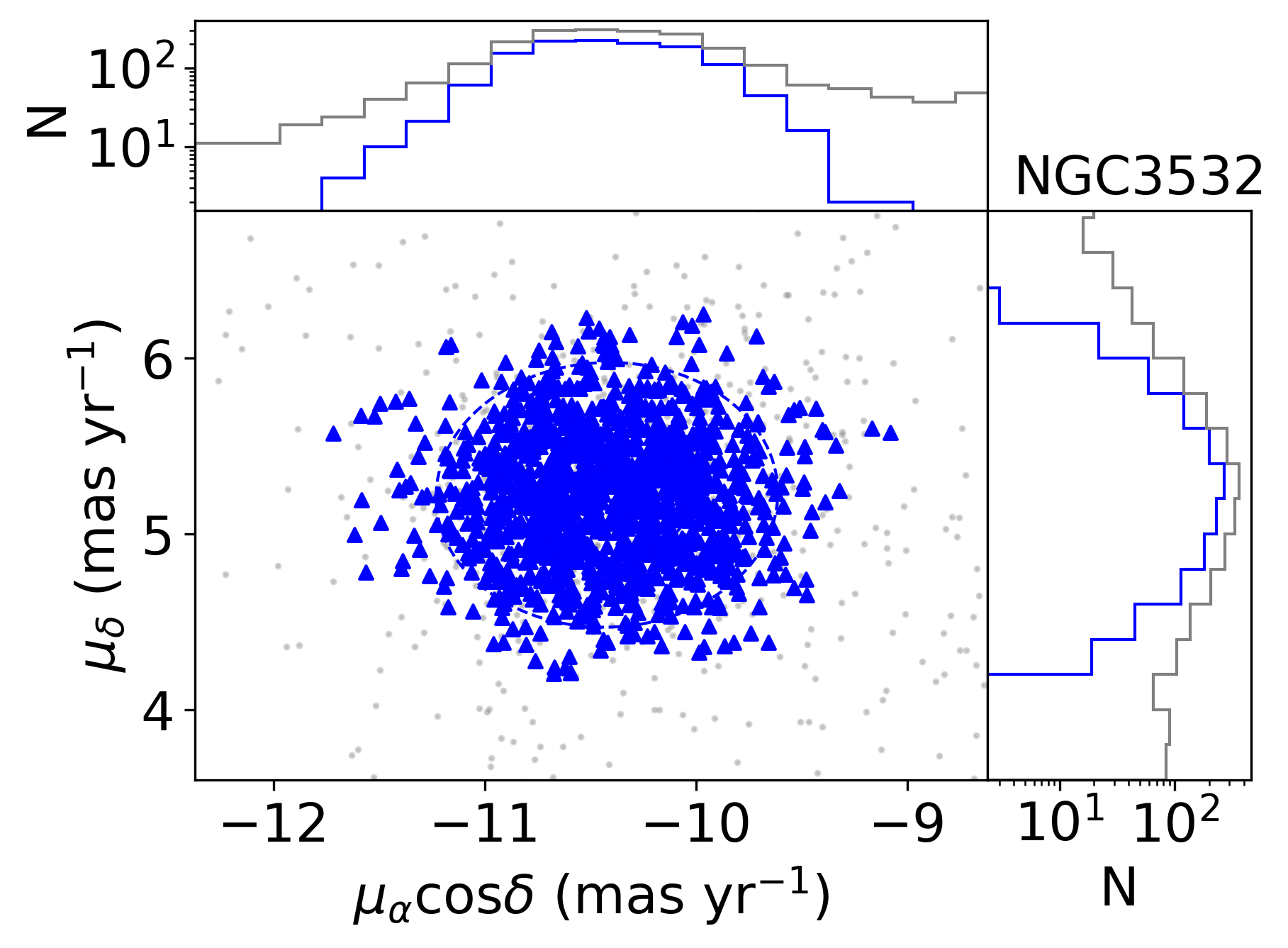}
\includegraphics[width=0.9\columnwidth]{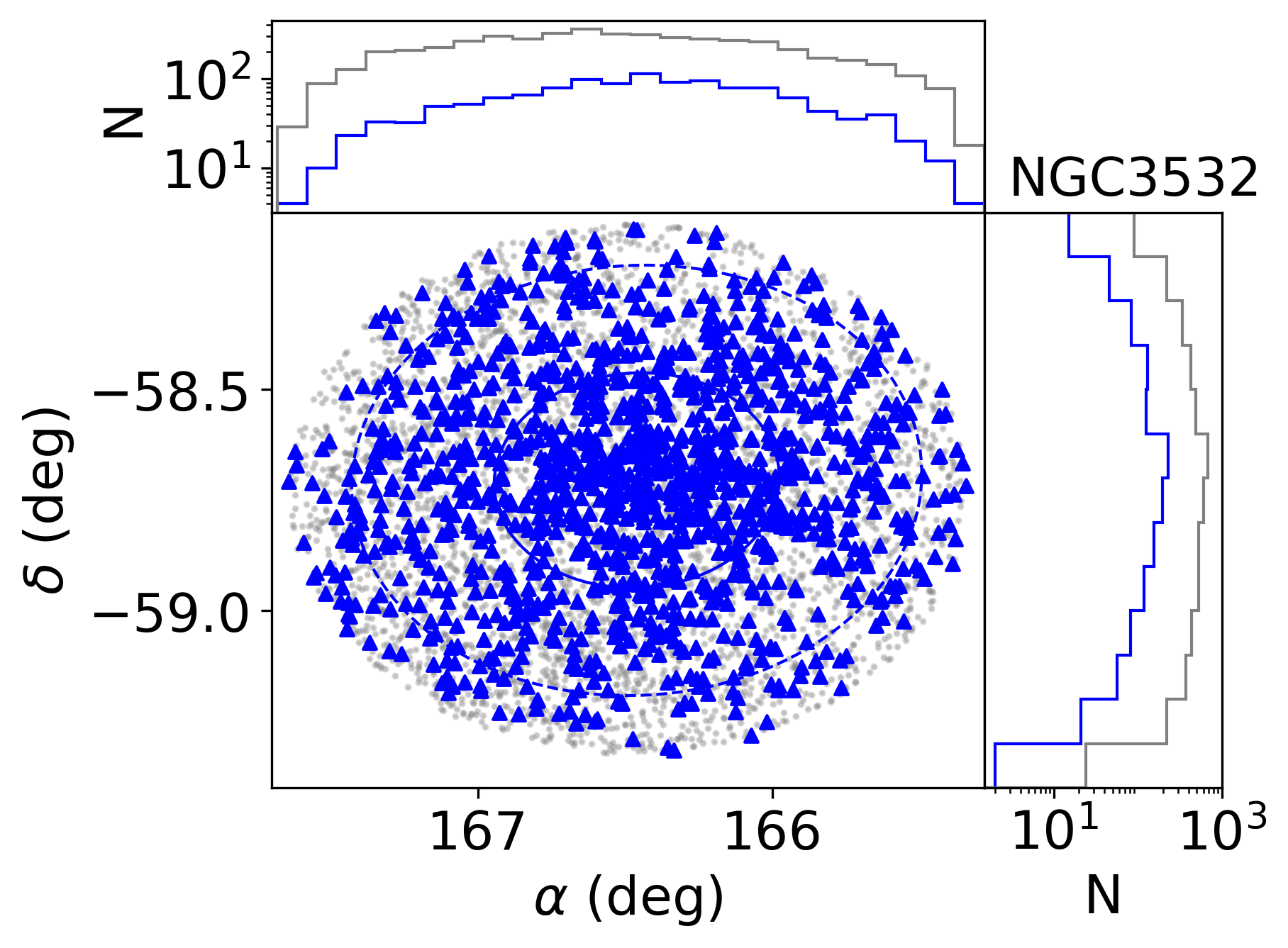}
\includegraphics[width=0.9\columnwidth]{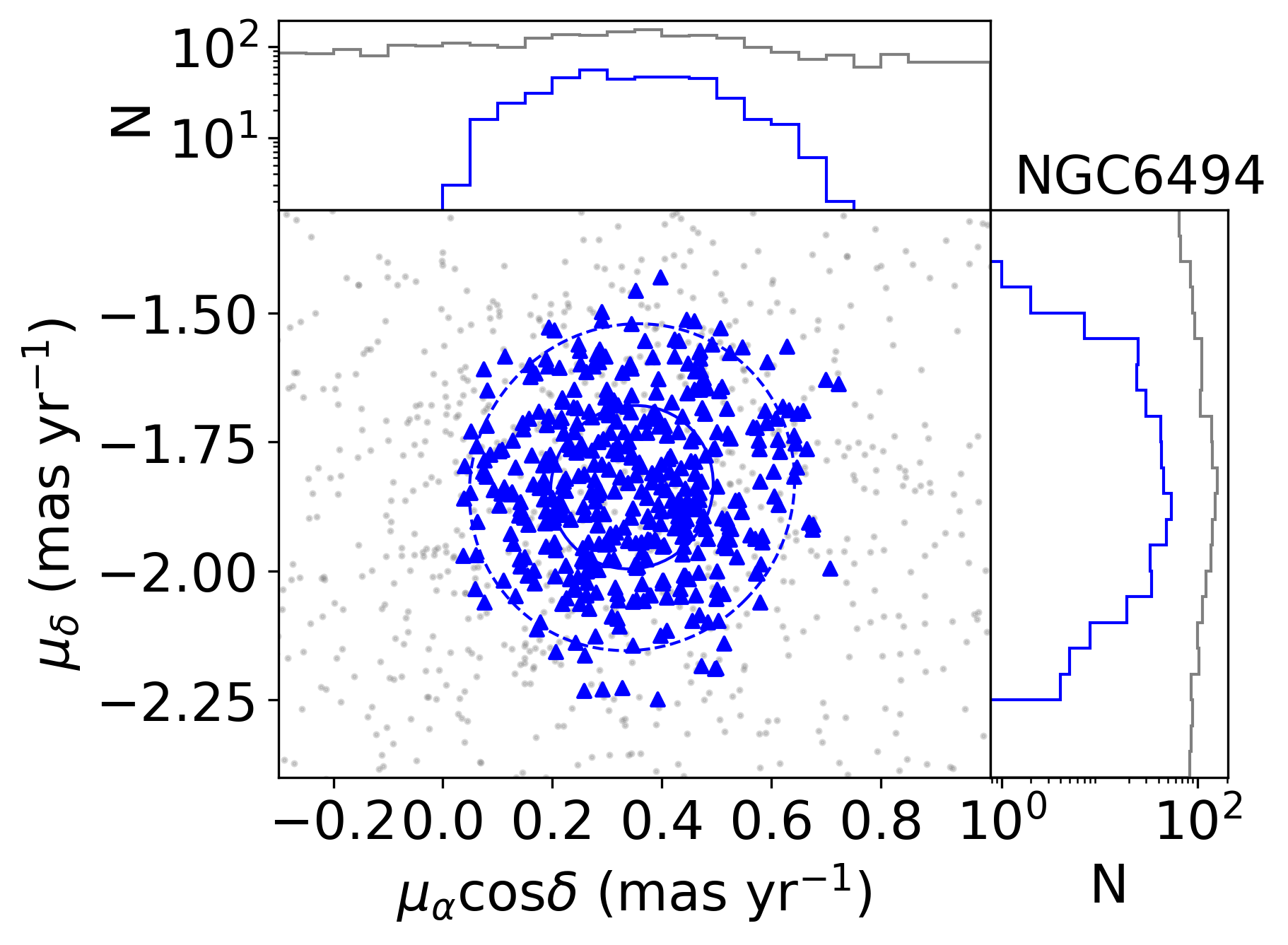}
\includegraphics[width=0.9\columnwidth]{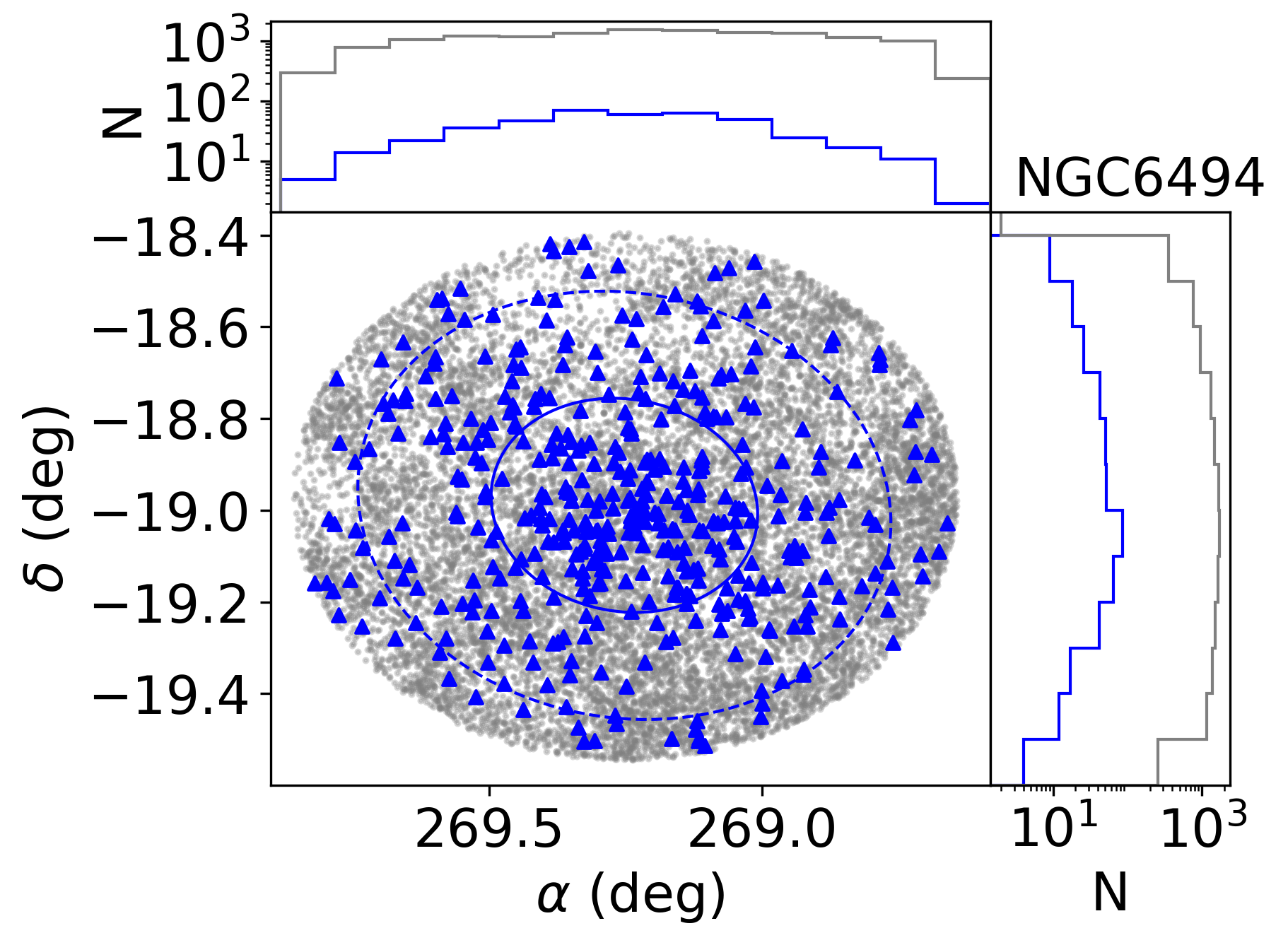}
\caption{The same as Fig. \ref{fig:member}.}
\label{fig:Amember1}
\end{figure*}
%%%% -----------------------------------------end Fig.B3

%%%%%%%--------------------------------- Fig.B4  Parallax as a function of G mag 
\begin{figure*}
\includegraphics[width=\columnwidth]{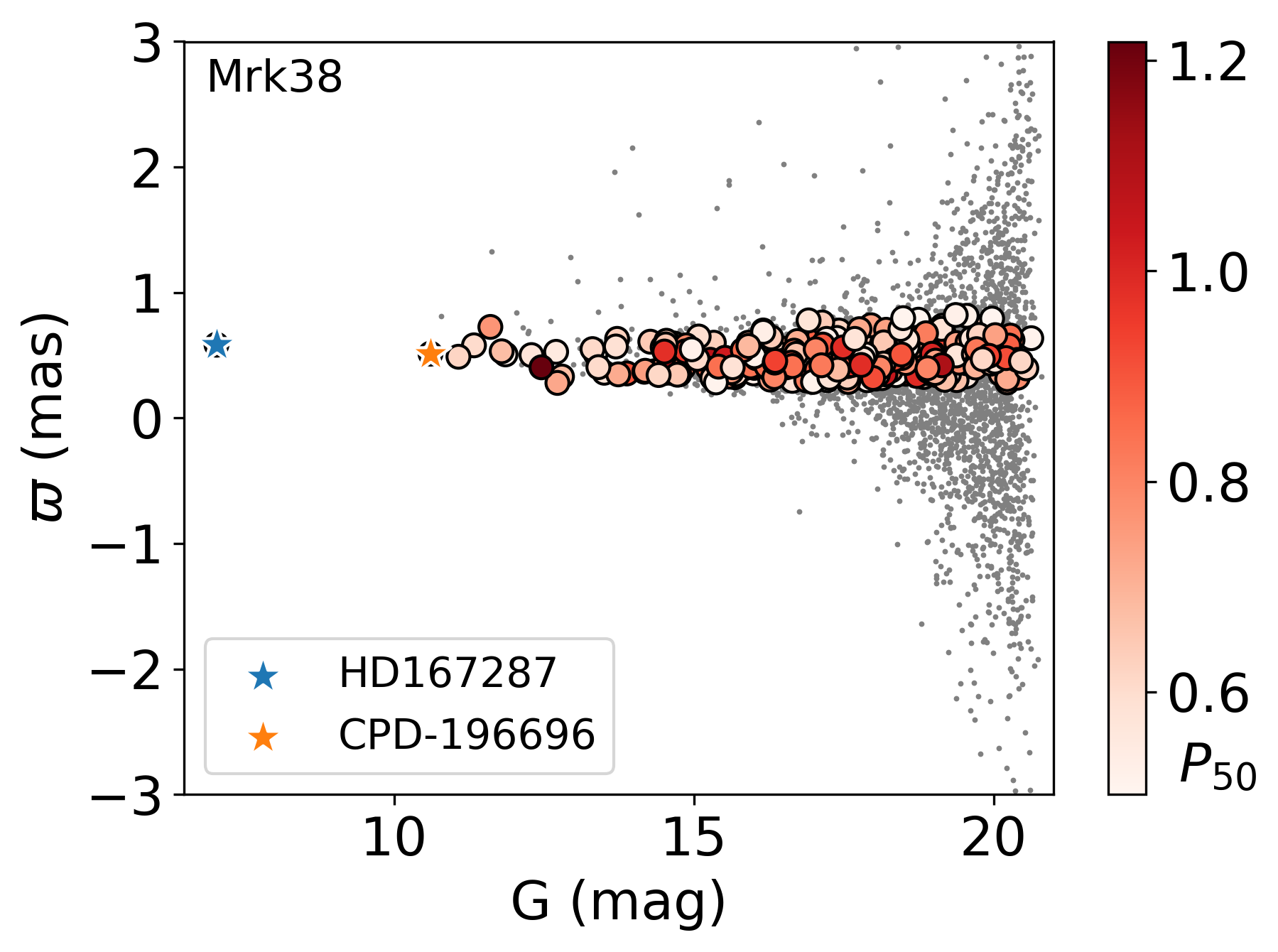}
\includegraphics[width=\columnwidth]{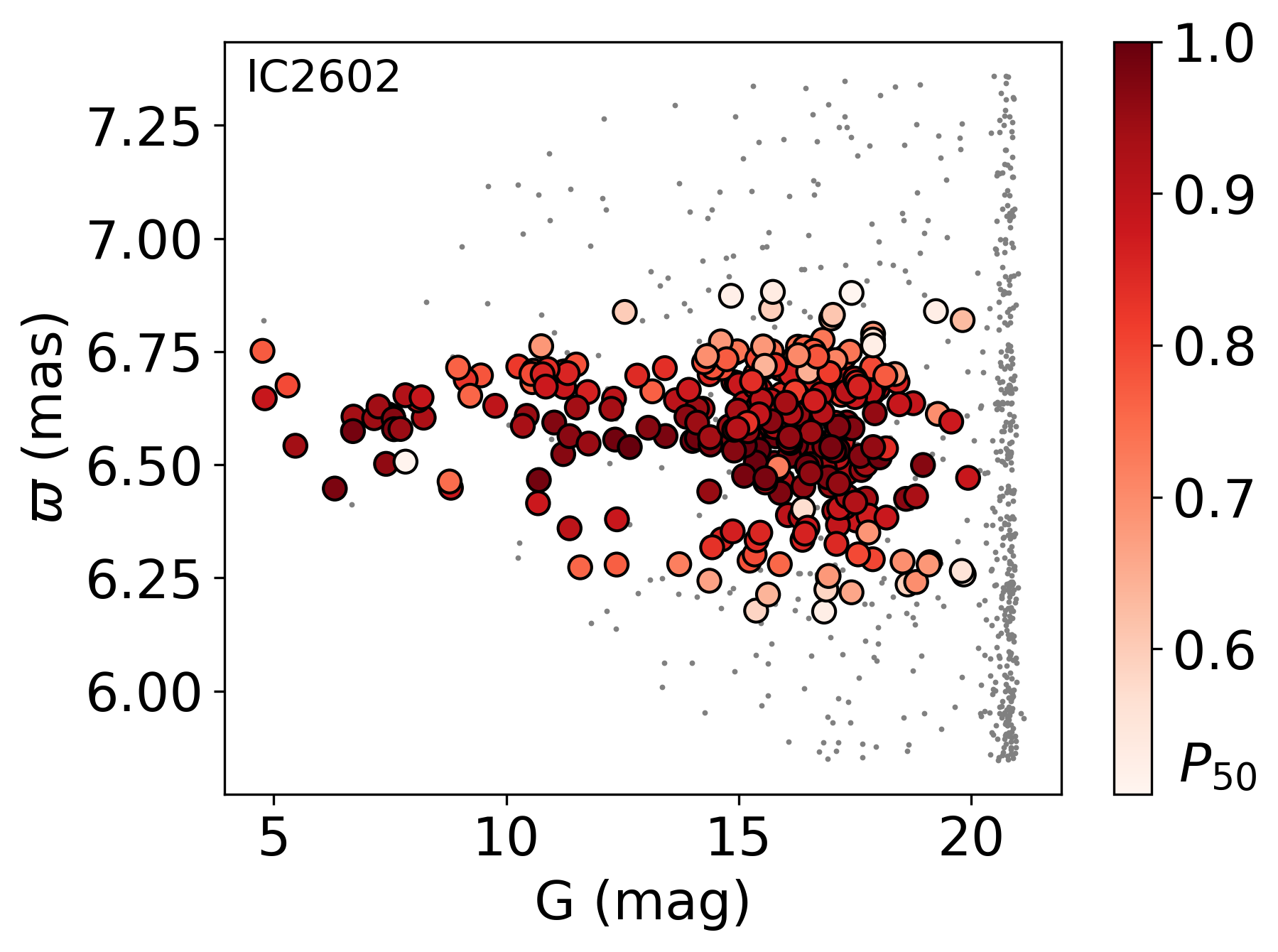}
\includegraphics[width=\columnwidth]{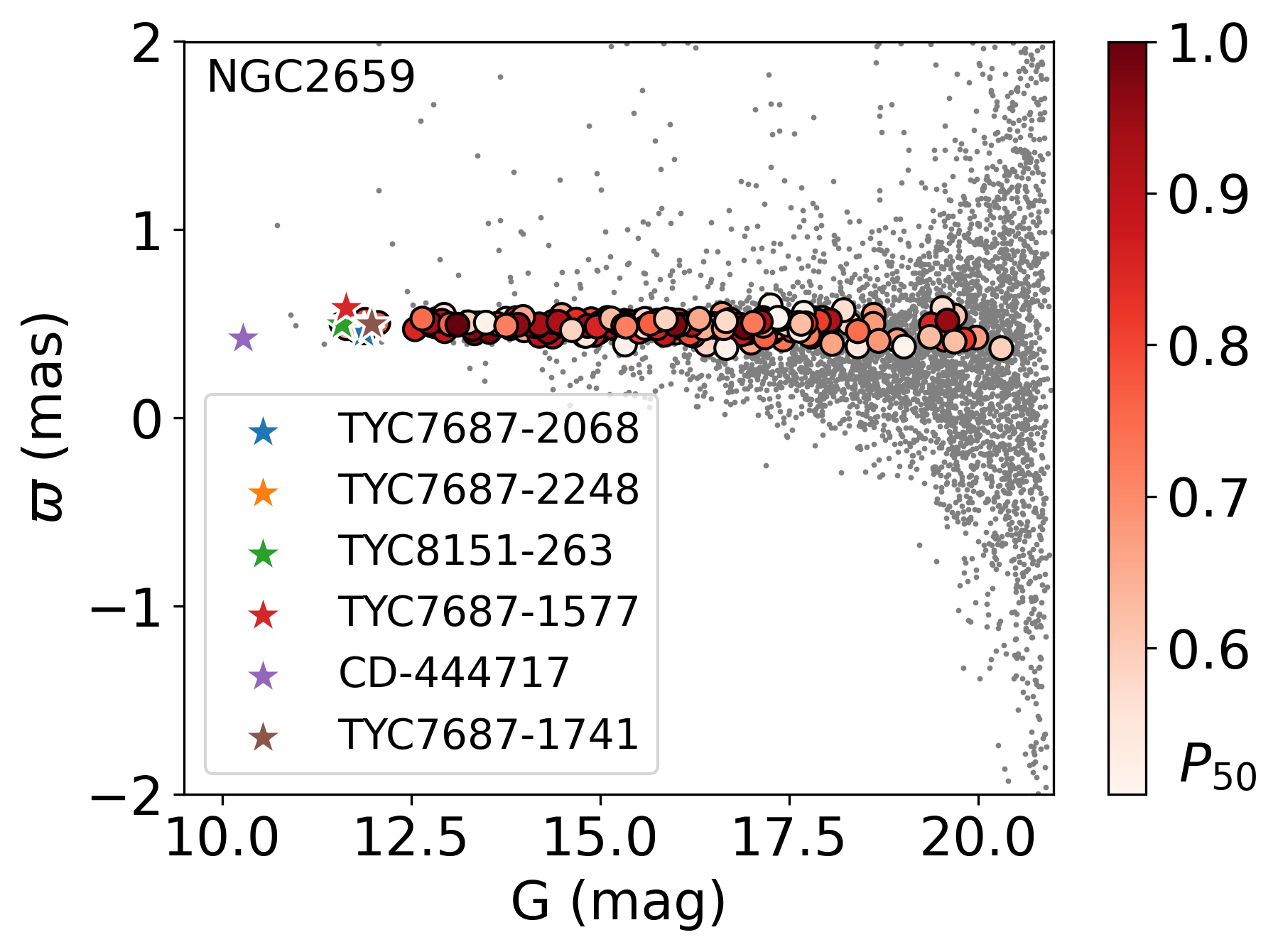}
\includegraphics[width=\columnwidth]{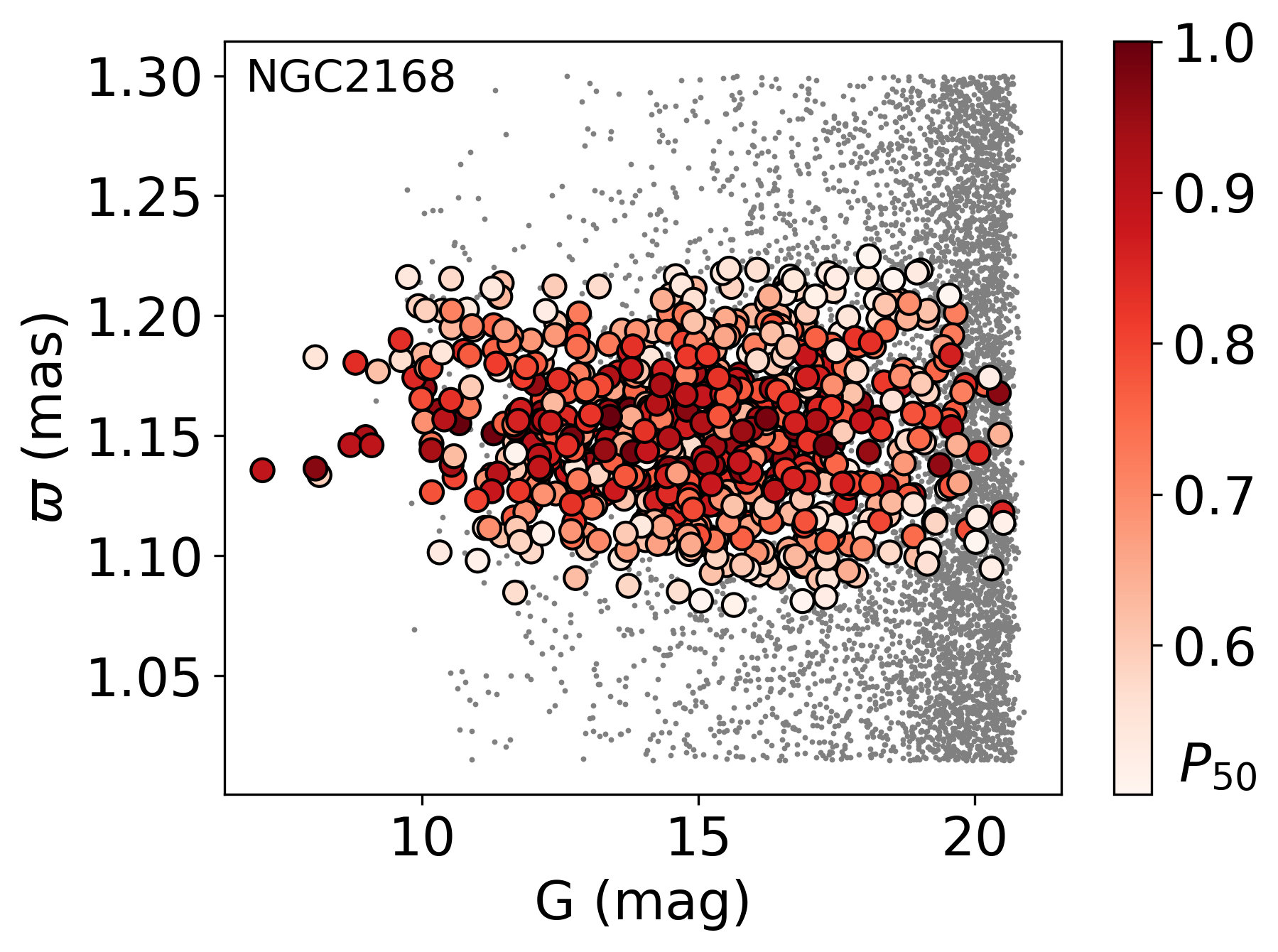}
\includegraphics[width=\columnwidth]{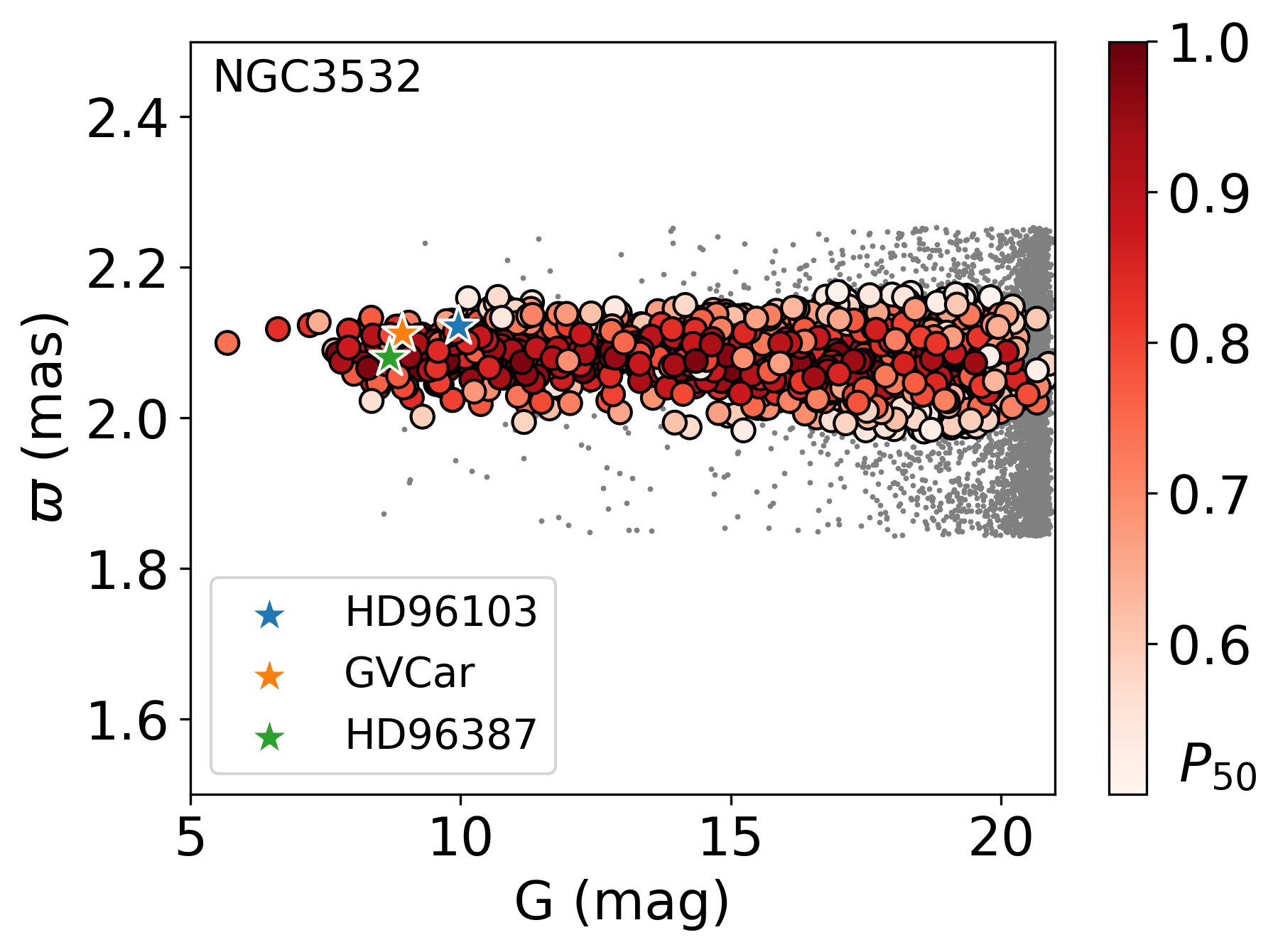}
\includegraphics[width=\columnwidth]{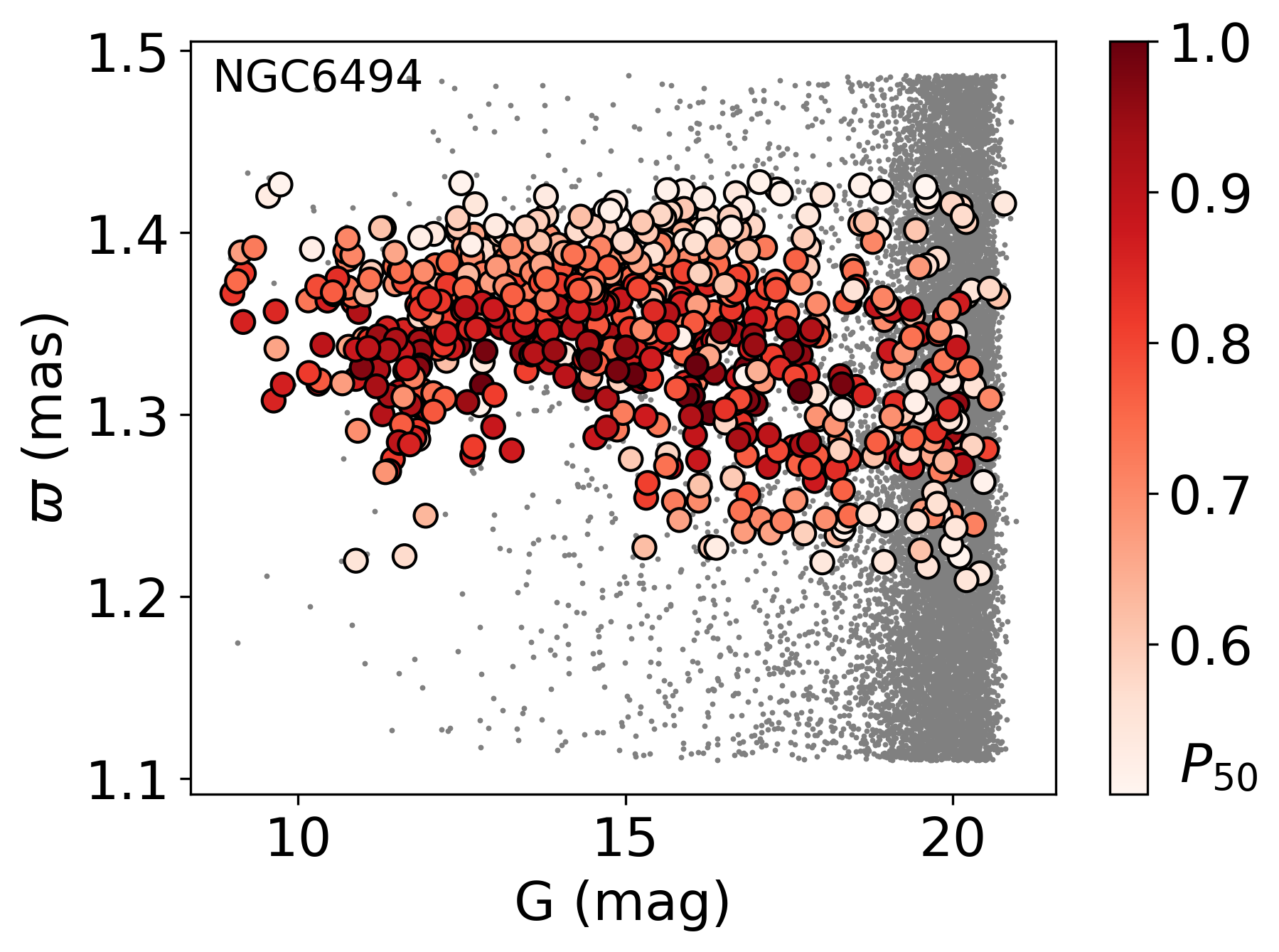}
\caption{The same as Fig. \ref{fig:G_pxl}.
}
\label{fig:AG_pxl}
\end{figure*}
%%%% -----------------------------------------end Fig.B4

%%--------new B5 Mass_Teff
\begin{figure*}
\includegraphics[width=\columnwidth]{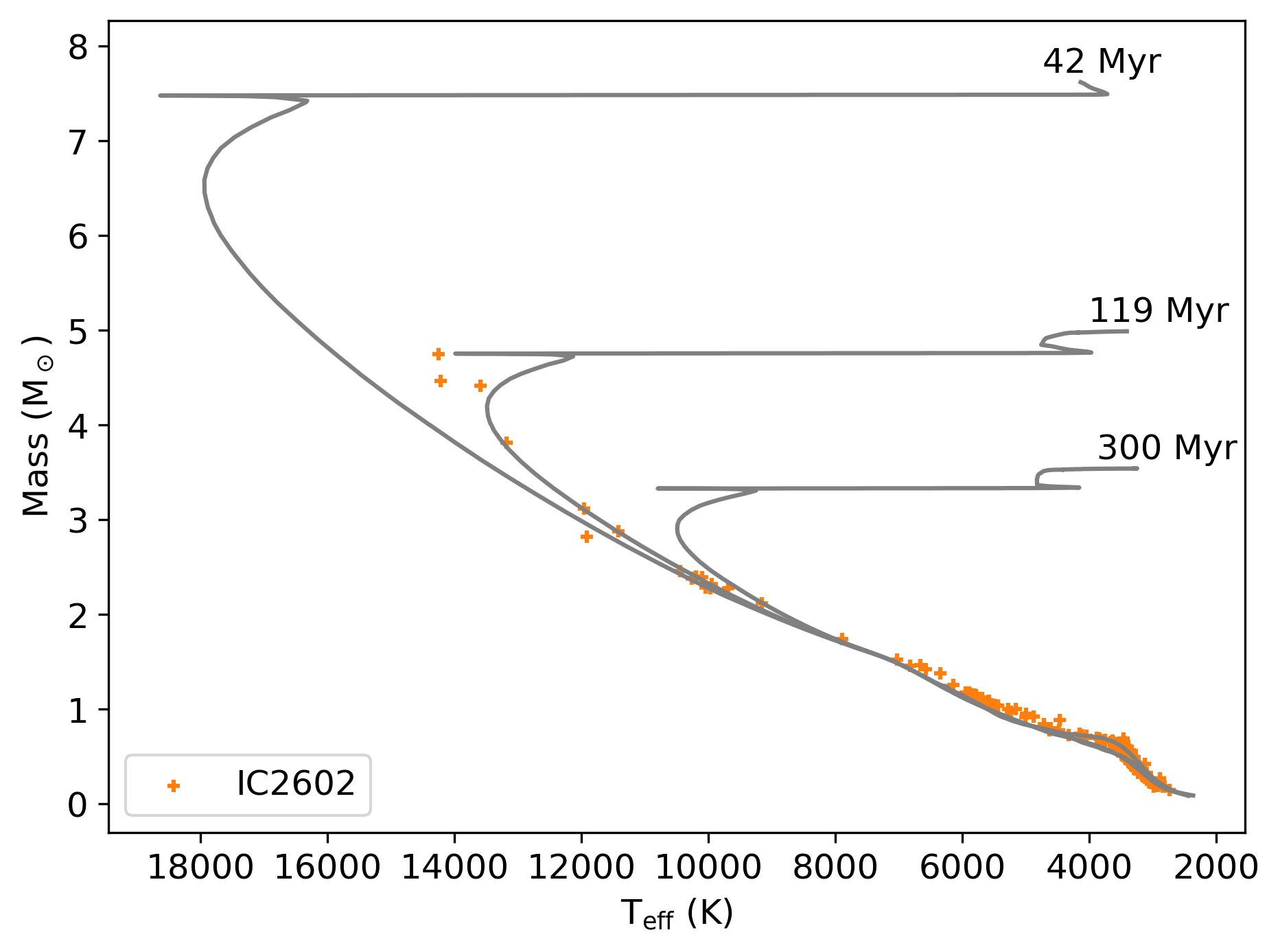}
\includegraphics[width=\columnwidth]{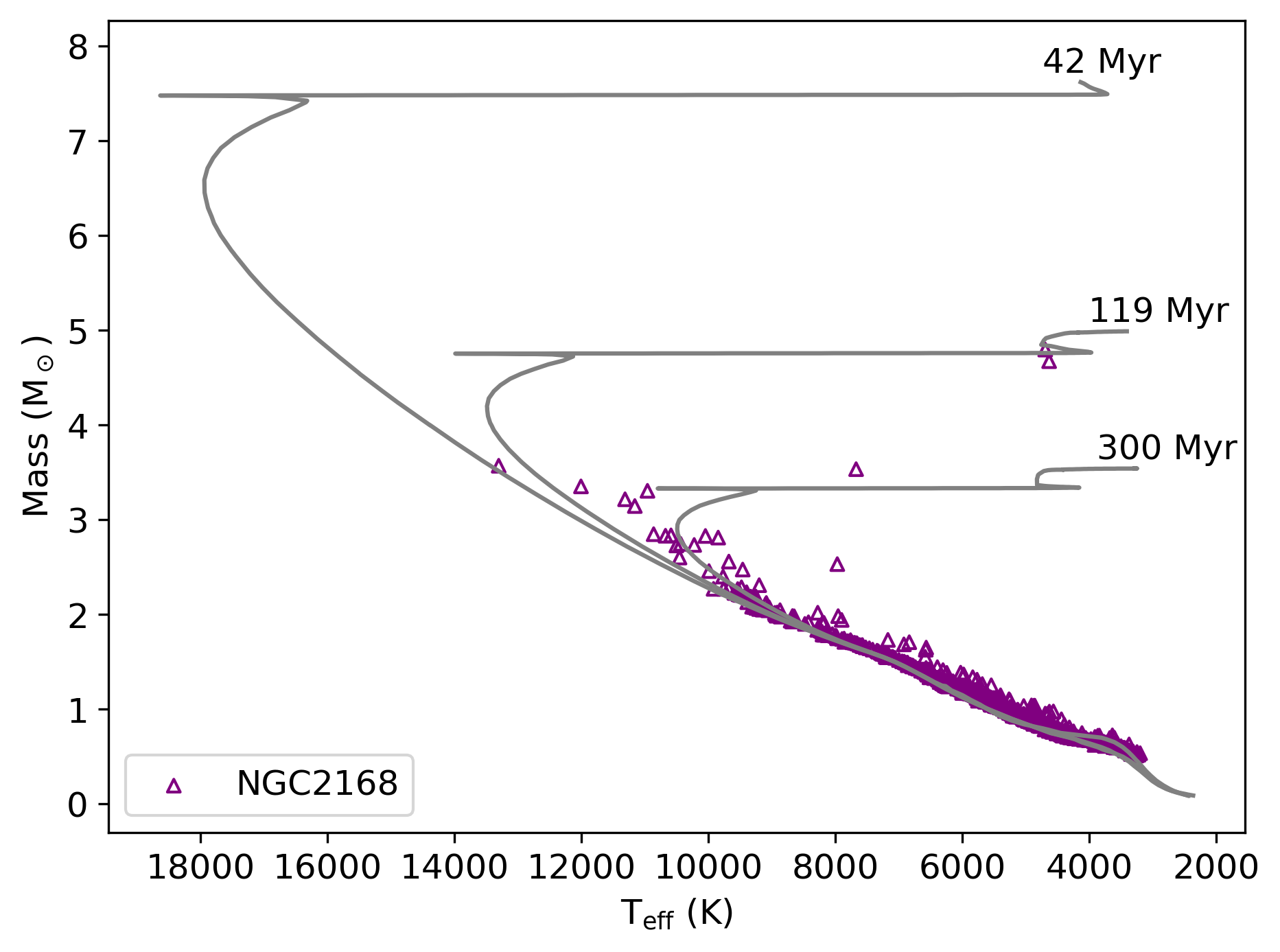}
\includegraphics[width=\columnwidth]{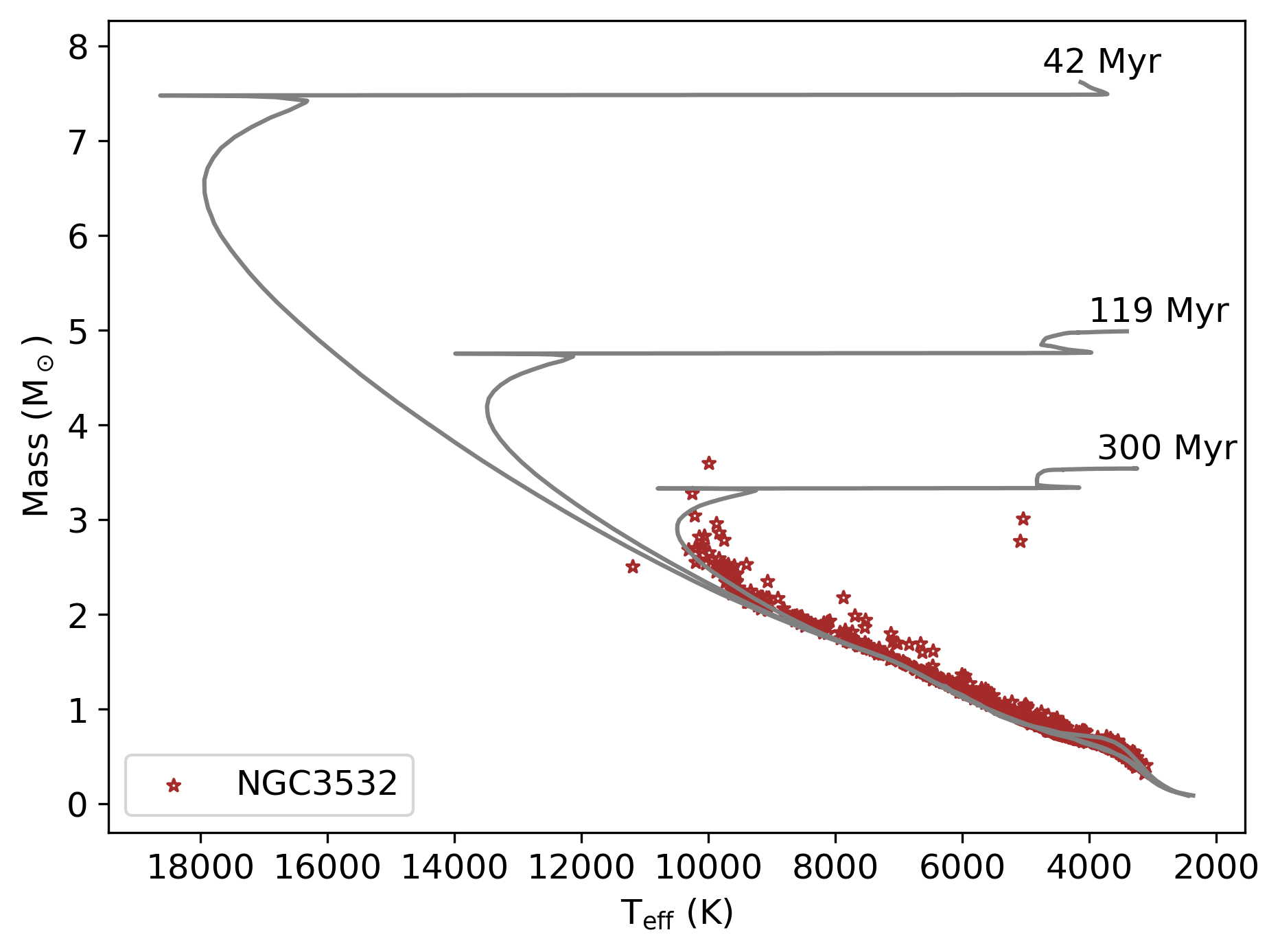}
\includegraphics[width=\columnwidth]{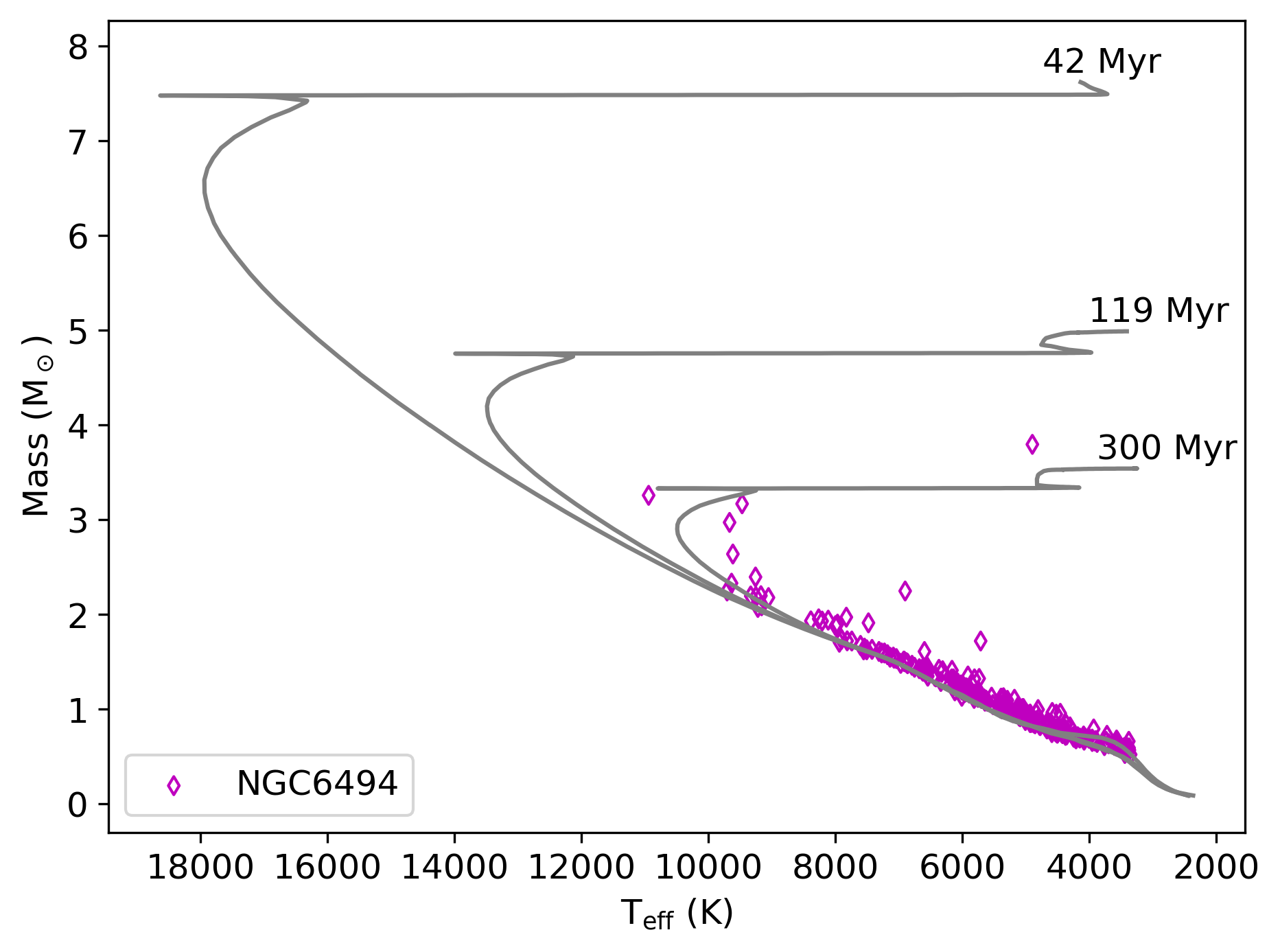}
 \caption{The same as Fig. \ref{fig:mteff}.}
    \label{fig:Amteff}
\end{figure*}
%%-----------------end Fig.6

%%%%%--------------------------------- Fig. B5 color-magnitude diagram 
\begin{figure*}
  \includegraphics[width=0.67\columnwidth]{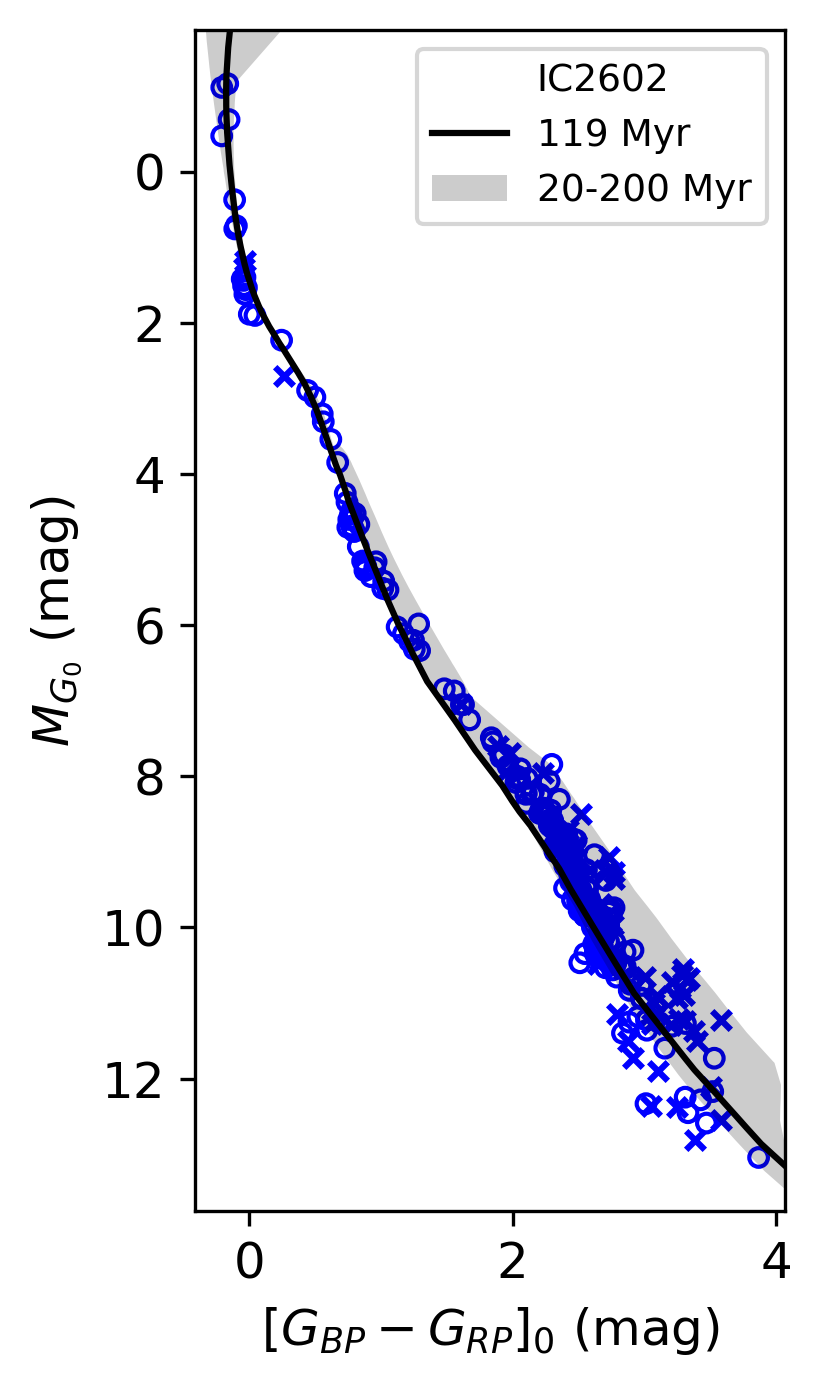}
\includegraphics[width=0.69\columnwidth]{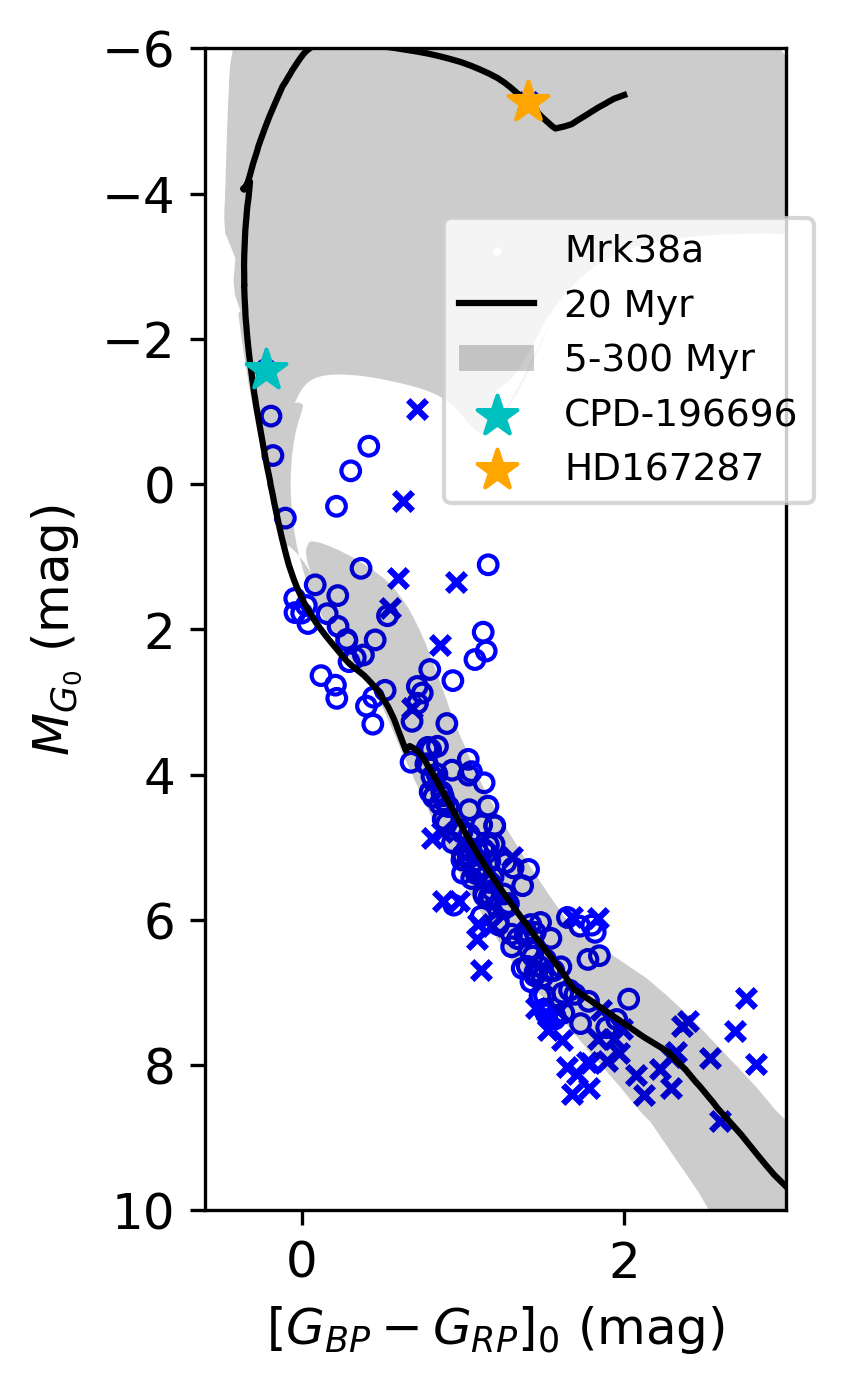}
\includegraphics[width=0.67\columnwidth]{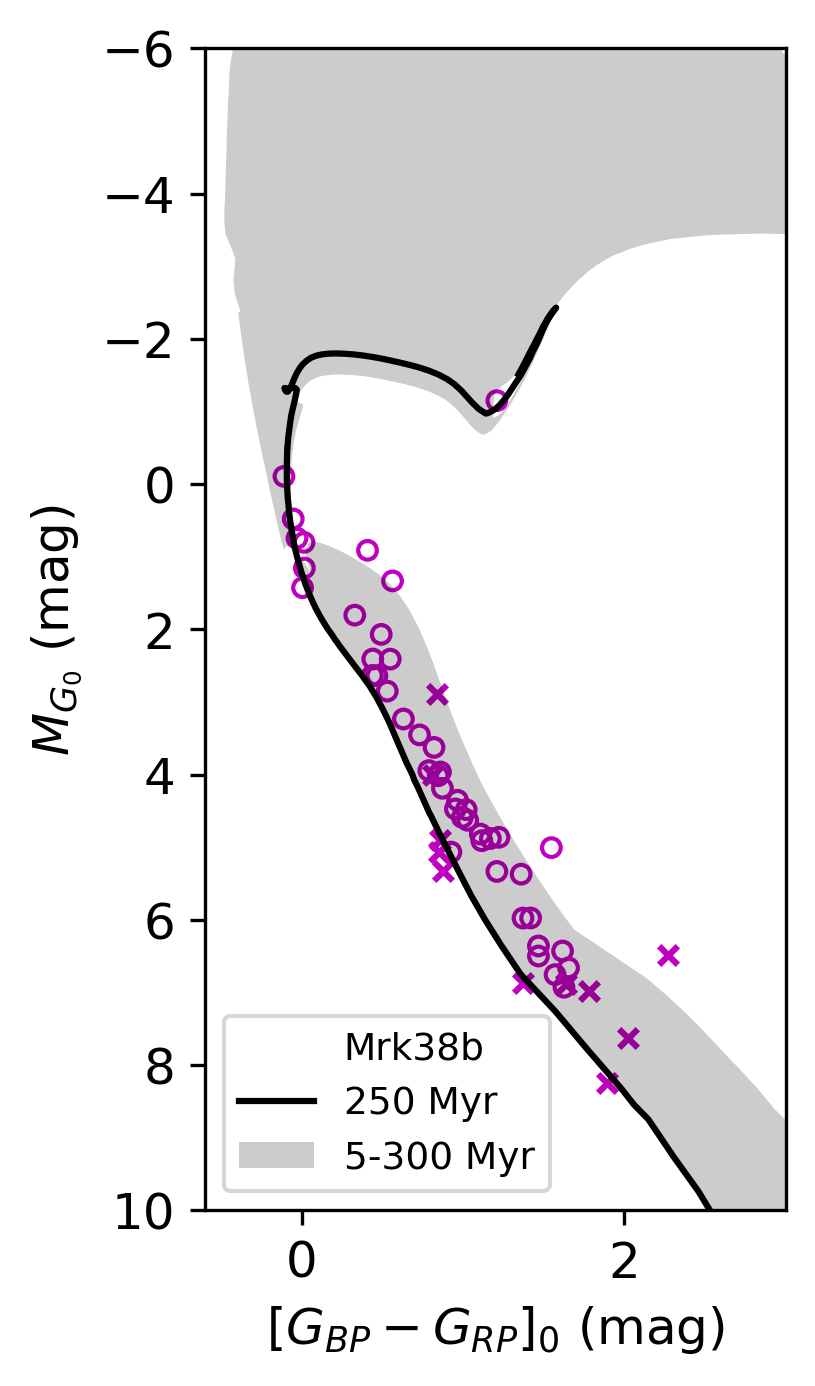}
   \includegraphics[width=0.68\columnwidth]{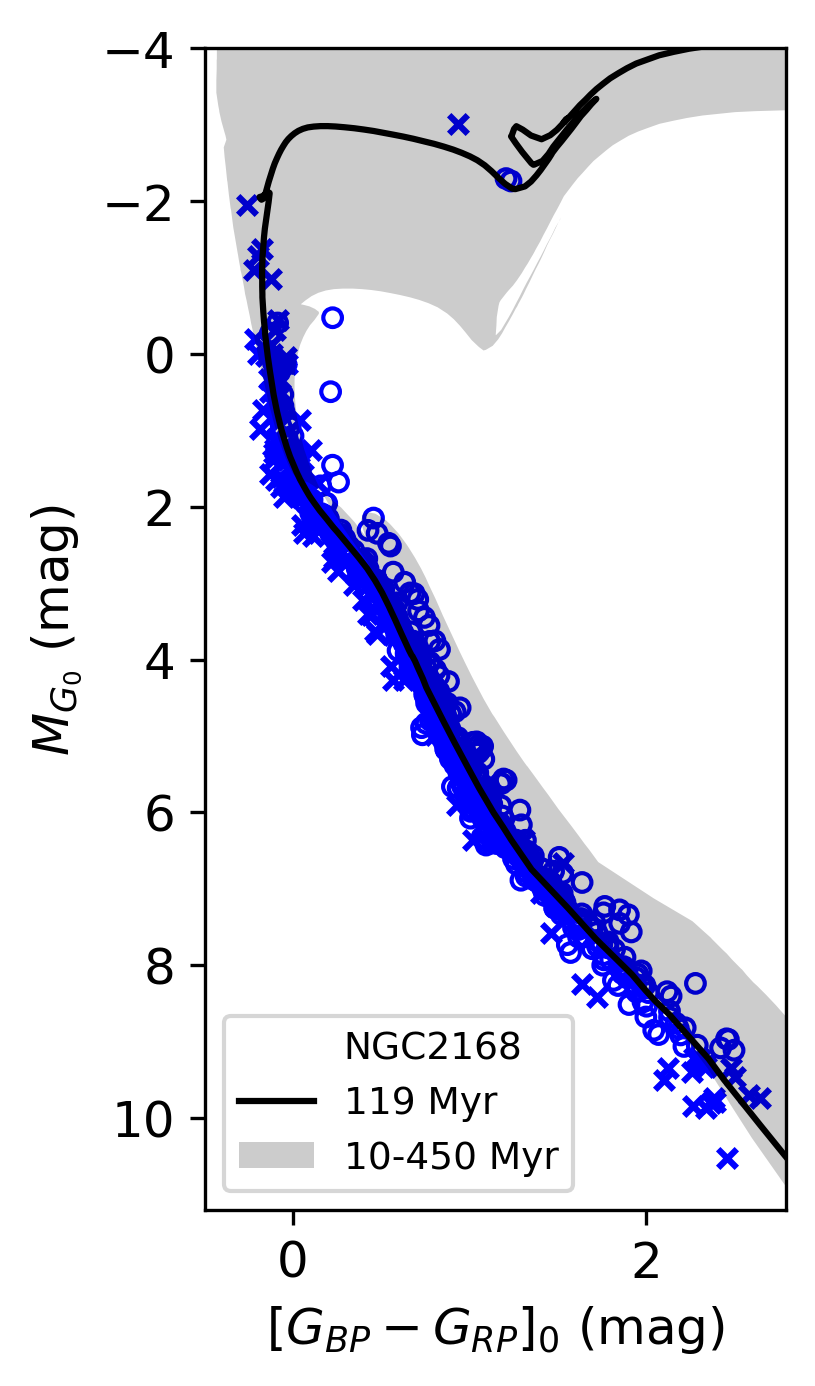}
     \includegraphics[width=0.68\columnwidth]{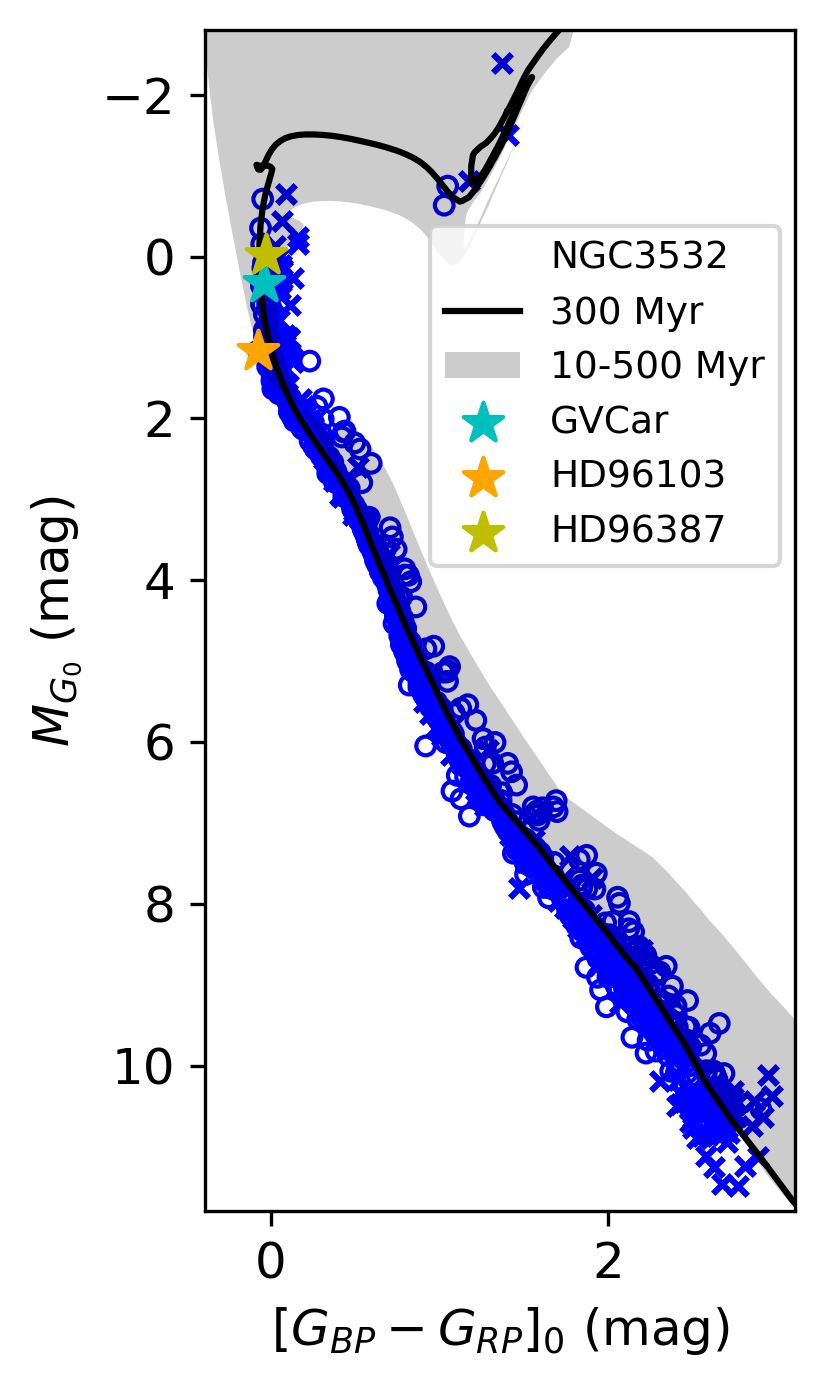}
   \includegraphics[width=0.68\columnwidth]{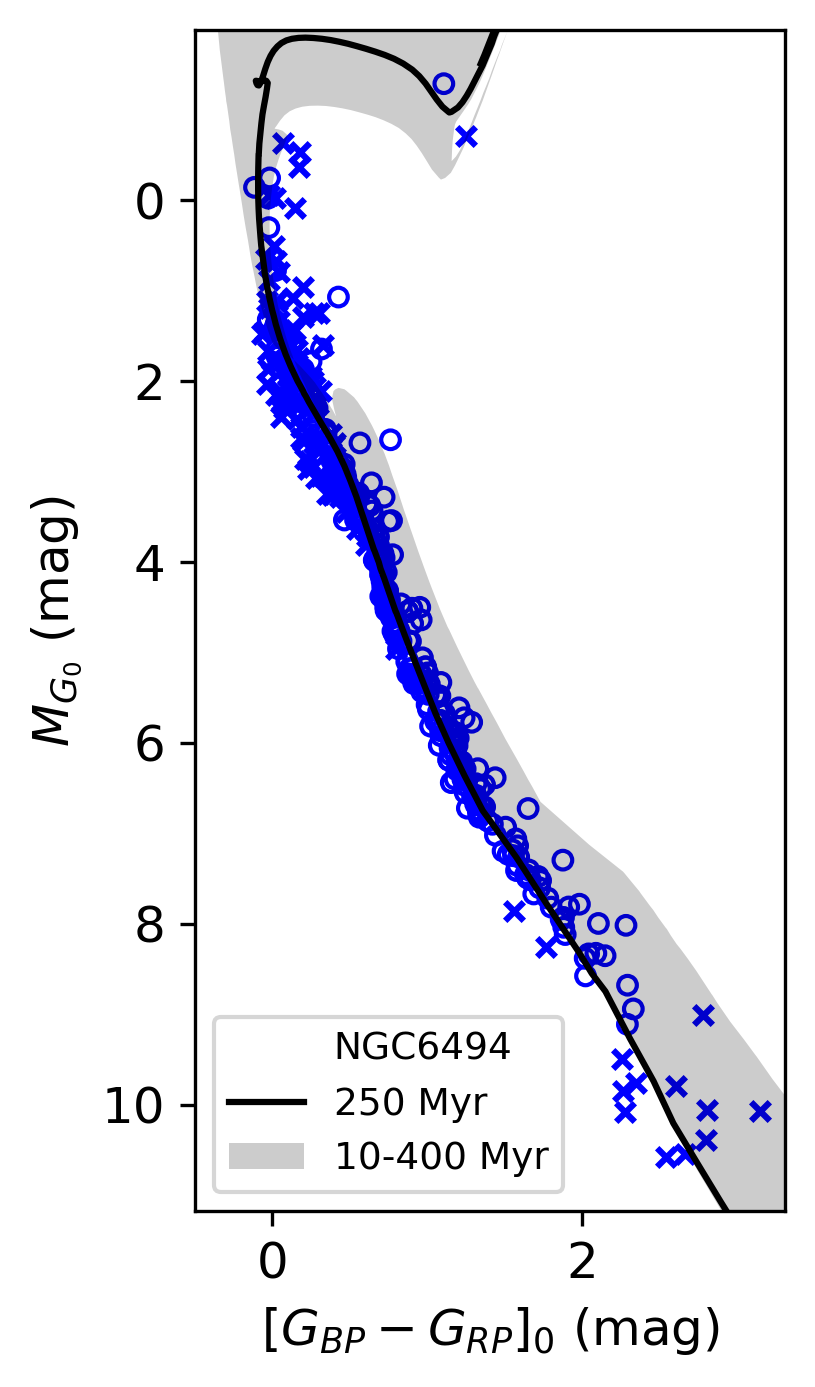}
   
 \caption{The same as Fig. \ref{fig:cormag}. }
    \label{fig:Acormag}
\end{figure*}

%%%%--------------------------------- Fig. B6  Lambda x N  + Sigma x Mass
\begin{figure*}
\includegraphics[width=0.65\columnwidth]{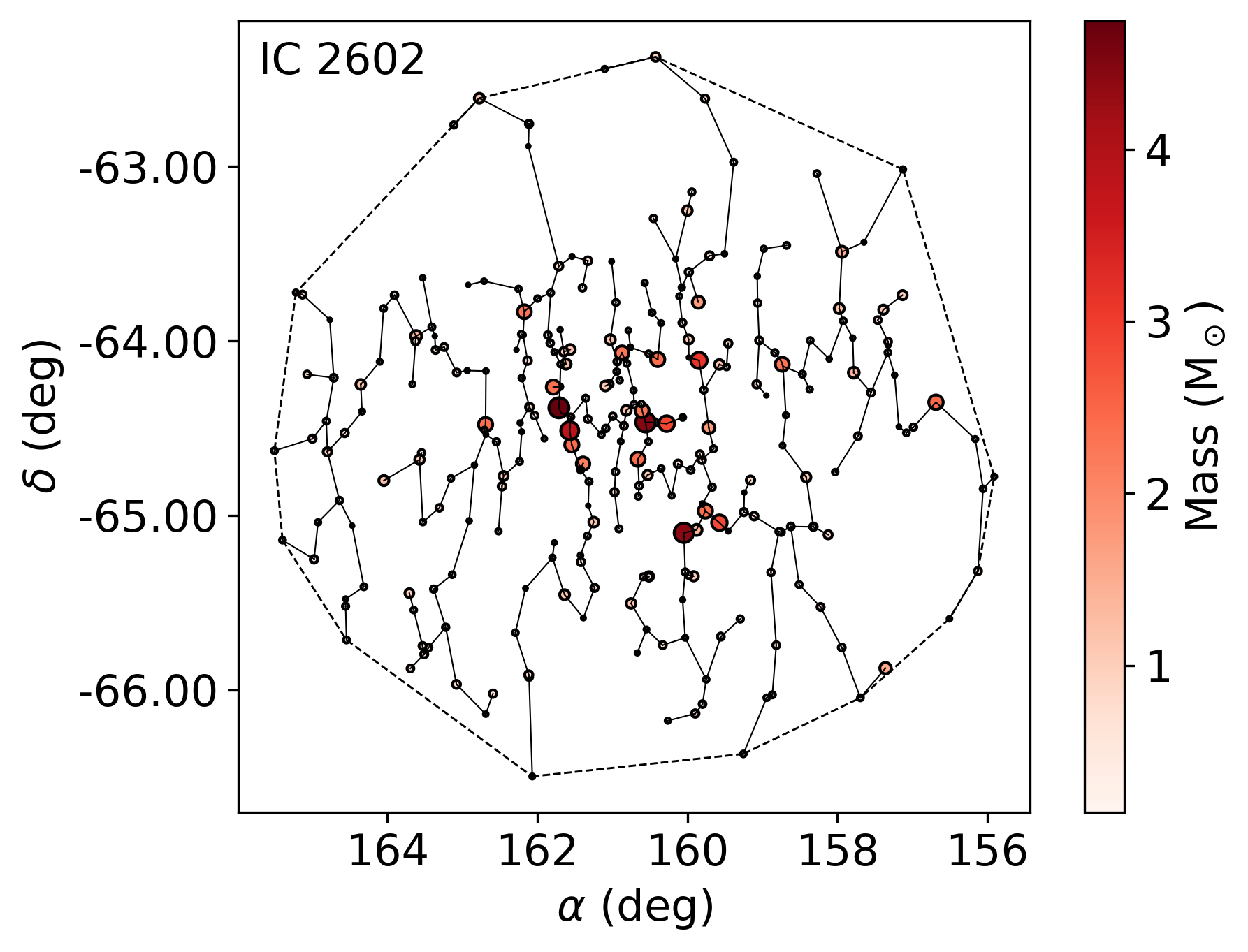}
\includegraphics[width=0.65\columnwidth]{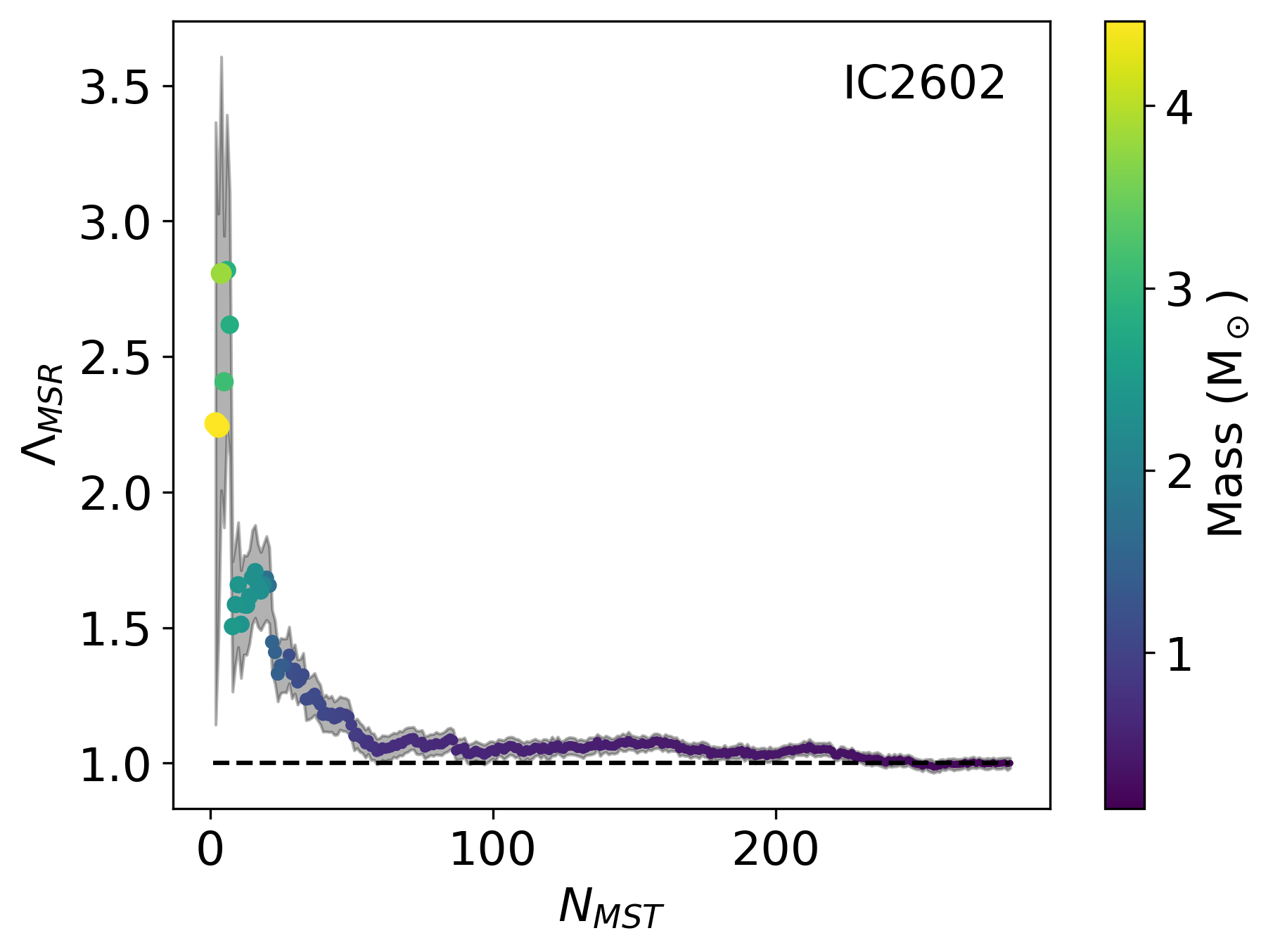}
\includegraphics[width=0.65\columnwidth]{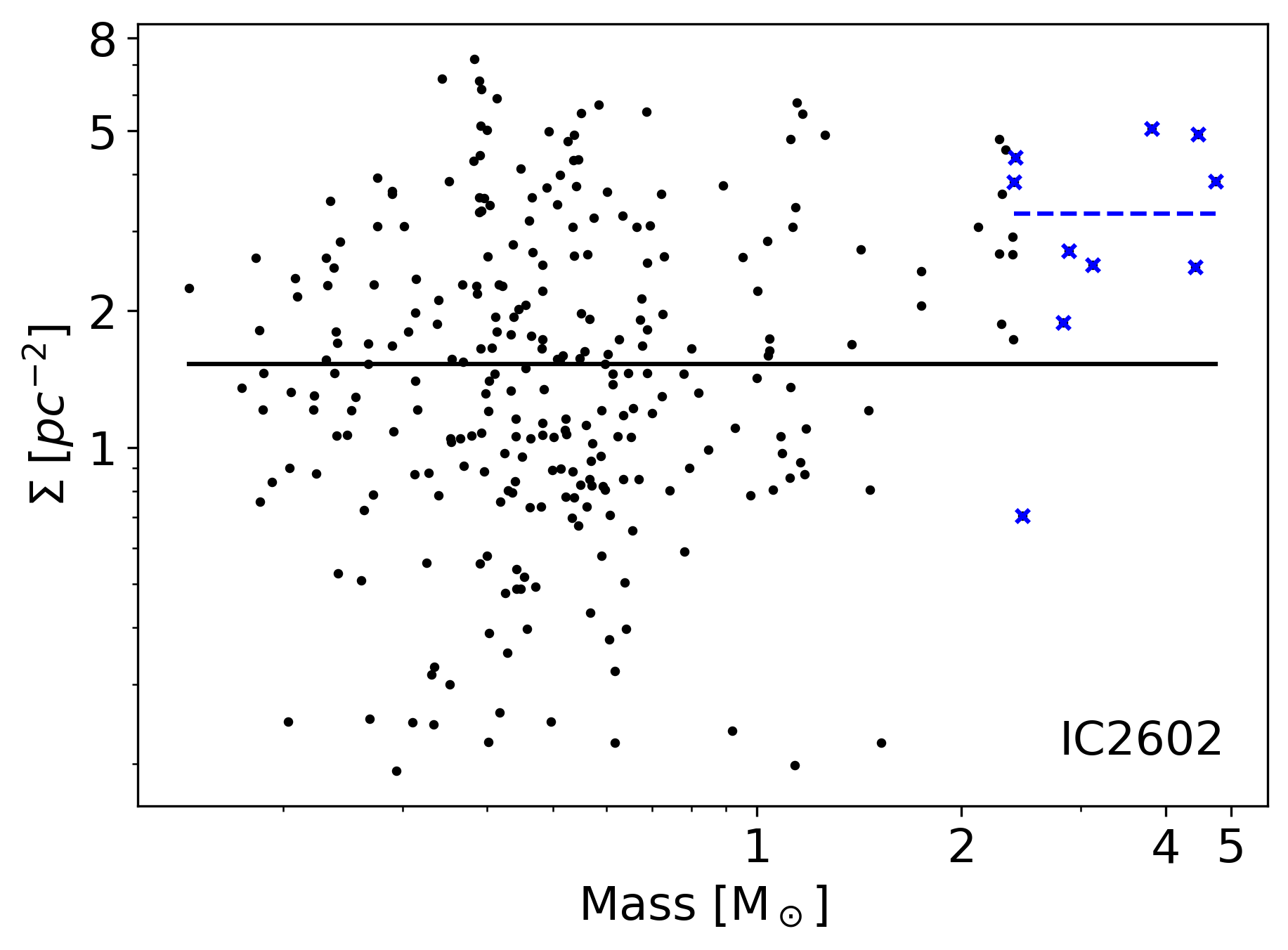}
\includegraphics[width=0.65\columnwidth]{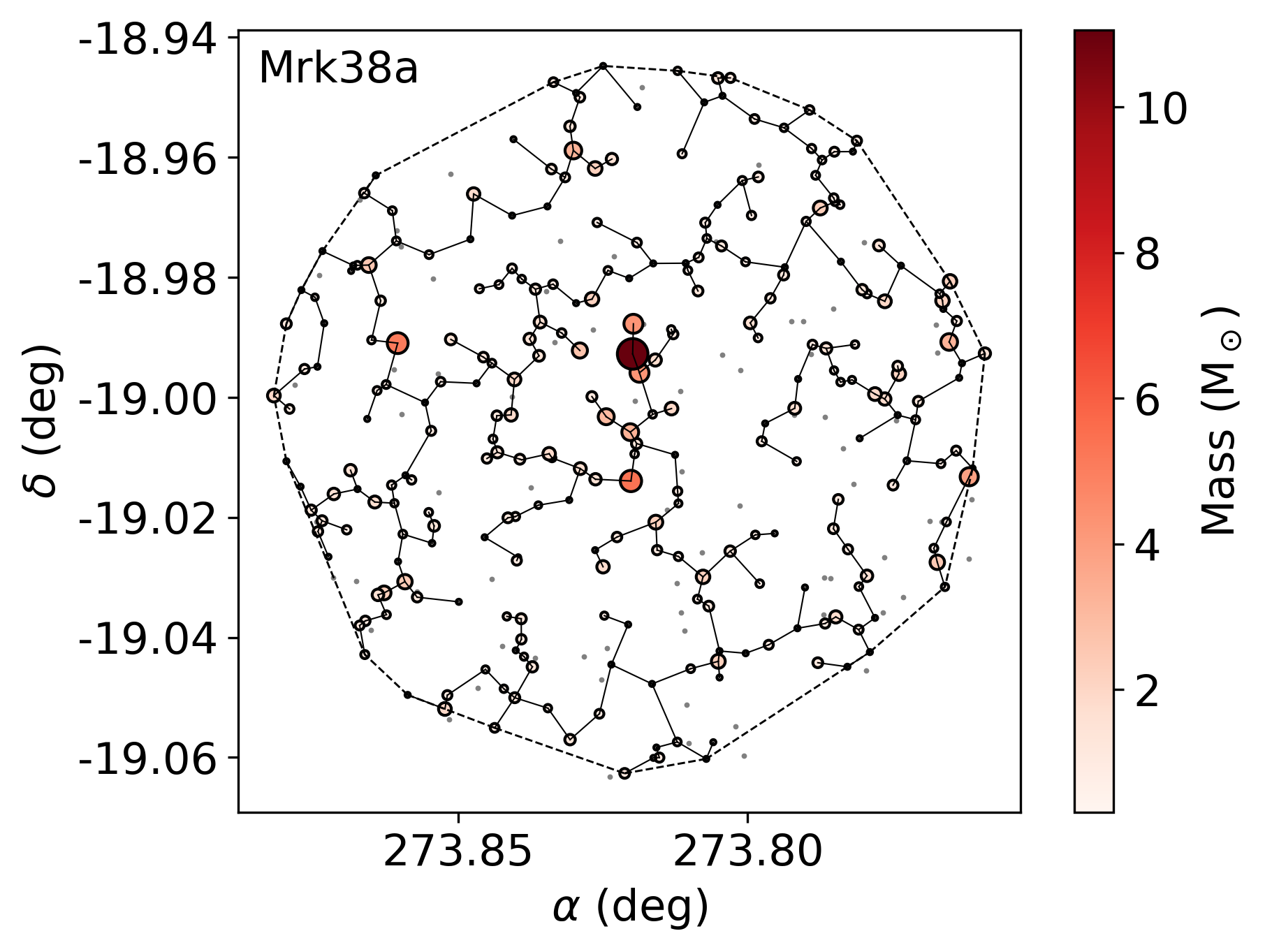}
\includegraphics[width=0.65\columnwidth]{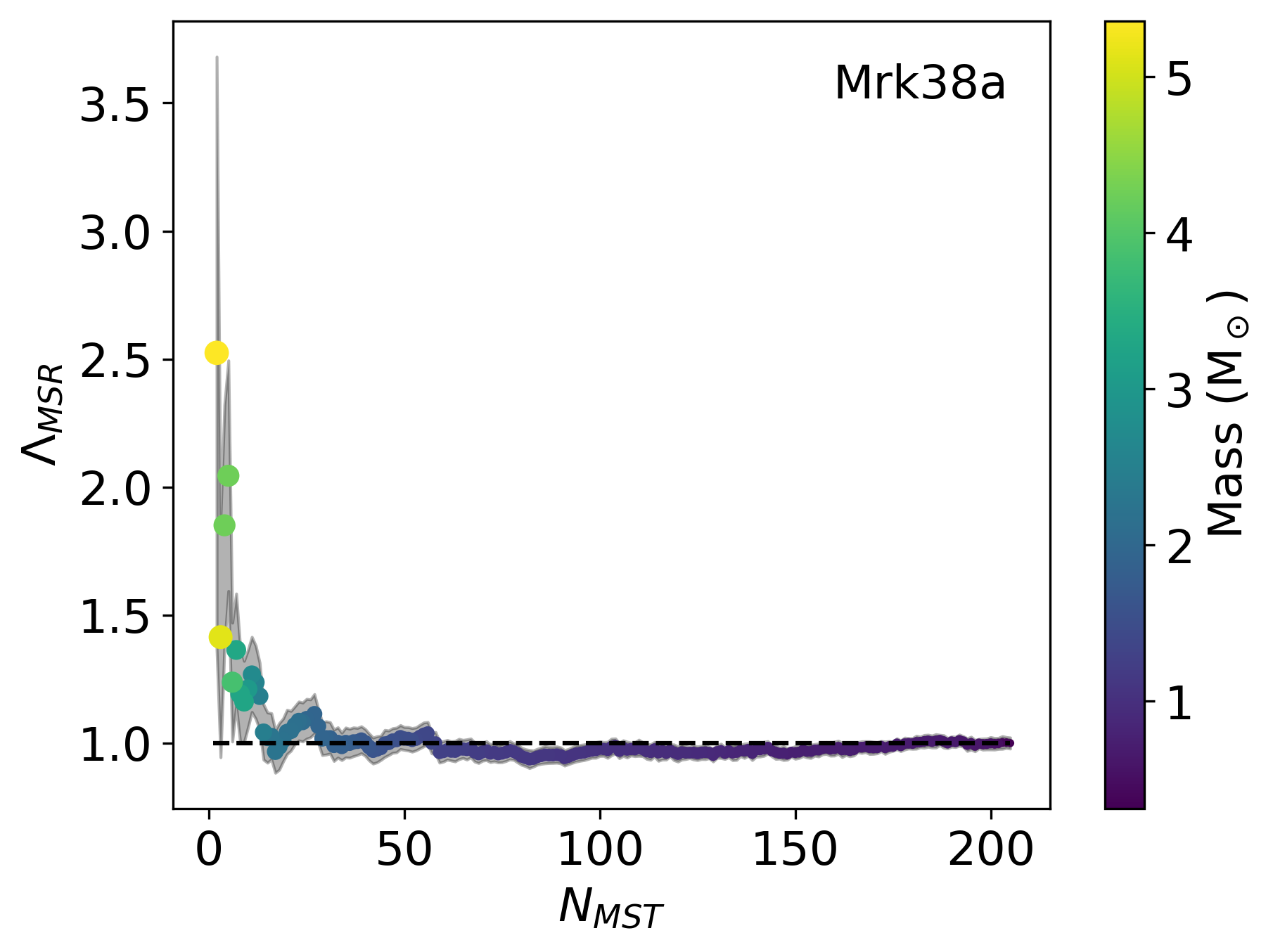}
\includegraphics[width=0.65\columnwidth]{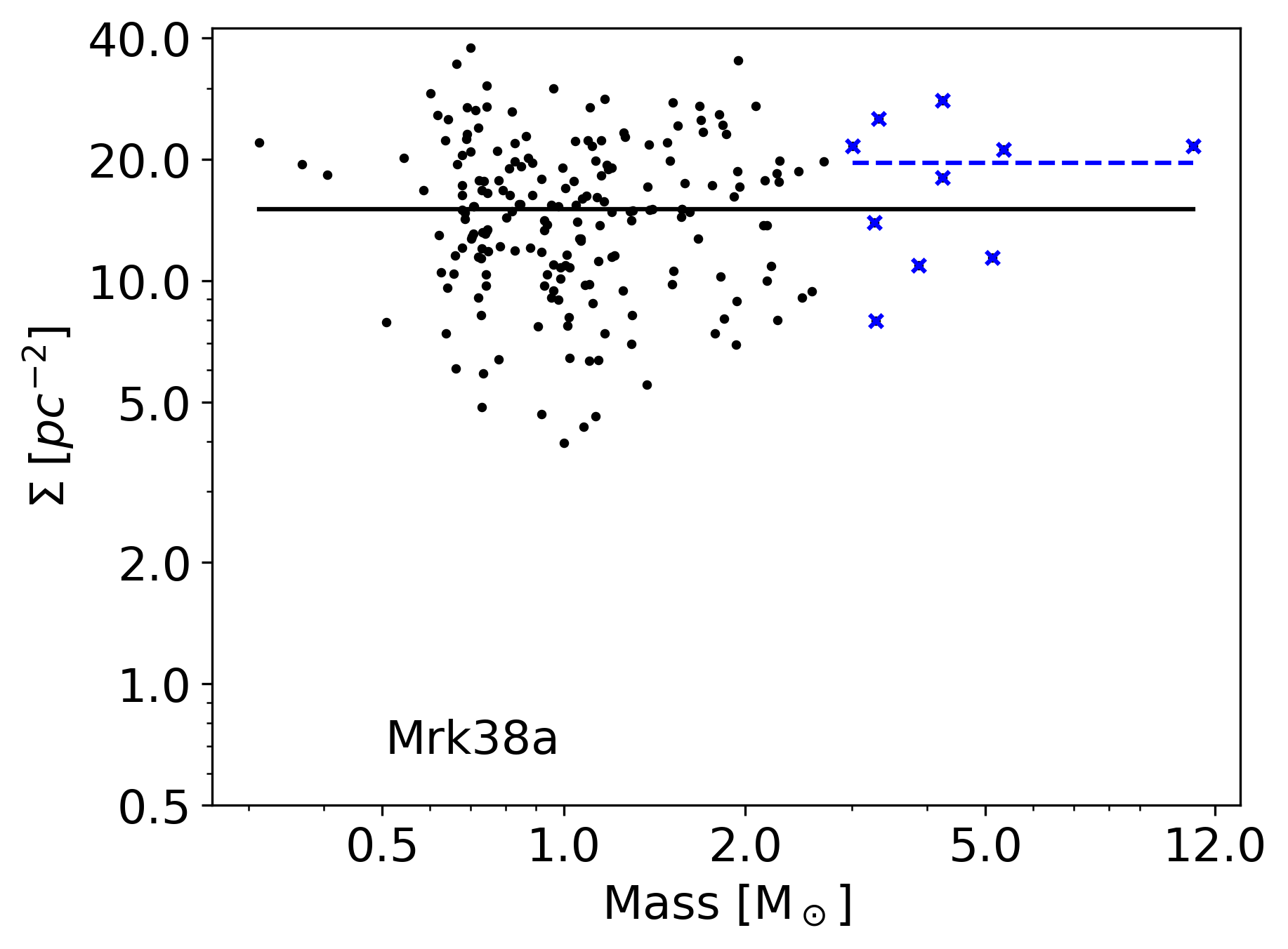}
\includegraphics[width=0.65\columnwidth]{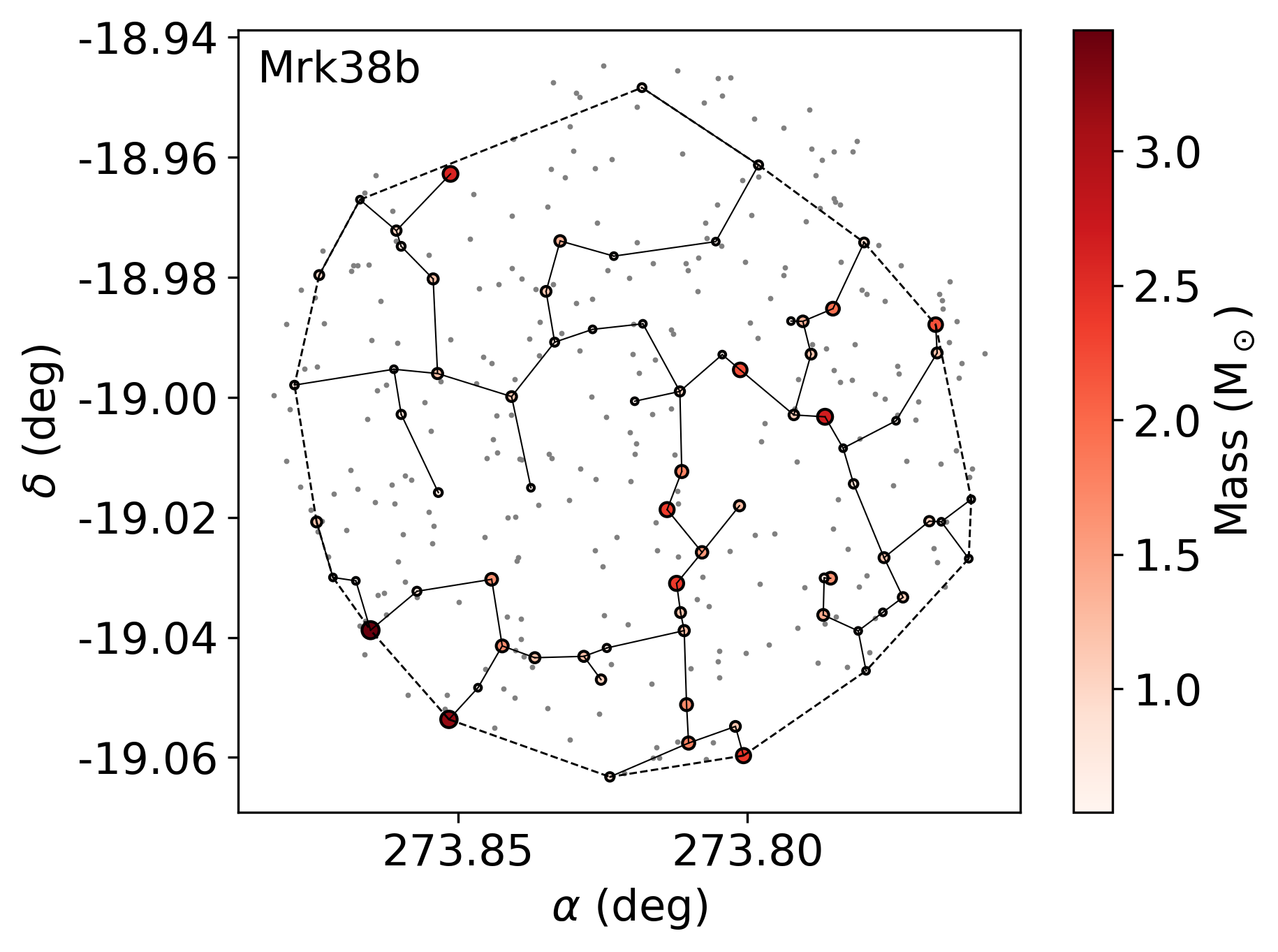}
\includegraphics[width=0.65\columnwidth]{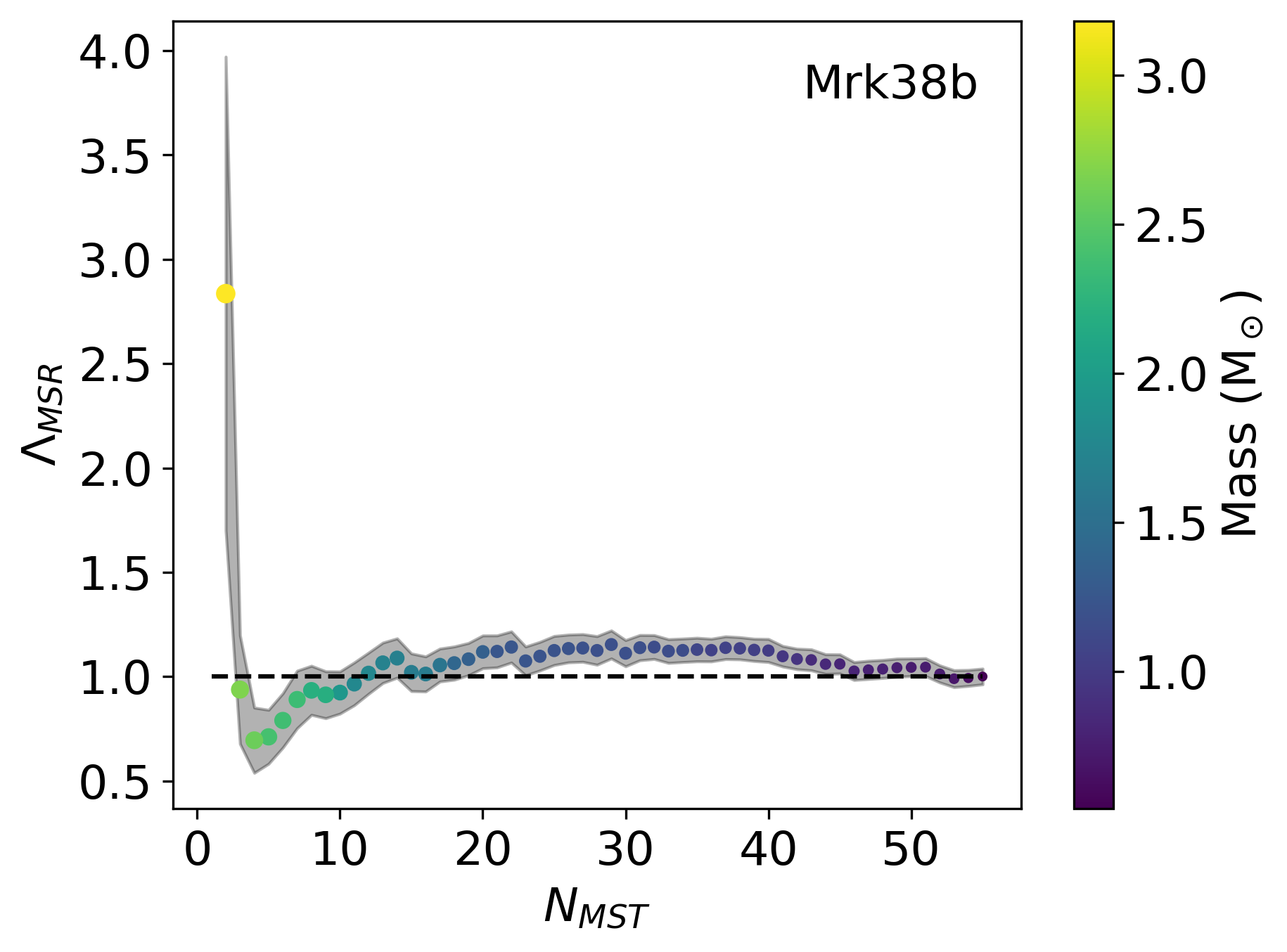}
\includegraphics[width=0.65\columnwidth]{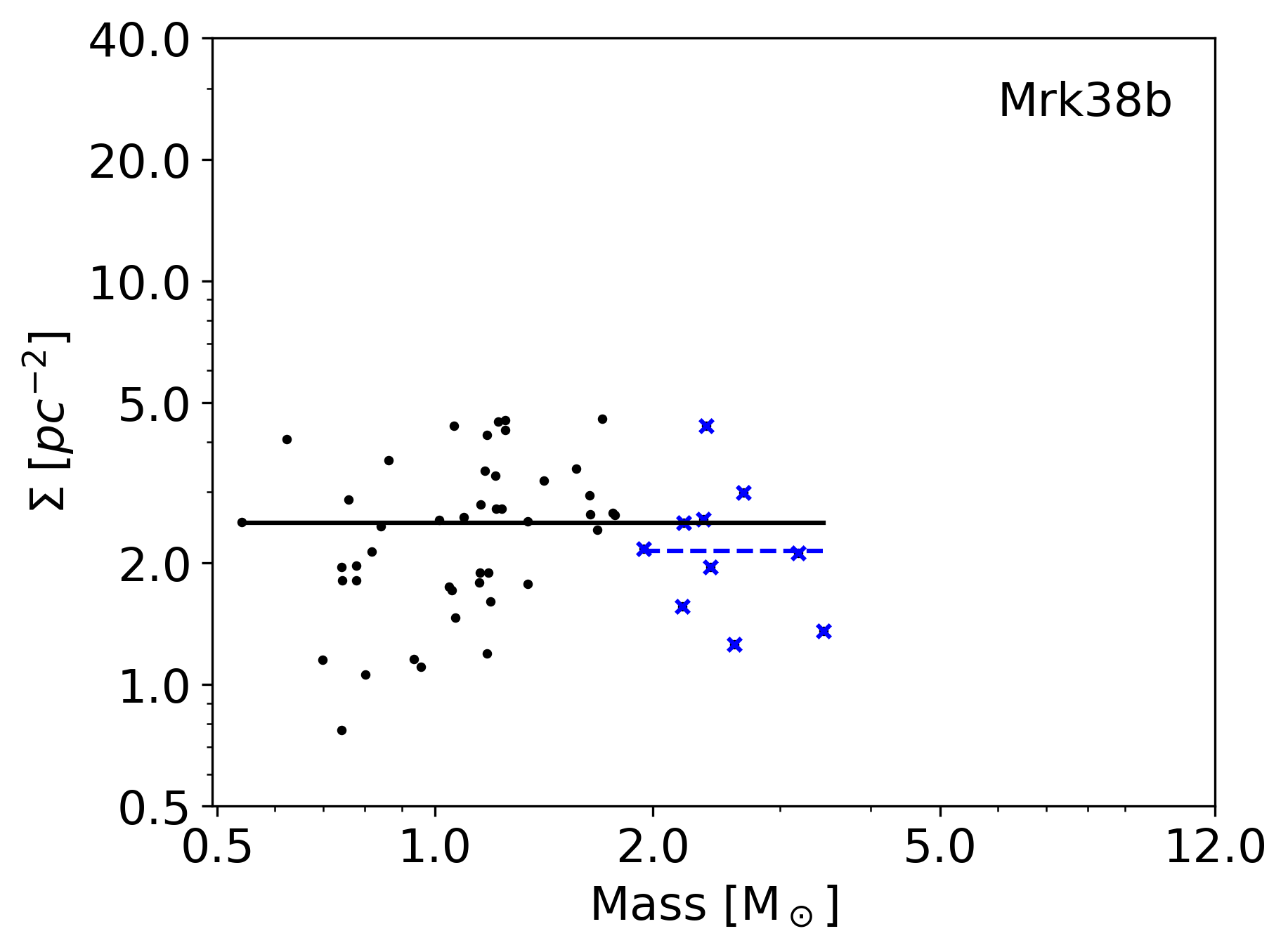}
\includegraphics[width=0.65\columnwidth]{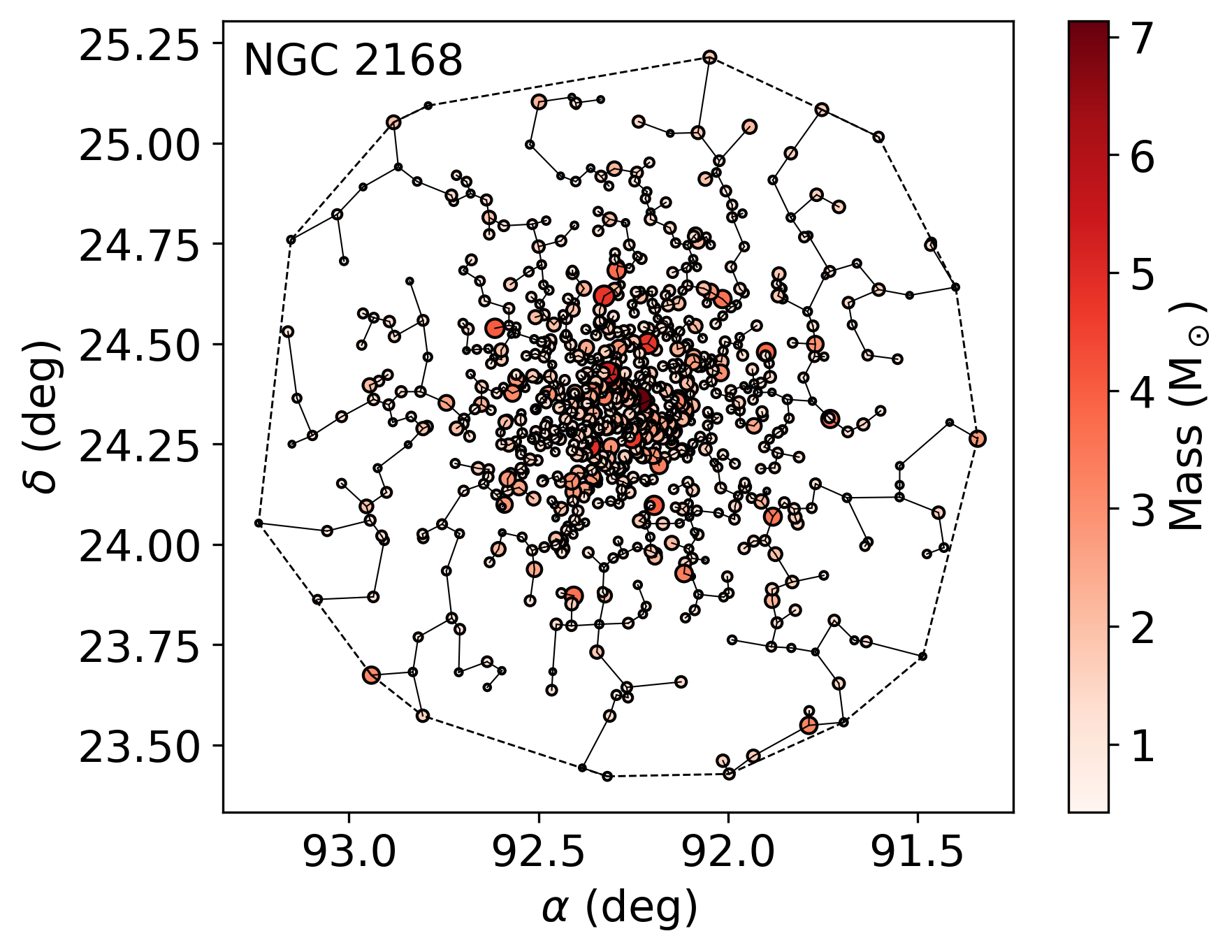}
\includegraphics[width=0.65\columnwidth]{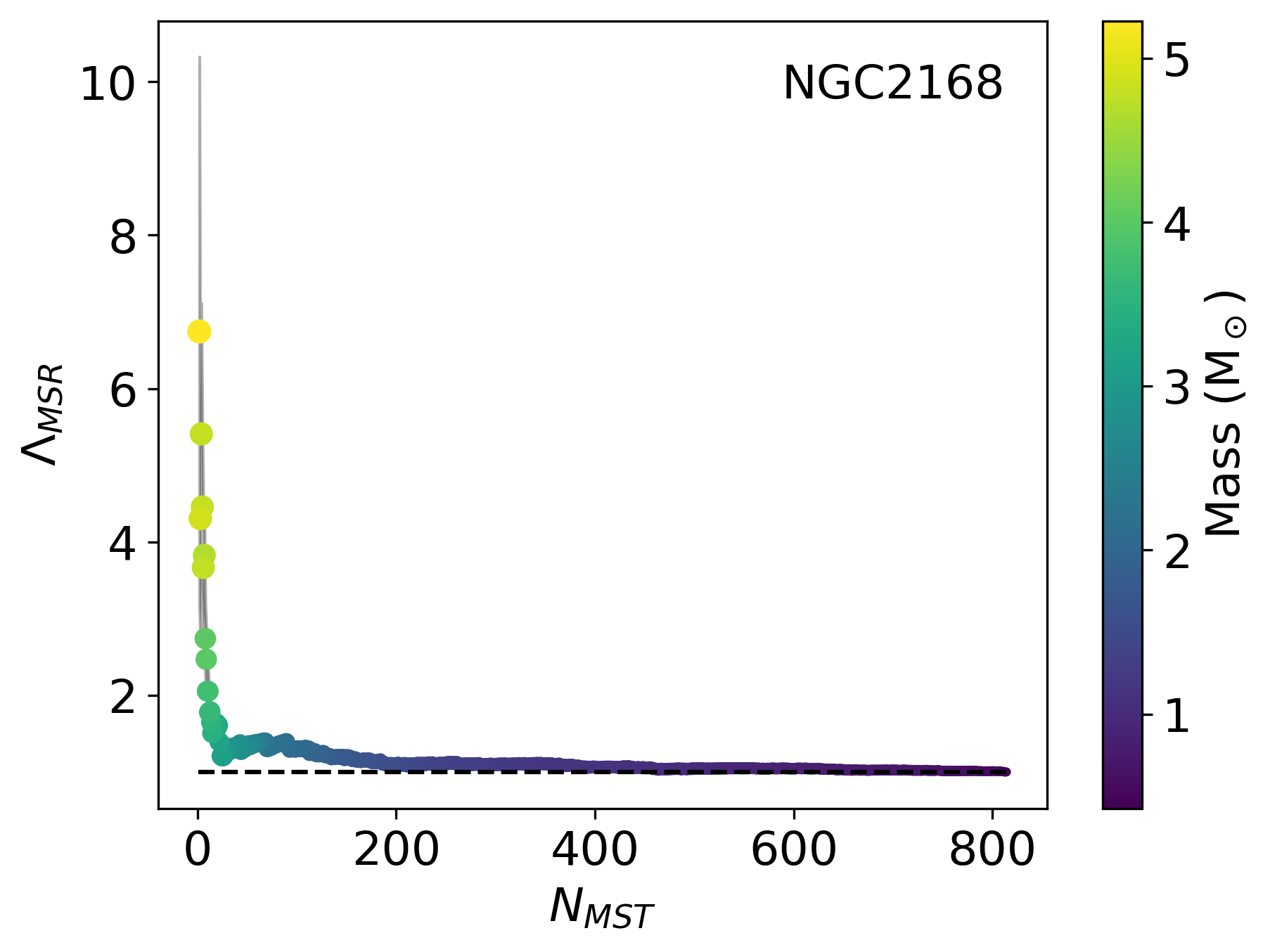}
\includegraphics[width=0.65\columnwidth]{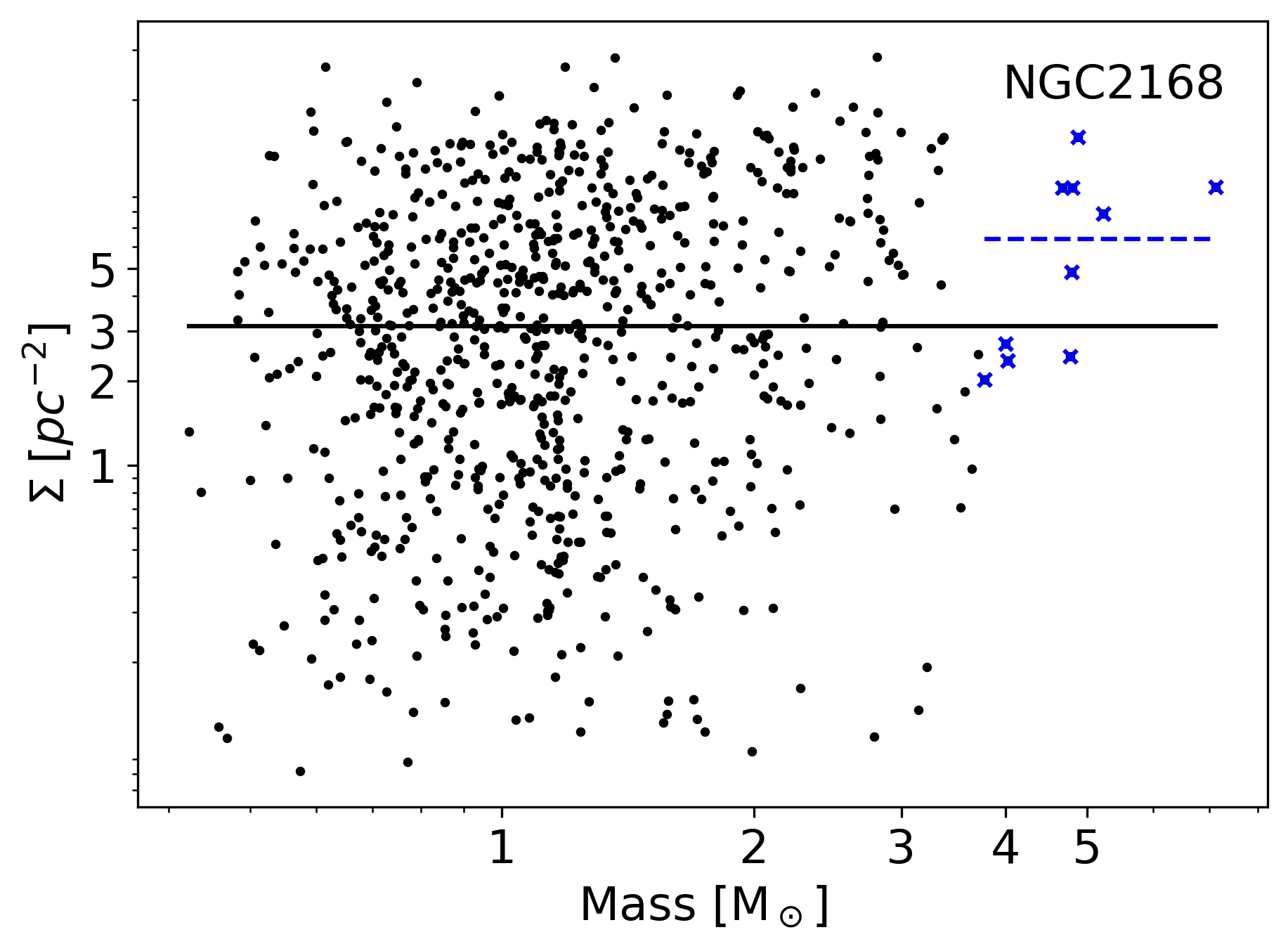}
\caption{The same as Fig. \ref{fig:Lambda}. For comparison, only the  $\Sigma - m$ plots for Mrk38a and Mrk38b are displayed at the same scale. To evidence the offset between the full line (average obtained for all the cluster members) and the dashed line (average estimated for the 10 most massive stars), the same scale is not adopted for all the other clusters since a clear separation between these lines would not appear in the plots.}
\label{fig:ALambda1}
\end{figure*}
%%-----------------------end 
%%%%--------------------------------- Fig. B7  Lambda x N  + Sigma x Mass
\begin{figure*}
\includegraphics[width=0.65\columnwidth]{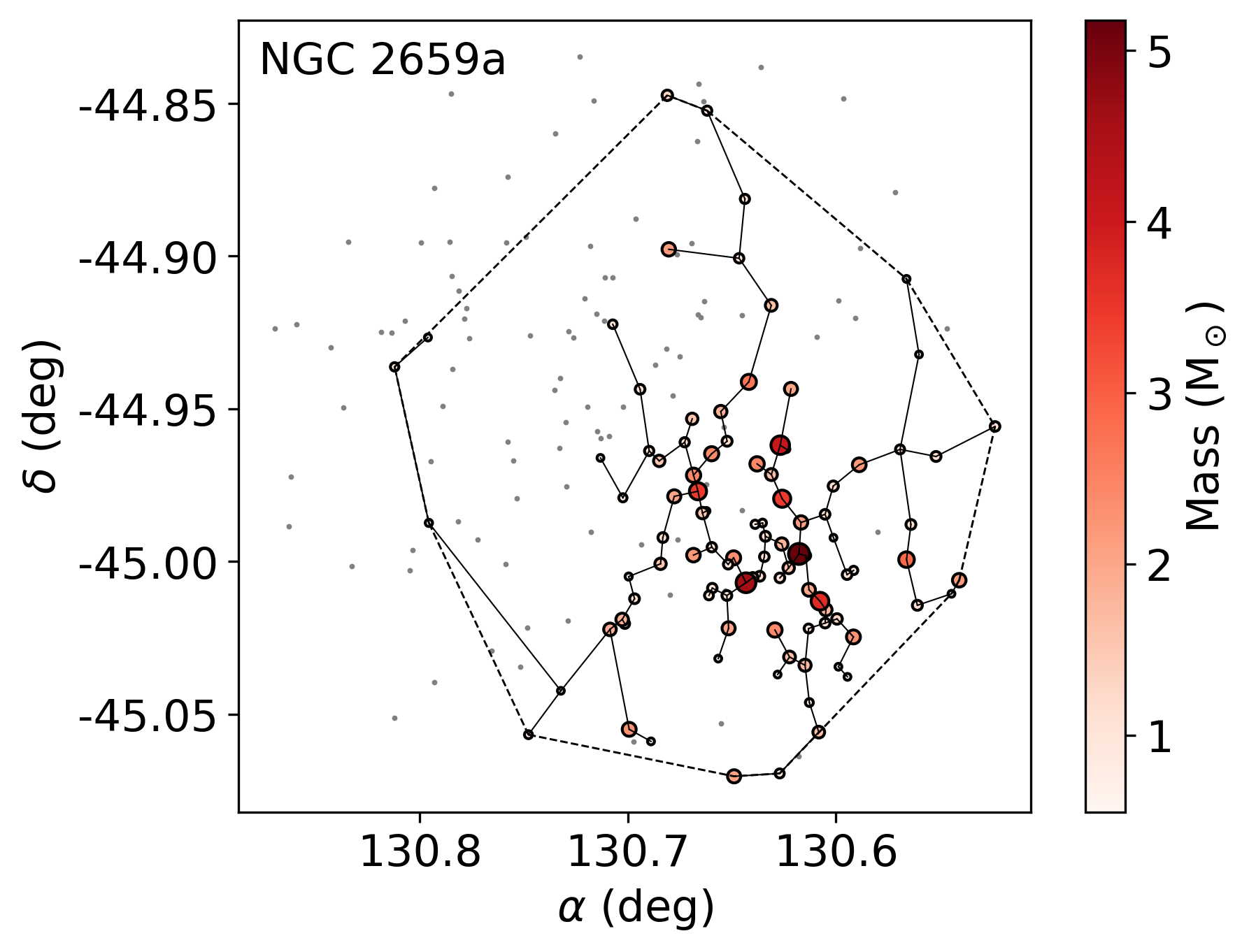}
\includegraphics[width=0.65\columnwidth]{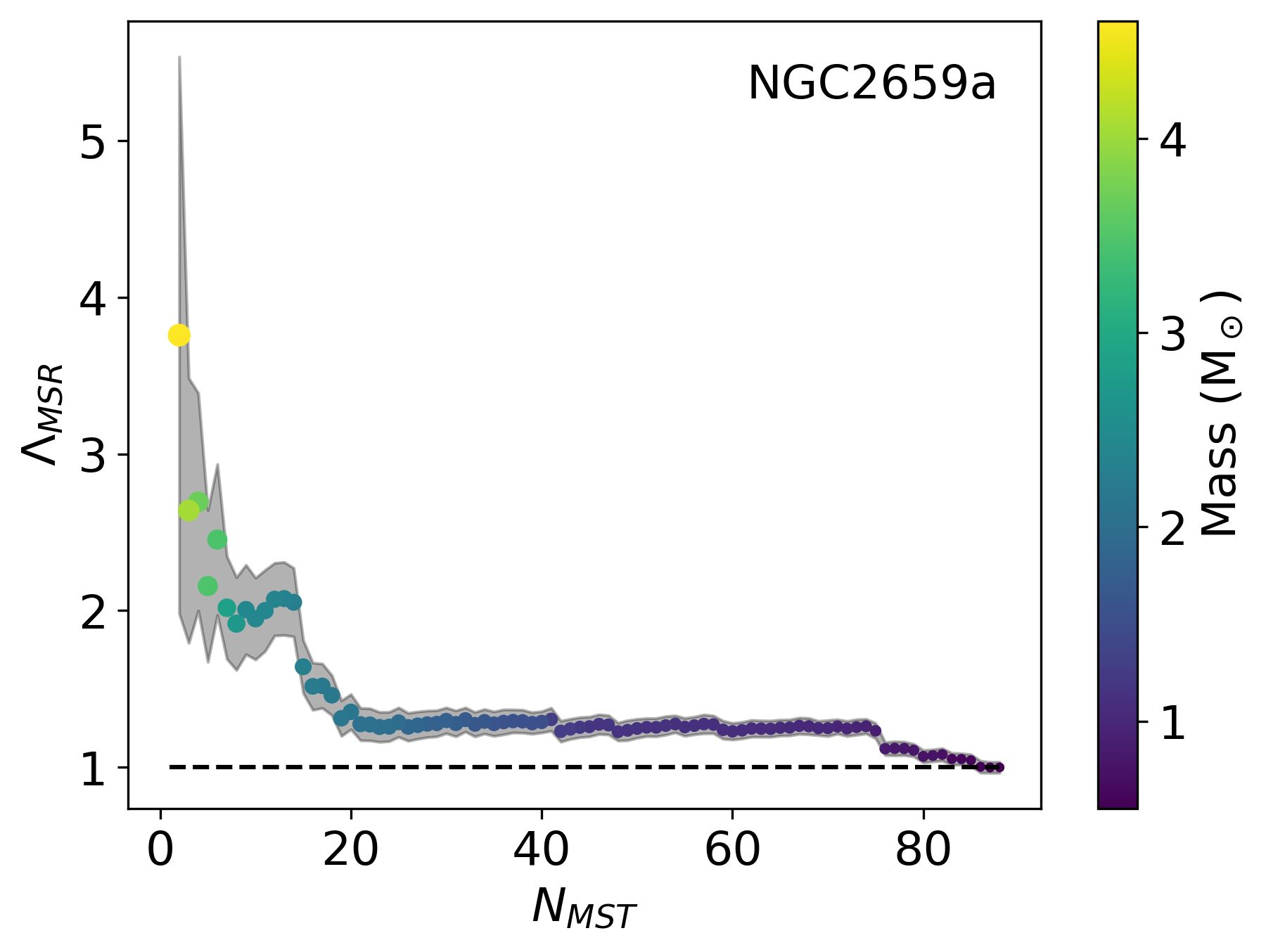}
\includegraphics[width=0.65\columnwidth]{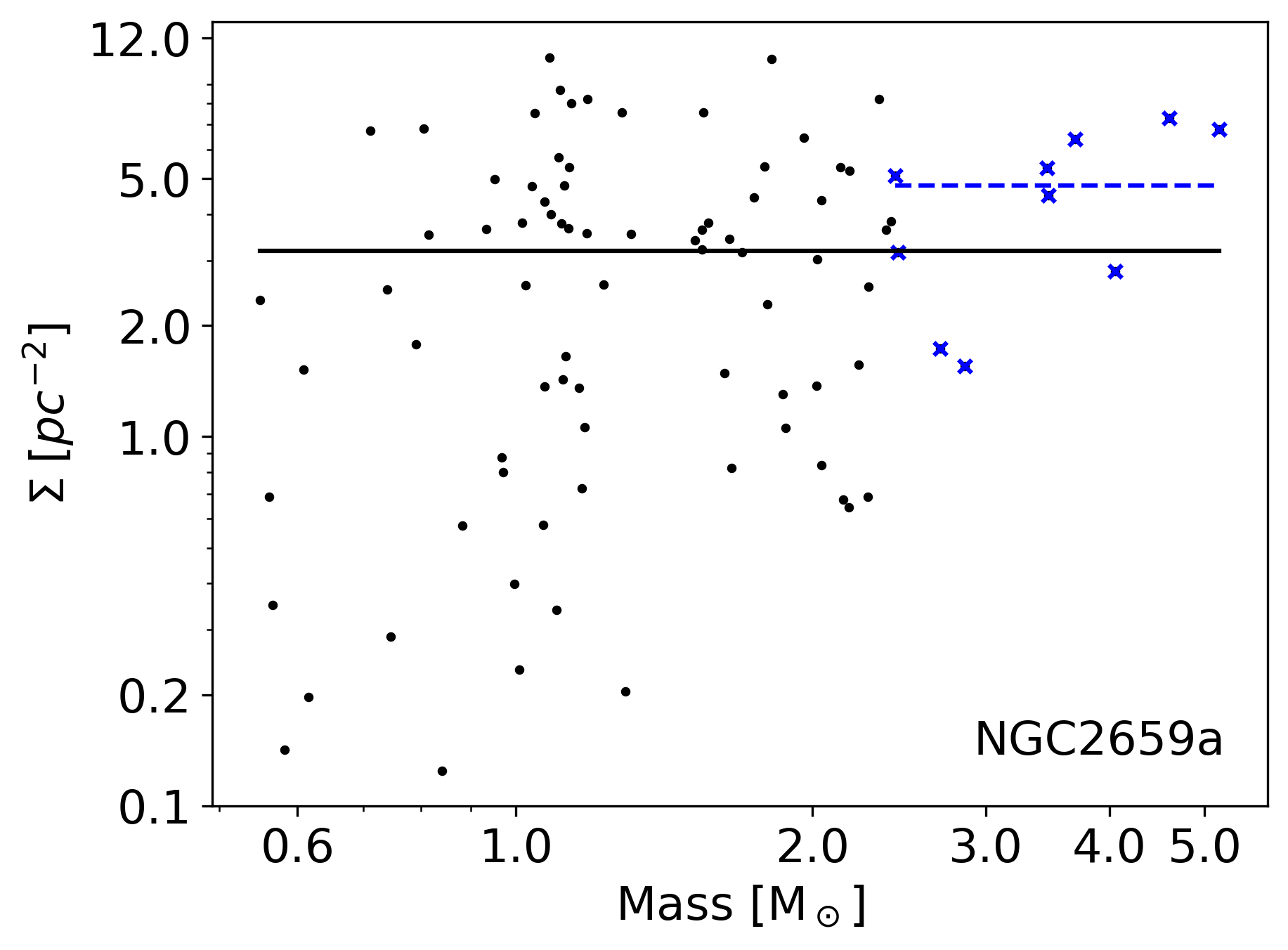}
\includegraphics[width=0.65\columnwidth]{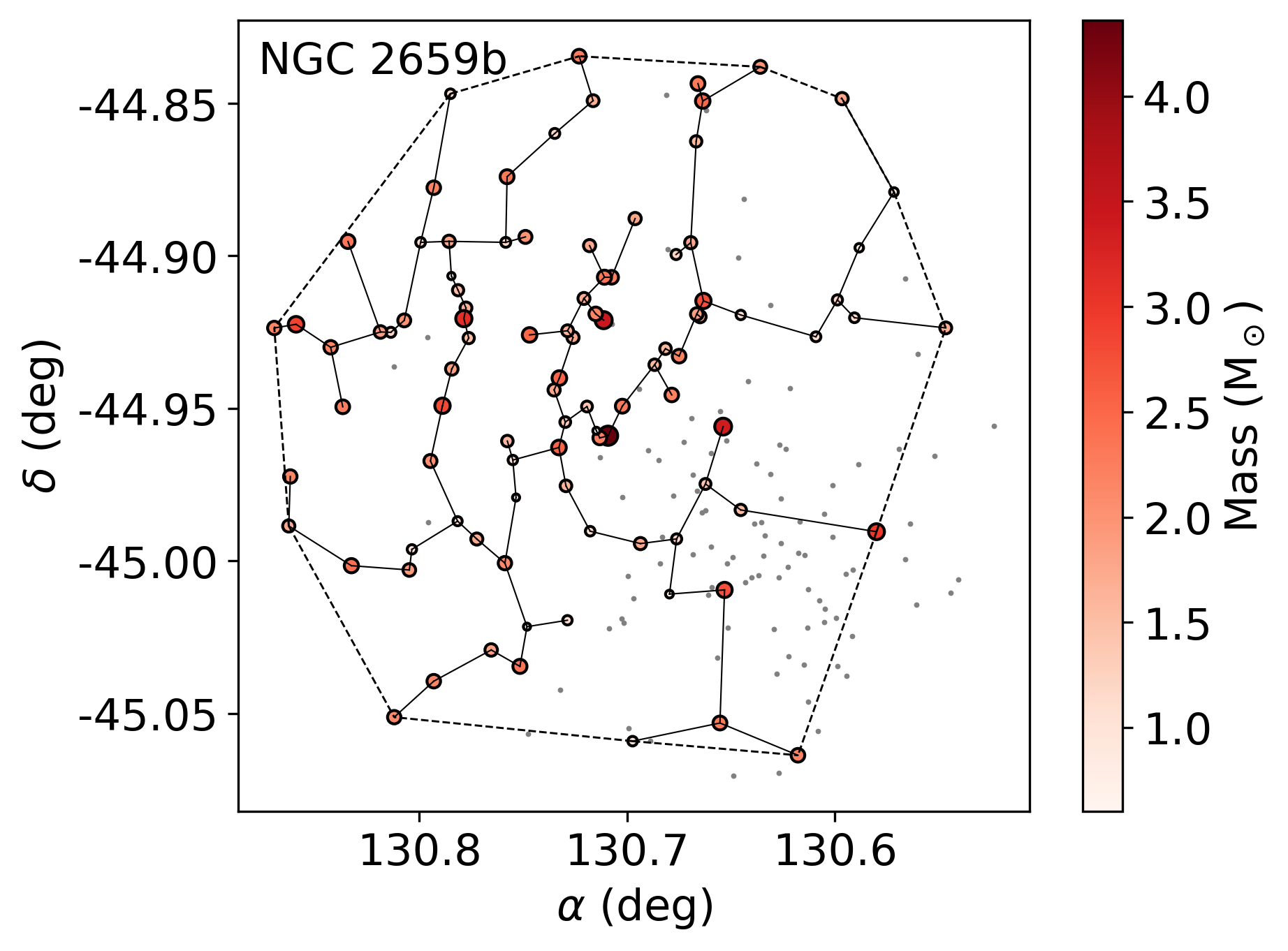}
\includegraphics[width=0.65\columnwidth]{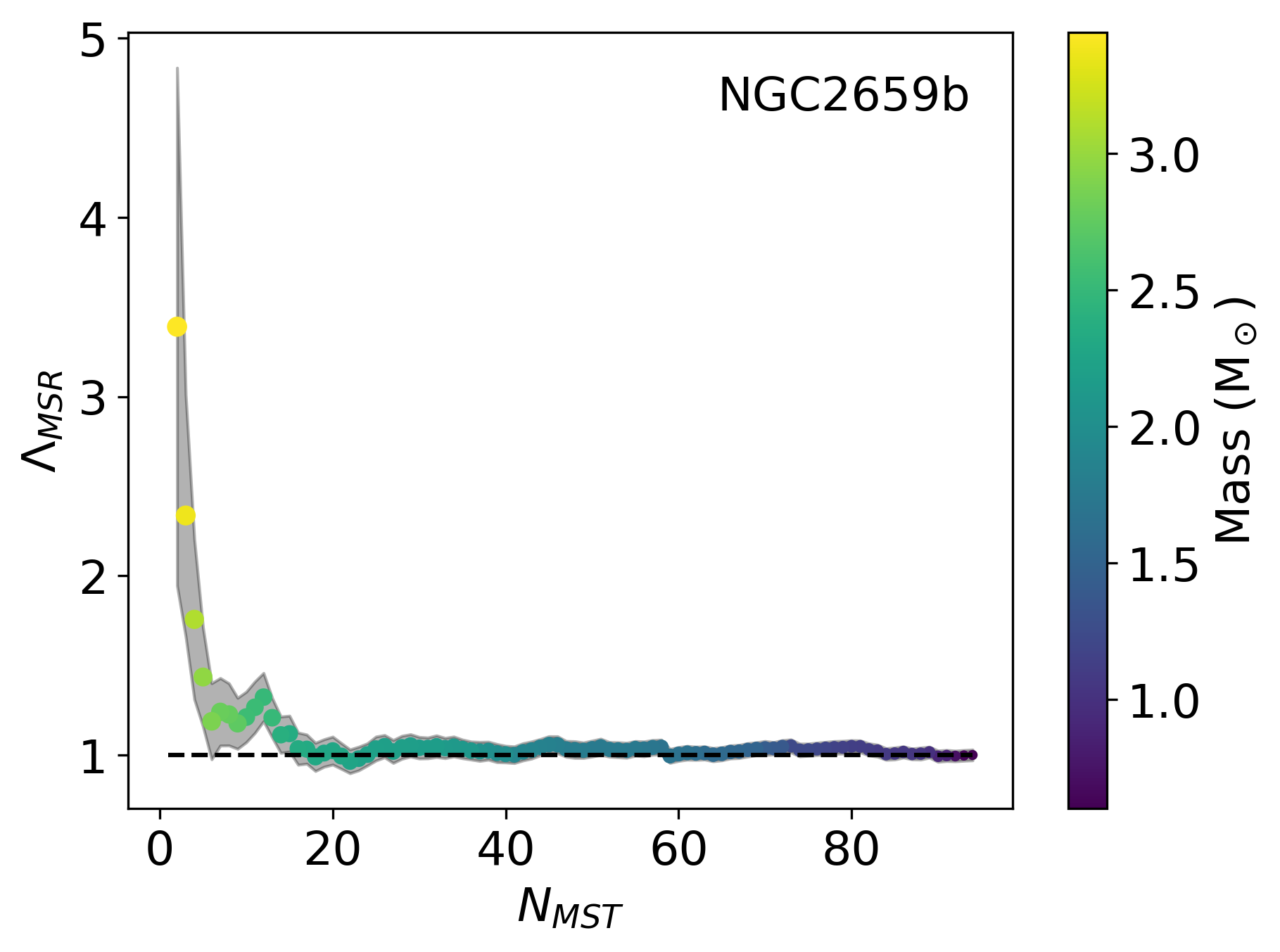}
\includegraphics[width=0.65\columnwidth]{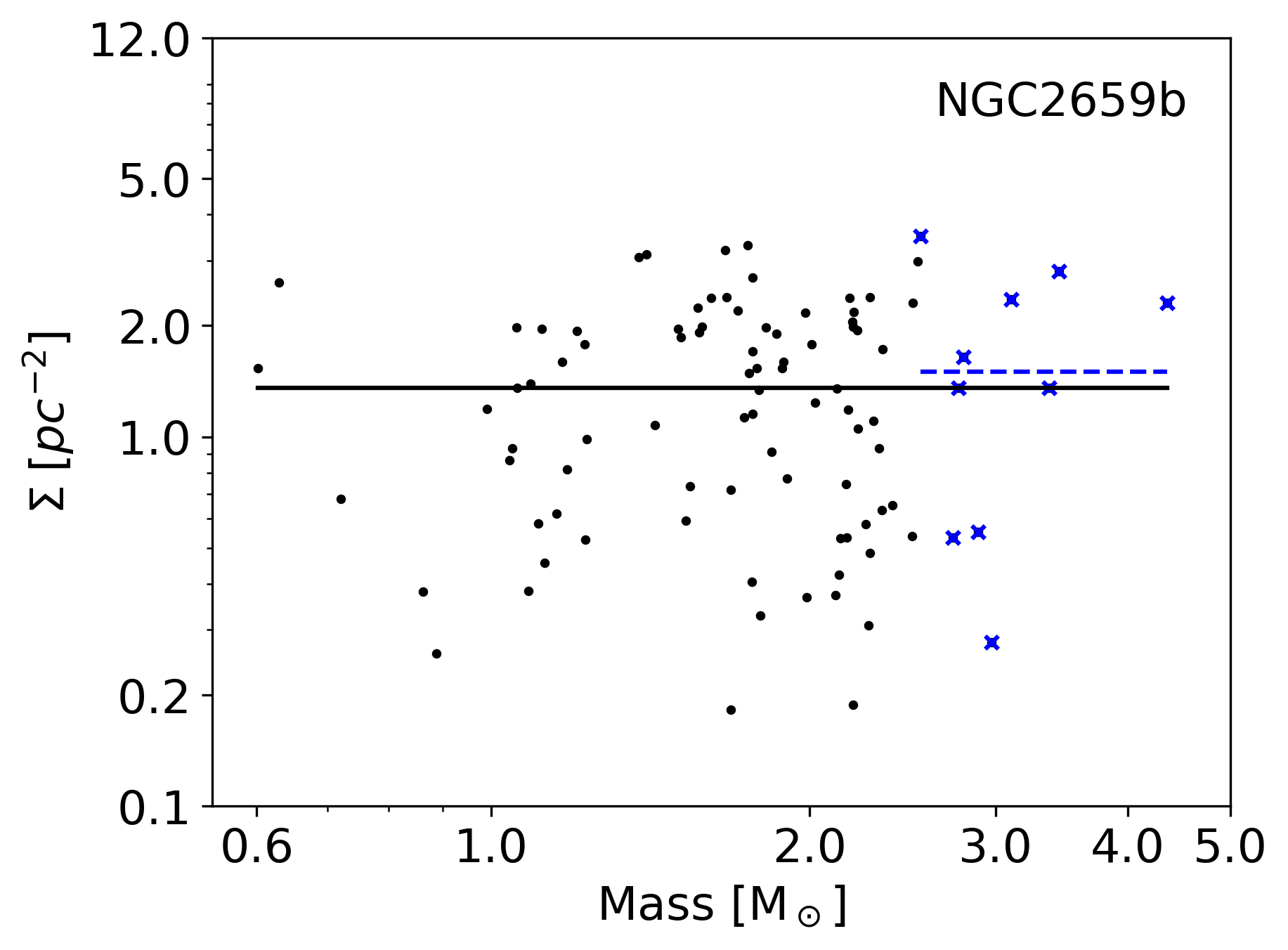}
\includegraphics[width=0.65\columnwidth]{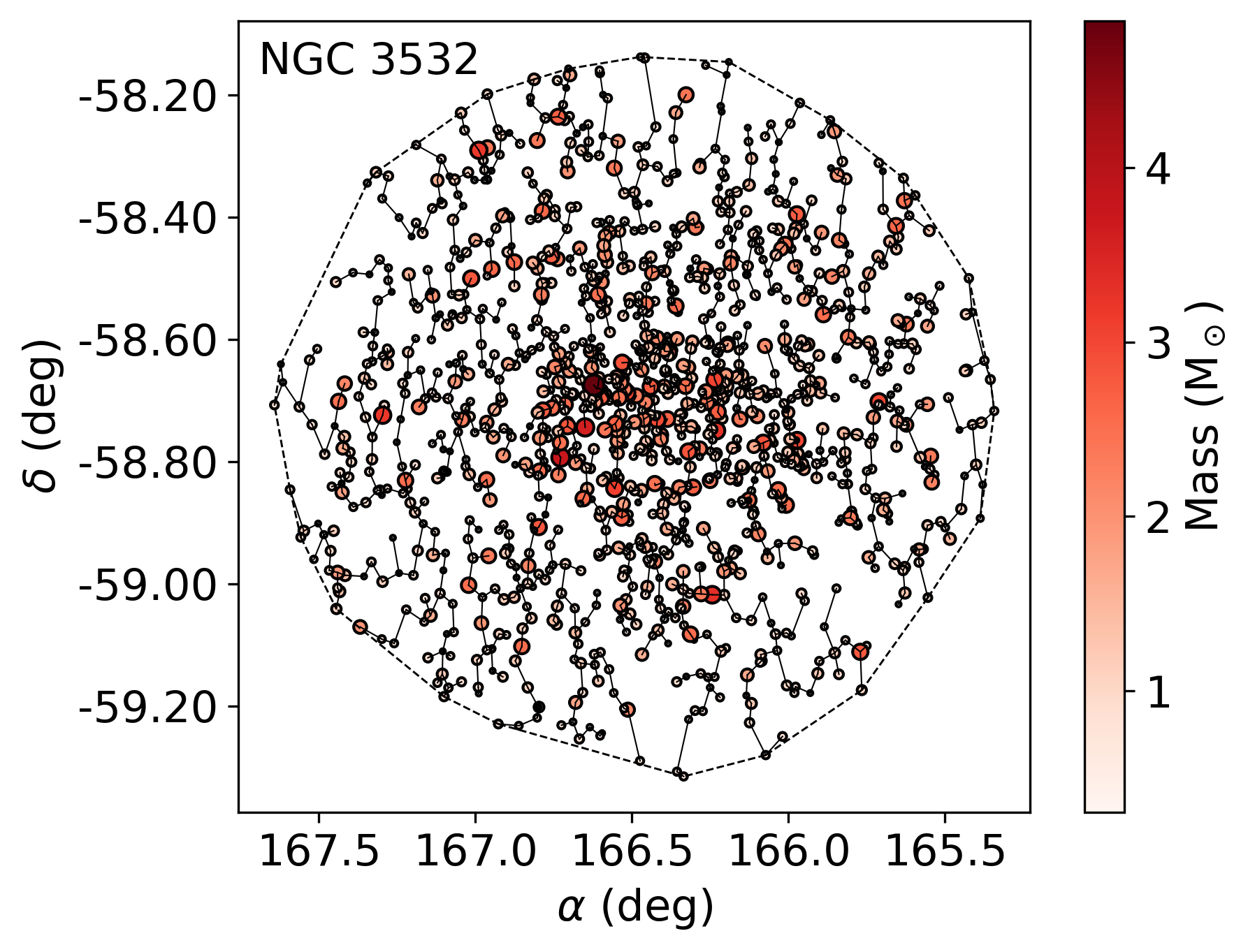}
\includegraphics[width=0.65\columnwidth]{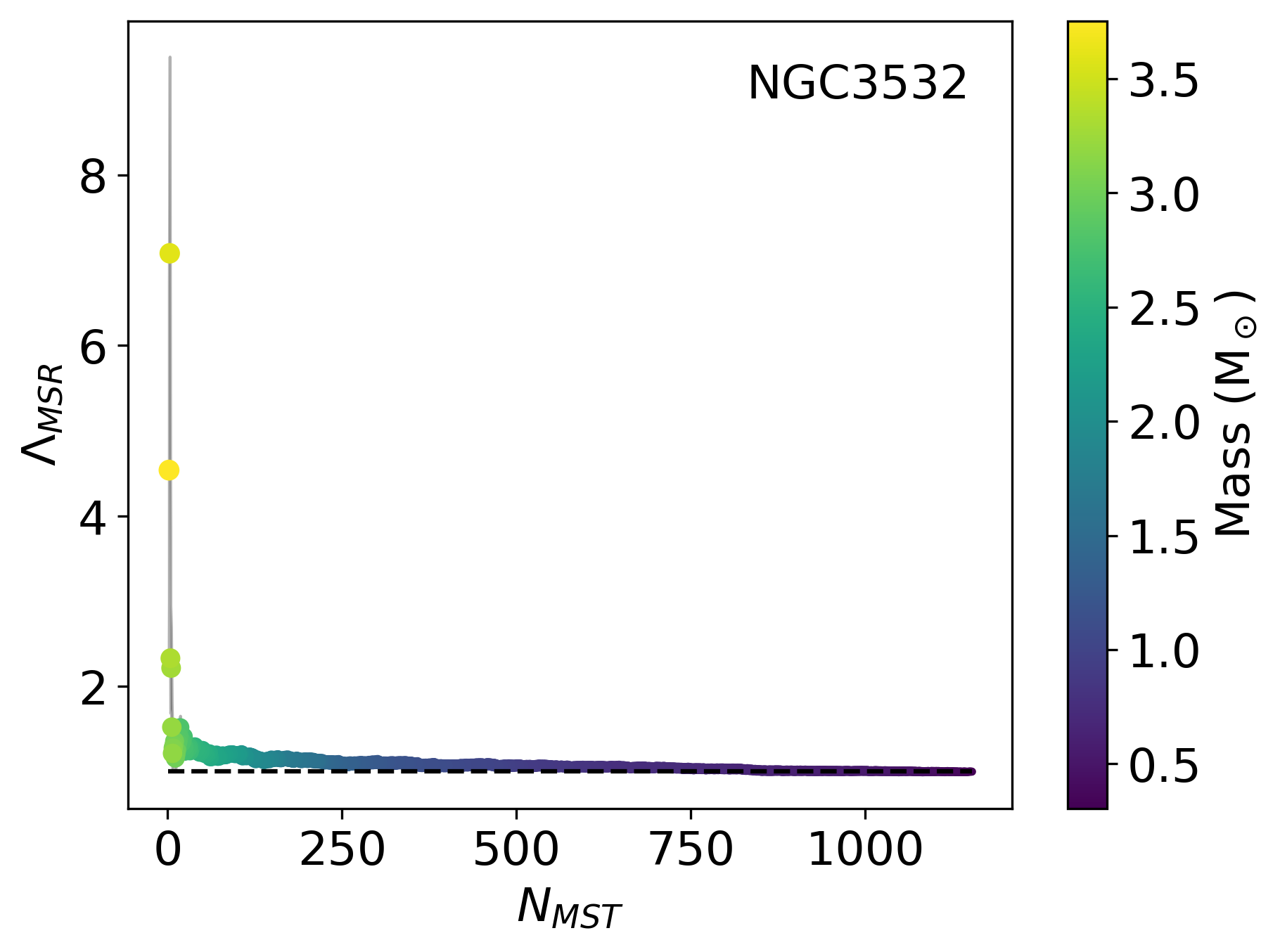}
\includegraphics[width=0.65\columnwidth]{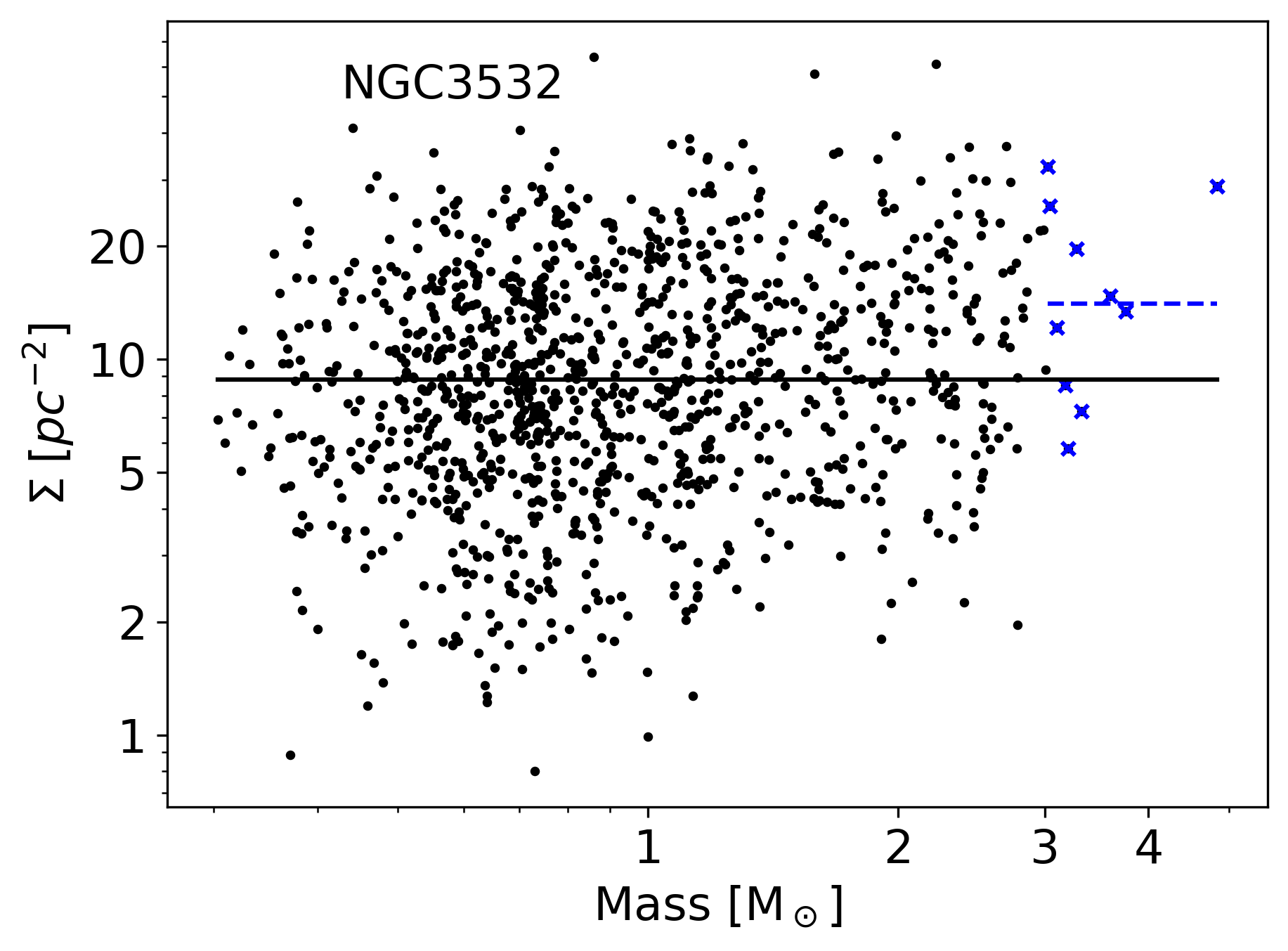}
\includegraphics[width=0.65\columnwidth]{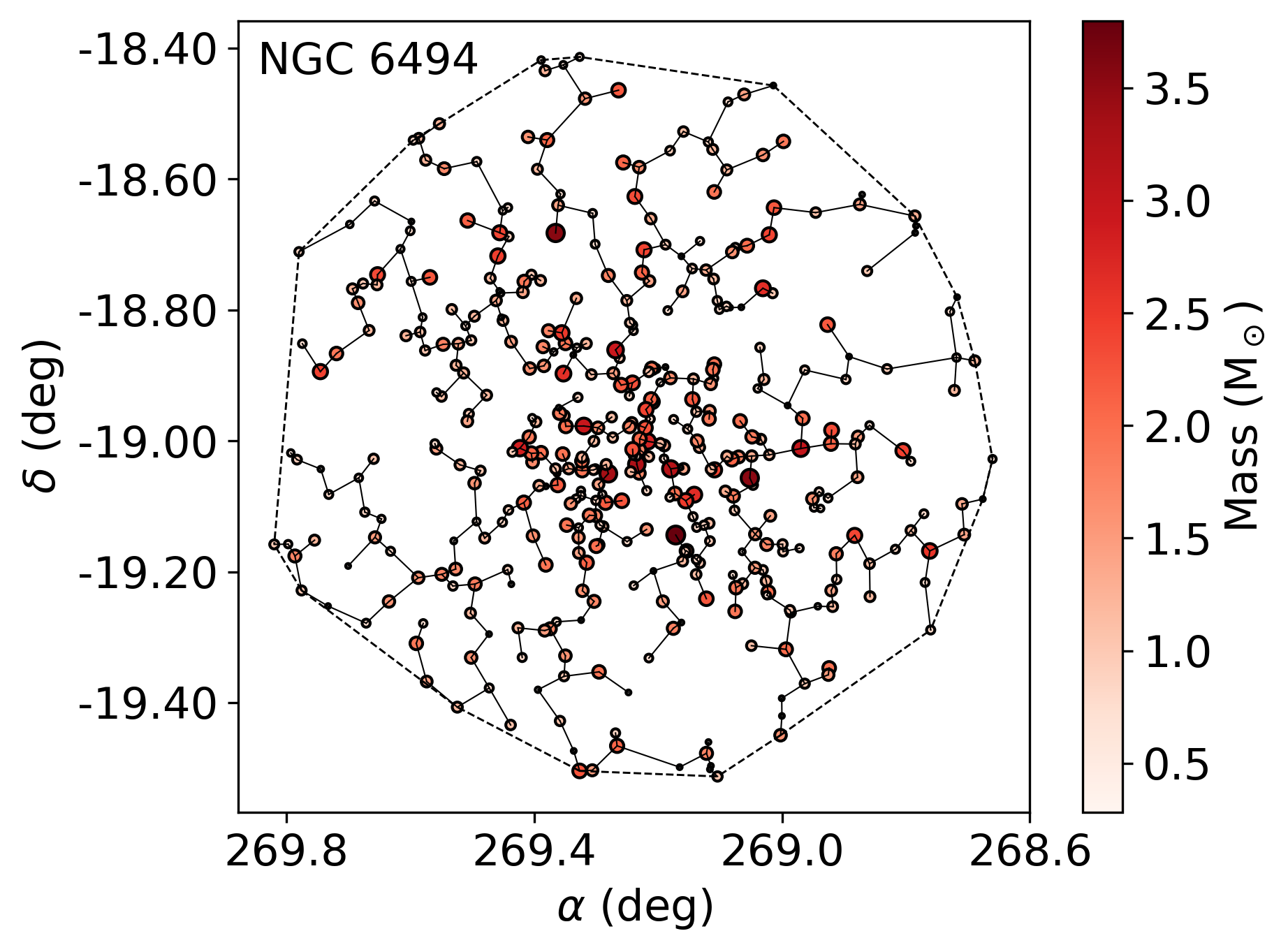}
\includegraphics[width=0.65\columnwidth]{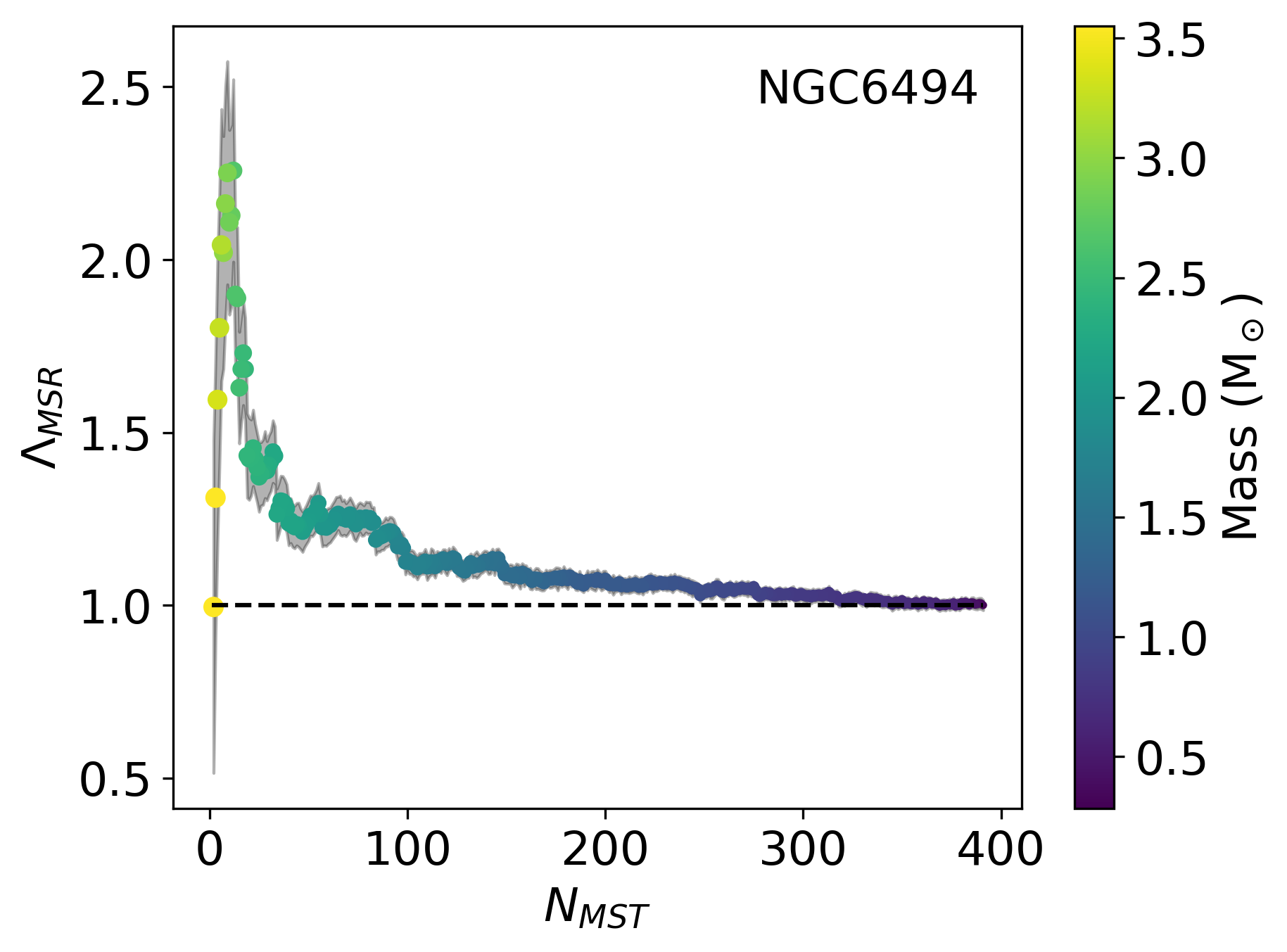}
\includegraphics[width=0.65\columnwidth]{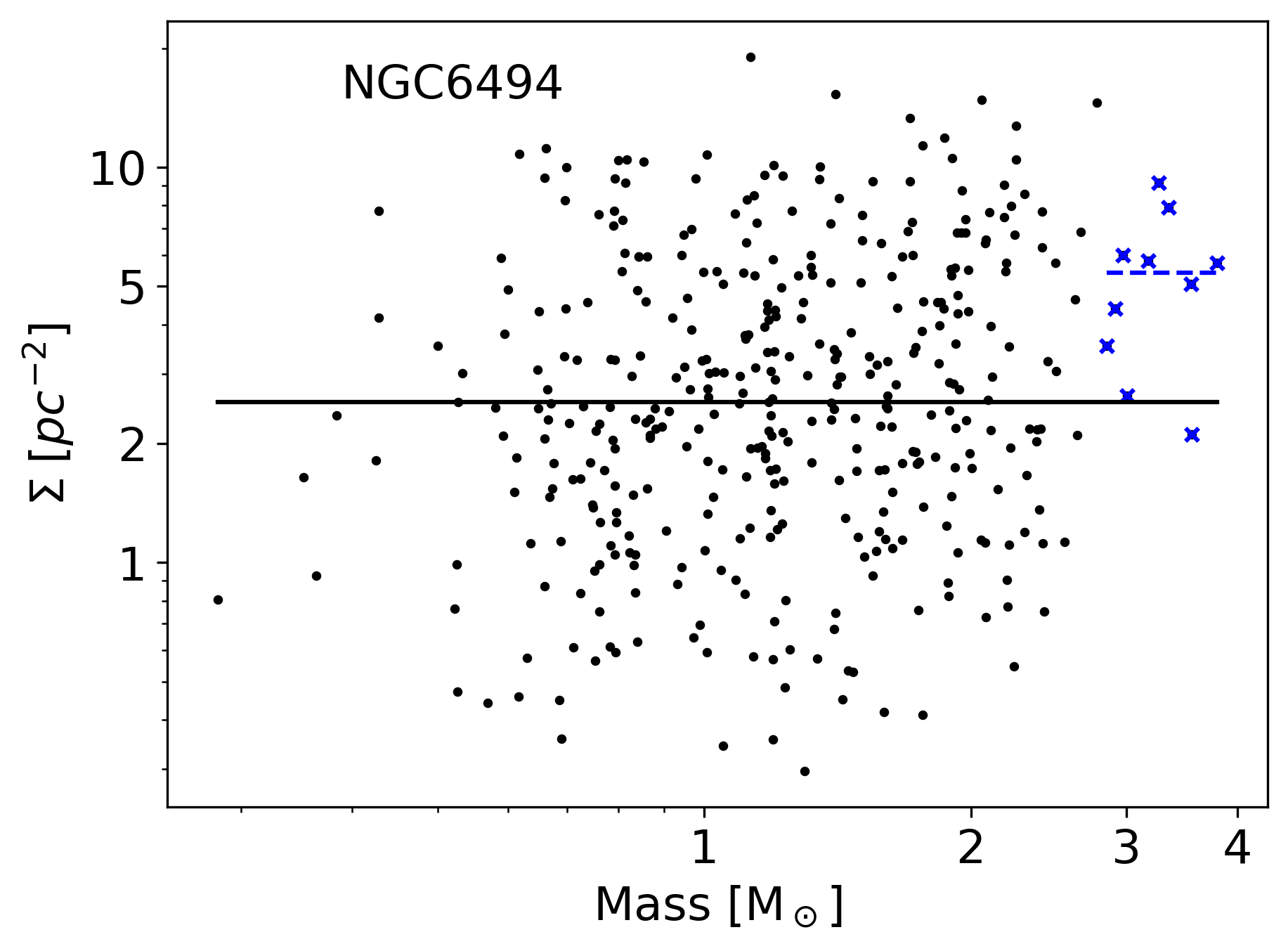}
\caption{Continuation of  Fig. \ref{fig:ALambda1}.  For comparison, only the $\Sigma - m$ plots for NGC~2659a and NGC~2659b are displayed at the same scale}.
\label{fig:ALambda2}
\end{figure*}
%%-----------------------end 

\end{document}